%% file: main.tex
\pdfoutput=1
\def\CAMERAREADY{}\def\PREPRINT{}
\documentclass[10pt,conference]{IEEEtran}
\IEEEoverridecommandlockouts

\newif\ifcameraready
\ifdefined\CAMERAREADY\camerareadytrue\fi
\newif\ifpreprint
\ifdefined\PREPRINT\preprinttrue\fi
\newenvironment{keeptogether}{\ifpreprint\else\samepage\fi}{}

\usepackage[T1]{fontenc}
\usepackage[utf8]{inputenc}
\usepackage{amsmath,amssymb}
\usepackage{graphicx}
\usepackage{booktabs}
\usepackage[table]{xcolor}
\usepackage{listings}
\usepackage{tikz}
\usetikzlibrary{arrows.meta}
\usepackage{fontawesome5}
\usepackage{capt-of}
\usepackage[nocompress]{cite}
\usepackage[hidelinks]{hyperref}
\usepackage{xurl}
\hypersetup{
  pdftitle={Do Coding Agents Still Need a Debugger? Frontier-Model-Based Fault Localization at the Accuracy Ceiling},
  pdfauthor={},
  pdflang={en-US}
}
\ifcameraready
  \hypersetup{pdfauthor={Christoph Daxerer, Johann Glock, Christian Macho, Rudolf Ramler, Martin Pinzger, Josef Pichler}}
\fi

\definecolor{draftmark}{RGB}{230,70,0}

\newif\ifshowresults
\showresultsfalse
\definecolor{resultmark}{RGB}{0,90,190}

\ExplSyntaxOn
\cs_generate_variant:Nn \tl_if_eq:NnTF { c }
\newcommand{\resultdef}[2]{\tl_gset:cn {resultvalue@#1} {#2}}
\DeclareRobustCommand{\result}[2]{
  \cs_if_exist:cTF {resultvalue@#1}
    {
      \tl_if_eq:cnTF {resultvalue@#1} {#2}
        {\ifshowresults{\color{resultmark}#2}\else#2\fi}
        {\PackageError{manuscript-results}{Result~'#1'~has~stale~shown~text}
          {Run~the~manuscript~synchronization~command~to~update~bound~values.}}
    }
    {\PackageError{manuscript-results}{Unknown~result~'#1'}
      {Generate~values.tex~and~synchronize~the~manuscript~before~building.}}
}
\ExplSyntaxOff
\InputIfFileExists{generated/values.tex}{}{}

\newcommand{\rqanswer}[2]{\par\smallskip\noindent\fbox{\parbox{\dimexpr\columnwidth-2\fboxsep-2\fboxrule\relax}{\textbf{Answer to RQ#1:} #2}}\par\smallskip}
\definecolor{faulthl}{RGB}{255,214,214}
\newcommand{\figcode}{\fontfamily{lmtt}\fontsize{6.4}{7.7}\selectfont}
\newcommand{\cln}[3]{{\color{gray}\tiny #1} & #2 & #3\\}

\title{Do Coding Agents Still Need a Debugger? Frontier-Model-Based Fault Localization\\at the Accuracy Ceiling}
\ifcameraready
\makeatletter
\newcommand{\linebreakand}{\end{@IEEEauthorhalign}\hfill\mbox{}\par\mbox{}\hfill\begin{@IEEEauthorhalign}}
\makeatother
\author{%
\IEEEauthorblockN{Christoph Daxerer}
\IEEEauthorblockA{\textit{Software Competence Center}\\
\textit{Hagenberg GmbH (SCCH)}\\
Hagenberg, Austria\\
0009-0002-6397-5959}
\and
\IEEEauthorblockN{Johann Glock}
\IEEEauthorblockA{\textit{Software Competence Center}\\
\textit{Hagenberg GmbH (SCCH)}\\
Hagenberg, Austria\\
0000-0002-0152-8611}
\and
\IEEEauthorblockN{Christian Macho}
\IEEEauthorblockA{\textit{Department of Informatics Systems}\\
\textit{University of Klagenfurt}\\
Klagenfurt, Austria\\
0000-0001-8182-7277}
\linebreakand
\IEEEauthorblockN{Rudolf Ramler}
\IEEEauthorblockA{\textit{Software Competence Center}\\
\textit{Hagenberg GmbH (SCCH)}\\
Hagenberg, Austria\\
0000-0001-9903-6107}
\and
\IEEEauthorblockN{Martin Pinzger}
\IEEEauthorblockA{\textit{Department of Informatics Systems}\\
\textit{University of Klagenfurt}\\
Klagenfurt, Austria\\
0000-0002-5536-3859}
\and
\IEEEauthorblockN{Josef Pichler}
\IEEEauthorblockA{\textit{School of Informatics, Communications and Media}\\
\textit{University of Applied Sciences Upper Austria}\\
Hagenberg, Austria\\
0009-0007-6908-0474}}
\else
\author{}
\fi
\newsavebox{\overviewbox}
\IEEEaftertitletext{%
  \setbox\overviewbox=\vbox{%
    \vspace*{-0.4\baselineskip}%
    {\centering\input{figures/study-overview}\par}%
    \vspace{0.3\baselineskip}%
    \captionof{figure}{Study design. From 150 LeetCodeDataset reference programs $P$, we seed one single-line fault per fault type to obtain faulty programs $M$ and rewrite $P$ and $M$ alike with five transformation recipes. Admitted pairs $(M, M')$ preserve the passing behavior of $P$, the failing assertion of $M$, and the restorable fault at line $L'$. Three LLM agents then localize every fault under three tool-use settings (Conditions A--F).}%
    \label{fig:study-overview}}%
  \nointerlineskip\box\overviewbox\vspace*{\baselineskip}}

\makeatletter
\AddToHookNext{shipout/after}{\global\let\@textbottom\relax\global\let\@texttop\relax}
\makeatother

\begin{document}
\maketitle

\begin{abstract}
\textbf{Background:} To localize a fault, a coding agent can reason from source code alone or use \emph{debugger assistance} by running the program and stepping through it in an interactive debugger. While earlier agents struggled with static execution tracing and debugger operation, frontier LLMs have improved at both.
\textbf{Objective:} We evaluate how accurately coding agents localize faults without tools across original and transformed programs, how much debugger assistance improves accuracy and at what cost, and how agents use it when optional.
\textbf{Methods:} We seed \result{programs.faulty.admitted}{610} single-line faults into \result{programs.reference.admitted}{144} LeetCodeDataset solutions and rewrite each faulty program with five recipes of behavior-preserving transformations. Agents based on Sol~6.1, Luna~6, and Qwen~3.8 localize each fault with no, optional, or encouraged debugger assistance.
\textbf{Results:} Without tools, the agents localized \result{primary.acc.untransformed.no-tools.min}{98.2} to \result{primary.acc.untransformed.no-tools.max}{99.9}\% of faults in original and \result{primary.acc.transformed.no-tools.min}{96.7} to \result{primary.acc.transformed.no-tools.max}{99.9}\% in transformed programs, losing accuracy mainly where recipes restructured loops and conditions. Encouraged assistance removed \result{primary.error-reduction.transformed.luna}{63}\% and \result{primary.error-reduction.transformed.qwen}{61}\% of Luna~6's and Qwen~3.8's remaining transformed errors (\result{primary.effect.transformed.luna}{+2.1} and \result{primary.effect.transformed.qwen}{+1.3} percentage points) at \result{cost.ratio.tokens.transformed.min}{14.7} to \result{cost.ratio.tokens.transformed.max}{27.0} times the token use, while Sol~6.1 stayed at \result{primary.acc.transformed.no-tools.sol}{99.9}\%. With optional assistance, Luna~6 and Qwen~3.8 still gained \result{primary.disc-minus-no.transformed.luna}{+1.5} and \result{primary.disc-minus-no.transformed.qwen}{+1.3} points, though the agents used the interactive debugger in only \result{tool.any-debugger.e.min}{0.2} to \result{tool.any-debugger.e.max}{6.2}\% of runs.
\textbf{Conclusions:} On short, precisely specified programs, current coding agents localize nearly all faults from source code alone, and debugger assistance removes most errors that remain. Given the choice, agents executed the program far more often than they stepped through it, yet kept most of the benefit.
\end{abstract}

\begin{IEEEkeywords}
agentic software engineering, fault localization, large language models, debugging
\end{IEEEkeywords}

\input{sections/01-introduction}

\input{sections/02-related-work}
\input{sections/03-study-design}
\input{sections/04-results}
\input{sections/05-discussion}
\input{sections/06-threats}
\input{sections/07-conclusion}
\input{sections/08-acknowledgment}

\ifpreprint\else\clearpage\fi
\bibliographystyle{IEEEtran}
\bibliography{references}

\end{document}

%% file: generated/values.tex
\resultdef{audit.exact-line.diagnosed}{59}
\resultdef{audit.exact-line.sampled}{60}
\resultdef{audit.statement-only.diagnosed}{133}
\resultdef{audit.statement-only.sampled}{135}
\resultdef{cost.forced-answer.a.luna}{0.0}
\resultdef{cost.forced-answer.a.max}{0.0}
\resultdef{cost.forced-answer.a.min}{0.0}
\resultdef{cost.forced-answer.a.qwen}{0.0}
\resultdef{cost.forced-answer.a.sol}{0.0}
\resultdef{cost.forced-answer.b.luna}{0.1}
\resultdef{cost.forced-answer.b.max}{0.1}
\resultdef{cost.forced-answer.b.min}{0.0}
\resultdef{cost.forced-answer.b.qwen}{0.0}
\resultdef{cost.forced-answer.b.sol}{0.0}
\resultdef{cost.forced-answer.c.luna}{0.0}
\resultdef{cost.forced-answer.c.max}{0.1}
\resultdef{cost.forced-answer.c.min}{0.0}
\resultdef{cost.forced-answer.c.qwen}{0.1}
\resultdef{cost.forced-answer.c.sol}{0.0}
\resultdef{cost.forced-answer.d.luna}{0.0}
\resultdef{cost.forced-answer.d.max}{0.0}
\resultdef{cost.forced-answer.d.min}{0.0}
\resultdef{cost.forced-answer.d.qwen}{0.0}
\resultdef{cost.forced-answer.d.sol}{0.0}
\resultdef{cost.forced-answer.e.luna}{0.1}
\resultdef{cost.forced-answer.e.max}{0.2}
\resultdef{cost.forced-answer.e.min}{0.0}
\resultdef{cost.forced-answer.e.qwen}{0.2}
\resultdef{cost.forced-answer.e.sol}{0.0}
\resultdef{cost.forced-answer.f.luna}{0.1}
\resultdef{cost.forced-answer.f.max}{0.3}
\resultdef{cost.forced-answer.f.min}{0.0}
\resultdef{cost.forced-answer.f.qwen}{0.3}
\resultdef{cost.forced-answer.f.sol}{0.0}
\resultdef{cost.no-answer.a.luna}{0.1}
\resultdef{cost.no-answer.a.max}{0.1}
\resultdef{cost.no-answer.a.min}{0.0}
\resultdef{cost.no-answer.a.qwen}{0.0}
\resultdef{cost.no-answer.a.sol}{0.0}
\resultdef{cost.no-answer.b.luna}{0.0}
\resultdef{cost.no-answer.b.max}{0.0}
\resultdef{cost.no-answer.b.min}{0.0}
\resultdef{cost.no-answer.b.qwen}{0.0}
\resultdef{cost.no-answer.b.sol}{0.0}
\resultdef{cost.no-answer.c.luna}{0.0}
\resultdef{cost.no-answer.c.max}{0.1}
\resultdef{cost.no-answer.c.min}{0.0}
\resultdef{cost.no-answer.c.qwen}{0.0}
\resultdef{cost.no-answer.c.sol}{0.1}
\resultdef{cost.no-answer.d.luna}{0.0}
\resultdef{cost.no-answer.d.max}{0.1}
\resultdef{cost.no-answer.d.min}{0.0}
\resultdef{cost.no-answer.d.qwen}{0.1}
\resultdef{cost.no-answer.d.sol}{0.0}
\resultdef{cost.no-answer.e.luna}{0.0}
\resultdef{cost.no-answer.e.max}{0.0}
\resultdef{cost.no-answer.e.min}{0.0}
\resultdef{cost.no-answer.e.qwen}{0.0}
\resultdef{cost.no-answer.e.sol}{0.0}
\resultdef{cost.no-answer.f.luna}{0.0}
\resultdef{cost.no-answer.f.max}{0.1}
\resultdef{cost.no-answer.f.min}{0.0}
\resultdef{cost.no-answer.f.qwen}{0.0}
\resultdef{cost.no-answer.f.sol}{0.1}
\resultdef{cost.ratio.tokens.transformed.luna}{27.0}
\resultdef{cost.ratio.tokens.transformed.max}{27.0}
\resultdef{cost.ratio.tokens.transformed.min}{14.7}
\resultdef{cost.ratio.tokens.transformed.qwen}{14.7}
\resultdef{cost.ratio.tokens.transformed.sol}{18.4}
\resultdef{cost.ratio.tokens.untransformed.luna}{29.7}
\resultdef{cost.ratio.tokens.untransformed.max}{29.7}
\resultdef{cost.ratio.tokens.untransformed.min}{11.2}
\resultdef{cost.ratio.tokens.untransformed.qwen}{11.2}
\resultdef{cost.ratio.tokens.untransformed.sol}{19.8}
\resultdef{cost.ratio.usd.transformed.luna}{5.0}
\resultdef{cost.ratio.usd.transformed.max}{5.0}
\resultdef{cost.ratio.usd.transformed.min}{3.4}
\resultdef{cost.ratio.usd.transformed.qwen}{3.4}
\resultdef{cost.ratio.usd.transformed.sol}{4.1}
\resultdef{cost.ratio.usd.untransformed.luna}{5.7}
\resultdef{cost.ratio.usd.untransformed.max}{5.7}
\resultdef{cost.ratio.usd.untransformed.min}{3.3}
\resultdef{cost.ratio.usd.untransformed.qwen}{3.3}
\resultdef{cost.ratio.usd.untransformed.sol}{4.1}
\resultdef{cost.ratio.wall.transformed.luna}{6.2}
\resultdef{cost.ratio.wall.transformed.max}{7.7}
\resultdef{cost.ratio.wall.transformed.min}{5.3}
\resultdef{cost.ratio.wall.transformed.qwen}{5.3}
\resultdef{cost.ratio.wall.transformed.sol}{7.7}
\resultdef{cost.ratio.wall.untransformed.luna}{7.0}
\resultdef{cost.ratio.wall.untransformed.max}{7.7}
\resultdef{cost.ratio.wall.untransformed.min}{5.6}
\resultdef{cost.ratio.wall.untransformed.qwen}{5.6}
\resultdef{cost.ratio.wall.untransformed.sol}{7.7}
\resultdef{cost.tokens-thousands.a.luna}{1.4}
\resultdef{cost.tokens-thousands.a.max}{3.3}
\resultdef{cost.tokens-thousands.a.min}{1.3}
\resultdef{cost.tokens-thousands.a.qwen}{3.3}
\resultdef{cost.tokens-thousands.a.sol}{1.3}
\resultdef{cost.tokens-thousands.b.luna}{2.2}
\resultdef{cost.tokens-thousands.b.max}{19.0}
\resultdef{cost.tokens-thousands.b.min}{2.2}
\resultdef{cost.tokens-thousands.b.qwen}{19.0}
\resultdef{cost.tokens-thousands.b.sol}{3.6}
\resultdef{cost.tokens-thousands.c.luna}{41.8}
\resultdef{cost.tokens-thousands.c.max}{41.8}
\resultdef{cost.tokens-thousands.c.min}{25.0}
\resultdef{cost.tokens-thousands.c.qwen}{37.4}
\resultdef{cost.tokens-thousands.c.sol}{25.0}
\resultdef{cost.tokens-thousands.d.luna}{2.0}
\resultdef{cost.tokens-thousands.d.max}{5.0}
\resultdef{cost.tokens-thousands.d.min}{1.7}
\resultdef{cost.tokens-thousands.d.qwen}{5.0}
\resultdef{cost.tokens-thousands.d.sol}{1.7}
\resultdef{cost.tokens-thousands.e.luna}{3.3}
\resultdef{cost.tokens-thousands.e.max}{38.3}
\resultdef{cost.tokens-thousands.e.min}{3.3}
\resultdef{cost.tokens-thousands.e.qwen}{38.3}
\resultdef{cost.tokens-thousands.e.sol}{4.9}
\resultdef{cost.tokens-thousands.f.luna}{55.0}
\resultdef{cost.tokens-thousands.f.max}{73.5}
\resultdef{cost.tokens-thousands.f.min}{32.2}
\resultdef{cost.tokens-thousands.f.qwen}{73.5}
\resultdef{cost.tokens-thousands.f.sol}{32.2}
\resultdef{cost.usd-per-1000.a.luna}{0.22}
\resultdef{cost.usd-per-1000.a.max}{4.76}
\resultdef{cost.usd-per-1000.a.min}{0.22}
\resultdef{cost.usd-per-1000.a.qwen}{4.76}
\resultdef{cost.usd-per-1000.a.sol}{2.86}
\resultdef{cost.usd-per-1000.b.luna}{0.25}
\resultdef{cost.usd-per-1000.b.max}{11.60}
\resultdef{cost.usd-per-1000.b.min}{0.25}
\resultdef{cost.usd-per-1000.b.qwen}{11.60}
\resultdef{cost.usd-per-1000.b.sol}{4.35}
\resultdef{cost.usd-per-1000.c.luna}{1.26}
\resultdef{cost.usd-per-1000.c.max}{15.68}
\resultdef{cost.usd-per-1000.c.min}{1.26}
\resultdef{cost.usd-per-1000.c.qwen}{15.68}
\resultdef{cost.usd-per-1000.c.sol}{11.81}
\resultdef{cost.usd-per-1000.d.luna}{0.32}
\resultdef{cost.usd-per-1000.d.max}{7.65}
\resultdef{cost.usd-per-1000.d.min}{0.32}
\resultdef{cost.usd-per-1000.d.qwen}{7.65}
\resultdef{cost.usd-per-1000.d.sol}{3.41}
\resultdef{cost.usd-per-1000.e.luna}{0.38}
\resultdef{cost.usd-per-1000.e.max}{18.92}
\resultdef{cost.usd-per-1000.e.min}{0.38}
\resultdef{cost.usd-per-1000.e.qwen}{18.92}
\resultdef{cost.usd-per-1000.e.sol}{5.30}
\resultdef{cost.usd-per-1000.f.luna}{1.60}
\resultdef{cost.usd-per-1000.f.max}{26.06}
\resultdef{cost.usd-per-1000.f.min}{1.60}
\resultdef{cost.usd-per-1000.f.qwen}{26.06}
\resultdef{cost.usd-per-1000.f.sol}{13.98}
\resultdef{cost.wall-seconds.a.luna}{5.9}
\resultdef{cost.wall-seconds.a.max}{37.9}
\resultdef{cost.wall-seconds.a.min}{4.9}
\resultdef{cost.wall-seconds.a.qwen}{37.9}
\resultdef{cost.wall-seconds.a.sol}{4.9}
\resultdef{cost.wall-seconds.b.luna}{4.5}
\resultdef{cost.wall-seconds.b.max}{136.1}
\resultdef{cost.wall-seconds.b.min}{4.5}
\resultdef{cost.wall-seconds.b.qwen}{136.1}
\resultdef{cost.wall-seconds.b.sol}{14.0}
\resultdef{cost.wall-seconds.c.luna}{41.2}
\resultdef{cost.wall-seconds.c.max}{213.6}
\resultdef{cost.wall-seconds.c.min}{37.8}
\resultdef{cost.wall-seconds.c.qwen}{213.6}
\resultdef{cost.wall-seconds.c.sol}{37.8}
\resultdef{cost.wall-seconds.d.luna}{7.5}
\resultdef{cost.wall-seconds.d.max}{66.4}
\resultdef{cost.wall-seconds.d.min}{5.2}
\resultdef{cost.wall-seconds.d.qwen}{66.4}
\resultdef{cost.wall-seconds.d.sol}{5.2}
\resultdef{cost.wall-seconds.e.luna}{6.4}
\resultdef{cost.wall-seconds.e.max}{245.0}
\resultdef{cost.wall-seconds.e.min}{6.4}
\resultdef{cost.wall-seconds.e.qwen}{245.0}
\resultdef{cost.wall-seconds.e.sol}{15.0}
\resultdef{cost.wall-seconds.f.luna}{47.0}
\resultdef{cost.wall-seconds.f.max}{350.4}
\resultdef{cost.wall-seconds.f.min}{39.9}
\resultdef{cost.wall-seconds.f.qwen}{350.4}
\resultdef{cost.wall-seconds.f.sol}{39.9}
\resultdef{count.cases.transformed.luna}{3{,}022}
\resultdef{count.cases.transformed.max}{3{,}022}
\resultdef{count.cases.transformed.min}{3{,}022}
\resultdef{count.cases.transformed.qwen}{3{,}022}
\resultdef{count.cases.transformed.sol}{3{,}022}
\resultdef{count.cases.untransformed.luna}{610}
\resultdef{count.cases.untransformed.max}{610}
\resultdef{count.cases.untransformed.min}{610}
\resultdef{count.cases.untransformed.qwen}{610}
\resultdef{count.cases.untransformed.sol}{610}
\resultdef{count.debugger-cases.transformed.luna}{17}
\resultdef{count.debugger-cases.transformed.max}{339}
\resultdef{count.debugger-cases.transformed.min}{11}
\resultdef{count.debugger-cases.transformed.qwen}{339}
\resultdef{count.debugger-cases.transformed.sol}{11}
\resultdef{count.debugger-cases.untransformed.luna}{0}
\resultdef{count.debugger-cases.untransformed.max}{11}
\resultdef{count.debugger-cases.untransformed.min}{0}
\resultdef{count.debugger-cases.untransformed.qwen}{11}
\resultdef{count.debugger-cases.untransformed.sol}{0}
\resultdef{count.debugger-programs.transformed.luna}{13}
\resultdef{count.debugger-programs.transformed.max}{111}
\resultdef{count.debugger-programs.transformed.min}{7}
\resultdef{count.debugger-programs.transformed.qwen}{111}
\resultdef{count.debugger-programs.transformed.sol}{7}
\resultdef{count.debugger-programs.untransformed.luna}{0}
\resultdef{count.debugger-programs.untransformed.max}{11}
\resultdef{count.debugger-programs.untransformed.min}{0}
\resultdef{count.debugger-programs.untransformed.qwen}{11}
\resultdef{count.debugger-programs.untransformed.sol}{0}
\resultdef{count.episodes.transformed.luna}{18{,}132}
\resultdef{count.episodes.transformed.max}{18{,}132}
\resultdef{count.episodes.transformed.min}{18{,}132}
\resultdef{count.episodes.transformed.qwen}{18{,}132}
\resultdef{count.episodes.transformed.sol}{18{,}132}
\resultdef{count.episodes.untransformed.luna}{3{,}660}
\resultdef{count.episodes.untransformed.max}{3{,}660}
\resultdef{count.episodes.untransformed.min}{3{,}660}
\resultdef{count.episodes.untransformed.qwen}{3{,}660}
\resultdef{count.episodes.untransformed.sol}{3{,}660}
\resultdef{count.excluded.pairs.luna}{0}
\resultdef{count.excluded.pairs.max}{0}
\resultdef{count.excluded.pairs.min}{0}
\resultdef{count.excluded.pairs.qwen}{0}
\resultdef{count.excluded.pairs.sol}{0}
\resultdef{count.excluded.transformed-cases.luna}{0}
\resultdef{count.excluded.transformed-cases.max}{0}
\resultdef{count.excluded.transformed-cases.min}{0}
\resultdef{count.excluded.transformed-cases.qwen}{0}
\resultdef{count.excluded.transformed-cases.sol}{0}
\resultdef{count.excluded.untransformed-cases.luna}{0}
\resultdef{count.excluded.untransformed-cases.max}{0}
\resultdef{count.excluded.untransformed-cases.min}{0}
\resultdef{count.excluded.untransformed-cases.qwen}{0}
\resultdef{count.excluded.untransformed-cases.sol}{0}
\resultdef{count.pairs.transformed.luna}{3{,}022}
\resultdef{count.pairs.transformed.max}{3{,}022}
\resultdef{count.pairs.transformed.min}{3{,}022}
\resultdef{count.pairs.transformed.qwen}{3{,}022}
\resultdef{count.pairs.transformed.sol}{3{,}022}
\resultdef{count.pairs.untransformed.luna}{3{,}022}
\resultdef{count.pairs.untransformed.max}{3{,}022}
\resultdef{count.pairs.untransformed.min}{3{,}022}
\resultdef{count.pairs.untransformed.qwen}{3{,}022}
\resultdef{count.pairs.untransformed.sol}{3{,}022}
\resultdef{count.probe-cases.transformed.luna}{596}
\resultdef{count.probe-cases.transformed.max}{3{,}020}
\resultdef{count.probe-cases.transformed.min}{596}
\resultdef{count.probe-cases.transformed.qwen}{3{,}020}
\resultdef{count.probe-cases.transformed.sol}{2{,}992}
\resultdef{count.probe-cases.untransformed.luna}{94}
\resultdef{count.probe-cases.untransformed.max}{610}
\resultdef{count.probe-cases.untransformed.min}{94}
\resultdef{count.probe-cases.untransformed.qwen}{610}
\resultdef{count.probe-cases.untransformed.sol}{541}
\resultdef{count.probe-programs.transformed.luna}{132}
\resultdef{count.probe-programs.transformed.max}{144}
\resultdef{count.probe-programs.transformed.min}{132}
\resultdef{count.probe-programs.transformed.qwen}{144}
\resultdef{count.probe-programs.transformed.sol}{144}
\resultdef{count.probe-programs.untransformed.luna}{66}
\resultdef{count.probe-programs.untransformed.max}{144}
\resultdef{count.probe-programs.untransformed.min}{66}
\resultdef{count.probe-programs.untransformed.qwen}{144}
\resultdef{count.probe-programs.untransformed.sol}{136}
\resultdef{count.reference-programs.transformed.luna}{144}
\resultdef{count.reference-programs.transformed.max}{144}
\resultdef{count.reference-programs.transformed.min}{144}
\resultdef{count.reference-programs.transformed.qwen}{144}
\resultdef{count.reference-programs.transformed.sol}{144}
\resultdef{count.reference-programs.untransformed.luna}{144}
\resultdef{count.reference-programs.untransformed.max}{144}
\resultdef{count.reference-programs.untransformed.min}{144}
\resultdef{count.reference-programs.untransformed.qwen}{144}
\resultdef{count.reference-programs.untransformed.sol}{144}
\resultdef{criterion.primary}{statement}
\resultdef{criterion.secondary}{line}
\resultdef{facts.cost.disc-over-enc.transformed.luna}{0.2}
\resultdef{facts.cost.disc-over-enc.transformed.max}{0.7}
\resultdef{facts.cost.disc-over-enc.transformed.min}{0.2}
\resultdef{facts.cost.disc-over-enc.transformed.qwen}{0.7}
\resultdef{facts.cost.disc-over-enc.transformed.sol}{0.4}
\resultdef{facts.cost.disc-over-enc.untransformed.luna}{0.2}
\resultdef{facts.cost.disc-over-enc.untransformed.max}{0.7}
\resultdef{facts.cost.disc-over-enc.untransformed.min}{0.2}
\resultdef{facts.cost.disc-over-enc.untransformed.qwen}{0.7}
\resultdef{facts.cost.disc-over-enc.untransformed.sol}{0.4}
\resultdef{facts.disc-use.episodes.transformed.luna}{19}
\resultdef{facts.disc-use.episodes.transformed.max}{373}
\resultdef{facts.disc-use.episodes.transformed.min}{12}
\resultdef{facts.disc-use.episodes.transformed.qwen}{373}
\resultdef{facts.disc-use.episodes.transformed.sol}{12}
\resultdef{facts.disc-use.episodes.untransformed.luna}{0}
\resultdef{facts.disc-use.episodes.untransformed.max}{11}
\resultdef{facts.disc-use.episodes.untransformed.min}{0}
\resultdef{facts.disc-use.episodes.untransformed.qwen}{11}
\resultdef{facts.disc-use.episodes.untransformed.sol}{0}
\resultdef{facts.disc-use.share.transformed.luna}{0.3}
\resultdef{facts.disc-use.share.transformed.max}{6.2}
\resultdef{facts.disc-use.share.transformed.min}{0.2}
\resultdef{facts.disc-use.share.transformed.qwen}{6.2}
\resultdef{facts.disc-use.share.transformed.sol}{0.2}
\resultdef{facts.disc-use.share.untransformed.luna}{0.0}
\resultdef{facts.disc-use.share.untransformed.max}{0.9}
\resultdef{facts.disc-use.share.untransformed.min}{0.0}
\resultdef{facts.disc-use.share.untransformed.qwen}{0.9}
\resultdef{facts.disc-use.share.untransformed.sol}{0.0}
\resultdef{facts.disc-use.total.transformed.luna}{6{,}044}
\resultdef{facts.disc-use.total.transformed.max}{6{,}044}
\resultdef{facts.disc-use.total.transformed.min}{6{,}044}
\resultdef{facts.disc-use.total.transformed.qwen}{6{,}044}
\resultdef{facts.disc-use.total.transformed.sol}{6{,}044}
\resultdef{facts.disc-use.total.untransformed.luna}{1{,}220}
\resultdef{facts.disc-use.total.untransformed.max}{1{,}220}
\resultdef{facts.disc-use.total.untransformed.min}{1{,}220}
\resultdef{facts.disc-use.total.untransformed.qwen}{1{,}220}
\resultdef{facts.disc-use.total.untransformed.sol}{1{,}220}
\resultdef{facts.episodes.kept}{65{,}376}
\resultdef{facts.episodes.kept.luna}{21{,}792}
\resultdef{facts.episodes.kept.max}{21{,}792}
\resultdef{facts.episodes.kept.min}{21{,}792}
\resultdef{facts.episodes.kept.qwen}{21{,}792}
\resultdef{facts.episodes.kept.sol}{21{,}792}
\resultdef{facts.episodes.missing-infrastructure}{0}
\resultdef{facts.episodes.missing-infrastructure.luna}{0}
\resultdef{facts.episodes.missing-infrastructure.max}{0}
\resultdef{facts.episodes.missing-infrastructure.min}{0}
\resultdef{facts.episodes.missing-infrastructure.qwen}{0}
\resultdef{facts.episodes.missing-infrastructure.sol}{0}
\resultdef{facts.episodes.missing-share}{0.0}
\resultdef{facts.episodes.missing-share.luna}{0.0}
\resultdef{facts.episodes.missing-share.max}{0.0}
\resultdef{facts.episodes.missing-share.min}{0.0}
\resultdef{facts.episodes.missing-share.qwen}{0.0}
\resultdef{facts.episodes.missing-share.sol}{0.0}
\resultdef{facts.episodes.missing-unattempted}{0}
\resultdef{facts.episodes.missing-unattempted.luna}{0}
\resultdef{facts.episodes.missing-unattempted.max}{0}
\resultdef{facts.episodes.missing-unattempted.min}{0}
\resultdef{facts.episodes.missing-unattempted.qwen}{0}
\resultdef{facts.episodes.missing-unattempted.sol}{0}
\resultdef{facts.episodes.planned}{65{,}376}
\resultdef{facts.episodes.planned.luna}{21{,}792}
\resultdef{facts.episodes.planned.max}{21{,}792}
\resultdef{facts.episodes.planned.min}{21{,}792}
\resultdef{facts.episodes.planned.qwen}{21{,}792}
\resultdef{facts.episodes.planned.sol}{21{,}792}
\resultdef{facts.foreign-access.discretionary-tools.luna}{0}
\resultdef{facts.foreign-access.discretionary-tools.max}{8}
\resultdef{facts.foreign-access.discretionary-tools.min}{0}
\resultdef{facts.foreign-access.discretionary-tools.qwen}{8}
\resultdef{facts.foreign-access.discretionary-tools.sol}{0}
\resultdef{facts.foreign-access.encouraged-tools.luna}{0}
\resultdef{facts.foreign-access.encouraged-tools.max}{6}
\resultdef{facts.foreign-access.encouraged-tools.min}{0}
\resultdef{facts.foreign-access.encouraged-tools.qwen}{6}
\resultdef{facts.foreign-access.encouraged-tools.sol}{0}
\resultdef{facts.foreign-access.no-tools.luna}{0}
\resultdef{facts.foreign-access.no-tools.max}{0}
\resultdef{facts.foreign-access.no-tools.min}{0}
\resultdef{facts.foreign-access.no-tools.qwen}{0}
\resultdef{facts.foreign-access.no-tools.sol}{0}
\resultdef{facts.position.earlier.discretionary-tools.luna}{4}
\resultdef{facts.position.earlier.discretionary-tools.max}{5}
\resultdef{facts.position.earlier.discretionary-tools.min}{0}
\resultdef{facts.position.earlier.discretionary-tools.qwen}{5}
\resultdef{facts.position.earlier.discretionary-tools.sol}{0}
\resultdef{facts.position.earlier.encouraged-tools.luna}{5}
\resultdef{facts.position.earlier.encouraged-tools.max}{5}
\resultdef{facts.position.earlier.encouraged-tools.min}{0}
\resultdef{facts.position.earlier.encouraged-tools.qwen}{5}
\resultdef{facts.position.earlier.encouraged-tools.sol}{0}
\resultdef{facts.position.earlier.no-tools.luna}{5}
\resultdef{facts.position.earlier.no-tools.max}{7}
\resultdef{facts.position.earlier.no-tools.min}{0}
\resultdef{facts.position.earlier.no-tools.qwen}{7}
\resultdef{facts.position.earlier.no-tools.sol}{0}
\resultdef{facts.position.fault.discretionary-tools.luna}{5{,}559}
\resultdef{facts.position.fault.discretionary-tools.max}{6{,}022}
\resultdef{facts.position.fault.discretionary-tools.min}{5{,}559}
\resultdef{facts.position.fault.discretionary-tools.qwen}{5{,}926}
\resultdef{facts.position.fault.discretionary-tools.sol}{6{,}022}
\resultdef{facts.position.fault.encouraged-tools.luna}{5{,}604}
\resultdef{facts.position.fault.encouraged-tools.max}{6{,}019}
\resultdef{facts.position.fault.encouraged-tools.min}{5{,}604}
\resultdef{facts.position.fault.encouraged-tools.qwen}{5{,}939}
\resultdef{facts.position.fault.encouraged-tools.sol}{6{,}019}
\resultdef{facts.position.fault.no-tools.luna}{5{,}401}
\resultdef{facts.position.fault.no-tools.max}{6{,}029}
\resultdef{facts.position.fault.no-tools.min}{5{,}401}
\resultdef{facts.position.fault.no-tools.qwen}{5{,}829}
\resultdef{facts.position.fault.no-tools.sol}{6{,}029}
\resultdef{facts.position.later-share.discretionary-tools.luna}{98.9}
\resultdef{facts.position.later-share.discretionary-tools.max}{100.0}
\resultdef{facts.position.later-share.discretionary-tools.min}{92.4}
\resultdef{facts.position.later-share.discretionary-tools.qwen}{92.4}
\resultdef{facts.position.later-share.discretionary-tools.sol}{100.0}
\resultdef{facts.position.later-share.encouraged-tools.luna}{98.6}
\resultdef{facts.position.later-share.encouraged-tools.max}{100.0}
\resultdef{facts.position.later-share.encouraged-tools.min}{90.9}
\resultdef{facts.position.later-share.encouraged-tools.qwen}{90.9}
\resultdef{facts.position.later-share.encouraged-tools.sol}{100.0}
\resultdef{facts.position.later-share.no-tools.luna}{98.9}
\resultdef{facts.position.later-share.no-tools.max}{100.0}
\resultdef{facts.position.later-share.no-tools.min}{92.0}
\resultdef{facts.position.later-share.no-tools.qwen}{92.0}
\resultdef{facts.position.later-share.no-tools.sol}{100.0}
\resultdef{facts.position.later.discretionary-tools.luna}{373}
\resultdef{facts.position.later.discretionary-tools.max}{373}
\resultdef{facts.position.later.discretionary-tools.min}{15}
\resultdef{facts.position.later.discretionary-tools.qwen}{61}
\resultdef{facts.position.later.discretionary-tools.sol}{15}
\resultdef{facts.position.later.encouraged-tools.luna}{362}
\resultdef{facts.position.later.encouraged-tools.max}{362}
\resultdef{facts.position.later.encouraged-tools.min}{21}
\resultdef{facts.position.later.encouraged-tools.qwen}{50}
\resultdef{facts.position.later.encouraged-tools.sol}{21}
\resultdef{facts.position.later.no-tools.luna}{438}
\resultdef{facts.position.later.no-tools.max}{438}
\resultdef{facts.position.later.no-tools.min}{9}
\resultdef{facts.position.later.no-tools.qwen}{80}
\resultdef{facts.position.later.no-tools.sol}{9}
\resultdef{facts.primary.failed-cases.luna}{153}
\resultdef{facts.primary.failed-cases.max}{153}
\resultdef{facts.primary.failed-cases.min}{4}
\resultdef{facts.primary.failed-cases.qwen}{104}
\resultdef{facts.primary.failed-cases.shared.luna-qwen}{39}
\resultdef{facts.primary.failed-cases.shared.luna-sol}{3}
\resultdef{facts.primary.failed-cases.shared.qwen-sol}{3}
\resultdef{facts.primary.failed-cases.sol}{4}
\resultdef{facts.primary.failed-episodes.0.transformed.discretionary-tools.luna}{2{,}938}
\resultdef{facts.primary.failed-episodes.0.transformed.discretionary-tools.max}{3{,}017}
\resultdef{facts.primary.failed-episodes.0.transformed.discretionary-tools.min}{2{,}938}
\resultdef{facts.primary.failed-episodes.0.transformed.discretionary-tools.qwen}{2{,}980}
\resultdef{facts.primary.failed-episodes.0.transformed.discretionary-tools.sol}{3{,}017}
\resultdef{facts.primary.failed-episodes.0.transformed.encouraged-tools.luna}{2{,}970}
\resultdef{facts.primary.failed-episodes.0.transformed.encouraged-tools.max}{3{,}020}
\resultdef{facts.primary.failed-episodes.0.transformed.encouraged-tools.min}{2{,}970}
\resultdef{facts.primary.failed-episodes.0.transformed.encouraged-tools.qwen}{2{,}980}
\resultdef{facts.primary.failed-episodes.0.transformed.encouraged-tools.sol}{3{,}020}
\resultdef{facts.primary.failed-episodes.0.transformed.no-tools.luna}{2{,}869}
\resultdef{facts.primary.failed-episodes.0.transformed.no-tools.max}{3{,}018}
\resultdef{facts.primary.failed-episodes.0.transformed.no-tools.min}{2{,}869}
\resultdef{facts.primary.failed-episodes.0.transformed.no-tools.qwen}{2{,}918}
\resultdef{facts.primary.failed-episodes.0.transformed.no-tools.sol}{3{,}018}
\resultdef{facts.primary.failed-episodes.0.untransformed.discretionary-tools.luna}{596}
\resultdef{facts.primary.failed-episodes.0.untransformed.discretionary-tools.max}{608}
\resultdef{facts.primary.failed-episodes.0.untransformed.discretionary-tools.min}{596}
\resultdef{facts.primary.failed-episodes.0.untransformed.discretionary-tools.qwen}{601}
\resultdef{facts.primary.failed-episodes.0.untransformed.discretionary-tools.sol}{608}
\resultdef{facts.primary.failed-episodes.0.untransformed.encouraged-tools.luna}{602}
\resultdef{facts.primary.failed-episodes.0.untransformed.encouraged-tools.max}{609}
\resultdef{facts.primary.failed-episodes.0.untransformed.encouraged-tools.min}{602}
\resultdef{facts.primary.failed-episodes.0.untransformed.encouraged-tools.qwen}{602}
\resultdef{facts.primary.failed-episodes.0.untransformed.encouraged-tools.sol}{609}
\resultdef{facts.primary.failed-episodes.0.untransformed.no-tools.luna}{594}
\resultdef{facts.primary.failed-episodes.0.untransformed.no-tools.max}{609}
\resultdef{facts.primary.failed-episodes.0.untransformed.no-tools.min}{594}
\resultdef{facts.primary.failed-episodes.0.untransformed.no-tools.qwen}{599}
\resultdef{facts.primary.failed-episodes.0.untransformed.no-tools.sol}{609}
\resultdef{facts.primary.failed-episodes.1.transformed.discretionary-tools.luna}{60}
\resultdef{facts.primary.failed-episodes.1.transformed.discretionary-tools.max}{60}
\resultdef{facts.primary.failed-episodes.1.transformed.discretionary-tools.min}{3}
\resultdef{facts.primary.failed-episodes.1.transformed.discretionary-tools.qwen}{32}
\resultdef{facts.primary.failed-episodes.1.transformed.discretionary-tools.sol}{3}
\resultdef{facts.primary.failed-episodes.1.transformed.encouraged-tools.luna}{31}
\resultdef{facts.primary.failed-episodes.1.transformed.encouraged-tools.max}{34}
\resultdef{facts.primary.failed-episodes.1.transformed.encouraged-tools.min}{31}
\resultdef{facts.primary.failed-episodes.1.transformed.encouraged-tools.qwen}{34}
\resultdef{facts.primary.failed-episodes.1.transformed.no-tools.luna}{106}
\resultdef{facts.primary.failed-episodes.1.transformed.no-tools.max}{106}
\resultdef{facts.primary.failed-episodes.1.transformed.no-tools.min}{2}
\resultdef{facts.primary.failed-episodes.1.transformed.no-tools.qwen}{80}
\resultdef{facts.primary.failed-episodes.1.transformed.no-tools.sol}{2}
\resultdef{facts.primary.failed-episodes.1.untransformed.discretionary-tools.luna}{5}
\resultdef{facts.primary.failed-episodes.1.untransformed.discretionary-tools.max}{5}
\resultdef{facts.primary.failed-episodes.1.untransformed.discretionary-tools.min}{1}
\resultdef{facts.primary.failed-episodes.1.untransformed.discretionary-tools.qwen}{5}
\resultdef{facts.primary.failed-episodes.1.untransformed.discretionary-tools.sol}{1}
\resultdef{facts.primary.failed-episodes.1.untransformed.encouraged-tools.luna}{3}
\resultdef{facts.primary.failed-episodes.1.untransformed.encouraged-tools.max}{5}
\resultdef{facts.primary.failed-episodes.1.untransformed.encouraged-tools.min}{3}
\resultdef{facts.primary.failed-episodes.1.untransformed.encouraged-tools.qwen}{5}
\resultdef{facts.primary.failed-episodes.1.untransformed.no-tools.luna}{10}
\resultdef{facts.primary.failed-episodes.1.untransformed.no-tools.max}{10}
\resultdef{facts.primary.failed-episodes.1.untransformed.no-tools.min}{1}
\resultdef{facts.primary.failed-episodes.1.untransformed.no-tools.qwen}{7}
\resultdef{facts.primary.failed-episodes.1.untransformed.no-tools.sol}{1}
\resultdef{facts.primary.failed-episodes.2.transformed.discretionary-tools.luna}{24}
\resultdef{facts.primary.failed-episodes.2.transformed.discretionary-tools.max}{24}
\resultdef{facts.primary.failed-episodes.2.transformed.discretionary-tools.min}{2}
\resultdef{facts.primary.failed-episodes.2.transformed.discretionary-tools.qwen}{10}
\resultdef{facts.primary.failed-episodes.2.transformed.discretionary-tools.sol}{2}
\resultdef{facts.primary.failed-episodes.2.transformed.encouraged-tools.luna}{21}
\resultdef{facts.primary.failed-episodes.2.transformed.encouraged-tools.max}{21}
\resultdef{facts.primary.failed-episodes.2.transformed.encouraged-tools.min}{2}
\resultdef{facts.primary.failed-episodes.2.transformed.encouraged-tools.qwen}{8}
\resultdef{facts.primary.failed-episodes.2.transformed.encouraged-tools.sol}{2}
\resultdef{facts.primary.failed-episodes.2.transformed.no-tools.luna}{47}
\resultdef{facts.primary.failed-episodes.2.transformed.no-tools.max}{47}
\resultdef{facts.primary.failed-episodes.2.transformed.no-tools.min}{2}
\resultdef{facts.primary.failed-episodes.2.transformed.no-tools.qwen}{24}
\resultdef{facts.primary.failed-episodes.2.transformed.no-tools.sol}{2}
\resultdef{facts.primary.failed-episodes.2.untransformed.discretionary-tools.luna}{9}
\resultdef{facts.primary.failed-episodes.2.untransformed.discretionary-tools.max}{9}
\resultdef{facts.primary.failed-episodes.2.untransformed.discretionary-tools.min}{1}
\resultdef{facts.primary.failed-episodes.2.untransformed.discretionary-tools.qwen}{4}
\resultdef{facts.primary.failed-episodes.2.untransformed.discretionary-tools.sol}{1}
\resultdef{facts.primary.failed-episodes.2.untransformed.encouraged-tools.luna}{5}
\resultdef{facts.primary.failed-episodes.2.untransformed.encouraged-tools.max}{5}
\resultdef{facts.primary.failed-episodes.2.untransformed.encouraged-tools.min}{1}
\resultdef{facts.primary.failed-episodes.2.untransformed.encouraged-tools.qwen}{3}
\resultdef{facts.primary.failed-episodes.2.untransformed.encouraged-tools.sol}{1}
\resultdef{facts.primary.failed-episodes.2.untransformed.no-tools.luna}{6}
\resultdef{facts.primary.failed-episodes.2.untransformed.no-tools.max}{6}
\resultdef{facts.primary.failed-episodes.2.untransformed.no-tools.min}{4}
\resultdef{facts.primary.failed-episodes.2.untransformed.no-tools.qwen}{4}
\resultdef{facts.primary.gain-by-use.discretionary-tools.debugger.transformed.luna}{+0.1}
\resultdef{facts.primary.gain-by-use.discretionary-tools.debugger.transformed.max}{+0.2}
\resultdef{facts.primary.gain-by-use.discretionary-tools.debugger.transformed.min}{+0.0}
\resultdef{facts.primary.gain-by-use.discretionary-tools.debugger.transformed.qwen}{+0.2}
\resultdef{facts.primary.gain-by-use.discretionary-tools.debugger.transformed.sol}{+0.0}
\resultdef{facts.primary.gain-by-use.discretionary-tools.debugger.untransformed.luna}{+0.0}
\resultdef{facts.primary.gain-by-use.discretionary-tools.debugger.untransformed.max}{+0.0}
\resultdef{facts.primary.gain-by-use.discretionary-tools.debugger.untransformed.min}{+0.0}
\resultdef{facts.primary.gain-by-use.discretionary-tools.debugger.untransformed.qwen}{+0.0}
\resultdef{facts.primary.gain-by-use.discretionary-tools.debugger.untransformed.sol}{+0.0}
\resultdef{facts.primary.gain-by-use.discretionary-tools.no-tool.transformed.luna}{+0.7}
\resultdef{facts.primary.gain-by-use.discretionary-tools.no-tool.transformed.max}{+0.7}
\resultdef{facts.primary.gain-by-use.discretionary-tools.no-tool.transformed.min}{+0.0}
\resultdef{facts.primary.gain-by-use.discretionary-tools.no-tool.transformed.qwen}{+0.0}
\resultdef{facts.primary.gain-by-use.discretionary-tools.no-tool.transformed.sol}{+0.0}
\resultdef{facts.primary.gain-by-use.discretionary-tools.no-tool.untransformed.luna}{$-$0.2}
\resultdef{facts.primary.gain-by-use.discretionary-tools.no-tool.untransformed.max}{+0.0}
\resultdef{facts.primary.gain-by-use.discretionary-tools.no-tool.untransformed.min}{$-$0.2}
\resultdef{facts.primary.gain-by-use.discretionary-tools.no-tool.untransformed.qwen}{+0.0}
\resultdef{facts.primary.gain-by-use.discretionary-tools.no-tool.untransformed.sol}{+0.0}
\resultdef{facts.primary.gain-by-use.discretionary-tools.probes-only.transformed.luna}{+0.7}
\resultdef{facts.primary.gain-by-use.discretionary-tools.probes-only.transformed.max}{+1.1}
\resultdef{facts.primary.gain-by-use.discretionary-tools.probes-only.transformed.min}{+0.0}
\resultdef{facts.primary.gain-by-use.discretionary-tools.probes-only.transformed.qwen}{+1.1}
\resultdef{facts.primary.gain-by-use.discretionary-tools.probes-only.transformed.sol}{+0.0}
\resultdef{facts.primary.gain-by-use.discretionary-tools.probes-only.untransformed.luna}{+0.1}
\resultdef{facts.primary.gain-by-use.discretionary-tools.probes-only.untransformed.max}{+0.2}
\resultdef{facts.primary.gain-by-use.discretionary-tools.probes-only.untransformed.min}{$-$0.2}
\resultdef{facts.primary.gain-by-use.discretionary-tools.probes-only.untransformed.qwen}{+0.2}
\resultdef{facts.primary.gain-by-use.discretionary-tools.probes-only.untransformed.sol}{$-$0.2}
\resultdef{facts.primary.gain-by-use.encouraged-tools.debugger.transformed.luna}{+2.1}
\resultdef{facts.primary.gain-by-use.encouraged-tools.debugger.transformed.max}{+2.1}
\resultdef{facts.primary.gain-by-use.encouraged-tools.debugger.transformed.min}{+0.0}
\resultdef{facts.primary.gain-by-use.encouraged-tools.debugger.transformed.qwen}{+0.4}
\resultdef{facts.primary.gain-by-use.encouraged-tools.debugger.transformed.sol}{+0.0}
\resultdef{facts.primary.gain-by-use.encouraged-tools.debugger.untransformed.luna}{+0.7}
\resultdef{facts.primary.gain-by-use.encouraged-tools.debugger.untransformed.max}{+0.7}
\resultdef{facts.primary.gain-by-use.encouraged-tools.debugger.untransformed.min}{$-$0.1}
\resultdef{facts.primary.gain-by-use.encouraged-tools.debugger.untransformed.qwen}{+0.0}
\resultdef{facts.primary.gain-by-use.encouraged-tools.debugger.untransformed.sol}{$-$0.1}
\resultdef{facts.primary.gain-by-use.encouraged-tools.no-tool.transformed.luna}{+0.0}
\resultdef{facts.primary.gain-by-use.encouraged-tools.no-tool.transformed.max}{+0.0}
\resultdef{facts.primary.gain-by-use.encouraged-tools.no-tool.transformed.min}{$-$0.1}
\resultdef{facts.primary.gain-by-use.encouraged-tools.no-tool.transformed.qwen}{$-$0.1}
\resultdef{facts.primary.gain-by-use.encouraged-tools.no-tool.transformed.sol}{+0.0}
\resultdef{facts.primary.gain-by-use.encouraged-tools.no-tool.untransformed.luna}{+0.0}
\resultdef{facts.primary.gain-by-use.encouraged-tools.no-tool.untransformed.max}{+0.0}
\resultdef{facts.primary.gain-by-use.encouraged-tools.no-tool.untransformed.min}{+0.0}
\resultdef{facts.primary.gain-by-use.encouraged-tools.no-tool.untransformed.qwen}{+0.0}
\resultdef{facts.primary.gain-by-use.encouraged-tools.no-tool.untransformed.sol}{+0.0}
\resultdef{facts.primary.gain-by-use.encouraged-tools.probes-only.transformed.luna}{+0.0}
\resultdef{facts.primary.gain-by-use.encouraged-tools.probes-only.transformed.max}{+1.0}
\resultdef{facts.primary.gain-by-use.encouraged-tools.probes-only.transformed.min}{+0.0}
\resultdef{facts.primary.gain-by-use.encouraged-tools.probes-only.transformed.qwen}{+1.0}
\resultdef{facts.primary.gain-by-use.encouraged-tools.probes-only.transformed.sol}{+0.0}
\resultdef{facts.primary.gain-by-use.encouraged-tools.probes-only.untransformed.luna}{+0.0}
\resultdef{facts.primary.gain-by-use.encouraged-tools.probes-only.untransformed.max}{+0.3}
\resultdef{facts.primary.gain-by-use.encouraged-tools.probes-only.untransformed.min}{+0.0}
\resultdef{facts.primary.gain-by-use.encouraged-tools.probes-only.untransformed.qwen}{+0.3}
\resultdef{facts.primary.gain-by-use.encouraged-tools.probes-only.untransformed.sol}{+0.0}
\resultdef{facts.primary.missing-as-right.diff.luna}{+1.4}
\resultdef{facts.primary.missing-as-right.diff.max}{+1.4}
\resultdef{facts.primary.missing-as-right.diff.min}{+0.1}
\resultdef{facts.primary.missing-as-right.diff.qwen}{+1.0}
\resultdef{facts.primary.missing-as-right.diff.sol}{+0.1}
\resultdef{facts.primary.missing-as-right.effect.transformed.luna}{+2.1}
\resultdef{facts.primary.missing-as-right.effect.transformed.max}{+2.1}
\resultdef{facts.primary.missing-as-right.effect.transformed.min}{+0.0}
\resultdef{facts.primary.missing-as-right.effect.transformed.qwen}{+1.3}
\resultdef{facts.primary.missing-as-right.effect.transformed.sol}{+0.0}
\resultdef{facts.primary.missing-as-wrong.diff.luna}{+1.4}
\resultdef{facts.primary.missing-as-wrong.diff.max}{+1.4}
\resultdef{facts.primary.missing-as-wrong.diff.min}{+0.1}
\resultdef{facts.primary.missing-as-wrong.diff.qwen}{+1.0}
\resultdef{facts.primary.missing-as-wrong.diff.sol}{+0.1}
\resultdef{facts.primary.missing-as-wrong.effect.transformed.luna}{+2.1}
\resultdef{facts.primary.missing-as-wrong.effect.transformed.max}{+2.1}
\resultdef{facts.primary.missing-as-wrong.effect.transformed.min}{+0.0}
\resultdef{facts.primary.missing-as-wrong.effect.transformed.qwen}{+1.3}
\resultdef{facts.primary.missing-as-wrong.effect.transformed.sol}{+0.0}
\resultdef{facts.primary.probing.not-probed.cases.transformed.luna}{2{,}426}
\resultdef{facts.primary.probing.not-probed.cases.transformed.max}{2{,}426}
\resultdef{facts.primary.probing.not-probed.cases.transformed.min}{2}
\resultdef{facts.primary.probing.not-probed.cases.transformed.qwen}{2}
\resultdef{facts.primary.probing.not-probed.cases.transformed.sol}{30}
\resultdef{facts.primary.probing.not-probed.cases.untransformed.luna}{516}
\resultdef{facts.primary.probing.not-probed.cases.untransformed.max}{516}
\resultdef{facts.primary.probing.not-probed.cases.untransformed.min}{0}
\resultdef{facts.primary.probing.not-probed.cases.untransformed.qwen}{0}
\resultdef{facts.primary.probing.not-probed.cases.untransformed.sol}{69}
\resultdef{facts.primary.probing.not-probed.effect.transformed.luna}{+0.8}
\resultdef{facts.primary.probing.not-probed.effect.transformed.max}{+0.8}
\resultdef{facts.primary.probing.not-probed.effect.transformed.min}{+0.0}
\resultdef{facts.primary.probing.not-probed.effect.transformed.qwen}{+0.0}
\resultdef{facts.primary.probing.not-probed.effect.transformed.sol}{+0.0}
\resultdef{facts.primary.probing.not-probed.effect.untransformed.luna}{+0.5}
\resultdef{facts.primary.probing.not-probed.effect.untransformed.max}{+0.5}
\resultdef{facts.primary.probing.not-probed.effect.untransformed.min}{+0.0}
\resultdef{facts.primary.probing.not-probed.effect.untransformed.sol}{+0.0}
\resultdef{facts.primary.probing.not-probed.no-tools.transformed.luna}{98.3}
\resultdef{facts.primary.probing.not-probed.no-tools.transformed.max}{100.0}
\resultdef{facts.primary.probing.not-probed.no-tools.transformed.min}{98.3}
\resultdef{facts.primary.probing.not-probed.no-tools.transformed.qwen}{100.0}
\resultdef{facts.primary.probing.not-probed.no-tools.transformed.sol}{100.0}
\resultdef{facts.primary.probing.not-probed.no-tools.untransformed.luna}{98.6}
\resultdef{facts.primary.probing.not-probed.no-tools.untransformed.max}{100.0}
\resultdef{facts.primary.probing.not-probed.no-tools.untransformed.min}{98.6}
\resultdef{facts.primary.probing.not-probed.no-tools.untransformed.sol}{100.0}
\resultdef{facts.primary.probing.probed.cases.transformed.luna}{596}
\resultdef{facts.primary.probing.probed.cases.transformed.max}{3{,}020}
\resultdef{facts.primary.probing.probed.cases.transformed.min}{596}
\resultdef{facts.primary.probing.probed.cases.transformed.qwen}{3{,}020}
\resultdef{facts.primary.probing.probed.cases.transformed.sol}{2{,}992}
\resultdef{facts.primary.probing.probed.cases.untransformed.luna}{94}
\resultdef{facts.primary.probing.probed.cases.untransformed.max}{610}
\resultdef{facts.primary.probing.probed.cases.untransformed.min}{94}
\resultdef{facts.primary.probing.probed.cases.untransformed.qwen}{610}
\resultdef{facts.primary.probing.probed.cases.untransformed.sol}{541}
\resultdef{facts.primary.probing.probed.effect.transformed.luna}{+7.5}
\resultdef{facts.primary.probing.probed.effect.transformed.max}{+7.5}
\resultdef{facts.primary.probing.probed.effect.transformed.min}{+0.0}
\resultdef{facts.primary.probing.probed.effect.transformed.qwen}{+1.3}
\resultdef{facts.primary.probing.probed.effect.transformed.sol}{+0.0}
\resultdef{facts.primary.probing.probed.effect.untransformed.luna}{+2.1}
\resultdef{facts.primary.probing.probed.effect.untransformed.max}{+2.1}
\resultdef{facts.primary.probing.probed.effect.untransformed.min}{$-$0.1}
\resultdef{facts.primary.probing.probed.effect.untransformed.qwen}{+0.3}
\resultdef{facts.primary.probing.probed.effect.untransformed.sol}{$-$0.1}
\resultdef{facts.primary.probing.probed.no-tools.transformed.luna}{90.0}
\resultdef{facts.primary.probing.probed.no-tools.transformed.max}{99.9}
\resultdef{facts.primary.probing.probed.no-tools.transformed.min}{90.0}
\resultdef{facts.primary.probing.probed.no-tools.transformed.qwen}{97.9}
\resultdef{facts.primary.probing.probed.no-tools.transformed.sol}{99.9}
\resultdef{facts.primary.probing.probed.no-tools.untransformed.luna}{95.7}
\resultdef{facts.primary.probing.probed.no-tools.untransformed.max}{99.9}
\resultdef{facts.primary.probing.probed.no-tools.untransformed.min}{95.7}
\resultdef{facts.primary.probing.probed.no-tools.untransformed.qwen}{98.8}
\resultdef{facts.primary.probing.probed.no-tools.untransformed.sol}{99.9}
\resultdef{facts.primary.probing.rest.cases.transformed.luna}{2{,}426}
\resultdef{facts.primary.probing.rest.cases.transformed.max}{2{,}712}
\resultdef{facts.primary.probing.rest.cases.transformed.min}{54}
\resultdef{facts.primary.probing.rest.cases.transformed.qwen}{2{,}712}
\resultdef{facts.primary.probing.rest.cases.transformed.sol}{54}
\resultdef{facts.primary.probing.rest.cases.untransformed.luna}{516}
\resultdef{facts.primary.probing.rest.cases.untransformed.max}{516}
\resultdef{facts.primary.probing.rest.cases.untransformed.min}{93}
\resultdef{facts.primary.probing.rest.cases.untransformed.qwen}{482}
\resultdef{facts.primary.probing.rest.cases.untransformed.sol}{93}
\resultdef{facts.primary.probing.rest.effect.transformed.luna}{+0.8}
\resultdef{facts.primary.probing.rest.effect.transformed.max}{+0.8}
\resultdef{facts.primary.probing.rest.effect.transformed.min}{+0.0}
\resultdef{facts.primary.probing.rest.effect.transformed.qwen}{+0.6}
\resultdef{facts.primary.probing.rest.effect.transformed.sol}{+0.0}
\resultdef{facts.primary.probing.rest.effect.untransformed.luna}{+0.5}
\resultdef{facts.primary.probing.rest.effect.untransformed.max}{+0.5}
\resultdef{facts.primary.probing.rest.effect.untransformed.min}{+0.0}
\resultdef{facts.primary.probing.rest.effect.untransformed.qwen}{+0.0}
\resultdef{facts.primary.probing.rest.effect.untransformed.sol}{+0.0}
\resultdef{facts.primary.probing.rest.mean-runs.transformed.luna}{0.0}
\resultdef{facts.primary.probing.rest.mean-runs.transformed.max}{1.8}
\resultdef{facts.primary.probing.rest.mean-runs.transformed.min}{0.0}
\resultdef{facts.primary.probing.rest.mean-runs.transformed.qwen}{1.8}
\resultdef{facts.primary.probing.rest.mean-runs.transformed.sol}{0.2}
\resultdef{facts.primary.probing.rest.mean-runs.untransformed.luna}{0.0}
\resultdef{facts.primary.probing.rest.mean-runs.untransformed.max}{1.5}
\resultdef{facts.primary.probing.rest.mean-runs.untransformed.min}{0.0}
\resultdef{facts.primary.probing.rest.mean-runs.untransformed.qwen}{1.5}
\resultdef{facts.primary.probing.rest.mean-runs.untransformed.sol}{0.1}
\resultdef{facts.primary.probing.rest.no-tools.transformed.luna}{98.3}
\resultdef{facts.primary.probing.rest.no-tools.transformed.max}{100.0}
\resultdef{facts.primary.probing.rest.no-tools.transformed.min}{98.3}
\resultdef{facts.primary.probing.rest.no-tools.transformed.qwen}{98.7}
\resultdef{facts.primary.probing.rest.no-tools.transformed.sol}{100.0}
\resultdef{facts.primary.probing.rest.no-tools.untransformed.luna}{98.6}
\resultdef{facts.primary.probing.rest.no-tools.untransformed.max}{100.0}
\resultdef{facts.primary.probing.rest.no-tools.untransformed.min}{98.6}
\resultdef{facts.primary.probing.rest.no-tools.untransformed.qwen}{99.5}
\resultdef{facts.primary.probing.rest.no-tools.untransformed.sol}{100.0}
\resultdef{facts.primary.probing.top-tenth.cases.transformed.luna}{596}
\resultdef{facts.primary.probing.top-tenth.cases.transformed.max}{2{,}968}
\resultdef{facts.primary.probing.top-tenth.cases.transformed.min}{310}
\resultdef{facts.primary.probing.top-tenth.cases.transformed.qwen}{310}
\resultdef{facts.primary.probing.top-tenth.cases.transformed.sol}{2{,}968}
\resultdef{facts.primary.probing.top-tenth.cases.untransformed.luna}{94}
\resultdef{facts.primary.probing.top-tenth.cases.untransformed.max}{517}
\resultdef{facts.primary.probing.top-tenth.cases.untransformed.min}{94}
\resultdef{facts.primary.probing.top-tenth.cases.untransformed.qwen}{128}
\resultdef{facts.primary.probing.top-tenth.cases.untransformed.sol}{517}
\resultdef{facts.primary.probing.top-tenth.effect.transformed.luna}{+7.5}
\resultdef{facts.primary.probing.top-tenth.effect.transformed.max}{+7.5}
\resultdef{facts.primary.probing.top-tenth.effect.transformed.min}{+0.0}
\resultdef{facts.primary.probing.top-tenth.effect.transformed.qwen}{+6.9}
\resultdef{facts.primary.probing.top-tenth.effect.transformed.sol}{+0.0}
\resultdef{facts.primary.probing.top-tenth.effect.untransformed.luna}{+2.1}
\resultdef{facts.primary.probing.top-tenth.effect.untransformed.max}{+2.1}
\resultdef{facts.primary.probing.top-tenth.effect.untransformed.min}{$-$0.1}
\resultdef{facts.primary.probing.top-tenth.effect.untransformed.qwen}{+1.6}
\resultdef{facts.primary.probing.top-tenth.effect.untransformed.sol}{$-$0.1}
\resultdef{facts.primary.probing.top-tenth.mean-runs.transformed.luna}{0.8}
\resultdef{facts.primary.probing.top-tenth.mean-runs.transformed.max}{5.7}
\resultdef{facts.primary.probing.top-tenth.mean-runs.transformed.min}{0.8}
\resultdef{facts.primary.probing.top-tenth.mean-runs.transformed.qwen}{5.7}
\resultdef{facts.primary.probing.top-tenth.mean-runs.transformed.sol}{1.0}
\resultdef{facts.primary.probing.top-tenth.mean-runs.untransformed.luna}{0.6}
\resultdef{facts.primary.probing.top-tenth.mean-runs.untransformed.max}{3.0}
\resultdef{facts.primary.probing.top-tenth.mean-runs.untransformed.min}{0.6}
\resultdef{facts.primary.probing.top-tenth.mean-runs.untransformed.qwen}{3.0}
\resultdef{facts.primary.probing.top-tenth.mean-runs.untransformed.sol}{1.0}
\resultdef{facts.primary.probing.top-tenth.no-tools.transformed.luna}{90.0}
\resultdef{facts.primary.probing.top-tenth.no-tools.transformed.max}{99.9}
\resultdef{facts.primary.probing.top-tenth.no-tools.transformed.min}{90.0}
\resultdef{facts.primary.probing.top-tenth.no-tools.transformed.qwen}{90.6}
\resultdef{facts.primary.probing.top-tenth.no-tools.transformed.sol}{99.9}
\resultdef{facts.primary.probing.top-tenth.no-tools.untransformed.luna}{95.7}
\resultdef{facts.primary.probing.top-tenth.no-tools.untransformed.max}{99.9}
\resultdef{facts.primary.probing.top-tenth.no-tools.untransformed.min}{95.7}
\resultdef{facts.primary.probing.top-tenth.no-tools.untransformed.qwen}{96.1}
\resultdef{facts.primary.probing.top-tenth.no-tools.untransformed.sol}{99.9}
\resultdef{facts.primary.reversal.all-to-none.transformed.luna}{1}
\resultdef{facts.primary.reversal.all-to-none.transformed.max}{1}
\resultdef{facts.primary.reversal.all-to-none.transformed.min}{0}
\resultdef{facts.primary.reversal.all-to-none.transformed.qwen}{0}
\resultdef{facts.primary.reversal.all-to-none.transformed.sol}{0}
\resultdef{facts.primary.reversal.all-to-none.untransformed.luna}{0}
\resultdef{facts.primary.reversal.all-to-none.untransformed.max}{0}
\resultdef{facts.primary.reversal.all-to-none.untransformed.min}{0}
\resultdef{facts.primary.reversal.all-to-none.untransformed.qwen}{0}
\resultdef{facts.primary.reversal.all-to-none.untransformed.sol}{0}
\resultdef{facts.primary.reversal.better.transformed.luna}{119}
\resultdef{facts.primary.reversal.better.transformed.max}{119}
\resultdef{facts.primary.reversal.better.transformed.min}{2}
\resultdef{facts.primary.reversal.better.transformed.qwen}{87}
\resultdef{facts.primary.reversal.better.transformed.sol}{2}
\resultdef{facts.primary.reversal.better.untransformed.luna}{11}
\resultdef{facts.primary.reversal.better.untransformed.max}{11}
\resultdef{facts.primary.reversal.better.untransformed.min}{0}
\resultdef{facts.primary.reversal.better.untransformed.qwen}{7}
\resultdef{facts.primary.reversal.better.untransformed.sol}{0}
\resultdef{facts.primary.reversal.none-to-all.transformed.luna}{26}
\resultdef{facts.primary.reversal.none-to-all.transformed.max}{26}
\resultdef{facts.primary.reversal.none-to-all.transformed.min}{1}
\resultdef{facts.primary.reversal.none-to-all.transformed.qwen}{12}
\resultdef{facts.primary.reversal.none-to-all.transformed.sol}{1}
\resultdef{facts.primary.reversal.none-to-all.untransformed.luna}{0}
\resultdef{facts.primary.reversal.none-to-all.untransformed.max}{2}
\resultdef{facts.primary.reversal.none-to-all.untransformed.min}{0}
\resultdef{facts.primary.reversal.none-to-all.untransformed.qwen}{2}
\resultdef{facts.primary.reversal.none-to-all.untransformed.sol}{0}
\resultdef{facts.primary.reversal.worse.transformed.luna}{17}
\resultdef{facts.primary.reversal.worse.transformed.max}{21}
\resultdef{facts.primary.reversal.worse.transformed.min}{1}
\resultdef{facts.primary.reversal.worse.transformed.qwen}{21}
\resultdef{facts.primary.reversal.worse.transformed.sol}{1}
\resultdef{facts.primary.reversal.worse.untransformed.luna}{2}
\resultdef{facts.primary.reversal.worse.untransformed.max}{5}
\resultdef{facts.primary.reversal.worse.untransformed.min}{1}
\resultdef{facts.primary.reversal.worse.untransformed.qwen}{5}
\resultdef{facts.primary.reversal.worse.untransformed.sol}{1}
\resultdef{facts.primary.wrong.counterpart-solved.combined.luna}{91}
\resultdef{facts.primary.wrong.counterpart-solved.combined.max}{91}
\resultdef{facts.primary.wrong.counterpart-solved.combined.min}{0}
\resultdef{facts.primary.wrong.counterpart-solved.combined.qwen}{66}
\resultdef{facts.primary.wrong.counterpart-solved.combined.sol}{0}
\resultdef{facts.primary.wrong.counterpart-solved.control-flow.luna}{36}
\resultdef{facts.primary.wrong.counterpart-solved.control-flow.max}{36}
\resultdef{facts.primary.wrong.counterpart-solved.control-flow.min}{0}
\resultdef{facts.primary.wrong.counterpart-solved.control-flow.qwen}{36}
\resultdef{facts.primary.wrong.counterpart-solved.control-flow.sol}{0}
\resultdef{facts.primary.wrong.counterpart-solved.maximal-mix.luna}{55}
\resultdef{facts.primary.wrong.counterpart-solved.maximal-mix.max}{55}
\resultdef{facts.primary.wrong.counterpart-solved.maximal-mix.min}{0}
\resultdef{facts.primary.wrong.counterpart-solved.maximal-mix.qwen}{30}
\resultdef{facts.primary.wrong.counterpart-solved.maximal-mix.sol}{0}
\resultdef{facts.primary.wrong.double-negation-share.combined.luna}{71.4}
\resultdef{facts.primary.wrong.double-negation-share.combined.max}{71.4}
\resultdef{facts.primary.wrong.double-negation-share.combined.min}{68.2}
\resultdef{facts.primary.wrong.double-negation-share.combined.qwen}{68.2}
\resultdef{facts.primary.wrong.double-negation-share.control-flow.luna}{72.2}
\resultdef{facts.primary.wrong.double-negation-share.control-flow.max}{72.2}
\resultdef{facts.primary.wrong.double-negation-share.control-flow.min}{66.7}
\resultdef{facts.primary.wrong.double-negation-share.control-flow.qwen}{66.7}
\resultdef{facts.primary.wrong.double-negation-share.maximal-mix.luna}{70.9}
\resultdef{facts.primary.wrong.double-negation-share.maximal-mix.max}{70.9}
\resultdef{facts.primary.wrong.double-negation-share.maximal-mix.min}{70.0}
\resultdef{facts.primary.wrong.double-negation-share.maximal-mix.qwen}{70.0}
\resultdef{facts.primary.wrong.double-negation.combined.luna}{65}
\resultdef{facts.primary.wrong.double-negation.combined.max}{65}
\resultdef{facts.primary.wrong.double-negation.combined.min}{0}
\resultdef{facts.primary.wrong.double-negation.combined.qwen}{45}
\resultdef{facts.primary.wrong.double-negation.combined.sol}{0}
\resultdef{facts.primary.wrong.double-negation.control-flow.luna}{26}
\resultdef{facts.primary.wrong.double-negation.control-flow.max}{26}
\resultdef{facts.primary.wrong.double-negation.control-flow.min}{0}
\resultdef{facts.primary.wrong.double-negation.control-flow.qwen}{24}
\resultdef{facts.primary.wrong.double-negation.control-flow.sol}{0}
\resultdef{facts.primary.wrong.double-negation.maximal-mix.luna}{39}
\resultdef{facts.primary.wrong.double-negation.maximal-mix.max}{39}
\resultdef{facts.primary.wrong.double-negation.maximal-mix.min}{0}
\resultdef{facts.primary.wrong.double-negation.maximal-mix.qwen}{21}
\resultdef{facts.primary.wrong.double-negation.maximal-mix.sol}{0}
\resultdef{facts.primary.wrong.reference-programs.combined.luna}{35}
\resultdef{facts.primary.wrong.reference-programs.combined.max}{35}
\resultdef{facts.primary.wrong.reference-programs.combined.min}{0}
\resultdef{facts.primary.wrong.reference-programs.combined.qwen}{33}
\resultdef{facts.primary.wrong.reference-programs.combined.sol}{0}
\resultdef{facts.primary.wrong.reference-programs.control-flow.luna}{21}
\resultdef{facts.primary.wrong.reference-programs.control-flow.max}{23}
\resultdef{facts.primary.wrong.reference-programs.control-flow.min}{0}
\resultdef{facts.primary.wrong.reference-programs.control-flow.qwen}{23}
\resultdef{facts.primary.wrong.reference-programs.control-flow.sol}{0}
\resultdef{facts.primary.wrong.reference-programs.maximal-mix.luna}{29}
\resultdef{facts.primary.wrong.reference-programs.maximal-mix.max}{29}
\resultdef{facts.primary.wrong.reference-programs.maximal-mix.min}{0}
\resultdef{facts.primary.wrong.reference-programs.maximal-mix.qwen}{21}
\resultdef{facts.primary.wrong.reference-programs.maximal-mix.sol}{0}
\resultdef{facts.primary.wrong.wrong-answers.combined.luna}{116}
\resultdef{facts.primary.wrong.wrong-answers.combined.max}{116}
\resultdef{facts.primary.wrong.wrong-answers.combined.min}{1}
\resultdef{facts.primary.wrong.wrong-answers.combined.qwen}{79}
\resultdef{facts.primary.wrong.wrong-answers.combined.sol}{1}
\resultdef{facts.primary.wrong.wrong-answers.control-flow.luna}{48}
\resultdef{facts.primary.wrong.wrong-answers.control-flow.max}{48}
\resultdef{facts.primary.wrong.wrong-answers.control-flow.min}{0}
\resultdef{facts.primary.wrong.wrong-answers.control-flow.qwen}{42}
\resultdef{facts.primary.wrong.wrong-answers.control-flow.sol}{0}
\resultdef{facts.primary.wrong.wrong-answers.maximal-mix.luna}{68}
\resultdef{facts.primary.wrong.wrong-answers.maximal-mix.max}{68}
\resultdef{facts.primary.wrong.wrong-answers.maximal-mix.min}{1}
\resultdef{facts.primary.wrong.wrong-answers.maximal-mix.qwen}{37}
\resultdef{facts.primary.wrong.wrong-answers.maximal-mix.sol}{1}
\resultdef{facts.probe.calls-program-share.transformed.discretionary-tools.luna}{85.6}
\resultdef{facts.probe.calls-program-share.transformed.discretionary-tools.max}{99.8}
\resultdef{facts.probe.calls-program-share.transformed.discretionary-tools.min}{57.9}
\resultdef{facts.probe.calls-program-share.transformed.discretionary-tools.qwen}{57.9}
\resultdef{facts.probe.calls-program-share.transformed.discretionary-tools.sol}{99.8}
\resultdef{facts.probe.calls-program-share.transformed.encouraged-tools.luna}{97.6}
\resultdef{facts.probe.calls-program-share.transformed.encouraged-tools.max}{98.2}
\resultdef{facts.probe.calls-program-share.transformed.encouraged-tools.min}{52.1}
\resultdef{facts.probe.calls-program-share.transformed.encouraged-tools.qwen}{52.1}
\resultdef{facts.probe.calls-program-share.transformed.encouraged-tools.sol}{98.2}
\resultdef{facts.probe.calls-program-share.untransformed.discretionary-tools.luna}{100.0}
\resultdef{facts.probe.calls-program-share.untransformed.discretionary-tools.max}{100.0}
\resultdef{facts.probe.calls-program-share.untransformed.discretionary-tools.min}{69.7}
\resultdef{facts.probe.calls-program-share.untransformed.discretionary-tools.qwen}{69.7}
\resultdef{facts.probe.calls-program-share.untransformed.discretionary-tools.sol}{100.0}
\resultdef{facts.probe.calls-program-share.untransformed.encouraged-tools.luna}{99.5}
\resultdef{facts.probe.calls-program-share.untransformed.encouraged-tools.max}{99.5}
\resultdef{facts.probe.calls-program-share.untransformed.encouraged-tools.min}{58.2}
\resultdef{facts.probe.calls-program-share.untransformed.encouraged-tools.qwen}{58.2}
\resultdef{facts.probe.calls-program-share.untransformed.encouraged-tools.sol}{98.5}
\resultdef{facts.probe.checking-episode-share.transformed.discretionary-tools.luna}{6.8}
\resultdef{facts.probe.checking-episode-share.transformed.discretionary-tools.max}{36.9}
\resultdef{facts.probe.checking-episode-share.transformed.discretionary-tools.min}{1.2}
\resultdef{facts.probe.checking-episode-share.transformed.discretionary-tools.qwen}{36.9}
\resultdef{facts.probe.checking-episode-share.transformed.discretionary-tools.sol}{1.2}
\resultdef{facts.probe.checking-episode-share.transformed.encouraged-tools.luna}{3.4}
\resultdef{facts.probe.checking-episode-share.transformed.encouraged-tools.max}{39.4}
\resultdef{facts.probe.checking-episode-share.transformed.encouraged-tools.min}{2.3}
\resultdef{facts.probe.checking-episode-share.transformed.encouraged-tools.qwen}{39.4}
\resultdef{facts.probe.checking-episode-share.transformed.encouraged-tools.sol}{2.3}
\resultdef{facts.probe.checking-episode-share.untransformed.discretionary-tools.luna}{3.7}
\resultdef{facts.probe.checking-episode-share.untransformed.discretionary-tools.max}{30.9}
\resultdef{facts.probe.checking-episode-share.untransformed.discretionary-tools.min}{1.3}
\resultdef{facts.probe.checking-episode-share.untransformed.discretionary-tools.qwen}{30.9}
\resultdef{facts.probe.checking-episode-share.untransformed.discretionary-tools.sol}{1.3}
\resultdef{facts.probe.checking-episode-share.untransformed.encouraged-tools.luna}{2.2}
\resultdef{facts.probe.checking-episode-share.untransformed.encouraged-tools.max}{32.6}
\resultdef{facts.probe.checking-episode-share.untransformed.encouraged-tools.min}{2.0}
\resultdef{facts.probe.checking-episode-share.untransformed.encouraged-tools.qwen}{32.6}
\resultdef{facts.probe.checking-episode-share.untransformed.encouraged-tools.sol}{2.0}
\resultdef{facts.probe.checks-expected-share.transformed.discretionary-tools.luna}{5.6}
\resultdef{facts.probe.checks-expected-share.transformed.discretionary-tools.max}{25.6}
\resultdef{facts.probe.checks-expected-share.transformed.discretionary-tools.min}{1.2}
\resultdef{facts.probe.checks-expected-share.transformed.discretionary-tools.qwen}{25.6}
\resultdef{facts.probe.checks-expected-share.transformed.discretionary-tools.sol}{1.2}
\resultdef{facts.probe.checks-expected-share.transformed.encouraged-tools.luna}{3.3}
\resultdef{facts.probe.checks-expected-share.transformed.encouraged-tools.max}{23.4}
\resultdef{facts.probe.checks-expected-share.transformed.encouraged-tools.min}{2.1}
\resultdef{facts.probe.checks-expected-share.transformed.encouraged-tools.qwen}{23.4}
\resultdef{facts.probe.checks-expected-share.transformed.encouraged-tools.sol}{2.1}
\resultdef{facts.probe.checks-expected-share.untransformed.discretionary-tools.luna}{4.6}
\resultdef{facts.probe.checks-expected-share.untransformed.discretionary-tools.max}{22.0}
\resultdef{facts.probe.checks-expected-share.untransformed.discretionary-tools.min}{1.3}
\resultdef{facts.probe.checks-expected-share.untransformed.discretionary-tools.qwen}{22.0}
\resultdef{facts.probe.checks-expected-share.untransformed.discretionary-tools.sol}{1.3}
\resultdef{facts.probe.checks-expected-share.untransformed.encouraged-tools.luna}{2.4}
\resultdef{facts.probe.checks-expected-share.untransformed.encouraged-tools.max}{19.2}
\resultdef{facts.probe.checks-expected-share.untransformed.encouraged-tools.min}{1.8}
\resultdef{facts.probe.checks-expected-share.untransformed.encouraged-tools.qwen}{19.2}
\resultdef{facts.probe.checks-expected-share.untransformed.encouraged-tools.sol}{1.8}
\resultdef{facts.probe.redefines-program-share.transformed.discretionary-tools.luna}{0.0}
\resultdef{facts.probe.redefines-program-share.transformed.discretionary-tools.max}{0.2}
\resultdef{facts.probe.redefines-program-share.transformed.discretionary-tools.min}{0.0}
\resultdef{facts.probe.redefines-program-share.transformed.discretionary-tools.qwen}{0.2}
\resultdef{facts.probe.redefines-program-share.transformed.discretionary-tools.sol}{0.0}
\resultdef{facts.probe.redefines-program-share.transformed.encouraged-tools.luna}{0.0}
\resultdef{facts.probe.redefines-program-share.transformed.encouraged-tools.max}{0.2}
\resultdef{facts.probe.redefines-program-share.transformed.encouraged-tools.min}{0.0}
\resultdef{facts.probe.redefines-program-share.transformed.encouraged-tools.qwen}{0.2}
\resultdef{facts.probe.redefines-program-share.transformed.encouraged-tools.sol}{0.0}
\resultdef{facts.probe.redefines-program-share.untransformed.discretionary-tools.luna}{0.0}
\resultdef{facts.probe.redefines-program-share.untransformed.discretionary-tools.max}{0.1}
\resultdef{facts.probe.redefines-program-share.untransformed.discretionary-tools.min}{0.0}
\resultdef{facts.probe.redefines-program-share.untransformed.discretionary-tools.qwen}{0.1}
\resultdef{facts.probe.redefines-program-share.untransformed.discretionary-tools.sol}{0.0}
\resultdef{facts.probe.redefines-program-share.untransformed.encouraged-tools.luna}{0.0}
\resultdef{facts.probe.redefines-program-share.untransformed.encouraged-tools.max}{0.3}
\resultdef{facts.probe.redefines-program-share.untransformed.encouraged-tools.min}{0.0}
\resultdef{facts.probe.redefines-program-share.untransformed.encouraged-tools.qwen}{0.3}
\resultdef{facts.probe.redefines-program-share.untransformed.encouraged-tools.sol}{0.0}
\resultdef{facts.probe.runs.transformed.discretionary-tools.luna}{972}
\resultdef{facts.probe.runs.transformed.discretionary-tools.max}{13{,}526}
\resultdef{facts.probe.runs.transformed.discretionary-tools.min}{972}
\resultdef{facts.probe.runs.transformed.discretionary-tools.qwen}{13{,}526}
\resultdef{facts.probe.runs.transformed.discretionary-tools.sol}{6{,}061}
\resultdef{facts.probe.runs.transformed.encouraged-tools.luna}{6{,}963}
\resultdef{facts.probe.runs.transformed.encouraged-tools.max}{16{,}510}
\resultdef{facts.probe.runs.transformed.encouraged-tools.min}{6{,}784}
\resultdef{facts.probe.runs.transformed.encouraged-tools.qwen}{16{,}510}
\resultdef{facts.probe.runs.transformed.encouraged-tools.sol}{6{,}784}
\resultdef{facts.probe.runs.transformed.no-tools.luna}{0}
\resultdef{facts.probe.runs.transformed.no-tools.max}{0}
\resultdef{facts.probe.runs.transformed.no-tools.min}{0}
\resultdef{facts.probe.runs.transformed.no-tools.qwen}{0}
\resultdef{facts.probe.runs.transformed.no-tools.sol}{0}
\resultdef{facts.probe.runs.untransformed.discretionary-tools.luna}{108}
\resultdef{facts.probe.runs.untransformed.discretionary-tools.max}{2{,}176}
\resultdef{facts.probe.runs.untransformed.discretionary-tools.min}{108}
\resultdef{facts.probe.runs.untransformed.discretionary-tools.qwen}{2{,}176}
\resultdef{facts.probe.runs.untransformed.discretionary-tools.sol}{1{,}074}
\resultdef{facts.probe.runs.untransformed.encouraged-tools.luna}{1{,}322}
\resultdef{facts.probe.runs.untransformed.encouraged-tools.max}{2{,}833}
\resultdef{facts.probe.runs.untransformed.encouraged-tools.min}{1{,}322}
\resultdef{facts.probe.runs.untransformed.encouraged-tools.qwen}{2{,}833}
\resultdef{facts.probe.runs.untransformed.encouraged-tools.sol}{1{,}343}
\resultdef{facts.probe.runs.untransformed.no-tools.luna}{0}
\resultdef{facts.probe.runs.untransformed.no-tools.max}{0}
\resultdef{facts.probe.runs.untransformed.no-tools.min}{0}
\resultdef{facts.probe.runs.untransformed.no-tools.qwen}{0}
\resultdef{facts.probe.runs.untransformed.no-tools.sol}{0}
\resultdef{facts.secondary.failed-cases.luna}{460}
\resultdef{facts.secondary.failed-cases.max}{460}
\resultdef{facts.secondary.failed-cases.min}{9}
\resultdef{facts.secondary.failed-cases.qwen}{181}
\resultdef{facts.secondary.failed-cases.shared.luna-qwen}{98}
\resultdef{facts.secondary.failed-cases.shared.luna-sol}{8}
\resultdef{facts.secondary.failed-cases.shared.qwen-sol}{5}
\resultdef{facts.secondary.failed-cases.sol}{9}
\resultdef{facts.secondary.failed-episodes.0.transformed.discretionary-tools.luna}{2{,}681}
\resultdef{facts.secondary.failed-episodes.0.transformed.discretionary-tools.max}{3{,}008}
\resultdef{facts.secondary.failed-episodes.0.transformed.discretionary-tools.min}{2{,}681}
\resultdef{facts.secondary.failed-episodes.0.transformed.discretionary-tools.qwen}{2{,}922}
\resultdef{facts.secondary.failed-episodes.0.transformed.discretionary-tools.sol}{3{,}008}
\resultdef{facts.secondary.failed-episodes.0.transformed.encouraged-tools.luna}{2{,}704}
\resultdef{facts.secondary.failed-episodes.0.transformed.encouraged-tools.max}{3{,}006}
\resultdef{facts.secondary.failed-episodes.0.transformed.encouraged-tools.min}{2{,}704}
\resultdef{facts.secondary.failed-episodes.0.transformed.encouraged-tools.qwen}{2{,}934}
\resultdef{facts.secondary.failed-episodes.0.transformed.encouraged-tools.sol}{3{,}006}
\resultdef{facts.secondary.failed-episodes.0.transformed.no-tools.luna}{2{,}562}
\resultdef{facts.secondary.failed-episodes.0.transformed.no-tools.max}{3{,}013}
\resultdef{facts.secondary.failed-episodes.0.transformed.no-tools.min}{2{,}562}
\resultdef{facts.secondary.failed-episodes.0.transformed.no-tools.qwen}{2{,}841}
\resultdef{facts.secondary.failed-episodes.0.transformed.no-tools.sol}{3{,}013}
\resultdef{facts.secondary.failed-episodes.0.untransformed.discretionary-tools.luna}{596}
\resultdef{facts.secondary.failed-episodes.0.untransformed.discretionary-tools.max}{608}
\resultdef{facts.secondary.failed-episodes.0.untransformed.discretionary-tools.min}{596}
\resultdef{facts.secondary.failed-episodes.0.untransformed.discretionary-tools.qwen}{601}
\resultdef{facts.secondary.failed-episodes.0.untransformed.discretionary-tools.sol}{608}
\resultdef{facts.secondary.failed-episodes.0.untransformed.encouraged-tools.luna}{602}
\resultdef{facts.secondary.failed-episodes.0.untransformed.encouraged-tools.max}{609}
\resultdef{facts.secondary.failed-episodes.0.untransformed.encouraged-tools.min}{602}
\resultdef{facts.secondary.failed-episodes.0.untransformed.encouraged-tools.qwen}{602}
\resultdef{facts.secondary.failed-episodes.0.untransformed.encouraged-tools.sol}{609}
\resultdef{facts.secondary.failed-episodes.0.untransformed.no-tools.luna}{594}
\resultdef{facts.secondary.failed-episodes.0.untransformed.no-tools.max}{609}
\resultdef{facts.secondary.failed-episodes.0.untransformed.no-tools.min}{594}
\resultdef{facts.secondary.failed-episodes.0.untransformed.no-tools.qwen}{599}
\resultdef{facts.secondary.failed-episodes.0.untransformed.no-tools.sol}{609}
\resultdef{facts.secondary.failed-episodes.1.transformed.discretionary-tools.luna}{197}
\resultdef{facts.secondary.failed-episodes.1.transformed.discretionary-tools.max}{197}
\resultdef{facts.secondary.failed-episodes.1.transformed.discretionary-tools.min}{6}
\resultdef{facts.secondary.failed-episodes.1.transformed.discretionary-tools.qwen}{82}
\resultdef{facts.secondary.failed-episodes.1.transformed.discretionary-tools.sol}{6}
\resultdef{facts.secondary.failed-episodes.1.transformed.encouraged-tools.luna}{196}
\resultdef{facts.secondary.failed-episodes.1.transformed.encouraged-tools.max}{196}
\resultdef{facts.secondary.failed-episodes.1.transformed.encouraged-tools.min}{7}
\resultdef{facts.secondary.failed-episodes.1.transformed.encouraged-tools.qwen}{71}
\resultdef{facts.secondary.failed-episodes.1.transformed.encouraged-tools.sol}{7}
\resultdef{facts.secondary.failed-episodes.1.transformed.no-tools.luna}{277}
\resultdef{facts.secondary.failed-episodes.1.transformed.no-tools.max}{277}
\resultdef{facts.secondary.failed-episodes.1.transformed.no-tools.min}{3}
\resultdef{facts.secondary.failed-episodes.1.transformed.no-tools.qwen}{147}
\resultdef{facts.secondary.failed-episodes.1.transformed.no-tools.sol}{3}
\resultdef{facts.secondary.failed-episodes.1.untransformed.discretionary-tools.luna}{5}
\resultdef{facts.secondary.failed-episodes.1.untransformed.discretionary-tools.max}{5}
\resultdef{facts.secondary.failed-episodes.1.untransformed.discretionary-tools.min}{1}
\resultdef{facts.secondary.failed-episodes.1.untransformed.discretionary-tools.qwen}{5}
\resultdef{facts.secondary.failed-episodes.1.untransformed.discretionary-tools.sol}{1}
\resultdef{facts.secondary.failed-episodes.1.untransformed.encouraged-tools.luna}{3}
\resultdef{facts.secondary.failed-episodes.1.untransformed.encouraged-tools.max}{5}
\resultdef{facts.secondary.failed-episodes.1.untransformed.encouraged-tools.min}{3}
\resultdef{facts.secondary.failed-episodes.1.untransformed.encouraged-tools.qwen}{5}
\resultdef{facts.secondary.failed-episodes.1.untransformed.no-tools.luna}{10}
\resultdef{facts.secondary.failed-episodes.1.untransformed.no-tools.max}{10}
\resultdef{facts.secondary.failed-episodes.1.untransformed.no-tools.min}{1}
\resultdef{facts.secondary.failed-episodes.1.untransformed.no-tools.qwen}{7}
\resultdef{facts.secondary.failed-episodes.1.untransformed.no-tools.sol}{1}
\resultdef{facts.secondary.failed-episodes.2.transformed.discretionary-tools.luna}{144}
\resultdef{facts.secondary.failed-episodes.2.transformed.discretionary-tools.max}{144}
\resultdef{facts.secondary.failed-episodes.2.transformed.discretionary-tools.min}{8}
\resultdef{facts.secondary.failed-episodes.2.transformed.discretionary-tools.qwen}{18}
\resultdef{facts.secondary.failed-episodes.2.transformed.discretionary-tools.sol}{8}
\resultdef{facts.secondary.failed-episodes.2.transformed.encouraged-tools.luna}{122}
\resultdef{facts.secondary.failed-episodes.2.transformed.encouraged-tools.max}{122}
\resultdef{facts.secondary.failed-episodes.2.transformed.encouraged-tools.min}{9}
\resultdef{facts.secondary.failed-episodes.2.transformed.encouraged-tools.qwen}{17}
\resultdef{facts.secondary.failed-episodes.2.transformed.encouraged-tools.sol}{9}
\resultdef{facts.secondary.failed-episodes.2.transformed.no-tools.luna}{183}
\resultdef{facts.secondary.failed-episodes.2.transformed.no-tools.max}{183}
\resultdef{facts.secondary.failed-episodes.2.transformed.no-tools.min}{6}
\resultdef{facts.secondary.failed-episodes.2.transformed.no-tools.qwen}{34}
\resultdef{facts.secondary.failed-episodes.2.transformed.no-tools.sol}{6}
\resultdef{facts.secondary.failed-episodes.2.untransformed.discretionary-tools.luna}{9}
\resultdef{facts.secondary.failed-episodes.2.untransformed.discretionary-tools.max}{9}
\resultdef{facts.secondary.failed-episodes.2.untransformed.discretionary-tools.min}{1}
\resultdef{facts.secondary.failed-episodes.2.untransformed.discretionary-tools.qwen}{4}
\resultdef{facts.secondary.failed-episodes.2.untransformed.discretionary-tools.sol}{1}
\resultdef{facts.secondary.failed-episodes.2.untransformed.encouraged-tools.luna}{5}
\resultdef{facts.secondary.failed-episodes.2.untransformed.encouraged-tools.max}{5}
\resultdef{facts.secondary.failed-episodes.2.untransformed.encouraged-tools.min}{1}
\resultdef{facts.secondary.failed-episodes.2.untransformed.encouraged-tools.qwen}{3}
\resultdef{facts.secondary.failed-episodes.2.untransformed.encouraged-tools.sol}{1}
\resultdef{facts.secondary.failed-episodes.2.untransformed.no-tools.luna}{6}
\resultdef{facts.secondary.failed-episodes.2.untransformed.no-tools.max}{6}
\resultdef{facts.secondary.failed-episodes.2.untransformed.no-tools.min}{4}
\resultdef{facts.secondary.failed-episodes.2.untransformed.no-tools.qwen}{4}
\resultdef{facts.secondary.gain-by-use.discretionary-tools.debugger.transformed.luna}{+0.1}
\resultdef{facts.secondary.gain-by-use.discretionary-tools.debugger.transformed.max}{+0.3}
\resultdef{facts.secondary.gain-by-use.discretionary-tools.debugger.transformed.min}{+0.0}
\resultdef{facts.secondary.gain-by-use.discretionary-tools.debugger.transformed.qwen}{+0.3}
\resultdef{facts.secondary.gain-by-use.discretionary-tools.debugger.transformed.sol}{+0.0}
\resultdef{facts.secondary.gain-by-use.discretionary-tools.debugger.untransformed.luna}{+0.0}
\resultdef{facts.secondary.gain-by-use.discretionary-tools.debugger.untransformed.max}{+0.0}
\resultdef{facts.secondary.gain-by-use.discretionary-tools.debugger.untransformed.min}{+0.0}
\resultdef{facts.secondary.gain-by-use.discretionary-tools.debugger.untransformed.qwen}{+0.0}
\resultdef{facts.secondary.gain-by-use.discretionary-tools.debugger.untransformed.sol}{+0.0}
\resultdef{facts.secondary.gain-by-use.discretionary-tools.no-tool.transformed.luna}{+1.7}
\resultdef{facts.secondary.gain-by-use.discretionary-tools.no-tool.transformed.max}{+1.7}
\resultdef{facts.secondary.gain-by-use.discretionary-tools.no-tool.transformed.min}{+0.0}
\resultdef{facts.secondary.gain-by-use.discretionary-tools.no-tool.transformed.qwen}{+0.0}
\resultdef{facts.secondary.gain-by-use.discretionary-tools.no-tool.transformed.sol}{+0.0}
\resultdef{facts.secondary.gain-by-use.discretionary-tools.no-tool.untransformed.luna}{$-$0.2}
\resultdef{facts.secondary.gain-by-use.discretionary-tools.no-tool.untransformed.max}{+0.0}
\resultdef{facts.secondary.gain-by-use.discretionary-tools.no-tool.untransformed.min}{$-$0.2}
\resultdef{facts.secondary.gain-by-use.discretionary-tools.no-tool.untransformed.qwen}{+0.0}
\resultdef{facts.secondary.gain-by-use.discretionary-tools.no-tool.untransformed.sol}{+0.0}
\resultdef{facts.secondary.gain-by-use.discretionary-tools.probes-only.transformed.luna}{+0.8}
\resultdef{facts.secondary.gain-by-use.discretionary-tools.probes-only.transformed.max}{+1.3}
\resultdef{facts.secondary.gain-by-use.discretionary-tools.probes-only.transformed.min}{$-$0.1}
\resultdef{facts.secondary.gain-by-use.discretionary-tools.probes-only.transformed.qwen}{+1.3}
\resultdef{facts.secondary.gain-by-use.discretionary-tools.probes-only.transformed.sol}{$-$0.1}
\resultdef{facts.secondary.gain-by-use.discretionary-tools.probes-only.untransformed.luna}{+0.1}
\resultdef{facts.secondary.gain-by-use.discretionary-tools.probes-only.untransformed.max}{+0.2}
\resultdef{facts.secondary.gain-by-use.discretionary-tools.probes-only.untransformed.min}{$-$0.2}
\resultdef{facts.secondary.gain-by-use.discretionary-tools.probes-only.untransformed.qwen}{+0.2}
\resultdef{facts.secondary.gain-by-use.discretionary-tools.probes-only.untransformed.sol}{$-$0.2}
\resultdef{facts.secondary.gain-by-use.encouraged-tools.debugger.transformed.luna}{+3.4}
\resultdef{facts.secondary.gain-by-use.encouraged-tools.debugger.transformed.max}{+3.4}
\resultdef{facts.secondary.gain-by-use.encouraged-tools.debugger.transformed.min}{$-$0.2}
\resultdef{facts.secondary.gain-by-use.encouraged-tools.debugger.transformed.qwen}{+0.7}
\resultdef{facts.secondary.gain-by-use.encouraged-tools.debugger.transformed.sol}{$-$0.2}
\resultdef{facts.secondary.gain-by-use.encouraged-tools.debugger.untransformed.luna}{+0.7}
\resultdef{facts.secondary.gain-by-use.encouraged-tools.debugger.untransformed.max}{+0.7}
\resultdef{facts.secondary.gain-by-use.encouraged-tools.debugger.untransformed.min}{$-$0.1}
\resultdef{facts.secondary.gain-by-use.encouraged-tools.debugger.untransformed.qwen}{+0.0}
\resultdef{facts.secondary.gain-by-use.encouraged-tools.debugger.untransformed.sol}{$-$0.1}
\resultdef{facts.secondary.gain-by-use.encouraged-tools.no-tool.transformed.luna}{+0.0}
\resultdef{facts.secondary.gain-by-use.encouraged-tools.no-tool.transformed.max}{+0.0}
\resultdef{facts.secondary.gain-by-use.encouraged-tools.no-tool.transformed.min}{$-$0.1}
\resultdef{facts.secondary.gain-by-use.encouraged-tools.no-tool.transformed.qwen}{$-$0.1}
\resultdef{facts.secondary.gain-by-use.encouraged-tools.no-tool.transformed.sol}{+0.0}
\resultdef{facts.secondary.gain-by-use.encouraged-tools.no-tool.untransformed.luna}{+0.0}
\resultdef{facts.secondary.gain-by-use.encouraged-tools.no-tool.untransformed.max}{+0.0}
\resultdef{facts.secondary.gain-by-use.encouraged-tools.no-tool.untransformed.min}{+0.0}
\resultdef{facts.secondary.gain-by-use.encouraged-tools.no-tool.untransformed.qwen}{+0.0}
\resultdef{facts.secondary.gain-by-use.encouraged-tools.no-tool.untransformed.sol}{+0.0}
\resultdef{facts.secondary.gain-by-use.encouraged-tools.probes-only.transformed.luna}{+0.0}
\resultdef{facts.secondary.gain-by-use.encouraged-tools.probes-only.transformed.max}{+1.2}
\resultdef{facts.secondary.gain-by-use.encouraged-tools.probes-only.transformed.min}{+0.0}
\resultdef{facts.secondary.gain-by-use.encouraged-tools.probes-only.transformed.qwen}{+1.2}
\resultdef{facts.secondary.gain-by-use.encouraged-tools.probes-only.transformed.sol}{+0.0}
\resultdef{facts.secondary.gain-by-use.encouraged-tools.probes-only.untransformed.luna}{+0.0}
\resultdef{facts.secondary.gain-by-use.encouraged-tools.probes-only.untransformed.max}{+0.3}
\resultdef{facts.secondary.gain-by-use.encouraged-tools.probes-only.untransformed.min}{+0.0}
\resultdef{facts.secondary.gain-by-use.encouraged-tools.probes-only.untransformed.qwen}{+0.3}
\resultdef{facts.secondary.gain-by-use.encouraged-tools.probes-only.untransformed.sol}{+0.0}
\resultdef{facts.secondary.missing-as-right.diff.luna}{+2.6}
\resultdef{facts.secondary.missing-as-right.diff.max}{+2.6}
\resultdef{facts.secondary.missing-as-right.diff.min}{$-$0.1}
\resultdef{facts.secondary.missing-as-right.diff.qwen}{+1.5}
\resultdef{facts.secondary.missing-as-right.diff.sol}{$-$0.1}
\resultdef{facts.secondary.missing-as-right.effect.transformed.luna}{+3.4}
\resultdef{facts.secondary.missing-as-right.effect.transformed.max}{+3.4}
\resultdef{facts.secondary.missing-as-right.effect.transformed.min}{$-$0.2}
\resultdef{facts.secondary.missing-as-right.effect.transformed.qwen}{+1.8}
\resultdef{facts.secondary.missing-as-right.effect.transformed.sol}{$-$0.2}
\resultdef{facts.secondary.missing-as-wrong.diff.luna}{+2.6}
\resultdef{facts.secondary.missing-as-wrong.diff.max}{+2.6}
\resultdef{facts.secondary.missing-as-wrong.diff.min}{$-$0.1}
\resultdef{facts.secondary.missing-as-wrong.diff.qwen}{+1.5}
\resultdef{facts.secondary.missing-as-wrong.diff.sol}{$-$0.1}
\resultdef{facts.secondary.missing-as-wrong.effect.transformed.luna}{+3.4}
\resultdef{facts.secondary.missing-as-wrong.effect.transformed.max}{+3.4}
\resultdef{facts.secondary.missing-as-wrong.effect.transformed.min}{$-$0.2}
\resultdef{facts.secondary.missing-as-wrong.effect.transformed.qwen}{+1.8}
\resultdef{facts.secondary.missing-as-wrong.effect.transformed.sol}{$-$0.2}
\resultdef{facts.secondary.probing.not-probed.cases.transformed.luna}{2{,}426}
\resultdef{facts.secondary.probing.not-probed.cases.transformed.max}{2{,}426}
\resultdef{facts.secondary.probing.not-probed.cases.transformed.min}{2}
\resultdef{facts.secondary.probing.not-probed.cases.transformed.qwen}{2}
\resultdef{facts.secondary.probing.not-probed.cases.transformed.sol}{30}
\resultdef{facts.secondary.probing.not-probed.cases.untransformed.luna}{516}
\resultdef{facts.secondary.probing.not-probed.cases.untransformed.max}{516}
\resultdef{facts.secondary.probing.not-probed.cases.untransformed.min}{0}
\resultdef{facts.secondary.probing.not-probed.cases.untransformed.qwen}{0}
\resultdef{facts.secondary.probing.not-probed.cases.untransformed.sol}{69}
\resultdef{facts.secondary.probing.not-probed.effect.transformed.luna}{+2.3}
\resultdef{facts.secondary.probing.not-probed.effect.transformed.max}{+2.3}
\resultdef{facts.secondary.probing.not-probed.effect.transformed.min}{+0.0}
\resultdef{facts.secondary.probing.not-probed.effect.transformed.qwen}{+0.0}
\resultdef{facts.secondary.probing.not-probed.effect.transformed.sol}{+0.0}
\resultdef{facts.secondary.probing.not-probed.effect.untransformed.luna}{+0.5}
\resultdef{facts.secondary.probing.not-probed.effect.untransformed.max}{+0.5}
\resultdef{facts.secondary.probing.not-probed.effect.untransformed.min}{+0.0}
\resultdef{facts.secondary.probing.not-probed.effect.untransformed.sol}{+0.0}
\resultdef{facts.secondary.probing.not-probed.no-tools.transformed.luna}{90.5}
\resultdef{facts.secondary.probing.not-probed.no-tools.transformed.max}{100.0}
\resultdef{facts.secondary.probing.not-probed.no-tools.transformed.min}{90.5}
\resultdef{facts.secondary.probing.not-probed.no-tools.transformed.qwen}{100.0}
\resultdef{facts.secondary.probing.not-probed.no-tools.transformed.sol}{100.0}
\resultdef{facts.secondary.probing.not-probed.no-tools.untransformed.luna}{98.6}
\resultdef{facts.secondary.probing.not-probed.no-tools.untransformed.max}{100.0}
\resultdef{facts.secondary.probing.not-probed.no-tools.untransformed.min}{98.6}
\resultdef{facts.secondary.probing.not-probed.no-tools.untransformed.sol}{100.0}
\resultdef{facts.secondary.probing.probed.cases.transformed.luna}{596}
\resultdef{facts.secondary.probing.probed.cases.transformed.max}{3{,}020}
\resultdef{facts.secondary.probing.probed.cases.transformed.min}{596}
\resultdef{facts.secondary.probing.probed.cases.transformed.qwen}{3{,}020}
\resultdef{facts.secondary.probing.probed.cases.transformed.sol}{2{,}992}
\resultdef{facts.secondary.probing.probed.cases.untransformed.luna}{94}
\resultdef{facts.secondary.probing.probed.cases.untransformed.max}{610}
\resultdef{facts.secondary.probing.probed.cases.untransformed.min}{94}
\resultdef{facts.secondary.probing.probed.cases.untransformed.qwen}{610}
\resultdef{facts.secondary.probing.probed.cases.untransformed.sol}{541}
\resultdef{facts.secondary.probing.probed.effect.transformed.luna}{+7.6}
\resultdef{facts.secondary.probing.probed.effect.transformed.max}{+7.6}
\resultdef{facts.secondary.probing.probed.effect.transformed.min}{$-$0.2}
\resultdef{facts.secondary.probing.probed.effect.transformed.qwen}{+1.8}
\resultdef{facts.secondary.probing.probed.effect.transformed.sol}{$-$0.2}
\resultdef{facts.secondary.probing.probed.effect.untransformed.luna}{+2.1}
\resultdef{facts.secondary.probing.probed.effect.untransformed.max}{+2.1}
\resultdef{facts.secondary.probing.probed.effect.untransformed.min}{$-$0.1}
\resultdef{facts.secondary.probing.probed.effect.untransformed.qwen}{+0.3}
\resultdef{facts.secondary.probing.probed.effect.untransformed.sol}{$-$0.1}
\resultdef{facts.secondary.probing.probed.no-tools.transformed.luna}{84.7}
\resultdef{facts.secondary.probing.probed.no-tools.transformed.max}{99.7}
\resultdef{facts.secondary.probing.probed.no-tools.transformed.min}{84.7}
\resultdef{facts.secondary.probing.probed.no-tools.transformed.qwen}{96.4}
\resultdef{facts.secondary.probing.probed.no-tools.transformed.sol}{99.7}
\resultdef{facts.secondary.probing.probed.no-tools.untransformed.luna}{95.7}
\resultdef{facts.secondary.probing.probed.no-tools.untransformed.max}{99.9}
\resultdef{facts.secondary.probing.probed.no-tools.untransformed.min}{95.7}
\resultdef{facts.secondary.probing.probed.no-tools.untransformed.qwen}{98.8}
\resultdef{facts.secondary.probing.probed.no-tools.untransformed.sol}{99.9}
\resultdef{facts.secondary.probing.rest.cases.transformed.luna}{2{,}426}
\resultdef{facts.secondary.probing.rest.cases.transformed.max}{2{,}712}
\resultdef{facts.secondary.probing.rest.cases.transformed.min}{54}
\resultdef{facts.secondary.probing.rest.cases.transformed.qwen}{2{,}712}
\resultdef{facts.secondary.probing.rest.cases.transformed.sol}{54}
\resultdef{facts.secondary.probing.rest.cases.untransformed.luna}{516}
\resultdef{facts.secondary.probing.rest.cases.untransformed.max}{516}
\resultdef{facts.secondary.probing.rest.cases.untransformed.min}{93}
\resultdef{facts.secondary.probing.rest.cases.untransformed.qwen}{482}
\resultdef{facts.secondary.probing.rest.cases.untransformed.sol}{93}
\resultdef{facts.secondary.probing.rest.effect.transformed.luna}{+2.3}
\resultdef{facts.secondary.probing.rest.effect.transformed.max}{+2.3}
\resultdef{facts.secondary.probing.rest.effect.transformed.min}{+0.0}
\resultdef{facts.secondary.probing.rest.effect.transformed.qwen}{+1.2}
\resultdef{facts.secondary.probing.rest.effect.transformed.sol}{+0.0}
\resultdef{facts.secondary.probing.rest.effect.untransformed.luna}{+0.5}
\resultdef{facts.secondary.probing.rest.effect.untransformed.max}{+0.5}
\resultdef{facts.secondary.probing.rest.effect.untransformed.min}{+0.0}
\resultdef{facts.secondary.probing.rest.effect.untransformed.qwen}{+0.0}
\resultdef{facts.secondary.probing.rest.effect.untransformed.sol}{+0.0}
\resultdef{facts.secondary.probing.rest.mean-runs.transformed.luna}{0.0}
\resultdef{facts.secondary.probing.rest.mean-runs.transformed.max}{1.8}
\resultdef{facts.secondary.probing.rest.mean-runs.transformed.min}{0.0}
\resultdef{facts.secondary.probing.rest.mean-runs.transformed.qwen}{1.8}
\resultdef{facts.secondary.probing.rest.mean-runs.transformed.sol}{0.2}
\resultdef{facts.secondary.probing.rest.mean-runs.untransformed.luna}{0.0}
\resultdef{facts.secondary.probing.rest.mean-runs.untransformed.max}{1.5}
\resultdef{facts.secondary.probing.rest.mean-runs.untransformed.min}{0.0}
\resultdef{facts.secondary.probing.rest.mean-runs.untransformed.qwen}{1.5}
\resultdef{facts.secondary.probing.rest.mean-runs.untransformed.sol}{0.1}
\resultdef{facts.secondary.probing.rest.no-tools.transformed.luna}{90.5}
\resultdef{facts.secondary.probing.rest.no-tools.transformed.max}{100.0}
\resultdef{facts.secondary.probing.rest.no-tools.transformed.min}{90.5}
\resultdef{facts.secondary.probing.rest.no-tools.transformed.qwen}{97.2}
\resultdef{facts.secondary.probing.rest.no-tools.transformed.sol}{100.0}
\resultdef{facts.secondary.probing.rest.no-tools.untransformed.luna}{98.6}
\resultdef{facts.secondary.probing.rest.no-tools.untransformed.max}{100.0}
\resultdef{facts.secondary.probing.rest.no-tools.untransformed.min}{98.6}
\resultdef{facts.secondary.probing.rest.no-tools.untransformed.qwen}{99.5}
\resultdef{facts.secondary.probing.rest.no-tools.untransformed.sol}{100.0}
\resultdef{facts.secondary.probing.top-tenth.cases.transformed.luna}{596}
\resultdef{facts.secondary.probing.top-tenth.cases.transformed.max}{2{,}968}
\resultdef{facts.secondary.probing.top-tenth.cases.transformed.min}{310}
\resultdef{facts.secondary.probing.top-tenth.cases.transformed.qwen}{310}
\resultdef{facts.secondary.probing.top-tenth.cases.transformed.sol}{2{,}968}
\resultdef{facts.secondary.probing.top-tenth.cases.untransformed.luna}{94}
\resultdef{facts.secondary.probing.top-tenth.cases.untransformed.max}{517}
\resultdef{facts.secondary.probing.top-tenth.cases.untransformed.min}{94}
\resultdef{facts.secondary.probing.top-tenth.cases.untransformed.qwen}{128}
\resultdef{facts.secondary.probing.top-tenth.cases.untransformed.sol}{517}
\resultdef{facts.secondary.probing.top-tenth.effect.transformed.luna}{+7.6}
\resultdef{facts.secondary.probing.top-tenth.effect.transformed.max}{+7.6}
\resultdef{facts.secondary.probing.top-tenth.effect.transformed.min}{$-$0.2}
\resultdef{facts.secondary.probing.top-tenth.effect.transformed.qwen}{+7.3}
\resultdef{facts.secondary.probing.top-tenth.effect.transformed.sol}{$-$0.2}
\resultdef{facts.secondary.probing.top-tenth.effect.untransformed.luna}{+2.1}
\resultdef{facts.secondary.probing.top-tenth.effect.untransformed.max}{+2.1}
\resultdef{facts.secondary.probing.top-tenth.effect.untransformed.min}{$-$0.1}
\resultdef{facts.secondary.probing.top-tenth.effect.untransformed.qwen}{+1.6}
\resultdef{facts.secondary.probing.top-tenth.effect.untransformed.sol}{$-$0.1}
\resultdef{facts.secondary.probing.top-tenth.mean-runs.transformed.luna}{0.8}
\resultdef{facts.secondary.probing.top-tenth.mean-runs.transformed.max}{5.7}
\resultdef{facts.secondary.probing.top-tenth.mean-runs.transformed.min}{0.8}
\resultdef{facts.secondary.probing.top-tenth.mean-runs.transformed.qwen}{5.7}
\resultdef{facts.secondary.probing.top-tenth.mean-runs.transformed.sol}{1.0}
\resultdef{facts.secondary.probing.top-tenth.mean-runs.untransformed.luna}{0.6}
\resultdef{facts.secondary.probing.top-tenth.mean-runs.untransformed.max}{3.0}
\resultdef{facts.secondary.probing.top-tenth.mean-runs.untransformed.min}{0.6}
\resultdef{facts.secondary.probing.top-tenth.mean-runs.untransformed.qwen}{3.0}
\resultdef{facts.secondary.probing.top-tenth.mean-runs.untransformed.sol}{1.0}
\resultdef{facts.secondary.probing.top-tenth.no-tools.transformed.luna}{84.7}
\resultdef{facts.secondary.probing.top-tenth.no-tools.transformed.max}{99.7}
\resultdef{facts.secondary.probing.top-tenth.no-tools.transformed.min}{84.7}
\resultdef{facts.secondary.probing.top-tenth.no-tools.transformed.qwen}{89.7}
\resultdef{facts.secondary.probing.top-tenth.no-tools.transformed.sol}{99.7}
\resultdef{facts.secondary.probing.top-tenth.no-tools.untransformed.luna}{95.7}
\resultdef{facts.secondary.probing.top-tenth.no-tools.untransformed.max}{99.9}
\resultdef{facts.secondary.probing.top-tenth.no-tools.untransformed.min}{95.7}
\resultdef{facts.secondary.probing.top-tenth.no-tools.untransformed.qwen}{96.1}
\resultdef{facts.secondary.probing.top-tenth.no-tools.untransformed.sol}{99.9}
\resultdef{facts.secondary.reversal.all-to-none.transformed.luna}{10}
\resultdef{facts.secondary.reversal.all-to-none.transformed.max}{10}
\resultdef{facts.secondary.reversal.all-to-none.transformed.min}{2}
\resultdef{facts.secondary.reversal.all-to-none.transformed.qwen}{2}
\resultdef{facts.secondary.reversal.all-to-none.transformed.sol}{3}
\resultdef{facts.secondary.reversal.all-to-none.untransformed.luna}{0}
\resultdef{facts.secondary.reversal.all-to-none.untransformed.max}{0}
\resultdef{facts.secondary.reversal.all-to-none.untransformed.min}{0}
\resultdef{facts.secondary.reversal.all-to-none.untransformed.qwen}{0}
\resultdef{facts.secondary.reversal.all-to-none.untransformed.sol}{0}
\resultdef{facts.secondary.reversal.better.transformed.luna}{266}
\resultdef{facts.secondary.reversal.better.transformed.max}{266}
\resultdef{facts.secondary.reversal.better.transformed.min}{3}
\resultdef{facts.secondary.reversal.better.transformed.qwen}{149}
\resultdef{facts.secondary.reversal.better.transformed.sol}{3}
\resultdef{facts.secondary.reversal.better.untransformed.luna}{11}
\resultdef{facts.secondary.reversal.better.untransformed.max}{11}
\resultdef{facts.secondary.reversal.better.untransformed.min}{0}
\resultdef{facts.secondary.reversal.better.untransformed.qwen}{7}
\resultdef{facts.secondary.reversal.better.untransformed.sol}{0}
\resultdef{facts.secondary.reversal.none-to-all.transformed.luna}{66}
\resultdef{facts.secondary.reversal.none-to-all.transformed.max}{66}
\resultdef{facts.secondary.reversal.none-to-all.transformed.min}{1}
\resultdef{facts.secondary.reversal.none-to-all.transformed.qwen}{16}
\resultdef{facts.secondary.reversal.none-to-all.transformed.sol}{1}
\resultdef{facts.secondary.reversal.none-to-all.untransformed.luna}{0}
\resultdef{facts.secondary.reversal.none-to-all.untransformed.max}{2}
\resultdef{facts.secondary.reversal.none-to-all.untransformed.min}{0}
\resultdef{facts.secondary.reversal.none-to-all.untransformed.qwen}{2}
\resultdef{facts.secondary.reversal.none-to-all.untransformed.sol}{0}
\resultdef{facts.secondary.reversal.worse.transformed.luna}{119}
\resultdef{facts.secondary.reversal.worse.transformed.max}{119}
\resultdef{facts.secondary.reversal.worse.transformed.min}{11}
\resultdef{facts.secondary.reversal.worse.transformed.qwen}{53}
\resultdef{facts.secondary.reversal.worse.transformed.sol}{11}
\resultdef{facts.secondary.reversal.worse.untransformed.luna}{2}
\resultdef{facts.secondary.reversal.worse.untransformed.max}{5}
\resultdef{facts.secondary.reversal.worse.untransformed.min}{1}
\resultdef{facts.secondary.reversal.worse.untransformed.qwen}{5}
\resultdef{facts.secondary.reversal.worse.untransformed.sol}{1}
\resultdef{facts.secondary.wrong.counterpart-solved.combined.luna}{404}
\resultdef{facts.secondary.wrong.counterpart-solved.combined.max}{404}
\resultdef{facts.secondary.wrong.counterpart-solved.combined.min}{7}
\resultdef{facts.secondary.wrong.counterpart-solved.combined.qwen}{128}
\resultdef{facts.secondary.wrong.counterpart-solved.combined.sol}{7}
\resultdef{facts.secondary.wrong.counterpart-solved.control-flow.luna}{200}
\resultdef{facts.secondary.wrong.counterpart-solved.control-flow.max}{200}
\resultdef{facts.secondary.wrong.counterpart-solved.control-flow.min}{5}
\resultdef{facts.secondary.wrong.counterpart-solved.control-flow.qwen}{58}
\resultdef{facts.secondary.wrong.counterpart-solved.control-flow.sol}{5}
\resultdef{facts.secondary.wrong.counterpart-solved.maximal-mix.luna}{204}
\resultdef{facts.secondary.wrong.counterpart-solved.maximal-mix.max}{204}
\resultdef{facts.secondary.wrong.counterpart-solved.maximal-mix.min}{2}
\resultdef{facts.secondary.wrong.counterpart-solved.maximal-mix.qwen}{70}
\resultdef{facts.secondary.wrong.counterpart-solved.maximal-mix.sol}{2}
\resultdef{facts.secondary.wrong.double-negation-share.combined.luna}{16.1}
\resultdef{facts.secondary.wrong.double-negation-share.combined.max}{35.2}
\resultdef{facts.secondary.wrong.double-negation-share.combined.min}{0.0}
\resultdef{facts.secondary.wrong.double-negation-share.combined.qwen}{35.2}
\resultdef{facts.secondary.wrong.double-negation-share.combined.sol}{0.0}
\resultdef{facts.secondary.wrong.double-negation-share.control-flow.luna}{13.0}
\resultdef{facts.secondary.wrong.double-negation-share.control-flow.max}{41.4}
\resultdef{facts.secondary.wrong.double-negation-share.control-flow.min}{0.0}
\resultdef{facts.secondary.wrong.double-negation-share.control-flow.qwen}{41.4}
\resultdef{facts.secondary.wrong.double-negation-share.control-flow.sol}{0.0}
\resultdef{facts.secondary.wrong.double-negation-share.maximal-mix.luna}{19.1}
\resultdef{facts.secondary.wrong.double-negation-share.maximal-mix.max}{30.0}
\resultdef{facts.secondary.wrong.double-negation-share.maximal-mix.min}{0.0}
\resultdef{facts.secondary.wrong.double-negation-share.maximal-mix.qwen}{30.0}
\resultdef{facts.secondary.wrong.double-negation-share.maximal-mix.sol}{0.0}
\resultdef{facts.secondary.wrong.double-negation.combined.luna}{65}
\resultdef{facts.secondary.wrong.double-negation.combined.max}{65}
\resultdef{facts.secondary.wrong.double-negation.combined.min}{0}
\resultdef{facts.secondary.wrong.double-negation.combined.qwen}{45}
\resultdef{facts.secondary.wrong.double-negation.combined.sol}{0}
\resultdef{facts.secondary.wrong.double-negation.control-flow.luna}{26}
\resultdef{facts.secondary.wrong.double-negation.control-flow.max}{26}
\resultdef{facts.secondary.wrong.double-negation.control-flow.min}{0}
\resultdef{facts.secondary.wrong.double-negation.control-flow.qwen}{24}
\resultdef{facts.secondary.wrong.double-negation.control-flow.sol}{0}
\resultdef{facts.secondary.wrong.double-negation.maximal-mix.luna}{39}
\resultdef{facts.secondary.wrong.double-negation.maximal-mix.max}{39}
\resultdef{facts.secondary.wrong.double-negation.maximal-mix.min}{0}
\resultdef{facts.secondary.wrong.double-negation.maximal-mix.qwen}{21}
\resultdef{facts.secondary.wrong.double-negation.maximal-mix.sol}{0}
\resultdef{facts.secondary.wrong.reference-programs.combined.luna}{117}
\resultdef{facts.secondary.wrong.reference-programs.combined.max}{117}
\resultdef{facts.secondary.wrong.reference-programs.combined.min}{3}
\resultdef{facts.secondary.wrong.reference-programs.combined.qwen}{70}
\resultdef{facts.secondary.wrong.reference-programs.combined.sol}{3}
\resultdef{facts.secondary.wrong.reference-programs.control-flow.luna}{91}
\resultdef{facts.secondary.wrong.reference-programs.control-flow.max}{91}
\resultdef{facts.secondary.wrong.reference-programs.control-flow.min}{3}
\resultdef{facts.secondary.wrong.reference-programs.control-flow.qwen}{42}
\resultdef{facts.secondary.wrong.reference-programs.control-flow.sol}{3}
\resultdef{facts.secondary.wrong.reference-programs.maximal-mix.luna}{94}
\resultdef{facts.secondary.wrong.reference-programs.maximal-mix.max}{94}
\resultdef{facts.secondary.wrong.reference-programs.maximal-mix.min}{1}
\resultdef{facts.secondary.wrong.reference-programs.maximal-mix.qwen}{53}
\resultdef{facts.secondary.wrong.reference-programs.maximal-mix.sol}{1}
\resultdef{facts.secondary.wrong.wrong-answers.combined.luna}{436}
\resultdef{facts.secondary.wrong.wrong-answers.combined.max}{436}
\resultdef{facts.secondary.wrong.wrong-answers.combined.min}{8}
\resultdef{facts.secondary.wrong.wrong-answers.combined.qwen}{141}
\resultdef{facts.secondary.wrong.wrong-answers.combined.sol}{8}
\resultdef{facts.secondary.wrong.wrong-answers.control-flow.luna}{216}
\resultdef{facts.secondary.wrong.wrong-answers.control-flow.max}{216}
\resultdef{facts.secondary.wrong.wrong-answers.control-flow.min}{5}
\resultdef{facts.secondary.wrong.wrong-answers.control-flow.qwen}{64}
\resultdef{facts.secondary.wrong.wrong-answers.control-flow.sol}{5}
\resultdef{facts.secondary.wrong.wrong-answers.maximal-mix.luna}{220}
\resultdef{facts.secondary.wrong.wrong-answers.maximal-mix.max}{220}
\resultdef{facts.secondary.wrong.wrong-answers.maximal-mix.min}{3}
\resultdef{facts.secondary.wrong.wrong-answers.maximal-mix.qwen}{77}
\resultdef{facts.secondary.wrong.wrong-answers.maximal-mix.sol}{3}
\resultdef{facts.statement-only.count.luna}{443}
\resultdef{facts.statement-only.count.max}{443}
\resultdef{facts.statement-only.count.min}{9}
\resultdef{facts.statement-only.count.qwen}{87}
\resultdef{facts.statement-only.count.sol}{9}
\resultdef{facts.statement-only.share.luna}{7.3}
\resultdef{facts.statement-only.share.max}{7.3}
\resultdef{facts.statement-only.share.min}{0.1}
\resultdef{facts.statement-only.share.qwen}{1.4}
\resultdef{facts.statement-only.share.sol}{0.1}
\resultdef{primary.acc.gap.discretionary-tools.ci.luna}{[$-$0.6, +0.9]}
\resultdef{primary.acc.gap.discretionary-tools.ci.qwen}{[$-$0.3, +0.8]}
\resultdef{primary.acc.gap.discretionary-tools.ci.sol}{[+0.0, +0.4]}
\resultdef{primary.acc.gap.discretionary-tools.control-flow.ci.luna}{[$-$0.7, +1.5]}
\resultdef{primary.acc.gap.discretionary-tools.control-flow.ci.qwen}{[$-$0.5, +0.9]}
\resultdef{primary.acc.gap.discretionary-tools.control-flow.ci.sol}{[+0.0, +0.7]}
\resultdef{primary.acc.gap.discretionary-tools.control-flow.luna}{+0.4}
\resultdef{primary.acc.gap.discretionary-tools.control-flow.max}{+0.4}
\resultdef{primary.acc.gap.discretionary-tools.control-flow.min}{+0.2}
\resultdef{primary.acc.gap.discretionary-tools.control-flow.qwen}{+0.2}
\resultdef{primary.acc.gap.discretionary-tools.control-flow.sol}{+0.2}
\resultdef{primary.acc.gap.discretionary-tools.distractors.ci.luna}{[$-$1.2, +0.6]}
\resultdef{primary.acc.gap.discretionary-tools.distractors.ci.qwen}{[$-$0.2, +1.0]}
\resultdef{primary.acc.gap.discretionary-tools.distractors.ci.sol}{[+0.0, +0.5]}
\resultdef{primary.acc.gap.discretionary-tools.distractors.luna}{$-$0.3}
\resultdef{primary.acc.gap.discretionary-tools.distractors.max}{+0.3}
\resultdef{primary.acc.gap.discretionary-tools.distractors.min}{$-$0.3}
\resultdef{primary.acc.gap.discretionary-tools.distractors.qwen}{+0.3}
\resultdef{primary.acc.gap.discretionary-tools.distractors.sol}{+0.2}
\resultdef{primary.acc.gap.discretionary-tools.luna}{+0.1}
\resultdef{primary.acc.gap.discretionary-tools.max}{+0.2}
\resultdef{primary.acc.gap.discretionary-tools.maximal-mix.ci.luna}{[$-$1.3, +1.0]}
\resultdef{primary.acc.gap.discretionary-tools.maximal-mix.ci.qwen}{[$-$0.4, +0.7]}
\resultdef{primary.acc.gap.discretionary-tools.maximal-mix.ci.sol}{[+0.0, +0.5]}
\resultdef{primary.acc.gap.discretionary-tools.maximal-mix.luna}{$-$0.2}
\resultdef{primary.acc.gap.discretionary-tools.maximal-mix.max}{+0.2}
\resultdef{primary.acc.gap.discretionary-tools.maximal-mix.min}{$-$0.2}
\resultdef{primary.acc.gap.discretionary-tools.maximal-mix.qwen}{+0.2}
\resultdef{primary.acc.gap.discretionary-tools.maximal-mix.sol}{+0.2}
\resultdef{primary.acc.gap.discretionary-tools.min}{+0.1}
\resultdef{primary.acc.gap.discretionary-tools.off-by-one.ci.luna}{[$-$2.9, +1.4]}
\resultdef{primary.acc.gap.discretionary-tools.off-by-one.ci.qwen}{[$-$0.2, +3.4]}
\resultdef{primary.acc.gap.discretionary-tools.off-by-one.ci.sol}{[+0.0, +1.8]}
\resultdef{primary.acc.gap.discretionary-tools.off-by-one.luna}{$-$0.9}
\resultdef{primary.acc.gap.discretionary-tools.off-by-one.max}{+1.2}
\resultdef{primary.acc.gap.discretionary-tools.off-by-one.min}{$-$0.9}
\resultdef{primary.acc.gap.discretionary-tools.off-by-one.qwen}{+1.2}
\resultdef{primary.acc.gap.discretionary-tools.off-by-one.sol}{+0.6}
\resultdef{primary.acc.gap.discretionary-tools.qwen}{+0.2}
\resultdef{primary.acc.gap.discretionary-tools.rename.ci.luna}{[$-$0.5, +1.2]}
\resultdef{primary.acc.gap.discretionary-tools.rename.ci.qwen}{[$-$0.2, +1.1]}
\resultdef{primary.acc.gap.discretionary-tools.rename.ci.sol}{[+0.0, +0.4]}
\resultdef{primary.acc.gap.discretionary-tools.rename.luna}{+0.3}
\resultdef{primary.acc.gap.discretionary-tools.rename.max}{+0.4}
\resultdef{primary.acc.gap.discretionary-tools.rename.min}{+0.2}
\resultdef{primary.acc.gap.discretionary-tools.rename.qwen}{+0.4}
\resultdef{primary.acc.gap.discretionary-tools.rename.sol}{+0.2}
\resultdef{primary.acc.gap.discretionary-tools.sol}{+0.1}
\resultdef{primary.acc.gap.discretionary-tools.state-and-literals.ci.luna}{[$-$0.6, +1.2]}
\resultdef{primary.acc.gap.discretionary-tools.state-and-literals.ci.qwen}{[$-$0.7, +0.7]}
\resultdef{primary.acc.gap.discretionary-tools.state-and-literals.ci.sol}{[$-$0.2, +0.0]}
\resultdef{primary.acc.gap.discretionary-tools.state-and-literals.luna}{+0.3}
\resultdef{primary.acc.gap.discretionary-tools.state-and-literals.max}{+0.3}
\resultdef{primary.acc.gap.discretionary-tools.state-and-literals.min}{$-$0.1}
\resultdef{primary.acc.gap.discretionary-tools.state-and-literals.qwen}{+0.0}
\resultdef{primary.acc.gap.discretionary-tools.state-and-literals.sol}{$-$0.1}
\resultdef{primary.acc.gap.discretionary-tools.static-return.ci.luna}{[$-$0.6, +2.2]}
\resultdef{primary.acc.gap.discretionary-tools.static-return.ci.qwen}{[$-$1.3, +0.7]}
\resultdef{primary.acc.gap.discretionary-tools.static-return.ci.sol}{[+0.0, +0.2]}
\resultdef{primary.acc.gap.discretionary-tools.static-return.luna}{+0.7}
\resultdef{primary.acc.gap.discretionary-tools.static-return.max}{+0.7}
\resultdef{primary.acc.gap.discretionary-tools.static-return.min}{$-$0.4}
\resultdef{primary.acc.gap.discretionary-tools.static-return.qwen}{$-$0.4}
\resultdef{primary.acc.gap.discretionary-tools.static-return.sol}{+0.1}
\resultdef{primary.acc.gap.discretionary-tools.wrong-arithmetic.ci.luna}{[$-$1.6, +0.0]}
\resultdef{primary.acc.gap.discretionary-tools.wrong-arithmetic.ci.qwen}{[$-$0.7, +0.0]}
\resultdef{primary.acc.gap.discretionary-tools.wrong-arithmetic.ci.sol}{[+0.0, +0.0]}
\resultdef{primary.acc.gap.discretionary-tools.wrong-arithmetic.luna}{$-$0.6}
\resultdef{primary.acc.gap.discretionary-tools.wrong-arithmetic.max}{+0.0}
\resultdef{primary.acc.gap.discretionary-tools.wrong-arithmetic.min}{$-$0.6}
\resultdef{primary.acc.gap.discretionary-tools.wrong-arithmetic.qwen}{$-$0.3}
\resultdef{primary.acc.gap.discretionary-tools.wrong-arithmetic.sol}{+0.0}
\resultdef{primary.acc.gap.discretionary-tools.wrong-logic.ci.luna}{[$-$1.9, +2.8]}
\resultdef{primary.acc.gap.discretionary-tools.wrong-logic.ci.qwen}{[$-$0.7, +0.8]}
\resultdef{primary.acc.gap.discretionary-tools.wrong-logic.ci.sol}{[+0.0, +0.0]}
\resultdef{primary.acc.gap.discretionary-tools.wrong-logic.luna}{+0.4}
\resultdef{primary.acc.gap.discretionary-tools.wrong-logic.max}{+0.4}
\resultdef{primary.acc.gap.discretionary-tools.wrong-logic.min}{+0.0}
\resultdef{primary.acc.gap.discretionary-tools.wrong-logic.qwen}{+0.0}
\resultdef{primary.acc.gap.discretionary-tools.wrong-logic.sol}{+0.0}
\resultdef{primary.acc.gap.discretionary-tools.wrong-variable.ci.luna}{[$-$0.8, +3.2]}
\resultdef{primary.acc.gap.discretionary-tools.wrong-variable.ci.qwen}{[$-$0.2, +1.4]}
\resultdef{primary.acc.gap.discretionary-tools.wrong-variable.ci.sol}{[+0.0, +0.0]}
\resultdef{primary.acc.gap.discretionary-tools.wrong-variable.luna}{+0.8}
\resultdef{primary.acc.gap.discretionary-tools.wrong-variable.max}{+0.8}
\resultdef{primary.acc.gap.discretionary-tools.wrong-variable.min}{+0.0}
\resultdef{primary.acc.gap.discretionary-tools.wrong-variable.qwen}{+0.5}
\resultdef{primary.acc.gap.discretionary-tools.wrong-variable.sol}{+0.0}
\resultdef{primary.acc.gap.encouraged-tools.ci.luna}{[$-$0.5, +0.3]}
\resultdef{primary.acc.gap.encouraged-tools.ci.qwen}{[$-$0.5, +0.8]}
\resultdef{primary.acc.gap.encouraged-tools.ci.sol}{[+0.0, +0.3]}
\resultdef{primary.acc.gap.encouraged-tools.control-flow.ci.luna}{[$-$0.4, +1.0]}
\resultdef{primary.acc.gap.encouraged-tools.control-flow.ci.qwen}{[$-$0.7, +0.9]}
\resultdef{primary.acc.gap.encouraged-tools.control-flow.ci.sol}{[+0.0, +0.5]}
\resultdef{primary.acc.gap.encouraged-tools.control-flow.luna}{+0.2}
\resultdef{primary.acc.gap.encouraged-tools.control-flow.max}{+0.2}
\resultdef{primary.acc.gap.encouraged-tools.control-flow.min}{+0.1}
\resultdef{primary.acc.gap.encouraged-tools.control-flow.qwen}{+0.1}
\resultdef{primary.acc.gap.encouraged-tools.control-flow.sol}{+0.2}
\resultdef{primary.acc.gap.encouraged-tools.distractors.ci.luna}{[$-$1.6, $-$0.3]}
\resultdef{primary.acc.gap.encouraged-tools.distractors.ci.qwen}{[$-$0.7, +0.7]}
\resultdef{primary.acc.gap.encouraged-tools.distractors.ci.sol}{[+0.0, +0.5]}
\resultdef{primary.acc.gap.encouraged-tools.distractors.luna}{$-$0.9}
\resultdef{primary.acc.gap.encouraged-tools.distractors.max}{+0.2}
\resultdef{primary.acc.gap.encouraged-tools.distractors.min}{$-$0.9}
\resultdef{primary.acc.gap.encouraged-tools.distractors.qwen}{+0.0}
\resultdef{primary.acc.gap.encouraged-tools.distractors.sol}{+0.2}
\resultdef{primary.acc.gap.encouraged-tools.luna}{$-$0.1}
\resultdef{primary.acc.gap.encouraged-tools.max}{+0.1}
\resultdef{primary.acc.gap.encouraged-tools.maximal-mix.ci.luna}{[$-$0.8, +0.5]}
\resultdef{primary.acc.gap.encouraged-tools.maximal-mix.ci.qwen}{[$-$1.1, +0.6]}
\resultdef{primary.acc.gap.encouraged-tools.maximal-mix.ci.sol}{[+0.0, +0.5]}
\resultdef{primary.acc.gap.encouraged-tools.maximal-mix.luna}{$-$0.2}
\resultdef{primary.acc.gap.encouraged-tools.maximal-mix.max}{+0.2}
\resultdef{primary.acc.gap.encouraged-tools.maximal-mix.min}{$-$0.2}
\resultdef{primary.acc.gap.encouraged-tools.maximal-mix.qwen}{$-$0.2}
\resultdef{primary.acc.gap.encouraged-tools.maximal-mix.sol}{+0.2}
\resultdef{primary.acc.gap.encouraged-tools.min}{$-$0.1}
\resultdef{primary.acc.gap.encouraged-tools.off-by-one.ci.luna}{[$-$2.1, +0.5]}
\resultdef{primary.acc.gap.encouraged-tools.off-by-one.ci.qwen}{[$-$1.3, +1.9]}
\resultdef{primary.acc.gap.encouraged-tools.off-by-one.ci.sol}{[+0.0, +1.5]}
\resultdef{primary.acc.gap.encouraged-tools.off-by-one.luna}{$-$0.6}
\resultdef{primary.acc.gap.encouraged-tools.off-by-one.max}{+0.5}
\resultdef{primary.acc.gap.encouraged-tools.off-by-one.min}{$-$0.6}
\resultdef{primary.acc.gap.encouraged-tools.off-by-one.qwen}{+0.1}
\resultdef{primary.acc.gap.encouraged-tools.off-by-one.sol}{+0.5}
\resultdef{primary.acc.gap.encouraged-tools.qwen}{+0.1}
\resultdef{primary.acc.gap.encouraged-tools.rename.ci.luna}{[$-$0.3, +0.5]}
\resultdef{primary.acc.gap.encouraged-tools.rename.ci.qwen}{[$-$0.2, +1.1]}
\resultdef{primary.acc.gap.encouraged-tools.rename.ci.sol}{[+0.0, +0.0]}
\resultdef{primary.acc.gap.encouraged-tools.rename.luna}{+0.1}
\resultdef{primary.acc.gap.encouraged-tools.rename.max}{+0.3}
\resultdef{primary.acc.gap.encouraged-tools.rename.min}{+0.0}
\resultdef{primary.acc.gap.encouraged-tools.rename.qwen}{+0.3}
\resultdef{primary.acc.gap.encouraged-tools.rename.sol}{+0.0}
\resultdef{primary.acc.gap.encouraged-tools.sol}{+0.1}
\resultdef{primary.acc.gap.encouraged-tools.state-and-literals.ci.luna}{[$-$0.3, +0.6]}
\resultdef{primary.acc.gap.encouraged-tools.state-and-literals.ci.qwen}{[$-$0.3, +1.0]}
\resultdef{primary.acc.gap.encouraged-tools.state-and-literals.ci.sol}{[+0.0, +0.0]}
\resultdef{primary.acc.gap.encouraged-tools.state-and-literals.luna}{+0.1}
\resultdef{primary.acc.gap.encouraged-tools.state-and-literals.max}{+0.2}
\resultdef{primary.acc.gap.encouraged-tools.state-and-literals.min}{+0.0}
\resultdef{primary.acc.gap.encouraged-tools.state-and-literals.qwen}{+0.2}
\resultdef{primary.acc.gap.encouraged-tools.state-and-literals.sol}{+0.0}
\resultdef{primary.acc.gap.encouraged-tools.static-return.ci.luna}{[$-$0.7, +0.0]}
\resultdef{primary.acc.gap.encouraged-tools.static-return.ci.qwen}{[$-$0.7, +1.0]}
\resultdef{primary.acc.gap.encouraged-tools.static-return.ci.sol}{[+0.0, +0.0]}
\resultdef{primary.acc.gap.encouraged-tools.static-return.luna}{$-$0.3}
\resultdef{primary.acc.gap.encouraged-tools.static-return.max}{+0.1}
\resultdef{primary.acc.gap.encouraged-tools.static-return.min}{$-$0.3}
\resultdef{primary.acc.gap.encouraged-tools.static-return.qwen}{+0.1}
\resultdef{primary.acc.gap.encouraged-tools.static-return.sol}{+0.0}
\resultdef{primary.acc.gap.encouraged-tools.wrong-arithmetic.ci.luna}{[$-$0.5, +0.0]}
\resultdef{primary.acc.gap.encouraged-tools.wrong-arithmetic.ci.qwen}{[$-$0.3, +1.5]}
\resultdef{primary.acc.gap.encouraged-tools.wrong-arithmetic.ci.sol}{[+0.0, +0.0]}
\resultdef{primary.acc.gap.encouraged-tools.wrong-arithmetic.luna}{$-$0.2}
\resultdef{primary.acc.gap.encouraged-tools.wrong-arithmetic.max}{+0.4}
\resultdef{primary.acc.gap.encouraged-tools.wrong-arithmetic.min}{$-$0.2}
\resultdef{primary.acc.gap.encouraged-tools.wrong-arithmetic.qwen}{+0.4}
\resultdef{primary.acc.gap.encouraged-tools.wrong-arithmetic.sol}{+0.0}
\resultdef{primary.acc.gap.encouraged-tools.wrong-logic.ci.luna}{[$-$1.3, +1.0]}
\resultdef{primary.acc.gap.encouraged-tools.wrong-logic.ci.qwen}{[$-$0.9, +2.4]}
\resultdef{primary.acc.gap.encouraged-tools.wrong-logic.ci.sol}{[+0.0, +0.0]}
\resultdef{primary.acc.gap.encouraged-tools.wrong-logic.luna}{$-$0.2}
\resultdef{primary.acc.gap.encouraged-tools.wrong-logic.max}{+0.6}
\resultdef{primary.acc.gap.encouraged-tools.wrong-logic.min}{$-$0.2}
\resultdef{primary.acc.gap.encouraged-tools.wrong-logic.qwen}{+0.6}
\resultdef{primary.acc.gap.encouraged-tools.wrong-logic.sol}{+0.0}
\resultdef{primary.acc.gap.encouraged-tools.wrong-variable.ci.luna}{[$-$0.4, +1.9]}
\resultdef{primary.acc.gap.encouraged-tools.wrong-variable.ci.qwen}{[$-$1.2, $-$0.1]}
\resultdef{primary.acc.gap.encouraged-tools.wrong-variable.ci.sol}{[+0.0, +0.0]}
\resultdef{primary.acc.gap.encouraged-tools.wrong-variable.luna}{+0.6}
\resultdef{primary.acc.gap.encouraged-tools.wrong-variable.max}{+0.6}
\resultdef{primary.acc.gap.encouraged-tools.wrong-variable.min}{$-$0.5}
\resultdef{primary.acc.gap.encouraged-tools.wrong-variable.qwen}{$-$0.5}
\resultdef{primary.acc.gap.encouraged-tools.wrong-variable.sol}{+0.0}
\resultdef{primary.acc.gap.no-tools.ci.luna}{[$-$2.3, $-$0.6]}
\resultdef{primary.acc.gap.no-tools.ci.qwen}{[$-$1.6, $-$0.1]}
\resultdef{primary.acc.gap.no-tools.ci.sol}{[$-$0.1, +0.0]}
\resultdef{primary.acc.gap.no-tools.control-flow.ci.luna}{[$-$3.7, $-$0.7]}
\resultdef{primary.acc.gap.no-tools.control-flow.ci.qwen}{[$-$4.0, $-$0.8]}
\resultdef{primary.acc.gap.no-tools.control-flow.ci.sol}{[+0.0, +0.3]}
\resultdef{primary.acc.gap.no-tools.control-flow.luna}{$-$2.1}
\resultdef{primary.acc.gap.no-tools.control-flow.max}{+0.1}
\resultdef{primary.acc.gap.no-tools.control-flow.min}{$-$2.4}
\resultdef{primary.acc.gap.no-tools.control-flow.qwen}{$-$2.4}
\resultdef{primary.acc.gap.no-tools.control-flow.sol}{+0.1}
\resultdef{primary.acc.gap.no-tools.distractors.ci.luna}{[$-$2.0, +0.0]}
\resultdef{primary.acc.gap.no-tools.distractors.ci.qwen}{[$-$0.9, +0.7]}
\resultdef{primary.acc.gap.no-tools.distractors.ci.sol}{[+0.0, +0.3]}
\resultdef{primary.acc.gap.no-tools.distractors.luna}{$-$1.0}
\resultdef{primary.acc.gap.no-tools.distractors.max}{+0.1}
\resultdef{primary.acc.gap.no-tools.distractors.min}{$-$1.0}
\resultdef{primary.acc.gap.no-tools.distractors.qwen}{$-$0.1}
\resultdef{primary.acc.gap.no-tools.distractors.sol}{+0.1}
\resultdef{primary.acc.gap.no-tools.luna}{$-$1.5}
\resultdef{primary.acc.gap.no-tools.max}{+0.0}
\resultdef{primary.acc.gap.no-tools.maximal-mix.ci.luna}{[$-$5.7, $-$1.9]}
\resultdef{primary.acc.gap.no-tools.maximal-mix.ci.qwen}{[$-$3.3, $-$0.7]}
\resultdef{primary.acc.gap.no-tools.maximal-mix.ci.sol}{[+0.0, +0.0]}
\resultdef{primary.acc.gap.no-tools.maximal-mix.luna}{$-$3.8}
\resultdef{primary.acc.gap.no-tools.maximal-mix.max}{+0.0}
\resultdef{primary.acc.gap.no-tools.maximal-mix.min}{$-$3.8}
\resultdef{primary.acc.gap.no-tools.maximal-mix.qwen}{$-$2.1}
\resultdef{primary.acc.gap.no-tools.maximal-mix.sol}{+0.0}
\resultdef{primary.acc.gap.no-tools.min}{$-$1.5}
\resultdef{primary.acc.gap.no-tools.off-by-one.ci.luna}{[$-$4.5, $-$0.6]}
\resultdef{primary.acc.gap.no-tools.off-by-one.ci.qwen}{[$-$1.6, +1.9]}
\resultdef{primary.acc.gap.no-tools.off-by-one.ci.sol}{[+0.0, +0.3]}
\resultdef{primary.acc.gap.no-tools.off-by-one.luna}{$-$2.6}
\resultdef{primary.acc.gap.no-tools.off-by-one.max}{+0.1}
\resultdef{primary.acc.gap.no-tools.off-by-one.min}{$-$2.6}
\resultdef{primary.acc.gap.no-tools.off-by-one.qwen}{+0.0}
\resultdef{primary.acc.gap.no-tools.off-by-one.sol}{+0.1}
\resultdef{primary.acc.gap.no-tools.qwen}{$-$0.9}
\resultdef{primary.acc.gap.no-tools.rename.ci.luna}{[$-$0.8, +0.7]}
\resultdef{primary.acc.gap.no-tools.rename.ci.qwen}{[$-$0.5, +0.7]}
\resultdef{primary.acc.gap.no-tools.rename.ci.sol}{[$-$0.3, +0.0]}
\resultdef{primary.acc.gap.no-tools.rename.luna}{+0.0}
\resultdef{primary.acc.gap.no-tools.rename.max}{+0.1}
\resultdef{primary.acc.gap.no-tools.rename.min}{$-$0.1}
\resultdef{primary.acc.gap.no-tools.rename.qwen}{+0.1}
\resultdef{primary.acc.gap.no-tools.rename.sol}{$-$0.1}
\resultdef{primary.acc.gap.no-tools.sol}{+0.0}
\resultdef{primary.acc.gap.no-tools.state-and-literals.ci.luna}{[$-$1.5, +0.5]}
\resultdef{primary.acc.gap.no-tools.state-and-literals.ci.qwen}{[$-$0.6, +0.8]}
\resultdef{primary.acc.gap.no-tools.state-and-literals.ci.sol}{[$-$0.5, +0.0]}
\resultdef{primary.acc.gap.no-tools.state-and-literals.luna}{$-$0.5}
\resultdef{primary.acc.gap.no-tools.state-and-literals.max}{+0.1}
\resultdef{primary.acc.gap.no-tools.state-and-literals.min}{$-$0.5}
\resultdef{primary.acc.gap.no-tools.state-and-literals.qwen}{+0.1}
\resultdef{primary.acc.gap.no-tools.state-and-literals.sol}{$-$0.2}
\resultdef{primary.acc.gap.no-tools.static-return.ci.luna}{[$-$2.6, +0.2]}
\resultdef{primary.acc.gap.no-tools.static-return.ci.qwen}{[$-$3.5, $-$1.0]}
\resultdef{primary.acc.gap.no-tools.static-return.ci.sol}{[$-$0.4, +0.0]}
\resultdef{primary.acc.gap.no-tools.static-return.luna}{$-$1.2}
\resultdef{primary.acc.gap.no-tools.static-return.max}{$-$0.1}
\resultdef{primary.acc.gap.no-tools.static-return.min}{$-$2.2}
\resultdef{primary.acc.gap.no-tools.static-return.qwen}{$-$2.2}
\resultdef{primary.acc.gap.no-tools.static-return.sol}{$-$0.1}
\resultdef{primary.acc.gap.no-tools.wrong-arithmetic.ci.luna}{[$-$1.7, $-$0.1]}
\resultdef{primary.acc.gap.no-tools.wrong-arithmetic.ci.qwen}{[$-$1.5, $-$0.1]}
\resultdef{primary.acc.gap.no-tools.wrong-arithmetic.ci.sol}{[+0.0, +0.0]}
\resultdef{primary.acc.gap.no-tools.wrong-arithmetic.luna}{$-$0.9}
\resultdef{primary.acc.gap.no-tools.wrong-arithmetic.max}{+0.0}
\resultdef{primary.acc.gap.no-tools.wrong-arithmetic.min}{$-$0.9}
\resultdef{primary.acc.gap.no-tools.wrong-arithmetic.qwen}{$-$0.7}
\resultdef{primary.acc.gap.no-tools.wrong-arithmetic.sol}{+0.0}
\resultdef{primary.acc.gap.no-tools.wrong-logic.ci.luna}{[$-$4.9, $-$0.8]}
\resultdef{primary.acc.gap.no-tools.wrong-logic.ci.qwen}{[$-$2.2, +2.5]}
\resultdef{primary.acc.gap.no-tools.wrong-logic.ci.sol}{[+0.0, +0.0]}
\resultdef{primary.acc.gap.no-tools.wrong-logic.luna}{$-$2.8}
\resultdef{primary.acc.gap.no-tools.wrong-logic.max}{+0.0}
\resultdef{primary.acc.gap.no-tools.wrong-logic.min}{$-$2.8}
\resultdef{primary.acc.gap.no-tools.wrong-logic.qwen}{$-$0.1}
\resultdef{primary.acc.gap.no-tools.wrong-logic.sol}{+0.0}
\resultdef{primary.acc.gap.no-tools.wrong-variable.ci.luna}{[$-$1.8, +2.1]}
\resultdef{primary.acc.gap.no-tools.wrong-variable.ci.qwen}{[$-$1.8, $-$0.5]}
\resultdef{primary.acc.gap.no-tools.wrong-variable.ci.sol}{[+0.0, +0.0]}
\resultdef{primary.acc.gap.no-tools.wrong-variable.luna}{$-$0.2}
\resultdef{primary.acc.gap.no-tools.wrong-variable.max}{+0.0}
\resultdef{primary.acc.gap.no-tools.wrong-variable.min}{$-$1.1}
\resultdef{primary.acc.gap.no-tools.wrong-variable.qwen}{$-$1.1}
\resultdef{primary.acc.gap.no-tools.wrong-variable.sol}{+0.0}
\resultdef{primary.acc.loss.discretionary-tools.control-flow.luna}{-0.4}
\resultdef{primary.acc.loss.discretionary-tools.control-flow.max}{-0.2}
\resultdef{primary.acc.loss.discretionary-tools.control-flow.min}{-0.4}
\resultdef{primary.acc.loss.discretionary-tools.control-flow.qwen}{-0.2}
\resultdef{primary.acc.loss.discretionary-tools.control-flow.sol}{-0.2}
\resultdef{primary.acc.loss.discretionary-tools.distractors.luna}{0.3}
\resultdef{primary.acc.loss.discretionary-tools.distractors.max}{0.3}
\resultdef{primary.acc.loss.discretionary-tools.distractors.min}{-0.3}
\resultdef{primary.acc.loss.discretionary-tools.distractors.qwen}{-0.3}
\resultdef{primary.acc.loss.discretionary-tools.distractors.sol}{-0.2}
\resultdef{primary.acc.loss.discretionary-tools.luna}{-0.1}
\resultdef{primary.acc.loss.discretionary-tools.max}{-0.1}
\resultdef{primary.acc.loss.discretionary-tools.maximal-mix.luna}{0.2}
\resultdef{primary.acc.loss.discretionary-tools.maximal-mix.max}{0.2}
\resultdef{primary.acc.loss.discretionary-tools.maximal-mix.min}{-0.2}
\resultdef{primary.acc.loss.discretionary-tools.maximal-mix.qwen}{-0.2}
\resultdef{primary.acc.loss.discretionary-tools.maximal-mix.sol}{-0.2}
\resultdef{primary.acc.loss.discretionary-tools.min}{-0.2}
\resultdef{primary.acc.loss.discretionary-tools.off-by-one.luna}{0.9}
\resultdef{primary.acc.loss.discretionary-tools.off-by-one.max}{0.9}
\resultdef{primary.acc.loss.discretionary-tools.off-by-one.min}{-1.2}
\resultdef{primary.acc.loss.discretionary-tools.off-by-one.qwen}{-1.2}
\resultdef{primary.acc.loss.discretionary-tools.off-by-one.sol}{-0.6}
\resultdef{primary.acc.loss.discretionary-tools.qwen}{-0.2}
\resultdef{primary.acc.loss.discretionary-tools.rename.luna}{-0.3}
\resultdef{primary.acc.loss.discretionary-tools.rename.max}{-0.2}
\resultdef{primary.acc.loss.discretionary-tools.rename.min}{-0.4}
\resultdef{primary.acc.loss.discretionary-tools.rename.qwen}{-0.4}
\resultdef{primary.acc.loss.discretionary-tools.rename.sol}{-0.2}
\resultdef{primary.acc.loss.discretionary-tools.sol}{-0.1}
\resultdef{primary.acc.loss.discretionary-tools.state-and-literals.luna}{-0.3}
\resultdef{primary.acc.loss.discretionary-tools.state-and-literals.max}{0.1}
\resultdef{primary.acc.loss.discretionary-tools.state-and-literals.min}{-0.3}
\resultdef{primary.acc.loss.discretionary-tools.state-and-literals.qwen}{0.0}
\resultdef{primary.acc.loss.discretionary-tools.state-and-literals.sol}{0.1}
\resultdef{primary.acc.loss.discretionary-tools.static-return.luna}{-0.7}
\resultdef{primary.acc.loss.discretionary-tools.static-return.max}{0.4}
\resultdef{primary.acc.loss.discretionary-tools.static-return.min}{-0.7}
\resultdef{primary.acc.loss.discretionary-tools.static-return.qwen}{0.4}
\resultdef{primary.acc.loss.discretionary-tools.static-return.sol}{-0.1}
\resultdef{primary.acc.loss.discretionary-tools.wrong-arithmetic.luna}{0.6}
\resultdef{primary.acc.loss.discretionary-tools.wrong-arithmetic.max}{0.6}
\resultdef{primary.acc.loss.discretionary-tools.wrong-arithmetic.min}{0.0}
\resultdef{primary.acc.loss.discretionary-tools.wrong-arithmetic.qwen}{0.3}
\resultdef{primary.acc.loss.discretionary-tools.wrong-arithmetic.sol}{0.0}
\resultdef{primary.acc.loss.discretionary-tools.wrong-logic.luna}{-0.4}
\resultdef{primary.acc.loss.discretionary-tools.wrong-logic.max}{0.0}
\resultdef{primary.acc.loss.discretionary-tools.wrong-logic.min}{-0.4}
\resultdef{primary.acc.loss.discretionary-tools.wrong-logic.qwen}{0.0}
\resultdef{primary.acc.loss.discretionary-tools.wrong-logic.sol}{0.0}
\resultdef{primary.acc.loss.discretionary-tools.wrong-variable.luna}{-0.8}
\resultdef{primary.acc.loss.discretionary-tools.wrong-variable.max}{0.0}
\resultdef{primary.acc.loss.discretionary-tools.wrong-variable.min}{-0.8}
\resultdef{primary.acc.loss.discretionary-tools.wrong-variable.qwen}{-0.5}
\resultdef{primary.acc.loss.discretionary-tools.wrong-variable.sol}{0.0}
\resultdef{primary.acc.loss.encouraged-tools.control-flow.luna}{-0.2}
\resultdef{primary.acc.loss.encouraged-tools.control-flow.max}{-0.1}
\resultdef{primary.acc.loss.encouraged-tools.control-flow.min}{-0.2}
\resultdef{primary.acc.loss.encouraged-tools.control-flow.qwen}{-0.1}
\resultdef{primary.acc.loss.encouraged-tools.control-flow.sol}{-0.2}
\resultdef{primary.acc.loss.encouraged-tools.distractors.luna}{0.9}
\resultdef{primary.acc.loss.encouraged-tools.distractors.max}{0.9}
\resultdef{primary.acc.loss.encouraged-tools.distractors.min}{-0.2}
\resultdef{primary.acc.loss.encouraged-tools.distractors.qwen}{0.0}
\resultdef{primary.acc.loss.encouraged-tools.distractors.sol}{-0.2}
\resultdef{primary.acc.loss.encouraged-tools.luna}{0.1}
\resultdef{primary.acc.loss.encouraged-tools.max}{0.1}
\resultdef{primary.acc.loss.encouraged-tools.maximal-mix.luna}{0.2}
\resultdef{primary.acc.loss.encouraged-tools.maximal-mix.max}{0.2}
\resultdef{primary.acc.loss.encouraged-tools.maximal-mix.min}{-0.2}
\resultdef{primary.acc.loss.encouraged-tools.maximal-mix.qwen}{0.2}
\resultdef{primary.acc.loss.encouraged-tools.maximal-mix.sol}{-0.2}
\resultdef{primary.acc.loss.encouraged-tools.min}{-0.1}
\resultdef{primary.acc.loss.encouraged-tools.off-by-one.luna}{0.6}
\resultdef{primary.acc.loss.encouraged-tools.off-by-one.max}{0.6}
\resultdef{primary.acc.loss.encouraged-tools.off-by-one.min}{-0.5}
\resultdef{primary.acc.loss.encouraged-tools.off-by-one.qwen}{-0.1}
\resultdef{primary.acc.loss.encouraged-tools.off-by-one.sol}{-0.5}
\resultdef{primary.acc.loss.encouraged-tools.qwen}{-0.1}
\resultdef{primary.acc.loss.encouraged-tools.rename.luna}{-0.1}
\resultdef{primary.acc.loss.encouraged-tools.rename.max}{0.0}
\resultdef{primary.acc.loss.encouraged-tools.rename.min}{-0.3}
\resultdef{primary.acc.loss.encouraged-tools.rename.qwen}{-0.3}
\resultdef{primary.acc.loss.encouraged-tools.rename.sol}{0.0}
\resultdef{primary.acc.loss.encouraged-tools.sol}{-0.1}
\resultdef{primary.acc.loss.encouraged-tools.state-and-literals.luna}{-0.1}
\resultdef{primary.acc.loss.encouraged-tools.state-and-literals.max}{0.0}
\resultdef{primary.acc.loss.encouraged-tools.state-and-literals.min}{-0.2}
\resultdef{primary.acc.loss.encouraged-tools.state-and-literals.qwen}{-0.2}
\resultdef{primary.acc.loss.encouraged-tools.state-and-literals.sol}{0.0}
\resultdef{primary.acc.loss.encouraged-tools.static-return.luna}{0.3}
\resultdef{primary.acc.loss.encouraged-tools.static-return.max}{0.3}
\resultdef{primary.acc.loss.encouraged-tools.static-return.min}{-0.1}
\resultdef{primary.acc.loss.encouraged-tools.static-return.qwen}{-0.1}
\resultdef{primary.acc.loss.encouraged-tools.static-return.sol}{0.0}
\resultdef{primary.acc.loss.encouraged-tools.wrong-arithmetic.luna}{0.2}
\resultdef{primary.acc.loss.encouraged-tools.wrong-arithmetic.max}{0.2}
\resultdef{primary.acc.loss.encouraged-tools.wrong-arithmetic.min}{-0.4}
\resultdef{primary.acc.loss.encouraged-tools.wrong-arithmetic.qwen}{-0.4}
\resultdef{primary.acc.loss.encouraged-tools.wrong-arithmetic.sol}{0.0}
\resultdef{primary.acc.loss.encouraged-tools.wrong-logic.luna}{0.2}
\resultdef{primary.acc.loss.encouraged-tools.wrong-logic.max}{0.2}
\resultdef{primary.acc.loss.encouraged-tools.wrong-logic.min}{-0.6}
\resultdef{primary.acc.loss.encouraged-tools.wrong-logic.qwen}{-0.6}
\resultdef{primary.acc.loss.encouraged-tools.wrong-logic.sol}{0.0}
\resultdef{primary.acc.loss.encouraged-tools.wrong-variable.luna}{-0.6}
\resultdef{primary.acc.loss.encouraged-tools.wrong-variable.max}{0.5}
\resultdef{primary.acc.loss.encouraged-tools.wrong-variable.min}{-0.6}
\resultdef{primary.acc.loss.encouraged-tools.wrong-variable.qwen}{0.5}
\resultdef{primary.acc.loss.encouraged-tools.wrong-variable.sol}{0.0}
\resultdef{primary.acc.loss.no-tools.control-flow.luna}{2.1}
\resultdef{primary.acc.loss.no-tools.control-flow.max}{2.4}
\resultdef{primary.acc.loss.no-tools.control-flow.min}{-0.1}
\resultdef{primary.acc.loss.no-tools.control-flow.qwen}{2.4}
\resultdef{primary.acc.loss.no-tools.control-flow.sol}{-0.1}
\resultdef{primary.acc.loss.no-tools.distractors.luna}{1.0}
\resultdef{primary.acc.loss.no-tools.distractors.max}{1.0}
\resultdef{primary.acc.loss.no-tools.distractors.min}{-0.1}
\resultdef{primary.acc.loss.no-tools.distractors.qwen}{0.1}
\resultdef{primary.acc.loss.no-tools.distractors.sol}{-0.1}
\resultdef{primary.acc.loss.no-tools.luna}{1.5}
\resultdef{primary.acc.loss.no-tools.max}{1.5}
\resultdef{primary.acc.loss.no-tools.maximal-mix.luna}{3.8}
\resultdef{primary.acc.loss.no-tools.maximal-mix.max}{3.8}
\resultdef{primary.acc.loss.no-tools.maximal-mix.min}{0.0}
\resultdef{primary.acc.loss.no-tools.maximal-mix.qwen}{2.1}
\resultdef{primary.acc.loss.no-tools.maximal-mix.sol}{0.0}
\resultdef{primary.acc.loss.no-tools.min}{0.0}
\resultdef{primary.acc.loss.no-tools.off-by-one.luna}{2.6}
\resultdef{primary.acc.loss.no-tools.off-by-one.max}{2.6}
\resultdef{primary.acc.loss.no-tools.off-by-one.min}{-0.1}
\resultdef{primary.acc.loss.no-tools.off-by-one.qwen}{0.0}
\resultdef{primary.acc.loss.no-tools.off-by-one.sol}{-0.1}
\resultdef{primary.acc.loss.no-tools.qwen}{0.9}
\resultdef{primary.acc.loss.no-tools.rename.luna}{0.0}
\resultdef{primary.acc.loss.no-tools.rename.max}{0.1}
\resultdef{primary.acc.loss.no-tools.rename.min}{-0.1}
\resultdef{primary.acc.loss.no-tools.rename.qwen}{-0.1}
\resultdef{primary.acc.loss.no-tools.rename.sol}{0.1}
\resultdef{primary.acc.loss.no-tools.sol}{0.0}
\resultdef{primary.acc.loss.no-tools.state-and-literals.luna}{0.5}
\resultdef{primary.acc.loss.no-tools.state-and-literals.max}{0.5}
\resultdef{primary.acc.loss.no-tools.state-and-literals.min}{-0.1}
\resultdef{primary.acc.loss.no-tools.state-and-literals.qwen}{-0.1}
\resultdef{primary.acc.loss.no-tools.state-and-literals.sol}{0.2}
\resultdef{primary.acc.loss.no-tools.static-return.luna}{1.2}
\resultdef{primary.acc.loss.no-tools.static-return.max}{2.2}
\resultdef{primary.acc.loss.no-tools.static-return.min}{0.1}
\resultdef{primary.acc.loss.no-tools.static-return.qwen}{2.2}
\resultdef{primary.acc.loss.no-tools.static-return.sol}{0.1}
\resultdef{primary.acc.loss.no-tools.wrong-arithmetic.luna}{0.9}
\resultdef{primary.acc.loss.no-tools.wrong-arithmetic.max}{0.9}
\resultdef{primary.acc.loss.no-tools.wrong-arithmetic.min}{0.0}
\resultdef{primary.acc.loss.no-tools.wrong-arithmetic.qwen}{0.7}
\resultdef{primary.acc.loss.no-tools.wrong-arithmetic.sol}{0.0}
\resultdef{primary.acc.loss.no-tools.wrong-logic.luna}{2.8}
\resultdef{primary.acc.loss.no-tools.wrong-logic.max}{2.8}
\resultdef{primary.acc.loss.no-tools.wrong-logic.min}{0.0}
\resultdef{primary.acc.loss.no-tools.wrong-logic.qwen}{0.1}
\resultdef{primary.acc.loss.no-tools.wrong-logic.sol}{0.0}
\resultdef{primary.acc.loss.no-tools.wrong-variable.luna}{0.2}
\resultdef{primary.acc.loss.no-tools.wrong-variable.max}{1.1}
\resultdef{primary.acc.loss.no-tools.wrong-variable.min}{0.0}
\resultdef{primary.acc.loss.no-tools.wrong-variable.qwen}{1.1}
\resultdef{primary.acc.loss.no-tools.wrong-variable.sol}{0.0}
\resultdef{primary.acc.transformed.discretionary-tools.control-flow.luna}{98.5}
\resultdef{primary.acc.transformed.discretionary-tools.control-flow.max}{100.0}
\resultdef{primary.acc.transformed.discretionary-tools.control-flow.min}{98.5}
\resultdef{primary.acc.transformed.discretionary-tools.control-flow.qwen}{99.1}
\resultdef{primary.acc.transformed.discretionary-tools.control-flow.sol}{100.0}
\resultdef{primary.acc.transformed.discretionary-tools.distractors.luna}{97.8}
\resultdef{primary.acc.transformed.discretionary-tools.distractors.max}{99.9}
\resultdef{primary.acc.transformed.discretionary-tools.distractors.min}{97.8}
\resultdef{primary.acc.transformed.discretionary-tools.distractors.qwen}{99.3}
\resultdef{primary.acc.transformed.discretionary-tools.distractors.sol}{99.9}
\resultdef{primary.acc.transformed.discretionary-tools.luna}{98.2}
\resultdef{primary.acc.transformed.discretionary-tools.max}{99.9}
\resultdef{primary.acc.transformed.discretionary-tools.maximal-mix.luna}{97.9}
\resultdef{primary.acc.transformed.discretionary-tools.maximal-mix.max}{99.9}
\resultdef{primary.acc.transformed.discretionary-tools.maximal-mix.min}{97.9}
\resultdef{primary.acc.transformed.discretionary-tools.maximal-mix.qwen}{99.1}
\resultdef{primary.acc.transformed.discretionary-tools.maximal-mix.sol}{99.9}
\resultdef{primary.acc.transformed.discretionary-tools.min}{98.2}
\resultdef{primary.acc.transformed.discretionary-tools.off-by-one.luna}{95.9}
\resultdef{primary.acc.transformed.discretionary-tools.off-by-one.max}{99.8}
\resultdef{primary.acc.transformed.discretionary-tools.off-by-one.min}{95.9}
\resultdef{primary.acc.transformed.discretionary-tools.off-by-one.qwen}{99.2}
\resultdef{primary.acc.transformed.discretionary-tools.off-by-one.sol}{99.8}
\resultdef{primary.acc.transformed.discretionary-tools.qwen}{99.1}
\resultdef{primary.acc.transformed.discretionary-tools.rename.luna}{98.4}
\resultdef{primary.acc.transformed.discretionary-tools.rename.max}{99.9}
\resultdef{primary.acc.transformed.discretionary-tools.rename.min}{98.4}
\resultdef{primary.acc.transformed.discretionary-tools.rename.qwen}{99.3}
\resultdef{primary.acc.transformed.discretionary-tools.rename.sol}{99.9}
\resultdef{primary.acc.transformed.discretionary-tools.sol}{99.9}
\resultdef{primary.acc.transformed.discretionary-tools.state-and-literals.luna}{98.4}
\resultdef{primary.acc.transformed.discretionary-tools.state-and-literals.max}{99.7}
\resultdef{primary.acc.transformed.discretionary-tools.state-and-literals.min}{98.4}
\resultdef{primary.acc.transformed.discretionary-tools.state-and-literals.qwen}{98.9}
\resultdef{primary.acc.transformed.discretionary-tools.state-and-literals.sol}{99.7}
\resultdef{primary.acc.transformed.discretionary-tools.static-return.luna}{99.2}
\resultdef{primary.acc.transformed.discretionary-tools.static-return.max}{99.7}
\resultdef{primary.acc.transformed.discretionary-tools.static-return.min}{98.6}
\resultdef{primary.acc.transformed.discretionary-tools.static-return.qwen}{98.6}
\resultdef{primary.acc.transformed.discretionary-tools.static-return.sol}{99.7}
\resultdef{primary.acc.transformed.discretionary-tools.subset-max.luna}{99.4}
\resultdef{primary.acc.transformed.discretionary-tools.subset-max.max}{100.0}
\resultdef{primary.acc.transformed.discretionary-tools.subset-max.min}{99.4}
\resultdef{primary.acc.transformed.discretionary-tools.subset-max.qwen}{99.7}
\resultdef{primary.acc.transformed.discretionary-tools.subset-max.sol}{100.0}
\resultdef{primary.acc.transformed.discretionary-tools.subset-min.luna}{95.9}
\resultdef{primary.acc.transformed.discretionary-tools.subset-min.max}{99.7}
\resultdef{primary.acc.transformed.discretionary-tools.subset-min.min}{95.9}
\resultdef{primary.acc.transformed.discretionary-tools.subset-min.qwen}{98.6}
\resultdef{primary.acc.transformed.discretionary-tools.subset-min.sol}{99.7}
\resultdef{primary.acc.transformed.discretionary-tools.wrong-arithmetic.luna}{99.4}
\resultdef{primary.acc.transformed.discretionary-tools.wrong-arithmetic.max}{100.0}
\resultdef{primary.acc.transformed.discretionary-tools.wrong-arithmetic.min}{99.4}
\resultdef{primary.acc.transformed.discretionary-tools.wrong-arithmetic.qwen}{99.7}
\resultdef{primary.acc.transformed.discretionary-tools.wrong-arithmetic.sol}{100.0}
\resultdef{primary.acc.transformed.discretionary-tools.wrong-logic.luna}{97.6}
\resultdef{primary.acc.transformed.discretionary-tools.wrong-logic.max}{100.0}
\resultdef{primary.acc.transformed.discretionary-tools.wrong-logic.min}{97.6}
\resultdef{primary.acc.transformed.discretionary-tools.wrong-logic.qwen}{98.6}
\resultdef{primary.acc.transformed.discretionary-tools.wrong-logic.sol}{100.0}
\resultdef{primary.acc.transformed.discretionary-tools.wrong-variable.luna}{98.9}
\resultdef{primary.acc.transformed.discretionary-tools.wrong-variable.max}{100.0}
\resultdef{primary.acc.transformed.discretionary-tools.wrong-variable.min}{98.9}
\resultdef{primary.acc.transformed.discretionary-tools.wrong-variable.qwen}{99.7}
\resultdef{primary.acc.transformed.discretionary-tools.wrong-variable.sol}{100.0}
\resultdef{primary.acc.transformed.encouraged-tools.control-flow.luna}{99.2}
\resultdef{primary.acc.transformed.encouraged-tools.control-flow.max}{100.0}
\resultdef{primary.acc.transformed.encouraged-tools.control-flow.min}{99.2}
\resultdef{primary.acc.transformed.encouraged-tools.control-flow.qwen}{99.2}
\resultdef{primary.acc.transformed.encouraged-tools.control-flow.sol}{100.0}
\resultdef{primary.acc.transformed.encouraged-tools.distractors.luna}{98.0}
\resultdef{primary.acc.transformed.encouraged-tools.distractors.max}{100.0}
\resultdef{primary.acc.transformed.encouraged-tools.distractors.min}{98.0}
\resultdef{primary.acc.transformed.encouraged-tools.distractors.qwen}{99.1}
\resultdef{primary.acc.transformed.encouraged-tools.distractors.sol}{100.0}
\resultdef{primary.acc.transformed.encouraged-tools.luna}{98.8}
\resultdef{primary.acc.transformed.encouraged-tools.max}{99.9}
\resultdef{primary.acc.transformed.encouraged-tools.maximal-mix.luna}{98.8}
\resultdef{primary.acc.transformed.encouraged-tools.maximal-mix.max}{100.0}
\resultdef{primary.acc.transformed.encouraged-tools.maximal-mix.min}{98.8}
\resultdef{primary.acc.transformed.encouraged-tools.maximal-mix.qwen}{98.8}
\resultdef{primary.acc.transformed.encouraged-tools.maximal-mix.sol}{100.0}
\resultdef{primary.acc.transformed.encouraged-tools.min}{98.8}
\resultdef{primary.acc.transformed.encouraged-tools.off-by-one.luna}{96.9}
\resultdef{primary.acc.transformed.encouraged-tools.off-by-one.max}{99.7}
\resultdef{primary.acc.transformed.encouraged-tools.off-by-one.min}{96.9}
\resultdef{primary.acc.transformed.encouraged-tools.off-by-one.qwen}{98.9}
\resultdef{primary.acc.transformed.encouraged-tools.off-by-one.sol}{99.7}
\resultdef{primary.acc.transformed.encouraged-tools.qwen}{99.2}
\resultdef{primary.acc.transformed.encouraged-tools.rename.luna}{99.0}
\resultdef{primary.acc.transformed.encouraged-tools.rename.max}{99.8}
\resultdef{primary.acc.transformed.encouraged-tools.rename.min}{99.0}
\resultdef{primary.acc.transformed.encouraged-tools.rename.qwen}{99.4}
\resultdef{primary.acc.transformed.encouraged-tools.rename.sol}{99.8}
\resultdef{primary.acc.transformed.encouraged-tools.sol}{99.9}
\resultdef{primary.acc.transformed.encouraged-tools.state-and-literals.luna}{99.0}
\resultdef{primary.acc.transformed.encouraged-tools.state-and-literals.max}{99.8}
\resultdef{primary.acc.transformed.encouraged-tools.state-and-literals.min}{99.0}
\resultdef{primary.acc.transformed.encouraged-tools.state-and-literals.qwen}{99.3}
\resultdef{primary.acc.transformed.encouraged-tools.state-and-literals.sol}{99.8}
\resultdef{primary.acc.transformed.encouraged-tools.static-return.luna}{99.7}
\resultdef{primary.acc.transformed.encouraged-tools.static-return.max}{100.0}
\resultdef{primary.acc.transformed.encouraged-tools.static-return.min}{99.0}
\resultdef{primary.acc.transformed.encouraged-tools.static-return.qwen}{99.0}
\resultdef{primary.acc.transformed.encouraged-tools.static-return.sol}{100.0}
\resultdef{primary.acc.transformed.encouraged-tools.subset-max.luna}{99.8}
\resultdef{primary.acc.transformed.encouraged-tools.subset-max.max}{100.0}
\resultdef{primary.acc.transformed.encouraged-tools.subset-max.min}{99.8}
\resultdef{primary.acc.transformed.encouraged-tools.subset-max.qwen}{99.9}
\resultdef{primary.acc.transformed.encouraged-tools.subset-max.sol}{100.0}
\resultdef{primary.acc.transformed.encouraged-tools.subset-min.luna}{96.9}
\resultdef{primary.acc.transformed.encouraged-tools.subset-min.max}{99.7}
\resultdef{primary.acc.transformed.encouraged-tools.subset-min.min}{96.9}
\resultdef{primary.acc.transformed.encouraged-tools.subset-min.qwen}{98.7}
\resultdef{primary.acc.transformed.encouraged-tools.subset-min.sol}{99.7}
\resultdef{primary.acc.transformed.encouraged-tools.wrong-arithmetic.luna}{99.8}
\resultdef{primary.acc.transformed.encouraged-tools.wrong-arithmetic.max}{100.0}
\resultdef{primary.acc.transformed.encouraged-tools.wrong-arithmetic.min}{99.8}
\resultdef{primary.acc.transformed.encouraged-tools.wrong-arithmetic.qwen}{99.9}
\resultdef{primary.acc.transformed.encouraged-tools.wrong-arithmetic.sol}{100.0}
\resultdef{primary.acc.transformed.encouraged-tools.wrong-logic.luna}{98.4}
\resultdef{primary.acc.transformed.encouraged-tools.wrong-logic.max}{100.0}
\resultdef{primary.acc.transformed.encouraged-tools.wrong-logic.min}{98.4}
\resultdef{primary.acc.transformed.encouraged-tools.wrong-logic.qwen}{98.7}
\resultdef{primary.acc.transformed.encouraged-tools.wrong-logic.sol}{100.0}
\resultdef{primary.acc.transformed.encouraged-tools.wrong-variable.luna}{99.1}
\resultdef{primary.acc.transformed.encouraged-tools.wrong-variable.max}{100.0}
\resultdef{primary.acc.transformed.encouraged-tools.wrong-variable.min}{99.1}
\resultdef{primary.acc.transformed.encouraged-tools.wrong-variable.qwen}{99.5}
\resultdef{primary.acc.transformed.encouraged-tools.wrong-variable.sol}{100.0}
\resultdef{primary.acc.transformed.no-tools.control-flow.luna}{96.1}
\resultdef{primary.acc.transformed.no-tools.control-flow.max}{100.0}
\resultdef{primary.acc.transformed.no-tools.control-flow.min}{96.1}
\resultdef{primary.acc.transformed.no-tools.control-flow.qwen}{96.4}
\resultdef{primary.acc.transformed.no-tools.control-flow.sol}{100.0}
\resultdef{primary.acc.transformed.no-tools.distractors.luna}{97.2}
\resultdef{primary.acc.transformed.no-tools.distractors.max}{100.0}
\resultdef{primary.acc.transformed.no-tools.distractors.min}{97.2}
\resultdef{primary.acc.transformed.no-tools.distractors.qwen}{98.7}
\resultdef{primary.acc.transformed.no-tools.distractors.sol}{100.0}
\resultdef{primary.acc.transformed.no-tools.luna}{96.7}
\resultdef{primary.acc.transformed.no-tools.max}{99.9}
\resultdef{primary.acc.transformed.no-tools.maximal-mix.luna}{94.4}
\resultdef{primary.acc.transformed.no-tools.maximal-mix.max}{99.9}
\resultdef{primary.acc.transformed.no-tools.maximal-mix.min}{94.4}
\resultdef{primary.acc.transformed.no-tools.maximal-mix.qwen}{96.7}
\resultdef{primary.acc.transformed.no-tools.maximal-mix.sol}{99.9}
\resultdef{primary.acc.transformed.no-tools.min}{96.7}
\resultdef{primary.acc.transformed.no-tools.off-by-one.luna}{93.8}
\resultdef{primary.acc.transformed.no-tools.off-by-one.max}{99.7}
\resultdef{primary.acc.transformed.no-tools.off-by-one.min}{93.8}
\resultdef{primary.acc.transformed.no-tools.off-by-one.qwen}{97.2}
\resultdef{primary.acc.transformed.no-tools.off-by-one.sol}{99.7}
\resultdef{primary.acc.transformed.no-tools.qwen}{97.9}
\resultdef{primary.acc.transformed.no-tools.rename.luna}{98.1}
\resultdef{primary.acc.transformed.no-tools.rename.max}{99.8}
\resultdef{primary.acc.transformed.no-tools.rename.min}{98.1}
\resultdef{primary.acc.transformed.no-tools.rename.qwen}{98.8}
\resultdef{primary.acc.transformed.no-tools.rename.sol}{99.8}
\resultdef{primary.acc.transformed.no-tools.sol}{99.9}
\resultdef{primary.acc.transformed.no-tools.state-and-literals.luna}{97.7}
\resultdef{primary.acc.transformed.no-tools.state-and-literals.max}{99.8}
\resultdef{primary.acc.transformed.no-tools.state-and-literals.min}{97.7}
\resultdef{primary.acc.transformed.no-tools.state-and-literals.qwen}{98.9}
\resultdef{primary.acc.transformed.no-tools.state-and-literals.sol}{99.8}
\resultdef{primary.acc.transformed.no-tools.static-return.luna}{97.7}
\resultdef{primary.acc.transformed.no-tools.static-return.max}{99.9}
\resultdef{primary.acc.transformed.no-tools.static-return.min}{97.0}
\resultdef{primary.acc.transformed.no-tools.static-return.qwen}{97.0}
\resultdef{primary.acc.transformed.no-tools.static-return.sol}{99.9}
\resultdef{primary.acc.transformed.no-tools.subset-max.luna}{98.6}
\resultdef{primary.acc.transformed.no-tools.subset-max.max}{100.0}
\resultdef{primary.acc.transformed.no-tools.subset-max.min}{98.6}
\resultdef{primary.acc.transformed.no-tools.subset-max.qwen}{99.3}
\resultdef{primary.acc.transformed.no-tools.subset-max.sol}{100.0}
\resultdef{primary.acc.transformed.no-tools.subset-min.luna}{93.8}
\resultdef{primary.acc.transformed.no-tools.subset-min.max}{99.7}
\resultdef{primary.acc.transformed.no-tools.subset-min.min}{93.8}
\resultdef{primary.acc.transformed.no-tools.subset-min.qwen}{96.4}
\resultdef{primary.acc.transformed.no-tools.subset-min.sol}{99.7}
\resultdef{primary.acc.transformed.no-tools.wrong-arithmetic.luna}{98.6}
\resultdef{primary.acc.transformed.no-tools.wrong-arithmetic.max}{100.0}
\resultdef{primary.acc.transformed.no-tools.wrong-arithmetic.min}{98.6}
\resultdef{primary.acc.transformed.no-tools.wrong-arithmetic.qwen}{99.3}
\resultdef{primary.acc.transformed.no-tools.wrong-arithmetic.sol}{100.0}
\resultdef{primary.acc.transformed.no-tools.wrong-logic.luna}{94.9}
\resultdef{primary.acc.transformed.no-tools.wrong-logic.max}{100.0}
\resultdef{primary.acc.transformed.no-tools.wrong-logic.min}{94.9}
\resultdef{primary.acc.transformed.no-tools.wrong-logic.qwen}{97.1}
\resultdef{primary.acc.transformed.no-tools.wrong-logic.sol}{100.0}
\resultdef{primary.acc.transformed.no-tools.wrong-variable.luna}{98.3}
\resultdef{primary.acc.transformed.no-tools.wrong-variable.max}{100.0}
\resultdef{primary.acc.transformed.no-tools.wrong-variable.min}{98.3}
\resultdef{primary.acc.transformed.no-tools.wrong-variable.qwen}{98.9}
\resultdef{primary.acc.transformed.no-tools.wrong-variable.sol}{100.0}
\resultdef{primary.acc.untransformed.discretionary-tools.control-flow.luna}{98.1}
\resultdef{primary.acc.untransformed.discretionary-tools.control-flow.max}{99.8}
\resultdef{primary.acc.untransformed.discretionary-tools.control-flow.min}{98.1}
\resultdef{primary.acc.untransformed.discretionary-tools.control-flow.qwen}{98.9}
\resultdef{primary.acc.untransformed.discretionary-tools.control-flow.sol}{99.8}
\resultdef{primary.acc.untransformed.discretionary-tools.distractors.luna}{98.1}
\resultdef{primary.acc.untransformed.discretionary-tools.distractors.max}{99.8}
\resultdef{primary.acc.untransformed.discretionary-tools.distractors.min}{98.1}
\resultdef{primary.acc.untransformed.discretionary-tools.distractors.qwen}{98.9}
\resultdef{primary.acc.untransformed.discretionary-tools.distractors.sol}{99.8}
\resultdef{primary.acc.untransformed.discretionary-tools.luna}{98.1}
\resultdef{primary.acc.untransformed.discretionary-tools.max}{99.8}
\resultdef{primary.acc.untransformed.discretionary-tools.maximal-mix.luna}{98.1}
\resultdef{primary.acc.untransformed.discretionary-tools.maximal-mix.max}{99.8}
\resultdef{primary.acc.untransformed.discretionary-tools.maximal-mix.min}{98.1}
\resultdef{primary.acc.untransformed.discretionary-tools.maximal-mix.qwen}{98.9}
\resultdef{primary.acc.untransformed.discretionary-tools.maximal-mix.sol}{99.8}
\resultdef{primary.acc.untransformed.discretionary-tools.min}{98.1}
\resultdef{primary.acc.untransformed.discretionary-tools.off-by-one.luna}{96.8}
\resultdef{primary.acc.untransformed.discretionary-tools.off-by-one.max}{99.2}
\resultdef{primary.acc.untransformed.discretionary-tools.off-by-one.min}{96.8}
\resultdef{primary.acc.untransformed.discretionary-tools.off-by-one.qwen}{98.0}
\resultdef{primary.acc.untransformed.discretionary-tools.off-by-one.sol}{99.2}
\resultdef{primary.acc.untransformed.discretionary-tools.qwen}{98.9}
\resultdef{primary.acc.untransformed.discretionary-tools.rename.luna}{98.0}
\resultdef{primary.acc.untransformed.discretionary-tools.rename.max}{99.7}
\resultdef{primary.acc.untransformed.discretionary-tools.rename.min}{98.0}
\resultdef{primary.acc.untransformed.discretionary-tools.rename.qwen}{98.9}
\resultdef{primary.acc.untransformed.discretionary-tools.rename.sol}{99.7}
\resultdef{primary.acc.untransformed.discretionary-tools.sol}{99.8}
\resultdef{primary.acc.untransformed.discretionary-tools.state-and-literals.luna}{98.1}
\resultdef{primary.acc.untransformed.discretionary-tools.state-and-literals.max}{99.8}
\resultdef{primary.acc.untransformed.discretionary-tools.state-and-literals.min}{98.1}
\resultdef{primary.acc.untransformed.discretionary-tools.state-and-literals.qwen}{98.9}
\resultdef{primary.acc.untransformed.discretionary-tools.state-and-literals.sol}{99.8}
\resultdef{primary.acc.untransformed.discretionary-tools.static-return.luna}{98.6}
\resultdef{primary.acc.untransformed.discretionary-tools.static-return.max}{99.6}
\resultdef{primary.acc.untransformed.discretionary-tools.static-return.min}{98.6}
\resultdef{primary.acc.untransformed.discretionary-tools.static-return.qwen}{98.9}
\resultdef{primary.acc.untransformed.discretionary-tools.static-return.sol}{99.6}
\resultdef{primary.acc.untransformed.discretionary-tools.subset-max.luna}{100.0}
\resultdef{primary.acc.untransformed.discretionary-tools.subset-max.max}{100.0}
\resultdef{primary.acc.untransformed.discretionary-tools.subset-max.min}{100.0}
\resultdef{primary.acc.untransformed.discretionary-tools.subset-max.qwen}{100.0}
\resultdef{primary.acc.untransformed.discretionary-tools.subset-max.sol}{100.0}
\resultdef{primary.acc.untransformed.discretionary-tools.subset-min.luna}{96.8}
\resultdef{primary.acc.untransformed.discretionary-tools.subset-min.max}{99.2}
\resultdef{primary.acc.untransformed.discretionary-tools.subset-min.min}{96.8}
\resultdef{primary.acc.untransformed.discretionary-tools.subset-min.qwen}{98.0}
\resultdef{primary.acc.untransformed.discretionary-tools.subset-min.sol}{99.2}
\resultdef{primary.acc.untransformed.discretionary-tools.wrong-arithmetic.luna}{100.0}
\resultdef{primary.acc.untransformed.discretionary-tools.wrong-arithmetic.max}{100.0}
\resultdef{primary.acc.untransformed.discretionary-tools.wrong-arithmetic.min}{100.0}
\resultdef{primary.acc.untransformed.discretionary-tools.wrong-arithmetic.qwen}{100.0}
\resultdef{primary.acc.untransformed.discretionary-tools.wrong-arithmetic.sol}{100.0}
\resultdef{primary.acc.untransformed.discretionary-tools.wrong-logic.luna}{97.2}
\resultdef{primary.acc.untransformed.discretionary-tools.wrong-logic.max}{100.0}
\resultdef{primary.acc.untransformed.discretionary-tools.wrong-logic.min}{97.2}
\resultdef{primary.acc.untransformed.discretionary-tools.wrong-logic.qwen}{98.6}
\resultdef{primary.acc.untransformed.discretionary-tools.wrong-logic.sol}{100.0}
\resultdef{primary.acc.untransformed.discretionary-tools.wrong-variable.luna}{98.1}
\resultdef{primary.acc.untransformed.discretionary-tools.wrong-variable.max}{100.0}
\resultdef{primary.acc.untransformed.discretionary-tools.wrong-variable.min}{98.1}
\resultdef{primary.acc.untransformed.discretionary-tools.wrong-variable.qwen}{99.2}
\resultdef{primary.acc.untransformed.discretionary-tools.wrong-variable.sol}{100.0}
\resultdef{primary.acc.untransformed.encouraged-tools.control-flow.luna}{98.9}
\resultdef{primary.acc.untransformed.encouraged-tools.control-flow.max}{99.8}
\resultdef{primary.acc.untransformed.encouraged-tools.control-flow.min}{98.9}
\resultdef{primary.acc.untransformed.encouraged-tools.control-flow.qwen}{99.1}
\resultdef{primary.acc.untransformed.encouraged-tools.control-flow.sol}{99.8}
\resultdef{primary.acc.untransformed.encouraged-tools.distractors.luna}{98.9}
\resultdef{primary.acc.untransformed.encouraged-tools.distractors.max}{99.8}
\resultdef{primary.acc.untransformed.encouraged-tools.distractors.min}{98.9}
\resultdef{primary.acc.untransformed.encouraged-tools.distractors.qwen}{99.1}
\resultdef{primary.acc.untransformed.encouraged-tools.distractors.sol}{99.8}
\resultdef{primary.acc.untransformed.encouraged-tools.luna}{98.9}
\resultdef{primary.acc.untransformed.encouraged-tools.max}{99.8}
\resultdef{primary.acc.untransformed.encouraged-tools.maximal-mix.luna}{98.9}
\resultdef{primary.acc.untransformed.encouraged-tools.maximal-mix.max}{99.8}
\resultdef{primary.acc.untransformed.encouraged-tools.maximal-mix.min}{98.9}
\resultdef{primary.acc.untransformed.encouraged-tools.maximal-mix.qwen}{99.1}
\resultdef{primary.acc.untransformed.encouraged-tools.maximal-mix.sol}{99.8}
\resultdef{primary.acc.untransformed.encouraged-tools.min}{98.9}
\resultdef{primary.acc.untransformed.encouraged-tools.off-by-one.luna}{97.6}
\resultdef{primary.acc.untransformed.encouraged-tools.off-by-one.max}{99.2}
\resultdef{primary.acc.untransformed.encouraged-tools.off-by-one.min}{97.6}
\resultdef{primary.acc.untransformed.encouraged-tools.off-by-one.qwen}{98.8}
\resultdef{primary.acc.untransformed.encouraged-tools.off-by-one.sol}{99.2}
\resultdef{primary.acc.untransformed.encouraged-tools.qwen}{99.1}
\resultdef{primary.acc.untransformed.encouraged-tools.rename.luna}{98.9}
\resultdef{primary.acc.untransformed.encouraged-tools.rename.max}{99.8}
\resultdef{primary.acc.untransformed.encouraged-tools.rename.min}{98.9}
\resultdef{primary.acc.untransformed.encouraged-tools.rename.qwen}{99.1}
\resultdef{primary.acc.untransformed.encouraged-tools.rename.sol}{99.8}
\resultdef{primary.acc.untransformed.encouraged-tools.sol}{99.8}
\resultdef{primary.acc.untransformed.encouraged-tools.state-and-literals.luna}{98.9}
\resultdef{primary.acc.untransformed.encouraged-tools.state-and-literals.max}{99.8}
\resultdef{primary.acc.untransformed.encouraged-tools.state-and-literals.min}{98.9}
\resultdef{primary.acc.untransformed.encouraged-tools.state-and-literals.qwen}{99.1}
\resultdef{primary.acc.untransformed.encouraged-tools.state-and-literals.sol}{99.8}
\resultdef{primary.acc.untransformed.encouraged-tools.static-return.luna}{100.0}
\resultdef{primary.acc.untransformed.encouraged-tools.static-return.max}{100.0}
\resultdef{primary.acc.untransformed.encouraged-tools.static-return.min}{98.9}
\resultdef{primary.acc.untransformed.encouraged-tools.static-return.qwen}{98.9}
\resultdef{primary.acc.untransformed.encouraged-tools.static-return.sol}{100.0}
\resultdef{primary.acc.untransformed.encouraged-tools.subset-max.luna}{100.0}
\resultdef{primary.acc.untransformed.encouraged-tools.subset-max.max}{100.0}
\resultdef{primary.acc.untransformed.encouraged-tools.subset-max.min}{100.0}
\resultdef{primary.acc.untransformed.encouraged-tools.subset-max.qwen}{100.0}
\resultdef{primary.acc.untransformed.encouraged-tools.subset-max.sol}{100.0}
\resultdef{primary.acc.untransformed.encouraged-tools.subset-min.luna}{97.6}
\resultdef{primary.acc.untransformed.encouraged-tools.subset-min.max}{99.2}
\resultdef{primary.acc.untransformed.encouraged-tools.subset-min.min}{97.6}
\resultdef{primary.acc.untransformed.encouraged-tools.subset-min.qwen}{98.1}
\resultdef{primary.acc.untransformed.encouraged-tools.subset-min.sol}{99.2}
\resultdef{primary.acc.untransformed.encouraged-tools.wrong-arithmetic.luna}{100.0}
\resultdef{primary.acc.untransformed.encouraged-tools.wrong-arithmetic.max}{100.0}
\resultdef{primary.acc.untransformed.encouraged-tools.wrong-arithmetic.min}{99.5}
\resultdef{primary.acc.untransformed.encouraged-tools.wrong-arithmetic.qwen}{99.5}
\resultdef{primary.acc.untransformed.encouraged-tools.wrong-arithmetic.sol}{100.0}
\resultdef{primary.acc.untransformed.encouraged-tools.wrong-logic.luna}{98.6}
\resultdef{primary.acc.untransformed.encouraged-tools.wrong-logic.max}{100.0}
\resultdef{primary.acc.untransformed.encouraged-tools.wrong-logic.min}{98.1}
\resultdef{primary.acc.untransformed.encouraged-tools.wrong-logic.qwen}{98.1}
\resultdef{primary.acc.untransformed.encouraged-tools.wrong-logic.sol}{100.0}
\resultdef{primary.acc.untransformed.encouraged-tools.wrong-variable.luna}{98.5}
\resultdef{primary.acc.untransformed.encouraged-tools.wrong-variable.max}{100.0}
\resultdef{primary.acc.untransformed.encouraged-tools.wrong-variable.min}{98.5}
\resultdef{primary.acc.untransformed.encouraged-tools.wrong-variable.qwen}{100.0}
\resultdef{primary.acc.untransformed.encouraged-tools.wrong-variable.sol}{100.0}
\resultdef{primary.acc.untransformed.no-tools.control-flow.luna}{98.2}
\resultdef{primary.acc.untransformed.no-tools.control-flow.max}{99.9}
\resultdef{primary.acc.untransformed.no-tools.control-flow.min}{98.2}
\resultdef{primary.acc.untransformed.no-tools.control-flow.qwen}{98.8}
\resultdef{primary.acc.untransformed.no-tools.control-flow.sol}{99.9}
\resultdef{primary.acc.untransformed.no-tools.distractors.luna}{98.2}
\resultdef{primary.acc.untransformed.no-tools.distractors.max}{99.9}
\resultdef{primary.acc.untransformed.no-tools.distractors.min}{98.2}
\resultdef{primary.acc.untransformed.no-tools.distractors.qwen}{98.8}
\resultdef{primary.acc.untransformed.no-tools.distractors.sol}{99.9}
\resultdef{primary.acc.untransformed.no-tools.luna}{98.2}
\resultdef{primary.acc.untransformed.no-tools.max}{99.9}
\resultdef{primary.acc.untransformed.no-tools.maximal-mix.luna}{98.2}
\resultdef{primary.acc.untransformed.no-tools.maximal-mix.max}{99.9}
\resultdef{primary.acc.untransformed.no-tools.maximal-mix.min}{98.2}
\resultdef{primary.acc.untransformed.no-tools.maximal-mix.qwen}{98.8}
\resultdef{primary.acc.untransformed.no-tools.maximal-mix.sol}{99.9}
\resultdef{primary.acc.untransformed.no-tools.min}{98.2}
\resultdef{primary.acc.untransformed.no-tools.off-by-one.luna}{96.4}
\resultdef{primary.acc.untransformed.no-tools.off-by-one.max}{99.6}
\resultdef{primary.acc.untransformed.no-tools.off-by-one.min}{96.4}
\resultdef{primary.acc.untransformed.no-tools.off-by-one.qwen}{97.2}
\resultdef{primary.acc.untransformed.no-tools.off-by-one.sol}{99.6}
\resultdef{primary.acc.untransformed.no-tools.qwen}{98.8}
\resultdef{primary.acc.untransformed.no-tools.rename.luna}{98.1}
\resultdef{primary.acc.untransformed.no-tools.rename.max}{99.9}
\resultdef{primary.acc.untransformed.no-tools.rename.min}{98.1}
\resultdef{primary.acc.untransformed.no-tools.rename.qwen}{98.7}
\resultdef{primary.acc.untransformed.no-tools.rename.sol}{99.9}
\resultdef{primary.acc.untransformed.no-tools.sol}{99.9}
\resultdef{primary.acc.untransformed.no-tools.state-and-literals.luna}{98.2}
\resultdef{primary.acc.untransformed.no-tools.state-and-literals.max}{99.9}
\resultdef{primary.acc.untransformed.no-tools.state-and-literals.min}{98.2}
\resultdef{primary.acc.untransformed.no-tools.state-and-literals.qwen}{98.8}
\resultdef{primary.acc.untransformed.no-tools.state-and-literals.sol}{99.9}
\resultdef{primary.acc.untransformed.no-tools.static-return.luna}{98.9}
\resultdef{primary.acc.untransformed.no-tools.static-return.max}{100.0}
\resultdef{primary.acc.untransformed.no-tools.static-return.min}{98.9}
\resultdef{primary.acc.untransformed.no-tools.static-return.qwen}{99.3}
\resultdef{primary.acc.untransformed.no-tools.static-return.sol}{100.0}
\resultdef{primary.acc.untransformed.no-tools.subset-max.luna}{99.5}
\resultdef{primary.acc.untransformed.no-tools.subset-max.max}{100.0}
\resultdef{primary.acc.untransformed.no-tools.subset-max.min}{99.5}
\resultdef{primary.acc.untransformed.no-tools.subset-max.qwen}{100.0}
\resultdef{primary.acc.untransformed.no-tools.subset-max.sol}{100.0}
\resultdef{primary.acc.untransformed.no-tools.subset-min.luna}{96.4}
\resultdef{primary.acc.untransformed.no-tools.subset-min.max}{99.6}
\resultdef{primary.acc.untransformed.no-tools.subset-min.min}{96.4}
\resultdef{primary.acc.untransformed.no-tools.subset-min.qwen}{97.2}
\resultdef{primary.acc.untransformed.no-tools.subset-min.sol}{99.6}
\resultdef{primary.acc.untransformed.no-tools.wrong-arithmetic.luna}{99.5}
\resultdef{primary.acc.untransformed.no-tools.wrong-arithmetic.max}{100.0}
\resultdef{primary.acc.untransformed.no-tools.wrong-arithmetic.min}{99.5}
\resultdef{primary.acc.untransformed.no-tools.wrong-arithmetic.qwen}{100.0}
\resultdef{primary.acc.untransformed.no-tools.wrong-arithmetic.sol}{100.0}
\resultdef{primary.acc.untransformed.no-tools.wrong-logic.luna}{97.7}
\resultdef{primary.acc.untransformed.no-tools.wrong-logic.max}{100.0}
\resultdef{primary.acc.untransformed.no-tools.wrong-logic.min}{97.2}
\resultdef{primary.acc.untransformed.no-tools.wrong-logic.qwen}{97.2}
\resultdef{primary.acc.untransformed.no-tools.wrong-logic.sol}{100.0}
\resultdef{primary.acc.untransformed.no-tools.wrong-variable.luna}{98.5}
\resultdef{primary.acc.untransformed.no-tools.wrong-variable.max}{100.0}
\resultdef{primary.acc.untransformed.no-tools.wrong-variable.min}{98.5}
\resultdef{primary.acc.untransformed.no-tools.wrong-variable.qwen}{100.0}
\resultdef{primary.acc.untransformed.no-tools.wrong-variable.sol}{100.0}
\resultdef{primary.diff.ci.luna}{[+0.6, +2.1]}
\resultdef{primary.diff.ci.qwen}{[+0.1, +1.7]}
\resultdef{primary.diff.ci.sol}{[+0.0, +0.3]}
\resultdef{primary.diff.control-flow.ci.luna}{[+1.0, +3.8]}
\resultdef{primary.diff.control-flow.ci.qwen}{[+0.8, +4.1]}
\resultdef{primary.diff.control-flow.ci.sol}{[+0.0, +0.3]}
\resultdef{primary.diff.control-flow.luna}{+2.4}
\resultdef{primary.diff.control-flow.max}{+2.5}
\resultdef{primary.diff.control-flow.min}{+0.1}
\resultdef{primary.diff.control-flow.qwen}{+2.5}
\resultdef{primary.diff.control-flow.sol}{+0.1}
\resultdef{primary.diff.distractors.ci.luna}{[$-$0.8, +1.0]}
\resultdef{primary.diff.distractors.ci.qwen}{[$-$0.7, +1.0]}
\resultdef{primary.diff.distractors.ci.sol}{[+0.0, +0.3]}
\resultdef{primary.diff.distractors.luna}{+0.1}
\resultdef{primary.diff.distractors.max}{+0.1}
\resultdef{primary.diff.distractors.min}{+0.1}
\resultdef{primary.diff.distractors.qwen}{+0.1}
\resultdef{primary.diff.distractors.sol}{+0.1}
\resultdef{primary.diff.luna}{+1.4}
\resultdef{primary.diff.max}{+1.4}
\resultdef{primary.diff.maximal-mix.ci.luna}{[+1.9, +5.3]}
\resultdef{primary.diff.maximal-mix.ci.qwen}{[+0.6, +3.0]}
\resultdef{primary.diff.maximal-mix.ci.sol}{[+0.0, +0.5]}
\resultdef{primary.diff.maximal-mix.luna}{+3.6}
\resultdef{primary.diff.maximal-mix.max}{+3.6}
\resultdef{primary.diff.maximal-mix.min}{+0.2}
\resultdef{primary.diff.maximal-mix.qwen}{+1.8}
\resultdef{primary.diff.maximal-mix.sol}{+0.2}
\resultdef{primary.diff.min}{+0.1}
\resultdef{primary.diff.off-by-one.ci.luna}{[$-$0.1, +4.0]}
\resultdef{primary.diff.off-by-one.ci.qwen}{[$-$1.8, +1.7]}
\resultdef{primary.diff.off-by-one.ci.sol}{[+0.0, +1.3]}
\resultdef{primary.diff.off-by-one.luna}{+1.9}
\resultdef{primary.diff.off-by-one.max}{+1.9}
\resultdef{primary.diff.off-by-one.min}{+0.1}
\resultdef{primary.diff.off-by-one.qwen}{+0.1}
\resultdef{primary.diff.off-by-one.sol}{+0.4}
\resultdef{primary.diff.qwen}{+1.0}
\resultdef{primary.diff.rename-distractors.subset-max.luna}{+0.1}
\resultdef{primary.diff.rename-distractors.subset-max.max}{+0.3}
\resultdef{primary.diff.rename-distractors.subset-max.min}{+0.1}
\resultdef{primary.diff.rename-distractors.subset-max.qwen}{+0.3}
\resultdef{primary.diff.rename-distractors.subset-max.sol}{+0.1}
\resultdef{primary.diff.rename-distractors.subset-min.luna}{+0.1}
\resultdef{primary.diff.rename-distractors.subset-min.max}{+0.1}
\resultdef{primary.diff.rename-distractors.subset-min.min}{+0.1}
\resultdef{primary.diff.rename-distractors.subset-min.qwen}{+0.1}
\resultdef{primary.diff.rename-distractors.subset-min.sol}{+0.1}
\resultdef{primary.diff.rename.ci.luna}{[$-$0.7, +0.9]}
\resultdef{primary.diff.rename.ci.qwen}{[$-$0.6, +1.1]}
\resultdef{primary.diff.rename.ci.sol}{[+0.0, +0.3]}
\resultdef{primary.diff.rename.luna}{+0.1}
\resultdef{primary.diff.rename.max}{+0.3}
\resultdef{primary.diff.rename.min}{+0.1}
\resultdef{primary.diff.rename.qwen}{+0.3}
\resultdef{primary.diff.rename.sol}{+0.1}
\resultdef{primary.diff.sol}{+0.1}
\resultdef{primary.diff.state-and-literals.ci.luna}{[$-$0.2, +1.5]}
\resultdef{primary.diff.state-and-literals.ci.qwen}{[$-$0.8, +1.1]}
\resultdef{primary.diff.state-and-literals.ci.sol}{[+0.0, +0.5]}
\resultdef{primary.diff.state-and-literals.luna}{+0.6}
\resultdef{primary.diff.state-and-literals.max}{+0.6}
\resultdef{primary.diff.state-and-literals.min}{+0.2}
\resultdef{primary.diff.state-and-literals.qwen}{+0.2}
\resultdef{primary.diff.state-and-literals.sol}{+0.2}
\resultdef{primary.diff.static-return.ci.luna}{[$-$0.4, +2.2]}
\resultdef{primary.diff.static-return.ci.qwen}{[+0.9, +3.9]}
\resultdef{primary.diff.static-return.ci.sol}{[+0.0, +0.4]}
\resultdef{primary.diff.static-return.luna}{+0.9}
\resultdef{primary.diff.static-return.max}{+2.3}
\resultdef{primary.diff.static-return.min}{+0.1}
\resultdef{primary.diff.static-return.qwen}{+2.3}
\resultdef{primary.diff.static-return.sol}{+0.1}
\resultdef{primary.diff.subset-max.luna}{+3.6}
\resultdef{primary.diff.subset-max.max}{+3.6}
\resultdef{primary.diff.subset-max.min}{+0.4}
\resultdef{primary.diff.subset-max.qwen}{+2.5}
\resultdef{primary.diff.subset-max.sol}{+0.4}
\resultdef{primary.diff.subset-min.luna}{+0.1}
\resultdef{primary.diff.subset-min.max}{+0.1}
\resultdef{primary.diff.subset-min.min}{+0.0}
\resultdef{primary.diff.subset-min.qwen}{+0.1}
\resultdef{primary.diff.subset-min.sol}{+0.0}
\resultdef{primary.diff.wrong-arithmetic.ci.luna}{[$-$0.2, +1.6]}
\resultdef{primary.diff.wrong-arithmetic.ci.qwen}{[+0.2, +2.3]}
\resultdef{primary.diff.wrong-arithmetic.ci.sol}{[+0.0, +0.0]}
\resultdef{primary.diff.wrong-arithmetic.luna}{+0.7}
\resultdef{primary.diff.wrong-arithmetic.max}{+1.1}
\resultdef{primary.diff.wrong-arithmetic.min}{+0.0}
\resultdef{primary.diff.wrong-arithmetic.qwen}{+1.1}
\resultdef{primary.diff.wrong-arithmetic.sol}{+0.0}
\resultdef{primary.diff.wrong-logic.ci.luna}{[+0.7, +4.5]}
\resultdef{primary.diff.wrong-logic.ci.qwen}{[$-$2.3, +3.2]}
\resultdef{primary.diff.wrong-logic.ci.sol}{[+0.0, +0.0]}
\resultdef{primary.diff.wrong-logic.luna}{+2.6}
\resultdef{primary.diff.wrong-logic.max}{+2.6}
\resultdef{primary.diff.wrong-logic.min}{+0.0}
\resultdef{primary.diff.wrong-logic.qwen}{+0.7}
\resultdef{primary.diff.wrong-logic.sol}{+0.0}
\resultdef{primary.diff.wrong-variable.ci.luna}{[$-$0.5, +1.9]}
\resultdef{primary.diff.wrong-variable.ci.qwen}{[$-$0.1, +1.2]}
\resultdef{primary.diff.wrong-variable.ci.sol}{[+0.0, +0.0]}
\resultdef{primary.diff.wrong-variable.luna}{+0.8}
\resultdef{primary.diff.wrong-variable.max}{+0.8}
\resultdef{primary.diff.wrong-variable.min}{+0.0}
\resultdef{primary.diff.wrong-variable.qwen}{+0.5}
\resultdef{primary.diff.wrong-variable.sol}{+0.0}
\resultdef{primary.disc-minus-enc.transformed.ci.luna}{[$-$1.0, $-$0.2]}
\resultdef{primary.disc-minus-enc.transformed.ci.qwen}{[$-$0.3, +0.2]}
\resultdef{primary.disc-minus-enc.transformed.ci.sol}{[$-$0.2, +0.0]}
\resultdef{primary.disc-minus-enc.transformed.luna}{$-$0.6}
\resultdef{primary.disc-minus-enc.transformed.max}{+0.0}
\resultdef{primary.disc-minus-enc.transformed.min}{$-$0.6}
\resultdef{primary.disc-minus-enc.transformed.qwen}{+0.0}
\resultdef{primary.disc-minus-enc.transformed.sol}{+0.0}
\resultdef{primary.disc-minus-enc.untransformed.ci.luna}{[$-$1.6, $-$0.2]}
\resultdef{primary.disc-minus-enc.untransformed.ci.qwen}{[$-$0.7, +0.3]}
\resultdef{primary.disc-minus-enc.untransformed.ci.sol}{[$-$0.2, +0.0]}
\resultdef{primary.disc-minus-enc.untransformed.luna}{$-$0.8}
\resultdef{primary.disc-minus-enc.untransformed.max}{$-$0.1}
\resultdef{primary.disc-minus-enc.untransformed.min}{$-$0.8}
\resultdef{primary.disc-minus-enc.untransformed.qwen}{$-$0.2}
\resultdef{primary.disc-minus-enc.untransformed.sol}{$-$0.1}
\resultdef{primary.disc-minus-no.transformed.ci.luna}{[+0.9, +2.2]}
\resultdef{primary.disc-minus-no.transformed.ci.qwen}{[+0.8, +1.8]}
\resultdef{primary.disc-minus-no.transformed.ci.sol}{[$-$0.1, +0.0]}
\resultdef{primary.disc-minus-no.transformed.luna}{+1.5}
\resultdef{primary.disc-minus-no.transformed.max}{+1.5}
\resultdef{primary.disc-minus-no.transformed.min}{+0.0}
\resultdef{primary.disc-minus-no.transformed.qwen}{+1.3}
\resultdef{primary.disc-minus-no.transformed.sol}{+0.0}
\resultdef{primary.disc-minus-no.untransformed.ci.luna}{[$-$1.0, +0.7]}
\resultdef{primary.disc-minus-no.untransformed.ci.qwen}{[$-$0.3, +0.7]}
\resultdef{primary.disc-minus-no.untransformed.ci.sol}{[$-$0.4, +0.0]}
\resultdef{primary.disc-minus-no.untransformed.luna}{$-$0.1}
\resultdef{primary.disc-minus-no.untransformed.max}{+0.2}
\resultdef{primary.disc-minus-no.untransformed.min}{$-$0.2}
\resultdef{primary.disc-minus-no.untransformed.qwen}{+0.2}
\resultdef{primary.disc-minus-no.untransformed.sol}{$-$0.2}
\resultdef{primary.effect.transformed.ci.luna}{[+1.4, +2.8]}
\resultdef{primary.effect.transformed.ci.qwen}{[+0.8, +1.8]}
\resultdef{primary.effect.transformed.ci.sol}{[+0.0, +0.1]}
\resultdef{primary.effect.transformed.control-flow.ci.luna}{[+1.9, +4.4]}
\resultdef{primary.effect.transformed.control-flow.ci.qwen}{[+1.5, +4.3]}
\resultdef{primary.effect.transformed.control-flow.ci.sol}{[+0.0, +0.0]}
\resultdef{primary.effect.transformed.control-flow.luna}{+3.1}
\resultdef{primary.effect.transformed.control-flow.max}{+3.1}
\resultdef{primary.effect.transformed.control-flow.min}{+0.0}
\resultdef{primary.effect.transformed.control-flow.qwen}{+2.8}
\resultdef{primary.effect.transformed.control-flow.sol}{+0.0}
\resultdef{primary.effect.transformed.distractors.ci.luna}{[+0.1, +1.7]}
\resultdef{primary.effect.transformed.distractors.ci.qwen}{[$-$0.3, +1.2]}
\resultdef{primary.effect.transformed.distractors.ci.sol}{[+0.0, +0.0]}
\resultdef{primary.effect.transformed.distractors.luna}{+0.8}
\resultdef{primary.effect.transformed.distractors.max}{+0.8}
\resultdef{primary.effect.transformed.distractors.min}{+0.0}
\resultdef{primary.effect.transformed.distractors.qwen}{+0.4}
\resultdef{primary.effect.transformed.distractors.sol}{+0.0}
\resultdef{primary.effect.transformed.luna}{+2.1}
\resultdef{primary.effect.transformed.max}{+2.1}
\resultdef{primary.effect.transformed.maximal-mix.ci.luna}{[+2.7, +6.1]}
\resultdef{primary.effect.transformed.maximal-mix.ci.qwen}{[+1.2, +3.2]}
\resultdef{primary.effect.transformed.maximal-mix.ci.sol}{[+0.0, +0.3]}
\resultdef{primary.effect.transformed.maximal-mix.luna}{+4.4}
\resultdef{primary.effect.transformed.maximal-mix.max}{+4.4}
\resultdef{primary.effect.transformed.maximal-mix.min}{+0.1}
\resultdef{primary.effect.transformed.maximal-mix.qwen}{+2.1}
\resultdef{primary.effect.transformed.maximal-mix.sol}{+0.1}
\resultdef{primary.effect.transformed.min}{+0.0}
\resultdef{primary.effect.transformed.off-by-one.ci.luna}{[+1.6, +4.8]}
\resultdef{primary.effect.transformed.off-by-one.ci.qwen}{[+0.7, +2.9]}
\resultdef{primary.effect.transformed.off-by-one.ci.sol}{[+0.0, +0.0]}
\resultdef{primary.effect.transformed.off-by-one.luna}{+3.1}
\resultdef{primary.effect.transformed.off-by-one.max}{+3.1}
\resultdef{primary.effect.transformed.off-by-one.min}{+0.0}
\resultdef{primary.effect.transformed.off-by-one.qwen}{+1.7}
\resultdef{primary.effect.transformed.off-by-one.sol}{+0.0}
\resultdef{primary.effect.transformed.qwen}{+1.3}
\resultdef{primary.effect.transformed.rename.ci.luna}{[+0.2, +1.6]}
\resultdef{primary.effect.transformed.rename.ci.qwen}{[+0.0, +1.3]}
\resultdef{primary.effect.transformed.rename.ci.sol}{[+0.0, +0.0]}
\resultdef{primary.effect.transformed.rename.luna}{+0.9}
\resultdef{primary.effect.transformed.rename.max}{+0.9}
\resultdef{primary.effect.transformed.rename.min}{+0.0}
\resultdef{primary.effect.transformed.rename.qwen}{+0.6}
\resultdef{primary.effect.transformed.rename.sol}{+0.0}
\resultdef{primary.effect.transformed.sol}{+0.0}
\resultdef{primary.effect.transformed.state-and-literals.ci.luna}{[+0.4, +2.2]}
\resultdef{primary.effect.transformed.state-and-literals.ci.qwen}{[$-$0.2, +1.2]}
\resultdef{primary.effect.transformed.state-and-literals.ci.sol}{[$-$0.2, +0.5]}
\resultdef{primary.effect.transformed.state-and-literals.luna}{+1.3}
\resultdef{primary.effect.transformed.state-and-literals.max}{+1.3}
\resultdef{primary.effect.transformed.state-and-literals.min}{+0.1}
\resultdef{primary.effect.transformed.state-and-literals.qwen}{+0.5}
\resultdef{primary.effect.transformed.state-and-literals.sol}{+0.1}
\resultdef{primary.effect.transformed.static-return.ci.luna}{[+0.9, +3.4]}
\resultdef{primary.effect.transformed.static-return.ci.qwen}{[+1.1, +2.9]}
\resultdef{primary.effect.transformed.static-return.ci.sol}{[+0.0, +0.4]}
\resultdef{primary.effect.transformed.static-return.luna}{+2.0}
\resultdef{primary.effect.transformed.static-return.max}{+2.0}
\resultdef{primary.effect.transformed.static-return.min}{+0.1}
\resultdef{primary.effect.transformed.static-return.qwen}{+2.0}
\resultdef{primary.effect.transformed.static-return.sol}{+0.1}
\resultdef{primary.effect.transformed.subset-max.luna}{+4.4}
\resultdef{primary.effect.transformed.subset-max.max}{+4.4}
\resultdef{primary.effect.transformed.subset-max.min}{+0.1}
\resultdef{primary.effect.transformed.subset-max.qwen}{+2.8}
\resultdef{primary.effect.transformed.subset-max.sol}{+0.1}
\resultdef{primary.effect.transformed.subset-min.luna}{+0.8}
\resultdef{primary.effect.transformed.subset-min.max}{+0.8}
\resultdef{primary.effect.transformed.subset-min.min}{+0.0}
\resultdef{primary.effect.transformed.subset-min.qwen}{+0.4}
\resultdef{primary.effect.transformed.subset-min.sol}{+0.0}
\resultdef{primary.effect.transformed.subsets-ci-positive.luna}{9}
\resultdef{primary.effect.transformed.subsets-ci-positive.max}{9}
\resultdef{primary.effect.transformed.subsets-ci-positive.min}{0}
\resultdef{primary.effect.transformed.subsets-ci-positive.qwen}{6}
\resultdef{primary.effect.transformed.subsets-ci-positive.sol}{0}
\resultdef{primary.effect.transformed.subsets-positive.luna}{10}
\resultdef{primary.effect.transformed.subsets-positive.max}{10}
\resultdef{primary.effect.transformed.subsets-positive.min}{3}
\resultdef{primary.effect.transformed.subsets-positive.qwen}{10}
\resultdef{primary.effect.transformed.subsets-positive.sol}{3}
\resultdef{primary.effect.transformed.wrong-arithmetic.ci.luna}{[+0.3, +2.2]}
\resultdef{primary.effect.transformed.wrong-arithmetic.ci.qwen}{[+0.1, +1.2]}
\resultdef{primary.effect.transformed.wrong-arithmetic.ci.sol}{[+0.0, +0.0]}
\resultdef{primary.effect.transformed.wrong-arithmetic.luna}{+1.2}
\resultdef{primary.effect.transformed.wrong-arithmetic.max}{+1.2}
\resultdef{primary.effect.transformed.wrong-arithmetic.min}{+0.0}
\resultdef{primary.effect.transformed.wrong-arithmetic.qwen}{+0.6}
\resultdef{primary.effect.transformed.wrong-arithmetic.sol}{+0.0}
\resultdef{primary.effect.transformed.wrong-logic.ci.luna}{[+2.2, +5.1]}
\resultdef{primary.effect.transformed.wrong-logic.ci.qwen}{[+0.5, +2.8]}
\resultdef{primary.effect.transformed.wrong-logic.ci.sol}{[+0.0, +0.0]}
\resultdef{primary.effect.transformed.wrong-logic.luna}{+3.5}
\resultdef{primary.effect.transformed.wrong-logic.max}{+3.5}
\resultdef{primary.effect.transformed.wrong-logic.min}{+0.0}
\resultdef{primary.effect.transformed.wrong-logic.qwen}{+1.6}
\resultdef{primary.effect.transformed.wrong-logic.sol}{+0.0}
\resultdef{primary.effect.transformed.wrong-variable.ci.luna}{[$-$0.2, +1.7]}
\resultdef{primary.effect.transformed.wrong-variable.ci.qwen}{[$-$0.1, +1.2]}
\resultdef{primary.effect.transformed.wrong-variable.ci.sol}{[+0.0, +0.0]}
\resultdef{primary.effect.transformed.wrong-variable.luna}{+0.8}
\resultdef{primary.effect.transformed.wrong-variable.max}{+0.8}
\resultdef{primary.effect.transformed.wrong-variable.min}{+0.0}
\resultdef{primary.effect.transformed.wrong-variable.qwen}{+0.5}
\resultdef{primary.effect.transformed.wrong-variable.sol}{+0.0}
\resultdef{primary.effect.untransformed.ci.luna}{[+0.2, +1.4]}
\resultdef{primary.effect.untransformed.ci.qwen}{[$-$0.3, +1.1]}
\resultdef{primary.effect.untransformed.ci.sol}{[$-$0.3, +0.0]}
\resultdef{primary.effect.untransformed.control-flow.ci.luna}{[+0.2, +1.3]}
\resultdef{primary.effect.untransformed.control-flow.ci.qwen}{[$-$0.3, +1.0]}
\resultdef{primary.effect.untransformed.control-flow.ci.sol}{[$-$0.3, +0.0]}
\resultdef{primary.effect.untransformed.control-flow.luna}{+0.7}
\resultdef{primary.effect.untransformed.control-flow.max}{+0.7}
\resultdef{primary.effect.untransformed.control-flow.min}{$-$0.1}
\resultdef{primary.effect.untransformed.control-flow.qwen}{+0.3}
\resultdef{primary.effect.untransformed.control-flow.sol}{$-$0.1}
\resultdef{primary.effect.untransformed.distractors.ci.luna}{[+0.2, +1.3]}
\resultdef{primary.effect.untransformed.distractors.ci.qwen}{[$-$0.3, +1.0]}
\resultdef{primary.effect.untransformed.distractors.ci.sol}{[$-$0.3, +0.0]}
\resultdef{primary.effect.untransformed.distractors.luna}{+0.7}
\resultdef{primary.effect.untransformed.distractors.max}{+0.7}
\resultdef{primary.effect.untransformed.distractors.min}{$-$0.1}
\resultdef{primary.effect.untransformed.distractors.qwen}{+0.3}
\resultdef{primary.effect.untransformed.distractors.sol}{$-$0.1}
\resultdef{primary.effect.untransformed.luna}{+0.7}
\resultdef{primary.effect.untransformed.max}{+0.7}
\resultdef{primary.effect.untransformed.maximal-mix.ci.luna}{[+0.2, +1.3]}
\resultdef{primary.effect.untransformed.maximal-mix.ci.qwen}{[$-$0.3, +1.1]}
\resultdef{primary.effect.untransformed.maximal-mix.ci.sol}{[$-$0.3, +0.0]}
\resultdef{primary.effect.untransformed.maximal-mix.luna}{+0.7}
\resultdef{primary.effect.untransformed.maximal-mix.max}{+0.7}
\resultdef{primary.effect.untransformed.maximal-mix.min}{$-$0.1}
\resultdef{primary.effect.untransformed.maximal-mix.qwen}{+0.3}
\resultdef{primary.effect.untransformed.maximal-mix.sol}{$-$0.1}
\resultdef{primary.effect.untransformed.min}{$-$0.1}
\resultdef{primary.effect.untransformed.off-by-one.ci.luna}{[+0.0, +2.7]}
\resultdef{primary.effect.untransformed.off-by-one.ci.qwen}{[+0.0, +3.7]}
\resultdef{primary.effect.untransformed.off-by-one.ci.sol}{[$-$1.3, +0.0]}
\resultdef{primary.effect.untransformed.off-by-one.luna}{+1.2}
\resultdef{primary.effect.untransformed.off-by-one.max}{+1.6}
\resultdef{primary.effect.untransformed.off-by-one.min}{$-$0.4}
\resultdef{primary.effect.untransformed.off-by-one.qwen}{+1.6}
\resultdef{primary.effect.untransformed.off-by-one.sol}{$-$0.4}
\resultdef{primary.effect.untransformed.qwen}{+0.3}
\resultdef{primary.effect.untransformed.rename.ci.luna}{[+0.2, +1.4]}
\resultdef{primary.effect.untransformed.rename.ci.qwen}{[$-$0.3, +1.1]}
\resultdef{primary.effect.untransformed.rename.ci.sol}{[$-$0.3, +0.0]}
\resultdef{primary.effect.untransformed.rename.luna}{+0.8}
\resultdef{primary.effect.untransformed.rename.max}{+0.8}
\resultdef{primary.effect.untransformed.rename.min}{$-$0.1}
\resultdef{primary.effect.untransformed.rename.qwen}{+0.3}
\resultdef{primary.effect.untransformed.rename.sol}{$-$0.1}
\resultdef{primary.effect.untransformed.sol}{$-$0.1}
\resultdef{primary.effect.untransformed.state-and-literals.ci.luna}{[+0.2, +1.3]}
\resultdef{primary.effect.untransformed.state-and-literals.ci.qwen}{[$-$0.3, +1.0]}
\resultdef{primary.effect.untransformed.state-and-literals.ci.sol}{[$-$0.3, +0.0]}
\resultdef{primary.effect.untransformed.state-and-literals.luna}{+0.7}
\resultdef{primary.effect.untransformed.state-and-literals.max}{+0.7}
\resultdef{primary.effect.untransformed.state-and-literals.min}{$-$0.1}
\resultdef{primary.effect.untransformed.state-and-literals.qwen}{+0.3}
\resultdef{primary.effect.untransformed.state-and-literals.sol}{$-$0.1}
\resultdef{primary.effect.untransformed.static-return.ci.luna}{[+0.0, +2.3]}
\resultdef{primary.effect.untransformed.static-return.ci.qwen}{[$-$1.5, +0.7]}
\resultdef{primary.effect.untransformed.static-return.ci.sol}{[+0.0, +0.0]}
\resultdef{primary.effect.untransformed.static-return.luna}{+1.1}
\resultdef{primary.effect.untransformed.static-return.max}{+1.1}
\resultdef{primary.effect.untransformed.static-return.min}{$-$0.4}
\resultdef{primary.effect.untransformed.static-return.qwen}{$-$0.4}
\resultdef{primary.effect.untransformed.static-return.sol}{+0.0}
\resultdef{primary.effect.untransformed.subset-max.luna}{+1.2}
\resultdef{primary.effect.untransformed.subset-max.max}{+1.6}
\resultdef{primary.effect.untransformed.subset-max.min}{+0.0}
\resultdef{primary.effect.untransformed.subset-max.qwen}{+1.6}
\resultdef{primary.effect.untransformed.subset-max.sol}{+0.0}
\resultdef{primary.effect.untransformed.subset-min.luna}{+0.0}
\resultdef{primary.effect.untransformed.subset-min.max}{+0.0}
\resultdef{primary.effect.untransformed.subset-min.min}{$-$0.5}
\resultdef{primary.effect.untransformed.subset-min.qwen}{$-$0.5}
\resultdef{primary.effect.untransformed.subset-min.sol}{$-$0.4}
\resultdef{primary.effect.untransformed.wrong-arithmetic.ci.luna}{[+0.0, +1.5]}
\resultdef{primary.effect.untransformed.wrong-arithmetic.ci.qwen}{[$-$1.6, +0.0]}
\resultdef{primary.effect.untransformed.wrong-arithmetic.ci.sol}{[+0.0, +0.0]}
\resultdef{primary.effect.untransformed.wrong-arithmetic.luna}{+0.5}
\resultdef{primary.effect.untransformed.wrong-arithmetic.max}{+0.5}
\resultdef{primary.effect.untransformed.wrong-arithmetic.min}{$-$0.5}
\resultdef{primary.effect.untransformed.wrong-arithmetic.qwen}{$-$0.5}
\resultdef{primary.effect.untransformed.wrong-arithmetic.sol}{+0.0}
\resultdef{primary.effect.untransformed.wrong-logic.ci.luna}{[$-$0.5, +2.9]}
\resultdef{primary.effect.untransformed.wrong-logic.ci.qwen}{[$-$1.4, +3.8]}
\resultdef{primary.effect.untransformed.wrong-logic.ci.sol}{[+0.0, +0.0]}
\resultdef{primary.effect.untransformed.wrong-logic.luna}{+0.9}
\resultdef{primary.effect.untransformed.wrong-logic.max}{+0.9}
\resultdef{primary.effect.untransformed.wrong-logic.min}{+0.0}
\resultdef{primary.effect.untransformed.wrong-logic.qwen}{+0.9}
\resultdef{primary.effect.untransformed.wrong-logic.sol}{+0.0}
\resultdef{primary.effect.untransformed.wrong-variable.ci.luna}{[$-$1.1, +1.1]}
\resultdef{primary.effect.untransformed.wrong-variable.ci.qwen}{[+0.0, +0.0]}
\resultdef{primary.effect.untransformed.wrong-variable.ci.sol}{[+0.0, +0.0]}
\resultdef{primary.effect.untransformed.wrong-variable.luna}{+0.0}
\resultdef{primary.effect.untransformed.wrong-variable.max}{+0.0}
\resultdef{primary.effect.untransformed.wrong-variable.min}{+0.0}
\resultdef{primary.effect.untransformed.wrong-variable.qwen}{+0.0}
\resultdef{primary.effect.untransformed.wrong-variable.sol}{+0.0}
\resultdef{primary.enc-minus-disc.transformed.luna}{+0.6}
\resultdef{primary.enc-minus-disc.transformed.max}{+0.6}
\resultdef{primary.enc-minus-disc.transformed.min}{+0.0}
\resultdef{primary.enc-minus-disc.transformed.qwen}{+0.0}
\resultdef{primary.enc-minus-disc.transformed.sol}{+0.0}
\resultdef{primary.enc-minus-disc.untransformed.luna}{+0.8}
\resultdef{primary.enc-minus-disc.untransformed.max}{+0.8}
\resultdef{primary.enc-minus-disc.untransformed.min}{+0.1}
\resultdef{primary.enc-minus-disc.untransformed.qwen}{+0.2}
\resultdef{primary.enc-minus-disc.untransformed.sol}{+0.1}
\resultdef{primary.error-reduction.transformed.control-flow.luna}{79}
\resultdef{primary.error-reduction.transformed.control-flow.max}{79}
\resultdef{primary.error-reduction.transformed.control-flow.min}{77}
\resultdef{primary.error-reduction.transformed.control-flow.qwen}{77}
\resultdef{primary.error-reduction.transformed.distractors.luna}{29}
\resultdef{primary.error-reduction.transformed.distractors.max}{31}
\resultdef{primary.error-reduction.transformed.distractors.min}{29}
\resultdef{primary.error-reduction.transformed.distractors.qwen}{31}
\resultdef{primary.error-reduction.transformed.luna}{63}
\resultdef{primary.error-reduction.transformed.max}{63}
\resultdef{primary.error-reduction.transformed.maximal-mix.luna}{78}
\resultdef{primary.error-reduction.transformed.maximal-mix.max}{100}
\resultdef{primary.error-reduction.transformed.maximal-mix.min}{65}
\resultdef{primary.error-reduction.transformed.maximal-mix.qwen}{65}
\resultdef{primary.error-reduction.transformed.maximal-mix.sol}{100}
\resultdef{primary.error-reduction.transformed.min}{33}
\resultdef{primary.error-reduction.transformed.off-by-one.luna}{51}
\resultdef{primary.error-reduction.transformed.off-by-one.max}{60}
\resultdef{primary.error-reduction.transformed.off-by-one.min}{0}
\resultdef{primary.error-reduction.transformed.off-by-one.qwen}{60}
\resultdef{primary.error-reduction.transformed.off-by-one.sol}{0}
\resultdef{primary.error-reduction.transformed.qwen}{61}
\resultdef{primary.error-reduction.transformed.rename.luna}{45}
\resultdef{primary.error-reduction.transformed.rename.max}{50}
\resultdef{primary.error-reduction.transformed.rename.min}{0}
\resultdef{primary.error-reduction.transformed.rename.qwen}{50}
\resultdef{primary.error-reduction.transformed.rename.sol}{0}
\resultdef{primary.error-reduction.transformed.sol}{33}
\resultdef{primary.error-reduction.transformed.state-and-literals.luna}{57}
\resultdef{primary.error-reduction.transformed.state-and-literals.max}{57}
\resultdef{primary.error-reduction.transformed.state-and-literals.min}{33}
\resultdef{primary.error-reduction.transformed.state-and-literals.qwen}{43}
\resultdef{primary.error-reduction.transformed.state-and-literals.sol}{33}
\resultdef{primary.error-reduction.transformed.static-return.luna}{87}
\resultdef{primary.error-reduction.transformed.static-return.max}{100}
\resultdef{primary.error-reduction.transformed.static-return.min}{66}
\resultdef{primary.error-reduction.transformed.static-return.qwen}{66}
\resultdef{primary.error-reduction.transformed.static-return.sol}{100}
\resultdef{primary.error-reduction.transformed.wrong-arithmetic.luna}{86}
\resultdef{primary.error-reduction.transformed.wrong-arithmetic.max}{86}
\resultdef{primary.error-reduction.transformed.wrong-arithmetic.min}{86}
\resultdef{primary.error-reduction.transformed.wrong-arithmetic.qwen}{86}
\resultdef{primary.error-reduction.transformed.wrong-logic.luna}{69}
\resultdef{primary.error-reduction.transformed.wrong-logic.max}{69}
\resultdef{primary.error-reduction.transformed.wrong-logic.min}{55}
\resultdef{primary.error-reduction.transformed.wrong-logic.qwen}{55}
\resultdef{primary.error-reduction.transformed.wrong-variable.luna}{45}
\resultdef{primary.error-reduction.transformed.wrong-variable.max}{50}
\resultdef{primary.error-reduction.transformed.wrong-variable.min}{45}
\resultdef{primary.error-reduction.transformed.wrong-variable.qwen}{50}
\resultdef{primary.error-reduction.untransformed.control-flow.luna}{41}
\resultdef{primary.error-reduction.untransformed.control-flow.max}{41}
\resultdef{primary.error-reduction.untransformed.control-flow.min}{-100}
\resultdef{primary.error-reduction.untransformed.control-flow.qwen}{27}
\resultdef{primary.error-reduction.untransformed.control-flow.sol}{-100}
\resultdef{primary.error-reduction.untransformed.distractors.luna}{41}
\resultdef{primary.error-reduction.untransformed.distractors.max}{41}
\resultdef{primary.error-reduction.untransformed.distractors.min}{-100}
\resultdef{primary.error-reduction.untransformed.distractors.qwen}{27}
\resultdef{primary.error-reduction.untransformed.distractors.sol}{-100}
\resultdef{primary.error-reduction.untransformed.luna}{41}
\resultdef{primary.error-reduction.untransformed.max}{41}
\resultdef{primary.error-reduction.untransformed.maximal-mix.luna}{41}
\resultdef{primary.error-reduction.untransformed.maximal-mix.max}{41}
\resultdef{primary.error-reduction.untransformed.maximal-mix.min}{-100}
\resultdef{primary.error-reduction.untransformed.maximal-mix.qwen}{27}
\resultdef{primary.error-reduction.untransformed.maximal-mix.sol}{-100}
\resultdef{primary.error-reduction.untransformed.min}{-100}
\resultdef{primary.error-reduction.untransformed.off-by-one.luna}{33}
\resultdef{primary.error-reduction.untransformed.off-by-one.max}{57}
\resultdef{primary.error-reduction.untransformed.off-by-one.min}{-100}
\resultdef{primary.error-reduction.untransformed.off-by-one.qwen}{57}
\resultdef{primary.error-reduction.untransformed.off-by-one.sol}{-100}
\resultdef{primary.error-reduction.untransformed.qwen}{27}
\resultdef{primary.error-reduction.untransformed.rename.luna}{41}
\resultdef{primary.error-reduction.untransformed.rename.max}{41}
\resultdef{primary.error-reduction.untransformed.rename.min}{-100}
\resultdef{primary.error-reduction.untransformed.rename.qwen}{27}
\resultdef{primary.error-reduction.untransformed.rename.sol}{-100}
\resultdef{primary.error-reduction.untransformed.sol}{-100}
\resultdef{primary.error-reduction.untransformed.state-and-literals.luna}{41}
\resultdef{primary.error-reduction.untransformed.state-and-literals.max}{41}
\resultdef{primary.error-reduction.untransformed.state-and-literals.min}{-100}
\resultdef{primary.error-reduction.untransformed.state-and-literals.qwen}{27}
\resultdef{primary.error-reduction.untransformed.state-and-literals.sol}{-100}
\resultdef{primary.error-reduction.untransformed.static-return.luna}{100}
\resultdef{primary.error-reduction.untransformed.static-return.max}{100}
\resultdef{primary.error-reduction.untransformed.static-return.min}{-50}
\resultdef{primary.error-reduction.untransformed.static-return.qwen}{-50}
\resultdef{primary.error-reduction.untransformed.wrong-arithmetic.luna}{100}
\resultdef{primary.error-reduction.untransformed.wrong-logic.luna}{40}
\resultdef{primary.error-reduction.untransformed.wrong-logic.max}{40}
\resultdef{primary.error-reduction.untransformed.wrong-logic.min}{33}
\resultdef{primary.error-reduction.untransformed.wrong-logic.qwen}{33}
\resultdef{primary.error-reduction.untransformed.wrong-variable.luna}{0}
\resultdef{primary.extra-correct.minutes.luna}{31.3}
\resultdef{primary.extra-correct.minutes.max}{366.7}
\resultdef{primary.extra-correct.minutes.min}{31.3}
\resultdef{primary.extra-correct.minutes.qwen}{366.7}
\resultdef{primary.extra-correct.usd.luna}{0.06}
\resultdef{primary.extra-correct.usd.max}{1.43}
\resultdef{primary.extra-correct.usd.min}{0.06}
\resultdef{primary.extra-correct.usd.qwen}{1.43}
\resultdef{primary.isolation.diff.ci.qwen}{[+0.1, +1.7]}
\resultdef{primary.isolation.diff.qwen}{+1.0}
\resultdef{primary.isolation.effect.transformed.ci.qwen}{[+0.8, +1.8]}
\resultdef{primary.isolation.effect.transformed.qwen}{+1.3}
\resultdef{primary.isolation.effect.untransformed.ci.qwen}{[$-$0.3, +1.1]}
\resultdef{primary.isolation.effect.untransformed.qwen}{+0.3}
\resultdef{primary.isolation.excluded.qwen}{14}
\resultdef{primary.isolation.slope.debugger.transformed.qwen}{+2.7}
\resultdef{primary.isolation.slope.probe.transformed.qwen}{+5.3}
\resultdef{primary.slope.debugger.transformed.ci.sol}{[$-$0.2, +0.0]}
\resultdef{primary.slope.debugger.transformed.luna}{+66.7}
\resultdef{primary.slope.debugger.transformed.max}{+66.7}
\resultdef{primary.slope.debugger.transformed.min}{$-$0.1}
\resultdef{primary.slope.debugger.transformed.placebo-miss.luna}{36.2}
\resultdef{primary.slope.debugger.transformed.placebo-miss.max}{36.2}
\resultdef{primary.slope.debugger.transformed.placebo-miss.min}{0.3}
\resultdef{primary.slope.debugger.transformed.placebo-miss.qwen}{6.7}
\resultdef{primary.slope.debugger.transformed.placebo-miss.sol}{0.3}
\resultdef{primary.slope.debugger.transformed.qwen}{+2.7}
\resultdef{primary.slope.debugger.transformed.sol}{$-$0.1}
\resultdef{primary.slope.debugger.untransformed.ci.qwen}{[$-$2.1, +0.7]}
\resultdef{primary.slope.debugger.untransformed.placebo-miss.qwen}{0.3}
\resultdef{primary.slope.debugger.untransformed.qwen}{$-$0.7}
\resultdef{primary.slope.probe.transformed.ci.qwen}{[+3.2, +7.5]}
\resultdef{primary.slope.probe.transformed.ci.sol}{[+0.0, +0.1]}
\resultdef{primary.slope.probe.transformed.luna}{+15.0}
\resultdef{primary.slope.probe.transformed.max}{+15.0}
\resultdef{primary.slope.probe.transformed.min}{+0.0}
\resultdef{primary.slope.probe.transformed.placebo-miss.luna}{6.6}
\resultdef{primary.slope.probe.transformed.placebo-miss.max}{6.6}
\resultdef{primary.slope.probe.transformed.placebo-miss.min}{4.2}
\resultdef{primary.slope.probe.transformed.placebo-miss.qwen}{4.2}
\resultdef{primary.slope.probe.transformed.placebo-miss.sol}{5.1}
\resultdef{primary.slope.probe.transformed.qwen}{+5.3}
\resultdef{primary.slope.probe.transformed.sol}{+0.0}
\resultdef{primary.slope.probe.untransformed.ci.qwen}{[$-$1.0, +5.5]}
\resultdef{primary.slope.probe.untransformed.luna}{+3.2}
\resultdef{primary.slope.probe.untransformed.max}{+3.2}
\resultdef{primary.slope.probe.untransformed.min}{$-$1.1}
\resultdef{primary.slope.probe.untransformed.placebo-miss.luna}{14.9}
\resultdef{primary.slope.probe.untransformed.placebo-miss.max}{71.2}
\resultdef{primary.slope.probe.untransformed.placebo-miss.min}{3.2}
\resultdef{primary.slope.probe.untransformed.placebo-miss.qwen}{3.2}
\resultdef{primary.slope.probe.untransformed.placebo-miss.sol}{71.2}
\resultdef{primary.slope.probe.untransformed.qwen}{+1.9}
\resultdef{primary.slope.probe.untransformed.sol}{$-$1.1}
\resultdef{programs.faulty.admitted}{610}
\resultdef{programs.faulty.admitted.off-by-one}{125}
\resultdef{programs.faulty.admitted.static-return}{141}
\resultdef{programs.faulty.admitted.wrong-arithmetic}{104}
\resultdef{programs.faulty.admitted.wrong-logic}{108}
\resultdef{programs.faulty.admitted.wrong-variable}{132}
\resultdef{programs.faulty.attempted}{750}
\resultdef{programs.faulty.cc.median.off-by-one}{5}
\resultdef{programs.faulty.cc.median.static-return}{4}
\resultdef{programs.faulty.cc.median.wrong-arithmetic}{5}
\resultdef{programs.faulty.cc.median.wrong-logic}{5}
\resultdef{programs.faulty.cc.median.wrong-variable}{4}
\resultdef{programs.faulty.loc.median.off-by-one}{13}
\resultdef{programs.faulty.loc.median.static-return}{12}
\resultdef{programs.faulty.loc.median.wrong-arithmetic}{13}
\resultdef{programs.faulty.loc.median.wrong-logic}{14}
\resultdef{programs.faulty.loc.median.wrong-variable}{12}
\resultdef{programs.faulty.multi-line-statements}{16}
\resultdef{programs.reference.admitted}{144}
\resultdef{programs.reference.admitted.easy}{50}
\resultdef{programs.reference.admitted.hard}{46}
\resultdef{programs.reference.admitted.medium}{48}
\resultdef{programs.reference.attempted}{150}
\resultdef{programs.reference.cc.median.easy}{3}
\resultdef{programs.reference.cc.median.hard}{8}
\resultdef{programs.reference.cc.median.medium}{4}
\resultdef{programs.reference.loc.median.easy}{8}
\resultdef{programs.reference.loc.median.hard}{19}
\resultdef{programs.reference.loc.median.medium}{11}
\resultdef{programs.sampled}{150}
\resultdef{programs.sampled.description-words.median}{176}
\resultdef{programs.sampled.multiple-functions-or-stateful.count}{31}
\resultdef{programs.sampled.multiple-functions-or-stateful.share}{20.7}
\resultdef{programs.sampled.release.after-split}{2}
\resultdef{programs.sampled.release.first}{August~7, 2015}
\resultdef{programs.sampled.release.last}{July~28, 2024}
\resultdef{programs.sampled.test-assertions.median}{99}
\resultdef{programs.transformed.admitted}{3{,}022}
\resultdef{programs.transformed.admitted.control-flow}{610}
\resultdef{programs.transformed.admitted.distractors}{610}
\resultdef{programs.transformed.admitted.maximal-mix}{608}
\resultdef{programs.transformed.admitted.rename}{584}
\resultdef{programs.transformed.admitted.state-and-literals}{610}
\resultdef{programs.transformed.attempted}{3{,}050}
\resultdef{programs.transformed.cc-change.median.control-flow}{1}
\resultdef{programs.transformed.cc-change.median.distractors}{11}
\resultdef{programs.transformed.cc-change.median.maximal-mix}{5}
\resultdef{programs.transformed.cc-change.median.rename}{0}
\resultdef{programs.transformed.cc-change.median.state-and-literals}{0}
\resultdef{programs.transformed.double-negation-share.control-flow}{7.8}
\resultdef{programs.transformed.double-negation-share.distractors}{0.0}
\resultdef{programs.transformed.double-negation-share.maximal-mix}{4.4}
\resultdef{programs.transformed.double-negation-share.rename}{0.0}
\resultdef{programs.transformed.double-negation-share.state-and-literals}{0.0}
\resultdef{programs.transformed.kept.median.control-flow}{51}
\resultdef{programs.transformed.kept.median.distractors}{71}
\resultdef{programs.transformed.kept.median.maximal-mix}{37}
\resultdef{programs.transformed.kept.median.rename}{40}
\resultdef{programs.transformed.kept.median.state-and-literals}{49}
\resultdef{programs.transformed.loc-change-percent.median.control-flow}{169}
\resultdef{programs.transformed.loc-change-percent.median.distractors}{400}
\resultdef{programs.transformed.loc-change-percent.median.maximal-mix}{450}
\resultdef{programs.transformed.loc-change-percent.median.rename}{0}
\resultdef{programs.transformed.loc-change-percent.median.state-and-literals}{71}
\resultdef{programs.transformed.spread-share}{47}
\resultdef{programs.transformed.spread-share.control-flow}{74}
\resultdef{programs.transformed.spread-share.distractors}{13}
\resultdef{programs.transformed.spread-share.maximal-mix}{75}
\resultdef{programs.transformed.spread-share.rename}{0}
\resultdef{programs.transformed.spread-share.state-and-literals}{70}
\resultdef{programs.transformed.statement-lines.max}{34}
\resultdef{secondary.acc.gap.discretionary-tools.ci.luna}{[$-$7.4, $-$4.8]}
\resultdef{secondary.acc.gap.discretionary-tools.ci.qwen}{[$-$1.5, $-$0.3]}
\resultdef{secondary.acc.gap.discretionary-tools.ci.sol}{[$-$0.4, +0.2]}
\resultdef{secondary.acc.gap.discretionary-tools.control-flow.ci.luna}{[$-$13.2, $-$8.1]}
\resultdef{secondary.acc.gap.discretionary-tools.control-flow.ci.qwen}{[$-$2.7, $-$0.5]}
\resultdef{secondary.acc.gap.discretionary-tools.control-flow.ci.sol}{[$-$0.6, +0.4]}
\resultdef{secondary.acc.gap.discretionary-tools.control-flow.luna}{$-$10.6}
\resultdef{secondary.acc.gap.discretionary-tools.control-flow.max}{$-$0.1}
\resultdef{secondary.acc.gap.discretionary-tools.control-flow.min}{$-$10.6}
\resultdef{secondary.acc.gap.discretionary-tools.control-flow.qwen}{$-$1.6}
\resultdef{secondary.acc.gap.discretionary-tools.control-flow.sol}{$-$0.1}
\resultdef{secondary.acc.gap.discretionary-tools.distractors.ci.luna}{[$-$2.9, $-$0.2]}
\resultdef{secondary.acc.gap.discretionary-tools.distractors.ci.qwen}{[$-$0.9, +0.6]}
\resultdef{secondary.acc.gap.discretionary-tools.distractors.ci.sol}{[+0.0, +0.5]}
\resultdef{secondary.acc.gap.discretionary-tools.distractors.luna}{$-$1.5}
\resultdef{secondary.acc.gap.discretionary-tools.distractors.max}{+0.2}
\resultdef{secondary.acc.gap.discretionary-tools.distractors.min}{$-$1.5}
\resultdef{secondary.acc.gap.discretionary-tools.distractors.qwen}{$-$0.2}
\resultdef{secondary.acc.gap.discretionary-tools.distractors.sol}{+0.2}
\resultdef{secondary.acc.gap.discretionary-tools.luna}{$-$6.1}
\resultdef{secondary.acc.gap.discretionary-tools.max}{$-$0.1}
\resultdef{secondary.acc.gap.discretionary-tools.maximal-mix.ci.luna}{[$-$13.8, $-$8.7]}
\resultdef{secondary.acc.gap.discretionary-tools.maximal-mix.ci.qwen}{[$-$2.6, $-$0.6]}
\resultdef{secondary.acc.gap.discretionary-tools.maximal-mix.ci.sol}{[$-$0.7, +0.3]}
\resultdef{secondary.acc.gap.discretionary-tools.maximal-mix.luna}{$-$11.3}
\resultdef{secondary.acc.gap.discretionary-tools.maximal-mix.max}{$-$0.2}
\resultdef{secondary.acc.gap.discretionary-tools.maximal-mix.min}{$-$11.3}
\resultdef{secondary.acc.gap.discretionary-tools.maximal-mix.qwen}{$-$1.6}
\resultdef{secondary.acc.gap.discretionary-tools.maximal-mix.sol}{$-$0.2}
\resultdef{secondary.acc.gap.discretionary-tools.min}{$-$6.1}
\resultdef{secondary.acc.gap.discretionary-tools.off-by-one.ci.luna}{[$-$11.0, $-$4.7]}
\resultdef{secondary.acc.gap.discretionary-tools.off-by-one.ci.qwen}{[$-$1.4, +2.6]}
\resultdef{secondary.acc.gap.discretionary-tools.off-by-one.ci.sol}{[$-$0.8, +1.5]}
\resultdef{secondary.acc.gap.discretionary-tools.off-by-one.luna}{$-$7.7}
\resultdef{secondary.acc.gap.discretionary-tools.off-by-one.max}{+0.3}
\resultdef{secondary.acc.gap.discretionary-tools.off-by-one.min}{$-$7.7}
\resultdef{secondary.acc.gap.discretionary-tools.off-by-one.qwen}{+0.3}
\resultdef{secondary.acc.gap.discretionary-tools.off-by-one.sol}{+0.2}
\resultdef{secondary.acc.gap.discretionary-tools.qwen}{$-$0.9}
\resultdef{secondary.acc.gap.discretionary-tools.rename.ci.luna}{[$-$0.5, +1.2]}
\resultdef{secondary.acc.gap.discretionary-tools.rename.ci.qwen}{[$-$0.2, +1.1]}
\resultdef{secondary.acc.gap.discretionary-tools.rename.ci.sol}{[+0.0, +0.4]}
\resultdef{secondary.acc.gap.discretionary-tools.rename.luna}{+0.3}
\resultdef{secondary.acc.gap.discretionary-tools.rename.max}{+0.4}
\resultdef{secondary.acc.gap.discretionary-tools.rename.min}{+0.2}
\resultdef{secondary.acc.gap.discretionary-tools.rename.qwen}{+0.4}
\resultdef{secondary.acc.gap.discretionary-tools.rename.sol}{+0.2}
\resultdef{secondary.acc.gap.discretionary-tools.sol}{$-$0.1}
\resultdef{secondary.acc.gap.discretionary-tools.state-and-literals.ci.luna}{[$-$9.4, $-$5.5]}
\resultdef{secondary.acc.gap.discretionary-tools.state-and-literals.ci.qwen}{[$-$2.5, $-$0.5]}
\resultdef{secondary.acc.gap.discretionary-tools.state-and-literals.ci.sol}{[$-$1.3, $-$0.2]}
\resultdef{secondary.acc.gap.discretionary-tools.state-and-literals.luna}{$-$7.4}
\resultdef{secondary.acc.gap.discretionary-tools.state-and-literals.max}{$-$0.7}
\resultdef{secondary.acc.gap.discretionary-tools.state-and-literals.min}{$-$7.4}
\resultdef{secondary.acc.gap.discretionary-tools.state-and-literals.qwen}{$-$1.5}
\resultdef{secondary.acc.gap.discretionary-tools.state-and-literals.sol}{$-$0.7}
\resultdef{secondary.acc.gap.discretionary-tools.static-return.ci.luna}{[$-$14.2, $-$8.8]}
\resultdef{secondary.acc.gap.discretionary-tools.static-return.ci.qwen}{[$-$4.8, $-$1.7]}
\resultdef{secondary.acc.gap.discretionary-tools.static-return.ci.sol}{[$-$0.2, +0.2]}
\resultdef{secondary.acc.gap.discretionary-tools.static-return.luna}{$-$11.4}
\resultdef{secondary.acc.gap.discretionary-tools.static-return.max}{+0.0}
\resultdef{secondary.acc.gap.discretionary-tools.static-return.min}{$-$11.4}
\resultdef{secondary.acc.gap.discretionary-tools.static-return.qwen}{$-$3.3}
\resultdef{secondary.acc.gap.discretionary-tools.static-return.sol}{+0.0}
\resultdef{secondary.acc.gap.discretionary-tools.wrong-arithmetic.ci.luna}{[$-$3.3, $-$0.7]}
\resultdef{secondary.acc.gap.discretionary-tools.wrong-arithmetic.ci.qwen}{[$-$1.0, $-$0.1]}
\resultdef{secondary.acc.gap.discretionary-tools.wrong-arithmetic.ci.sol}{[$-$1.8, +0.0]}
\resultdef{secondary.acc.gap.discretionary-tools.wrong-arithmetic.luna}{$-$1.8}
\resultdef{secondary.acc.gap.discretionary-tools.wrong-arithmetic.max}{$-$0.5}
\resultdef{secondary.acc.gap.discretionary-tools.wrong-arithmetic.min}{$-$1.8}
\resultdef{secondary.acc.gap.discretionary-tools.wrong-arithmetic.qwen}{$-$0.5}
\resultdef{secondary.acc.gap.discretionary-tools.wrong-arithmetic.sol}{$-$0.7}
\resultdef{secondary.acc.gap.discretionary-tools.wrong-logic.ci.luna}{[$-$1.9, +2.8]}
\resultdef{secondary.acc.gap.discretionary-tools.wrong-logic.ci.qwen}{[$-$0.7, +0.8]}
\resultdef{secondary.acc.gap.discretionary-tools.wrong-logic.ci.sol}{[+0.0, +0.0]}
\resultdef{secondary.acc.gap.discretionary-tools.wrong-logic.luna}{+0.4}
\resultdef{secondary.acc.gap.discretionary-tools.wrong-logic.max}{+0.4}
\resultdef{secondary.acc.gap.discretionary-tools.wrong-logic.min}{+0.0}
\resultdef{secondary.acc.gap.discretionary-tools.wrong-logic.qwen}{+0.0}
\resultdef{secondary.acc.gap.discretionary-tools.wrong-logic.sol}{+0.0}
\resultdef{secondary.acc.gap.discretionary-tools.wrong-variable.ci.luna}{[$-$11.5, $-$4.0]}
\resultdef{secondary.acc.gap.discretionary-tools.wrong-variable.ci.qwen}{[$-$1.7, +0.7]}
\resultdef{secondary.acc.gap.discretionary-tools.wrong-variable.ci.sol}{[$-$0.5, +0.0]}
\resultdef{secondary.acc.gap.discretionary-tools.wrong-variable.luna}{$-$7.7}
\resultdef{secondary.acc.gap.discretionary-tools.wrong-variable.max}{$-$0.2}
\resultdef{secondary.acc.gap.discretionary-tools.wrong-variable.min}{$-$7.7}
\resultdef{secondary.acc.gap.discretionary-tools.wrong-variable.qwen}{$-$0.5}
\resultdef{secondary.acc.gap.discretionary-tools.wrong-variable.sol}{$-$0.2}
\resultdef{secondary.acc.gap.encouraged-tools.ci.luna}{[$-$7.4, $-$5.0]}
\resultdef{secondary.acc.gap.encouraged-tools.ci.qwen}{[$-$1.5, +0.0]}
\resultdef{secondary.acc.gap.encouraged-tools.ci.sol}{[$-$0.6, +0.1]}
\resultdef{secondary.acc.gap.encouraged-tools.control-flow.ci.luna}{[$-$13.0, $-$8.5]}
\resultdef{secondary.acc.gap.encouraged-tools.control-flow.ci.qwen}{[$-$2.0, $-$0.1]}
\resultdef{secondary.acc.gap.encouraged-tools.control-flow.ci.sol}{[$-$1.2, +0.2]}
\resultdef{secondary.acc.gap.encouraged-tools.control-flow.luna}{$-$10.7}
\resultdef{secondary.acc.gap.encouraged-tools.control-flow.max}{$-$0.5}
\resultdef{secondary.acc.gap.encouraged-tools.control-flow.min}{$-$10.7}
\resultdef{secondary.acc.gap.encouraged-tools.control-flow.qwen}{$-$1.1}
\resultdef{secondary.acc.gap.encouraged-tools.control-flow.sol}{$-$0.5}
\resultdef{secondary.acc.gap.encouraged-tools.distractors.ci.luna}{[$-$3.3, $-$1.0]}
\resultdef{secondary.acc.gap.encouraged-tools.distractors.ci.qwen}{[$-$1.2, +0.5]}
\resultdef{secondary.acc.gap.encouraged-tools.distractors.ci.sol}{[$-$0.2, +0.5]}
\resultdef{secondary.acc.gap.encouraged-tools.distractors.luna}{$-$2.0}
\resultdef{secondary.acc.gap.encouraged-tools.distractors.max}{+0.1}
\resultdef{secondary.acc.gap.encouraged-tools.distractors.min}{$-$2.0}
\resultdef{secondary.acc.gap.encouraged-tools.distractors.qwen}{$-$0.3}
\resultdef{secondary.acc.gap.encouraged-tools.distractors.sol}{+0.1}
\resultdef{secondary.acc.gap.encouraged-tools.luna}{$-$6.2}
\resultdef{secondary.acc.gap.encouraged-tools.max}{$-$0.2}
\resultdef{secondary.acc.gap.encouraged-tools.maximal-mix.ci.luna}{[$-$13.4, $-$8.8]}
\resultdef{secondary.acc.gap.encouraged-tools.maximal-mix.ci.qwen}{[$-$3.8, $-$1.2]}
\resultdef{secondary.acc.gap.encouraged-tools.maximal-mix.ci.sol}{[$-$1.1, +0.2]}
\resultdef{secondary.acc.gap.encouraged-tools.maximal-mix.luna}{$-$11.1}
\resultdef{secondary.acc.gap.encouraged-tools.maximal-mix.max}{$-$0.5}
\resultdef{secondary.acc.gap.encouraged-tools.maximal-mix.min}{$-$11.1}
\resultdef{secondary.acc.gap.encouraged-tools.maximal-mix.qwen}{$-$2.5}
\resultdef{secondary.acc.gap.encouraged-tools.maximal-mix.sol}{$-$0.5}
\resultdef{secondary.acc.gap.encouraged-tools.min}{$-$6.2}
\resultdef{secondary.acc.gap.encouraged-tools.off-by-one.ci.luna}{[$-$9.9, $-$5.1]}
\resultdef{secondary.acc.gap.encouraged-tools.off-by-one.ci.qwen}{[$-$2.1, +1.4]}
\resultdef{secondary.acc.gap.encouraged-tools.off-by-one.ci.sol}{[$-$0.8, +1.4]}
\resultdef{secondary.acc.gap.encouraged-tools.off-by-one.luna}{$-$7.3}
\resultdef{secondary.acc.gap.encouraged-tools.off-by-one.max}{+0.2}
\resultdef{secondary.acc.gap.encouraged-tools.off-by-one.min}{$-$7.3}
\resultdef{secondary.acc.gap.encouraged-tools.off-by-one.qwen}{$-$0.6}
\resultdef{secondary.acc.gap.encouraged-tools.off-by-one.sol}{+0.2}
\resultdef{secondary.acc.gap.encouraged-tools.qwen}{$-$0.8}
\resultdef{secondary.acc.gap.encouraged-tools.rename.ci.luna}{[$-$0.3, +0.5]}
\resultdef{secondary.acc.gap.encouraged-tools.rename.ci.qwen}{[$-$0.2, +1.1]}
\resultdef{secondary.acc.gap.encouraged-tools.rename.ci.sol}{[+0.0, +0.0]}
\resultdef{secondary.acc.gap.encouraged-tools.rename.luna}{+0.1}
\resultdef{secondary.acc.gap.encouraged-tools.rename.max}{+0.3}
\resultdef{secondary.acc.gap.encouraged-tools.rename.min}{+0.0}
\resultdef{secondary.acc.gap.encouraged-tools.rename.qwen}{+0.3}
\resultdef{secondary.acc.gap.encouraged-tools.rename.sol}{+0.0}
\resultdef{secondary.acc.gap.encouraged-tools.sol}{$-$0.2}
\resultdef{secondary.acc.gap.encouraged-tools.state-and-literals.ci.luna}{[$-$8.7, $-$5.2]}
\resultdef{secondary.acc.gap.encouraged-tools.state-and-literals.ci.qwen}{[$-$1.3, +0.2]}
\resultdef{secondary.acc.gap.encouraged-tools.state-and-literals.ci.sol}{[$-$0.8, +0.0]}
\resultdef{secondary.acc.gap.encouraged-tools.state-and-literals.luna}{$-$7.0}
\resultdef{secondary.acc.gap.encouraged-tools.state-and-literals.max}{$-$0.3}
\resultdef{secondary.acc.gap.encouraged-tools.state-and-literals.min}{$-$7.0}
\resultdef{secondary.acc.gap.encouraged-tools.state-and-literals.qwen}{$-$0.6}
\resultdef{secondary.acc.gap.encouraged-tools.state-and-literals.sol}{$-$0.3}
\resultdef{secondary.acc.gap.encouraged-tools.static-return.ci.luna}{[$-$12.9, $-$8.5]}
\resultdef{secondary.acc.gap.encouraged-tools.static-return.ci.qwen}{[$-$2.4, $-$0.3]}
\resultdef{secondary.acc.gap.encouraged-tools.static-return.ci.sol}{[$-$0.4, +0.0]}
\resultdef{secondary.acc.gap.encouraged-tools.static-return.luna}{$-$10.6}
\resultdef{secondary.acc.gap.encouraged-tools.static-return.max}{$-$0.1}
\resultdef{secondary.acc.gap.encouraged-tools.static-return.min}{$-$10.6}
\resultdef{secondary.acc.gap.encouraged-tools.static-return.qwen}{$-$1.4}
\resultdef{secondary.acc.gap.encouraged-tools.static-return.sol}{$-$0.1}
\resultdef{secondary.acc.gap.encouraged-tools.wrong-arithmetic.ci.luna}{[$-$2.4, $-$0.4]}
\resultdef{secondary.acc.gap.encouraged-tools.wrong-arithmetic.ci.qwen}{[$-$0.6, +1.4]}
\resultdef{secondary.acc.gap.encouraged-tools.wrong-arithmetic.ci.sol}{[$-$2.1, +0.0]}
\resultdef{secondary.acc.gap.encouraged-tools.wrong-arithmetic.luna}{$-$1.3}
\resultdef{secondary.acc.gap.encouraged-tools.wrong-arithmetic.max}{+0.2}
\resultdef{secondary.acc.gap.encouraged-tools.wrong-arithmetic.min}{$-$1.3}
\resultdef{secondary.acc.gap.encouraged-tools.wrong-arithmetic.qwen}{+0.2}
\resultdef{secondary.acc.gap.encouraged-tools.wrong-arithmetic.sol}{$-$0.9}
\resultdef{secondary.acc.gap.encouraged-tools.wrong-logic.ci.luna}{[$-$1.3, +1.0]}
\resultdef{secondary.acc.gap.encouraged-tools.wrong-logic.ci.qwen}{[$-$0.9, +2.4]}
\resultdef{secondary.acc.gap.encouraged-tools.wrong-logic.ci.sol}{[+0.0, +0.0]}
\resultdef{secondary.acc.gap.encouraged-tools.wrong-logic.luna}{$-$0.2}
\resultdef{secondary.acc.gap.encouraged-tools.wrong-logic.max}{+0.6}
\resultdef{secondary.acc.gap.encouraged-tools.wrong-logic.min}{$-$0.2}
\resultdef{secondary.acc.gap.encouraged-tools.wrong-logic.qwen}{+0.6}
\resultdef{secondary.acc.gap.encouraged-tools.wrong-logic.sol}{+0.0}
\resultdef{secondary.acc.gap.encouraged-tools.wrong-variable.ci.luna}{[$-$12.8, $-$6.0]}
\resultdef{secondary.acc.gap.encouraged-tools.wrong-variable.ci.qwen}{[$-$4.1, $-$1.0]}
\resultdef{secondary.acc.gap.encouraged-tools.wrong-variable.ci.sol}{[$-$1.3, +0.0]}
\resultdef{secondary.acc.gap.encouraged-tools.wrong-variable.luna}{$-$9.4}
\resultdef{secondary.acc.gap.encouraged-tools.wrong-variable.max}{$-$0.5}
\resultdef{secondary.acc.gap.encouraged-tools.wrong-variable.min}{$-$9.4}
\resultdef{secondary.acc.gap.encouraged-tools.wrong-variable.qwen}{$-$2.4}
\resultdef{secondary.acc.gap.encouraged-tools.wrong-variable.sol}{$-$0.5}
\resultdef{secondary.acc.gap.no-tools.ci.luna}{[$-$10.1, $-$7.5]}
\resultdef{secondary.acc.gap.no-tools.ci.qwen}{[$-$3.1, $-$1.4]}
\resultdef{secondary.acc.gap.no-tools.ci.sol}{[$-$0.4, +0.0]}
\resultdef{secondary.acc.gap.no-tools.control-flow.ci.luna}{[$-$18.5, $-$13.2]}
\resultdef{secondary.acc.gap.no-tools.control-flow.ci.qwen}{[$-$5.9, $-$2.5]}
\resultdef{secondary.acc.gap.no-tools.control-flow.ci.sol}{[$-$0.8, +0.1]}
\resultdef{secondary.acc.gap.no-tools.control-flow.luna}{$-$15.9}
\resultdef{secondary.acc.gap.no-tools.control-flow.max}{$-$0.3}
\resultdef{secondary.acc.gap.no-tools.control-flow.min}{$-$15.9}
\resultdef{secondary.acc.gap.no-tools.control-flow.qwen}{$-$4.2}
\resultdef{secondary.acc.gap.no-tools.control-flow.sol}{$-$0.3}
\resultdef{secondary.acc.gap.no-tools.distractors.ci.luna}{[$-$4.4, $-$1.3]}
\resultdef{secondary.acc.gap.no-tools.distractors.ci.qwen}{[$-$1.7, +0.2]}
\resultdef{secondary.acc.gap.no-tools.distractors.ci.sol}{[+0.0, +0.3]}
\resultdef{secondary.acc.gap.no-tools.distractors.luna}{$-$2.8}
\resultdef{secondary.acc.gap.no-tools.distractors.max}{+0.1}
\resultdef{secondary.acc.gap.no-tools.distractors.min}{$-$2.8}
\resultdef{secondary.acc.gap.no-tools.distractors.qwen}{$-$0.7}
\resultdef{secondary.acc.gap.no-tools.distractors.sol}{+0.1}
\resultdef{secondary.acc.gap.no-tools.luna}{$-$8.8}
\resultdef{secondary.acc.gap.no-tools.max}{$-$0.2}
\resultdef{secondary.acc.gap.no-tools.maximal-mix.ci.luna}{[$-$18.9, $-$13.7]}
\resultdef{secondary.acc.gap.no-tools.maximal-mix.ci.qwen}{[$-$7.0, $-$3.7]}
\resultdef{secondary.acc.gap.no-tools.maximal-mix.ci.sol}{[$-$0.5, +0.0]}
\resultdef{secondary.acc.gap.no-tools.maximal-mix.luna}{$-$16.3}
\resultdef{secondary.acc.gap.no-tools.maximal-mix.max}{$-$0.2}
\resultdef{secondary.acc.gap.no-tools.maximal-mix.min}{$-$16.3}
\resultdef{secondary.acc.gap.no-tools.maximal-mix.qwen}{$-$5.3}
\resultdef{secondary.acc.gap.no-tools.maximal-mix.sol}{$-$0.2}
\resultdef{secondary.acc.gap.no-tools.min}{$-$8.8}
\resultdef{secondary.acc.gap.no-tools.off-by-one.ci.luna}{[$-$12.0, $-$6.8]}
\resultdef{secondary.acc.gap.no-tools.off-by-one.ci.qwen}{[$-$3.2, +0.8]}
\resultdef{secondary.acc.gap.no-tools.off-by-one.ci.sol}{[$-$0.5, +0.2]}
\resultdef{secondary.acc.gap.no-tools.off-by-one.luna}{$-$9.3}
\resultdef{secondary.acc.gap.no-tools.off-by-one.max}{$-$0.1}
\resultdef{secondary.acc.gap.no-tools.off-by-one.min}{$-$9.3}
\resultdef{secondary.acc.gap.no-tools.off-by-one.qwen}{$-$1.4}
\resultdef{secondary.acc.gap.no-tools.off-by-one.sol}{$-$0.1}
\resultdef{secondary.acc.gap.no-tools.qwen}{$-$2.3}
\resultdef{secondary.acc.gap.no-tools.rename.ci.luna}{[$-$0.8, +0.7]}
\resultdef{secondary.acc.gap.no-tools.rename.ci.qwen}{[$-$0.5, +0.7]}
\resultdef{secondary.acc.gap.no-tools.rename.ci.sol}{[$-$0.3, +0.0]}
\resultdef{secondary.acc.gap.no-tools.rename.luna}{+0.0}
\resultdef{secondary.acc.gap.no-tools.rename.max}{+0.1}
\resultdef{secondary.acc.gap.no-tools.rename.min}{$-$0.1}
\resultdef{secondary.acc.gap.no-tools.rename.qwen}{+0.1}
\resultdef{secondary.acc.gap.no-tools.rename.sol}{$-$0.1}
\resultdef{secondary.acc.gap.no-tools.sol}{$-$0.2}
\resultdef{secondary.acc.gap.no-tools.state-and-literals.ci.luna}{[$-$10.8, $-$6.8]}
\resultdef{secondary.acc.gap.no-tools.state-and-literals.ci.qwen}{[$-$2.4, $-$0.4]}
\resultdef{secondary.acc.gap.no-tools.state-and-literals.ci.sol}{[$-$0.8, +0.0]}
\resultdef{secondary.acc.gap.no-tools.state-and-literals.luna}{$-$8.8}
\resultdef{secondary.acc.gap.no-tools.state-and-literals.max}{$-$0.3}
\resultdef{secondary.acc.gap.no-tools.state-and-literals.min}{$-$8.8}
\resultdef{secondary.acc.gap.no-tools.state-and-literals.qwen}{$-$1.4}
\resultdef{secondary.acc.gap.no-tools.state-and-literals.sol}{$-$0.3}
\resultdef{secondary.acc.gap.no-tools.static-return.ci.luna}{[$-$20.9, $-$15.7]}
\resultdef{secondary.acc.gap.no-tools.static-return.ci.qwen}{[$-$7.2, $-$3.9]}
\resultdef{secondary.acc.gap.no-tools.static-return.ci.sol}{[$-$0.4, +0.0]}
\resultdef{secondary.acc.gap.no-tools.static-return.luna}{$-$18.3}
\resultdef{secondary.acc.gap.no-tools.static-return.max}{$-$0.1}
\resultdef{secondary.acc.gap.no-tools.static-return.min}{$-$18.3}
\resultdef{secondary.acc.gap.no-tools.static-return.qwen}{$-$5.6}
\resultdef{secondary.acc.gap.no-tools.static-return.sol}{$-$0.1}
\resultdef{secondary.acc.gap.no-tools.wrong-arithmetic.ci.luna}{[$-$3.5, $-$0.9]}
\resultdef{secondary.acc.gap.no-tools.wrong-arithmetic.ci.qwen}{[$-$1.6, $-$0.2]}
\resultdef{secondary.acc.gap.no-tools.wrong-arithmetic.ci.sol}{[$-$1.4, +0.0]}
\resultdef{secondary.acc.gap.no-tools.wrong-arithmetic.luna}{$-$2.1}
\resultdef{secondary.acc.gap.no-tools.wrong-arithmetic.max}{$-$0.5}
\resultdef{secondary.acc.gap.no-tools.wrong-arithmetic.min}{$-$2.1}
\resultdef{secondary.acc.gap.no-tools.wrong-arithmetic.qwen}{$-$0.8}
\resultdef{secondary.acc.gap.no-tools.wrong-arithmetic.sol}{$-$0.5}
\resultdef{secondary.acc.gap.no-tools.wrong-logic.ci.luna}{[$-$4.9, $-$0.8]}
\resultdef{secondary.acc.gap.no-tools.wrong-logic.ci.qwen}{[$-$2.2, +2.5]}
\resultdef{secondary.acc.gap.no-tools.wrong-logic.ci.sol}{[+0.0, +0.0]}
\resultdef{secondary.acc.gap.no-tools.wrong-logic.luna}{$-$2.8}
\resultdef{secondary.acc.gap.no-tools.wrong-logic.max}{+0.0}
\resultdef{secondary.acc.gap.no-tools.wrong-logic.min}{$-$2.8}
\resultdef{secondary.acc.gap.no-tools.wrong-logic.qwen}{$-$0.1}
\resultdef{secondary.acc.gap.no-tools.wrong-logic.sol}{+0.0}
\resultdef{secondary.acc.gap.no-tools.wrong-variable.ci.luna}{[$-$12.1, $-$5.1]}
\resultdef{secondary.acc.gap.no-tools.wrong-variable.ci.qwen}{[$-$4.4, $-$1.6]}
\resultdef{secondary.acc.gap.no-tools.wrong-variable.ci.sol}{[$-$0.5, +0.0]}
\resultdef{secondary.acc.gap.no-tools.wrong-variable.luna}{$-$8.6}
\resultdef{secondary.acc.gap.no-tools.wrong-variable.max}{$-$0.2}
\resultdef{secondary.acc.gap.no-tools.wrong-variable.min}{$-$8.6}
\resultdef{secondary.acc.gap.no-tools.wrong-variable.qwen}{$-$2.8}
\resultdef{secondary.acc.gap.no-tools.wrong-variable.sol}{$-$0.2}
\resultdef{secondary.acc.loss.discretionary-tools.control-flow.luna}{10.6}
\resultdef{secondary.acc.loss.discretionary-tools.control-flow.max}{10.6}
\resultdef{secondary.acc.loss.discretionary-tools.control-flow.min}{0.1}
\resultdef{secondary.acc.loss.discretionary-tools.control-flow.qwen}{1.6}
\resultdef{secondary.acc.loss.discretionary-tools.control-flow.sol}{0.1}
\resultdef{secondary.acc.loss.discretionary-tools.distractors.luna}{1.5}
\resultdef{secondary.acc.loss.discretionary-tools.distractors.max}{1.5}
\resultdef{secondary.acc.loss.discretionary-tools.distractors.min}{-0.2}
\resultdef{secondary.acc.loss.discretionary-tools.distractors.qwen}{0.2}
\resultdef{secondary.acc.loss.discretionary-tools.distractors.sol}{-0.2}
\resultdef{secondary.acc.loss.discretionary-tools.luna}{6.1}
\resultdef{secondary.acc.loss.discretionary-tools.max}{6.1}
\resultdef{secondary.acc.loss.discretionary-tools.maximal-mix.luna}{11.3}
\resultdef{secondary.acc.loss.discretionary-tools.maximal-mix.max}{11.3}
\resultdef{secondary.acc.loss.discretionary-tools.maximal-mix.min}{0.2}
\resultdef{secondary.acc.loss.discretionary-tools.maximal-mix.qwen}{1.6}
\resultdef{secondary.acc.loss.discretionary-tools.maximal-mix.sol}{0.2}
\resultdef{secondary.acc.loss.discretionary-tools.min}{0.1}
\resultdef{secondary.acc.loss.discretionary-tools.off-by-one.luna}{7.7}
\resultdef{secondary.acc.loss.discretionary-tools.off-by-one.max}{7.7}
\resultdef{secondary.acc.loss.discretionary-tools.off-by-one.min}{-0.3}
\resultdef{secondary.acc.loss.discretionary-tools.off-by-one.qwen}{-0.3}
\resultdef{secondary.acc.loss.discretionary-tools.off-by-one.sol}{-0.2}
\resultdef{secondary.acc.loss.discretionary-tools.qwen}{0.9}
\resultdef{secondary.acc.loss.discretionary-tools.rename.luna}{-0.3}
\resultdef{secondary.acc.loss.discretionary-tools.rename.max}{-0.2}
\resultdef{secondary.acc.loss.discretionary-tools.rename.min}{-0.4}
\resultdef{secondary.acc.loss.discretionary-tools.rename.qwen}{-0.4}
\resultdef{secondary.acc.loss.discretionary-tools.rename.sol}{-0.2}
\resultdef{secondary.acc.loss.discretionary-tools.sol}{0.1}
\resultdef{secondary.acc.loss.discretionary-tools.state-and-literals.luna}{7.4}
\resultdef{secondary.acc.loss.discretionary-tools.state-and-literals.max}{7.4}
\resultdef{secondary.acc.loss.discretionary-tools.state-and-literals.min}{0.7}
\resultdef{secondary.acc.loss.discretionary-tools.state-and-literals.qwen}{1.5}
\resultdef{secondary.acc.loss.discretionary-tools.state-and-literals.sol}{0.7}
\resultdef{secondary.acc.loss.discretionary-tools.static-return.luna}{11.4}
\resultdef{secondary.acc.loss.discretionary-tools.static-return.max}{11.4}
\resultdef{secondary.acc.loss.discretionary-tools.static-return.min}{0.0}
\resultdef{secondary.acc.loss.discretionary-tools.static-return.qwen}{3.3}
\resultdef{secondary.acc.loss.discretionary-tools.static-return.sol}{0.0}
\resultdef{secondary.acc.loss.discretionary-tools.wrong-arithmetic.luna}{1.8}
\resultdef{secondary.acc.loss.discretionary-tools.wrong-arithmetic.max}{1.8}
\resultdef{secondary.acc.loss.discretionary-tools.wrong-arithmetic.min}{0.5}
\resultdef{secondary.acc.loss.discretionary-tools.wrong-arithmetic.qwen}{0.5}
\resultdef{secondary.acc.loss.discretionary-tools.wrong-arithmetic.sol}{0.7}
\resultdef{secondary.acc.loss.discretionary-tools.wrong-logic.luna}{-0.4}
\resultdef{secondary.acc.loss.discretionary-tools.wrong-logic.max}{0.0}
\resultdef{secondary.acc.loss.discretionary-tools.wrong-logic.min}{-0.4}
\resultdef{secondary.acc.loss.discretionary-tools.wrong-logic.qwen}{0.0}
\resultdef{secondary.acc.loss.discretionary-tools.wrong-logic.sol}{0.0}
\resultdef{secondary.acc.loss.discretionary-tools.wrong-variable.luna}{7.7}
\resultdef{secondary.acc.loss.discretionary-tools.wrong-variable.max}{7.7}
\resultdef{secondary.acc.loss.discretionary-tools.wrong-variable.min}{0.2}
\resultdef{secondary.acc.loss.discretionary-tools.wrong-variable.qwen}{0.5}
\resultdef{secondary.acc.loss.discretionary-tools.wrong-variable.sol}{0.2}
\resultdef{secondary.acc.loss.encouraged-tools.control-flow.luna}{10.7}
\resultdef{secondary.acc.loss.encouraged-tools.control-flow.max}{10.7}
\resultdef{secondary.acc.loss.encouraged-tools.control-flow.min}{0.5}
\resultdef{secondary.acc.loss.encouraged-tools.control-flow.qwen}{1.1}
\resultdef{secondary.acc.loss.encouraged-tools.control-flow.sol}{0.5}
\resultdef{secondary.acc.loss.encouraged-tools.distractors.luna}{2.0}
\resultdef{secondary.acc.loss.encouraged-tools.distractors.max}{2.0}
\resultdef{secondary.acc.loss.encouraged-tools.distractors.min}{-0.1}
\resultdef{secondary.acc.loss.encouraged-tools.distractors.qwen}{0.3}
\resultdef{secondary.acc.loss.encouraged-tools.distractors.sol}{-0.1}
\resultdef{secondary.acc.loss.encouraged-tools.luna}{6.2}
\resultdef{secondary.acc.loss.encouraged-tools.max}{6.2}
\resultdef{secondary.acc.loss.encouraged-tools.maximal-mix.luna}{11.1}
\resultdef{secondary.acc.loss.encouraged-tools.maximal-mix.max}{11.1}
\resultdef{secondary.acc.loss.encouraged-tools.maximal-mix.min}{0.5}
\resultdef{secondary.acc.loss.encouraged-tools.maximal-mix.qwen}{2.5}
\resultdef{secondary.acc.loss.encouraged-tools.maximal-mix.sol}{0.5}
\resultdef{secondary.acc.loss.encouraged-tools.min}{0.2}
\resultdef{secondary.acc.loss.encouraged-tools.off-by-one.luna}{7.3}
\resultdef{secondary.acc.loss.encouraged-tools.off-by-one.max}{7.3}
\resultdef{secondary.acc.loss.encouraged-tools.off-by-one.min}{-0.2}
\resultdef{secondary.acc.loss.encouraged-tools.off-by-one.qwen}{0.6}
\resultdef{secondary.acc.loss.encouraged-tools.off-by-one.sol}{-0.2}
\resultdef{secondary.acc.loss.encouraged-tools.qwen}{0.8}
\resultdef{secondary.acc.loss.encouraged-tools.rename.luna}{-0.1}
\resultdef{secondary.acc.loss.encouraged-tools.rename.max}{0.0}
\resultdef{secondary.acc.loss.encouraged-tools.rename.min}{-0.3}
\resultdef{secondary.acc.loss.encouraged-tools.rename.qwen}{-0.3}
\resultdef{secondary.acc.loss.encouraged-tools.rename.sol}{0.0}
\resultdef{secondary.acc.loss.encouraged-tools.sol}{0.2}
\resultdef{secondary.acc.loss.encouraged-tools.state-and-literals.luna}{7.0}
\resultdef{secondary.acc.loss.encouraged-tools.state-and-literals.max}{7.0}
\resultdef{secondary.acc.loss.encouraged-tools.state-and-literals.min}{0.3}
\resultdef{secondary.acc.loss.encouraged-tools.state-and-literals.qwen}{0.6}
\resultdef{secondary.acc.loss.encouraged-tools.state-and-literals.sol}{0.3}
\resultdef{secondary.acc.loss.encouraged-tools.static-return.luna}{10.6}
\resultdef{secondary.acc.loss.encouraged-tools.static-return.max}{10.6}
\resultdef{secondary.acc.loss.encouraged-tools.static-return.min}{0.1}
\resultdef{secondary.acc.loss.encouraged-tools.static-return.qwen}{1.4}
\resultdef{secondary.acc.loss.encouraged-tools.static-return.sol}{0.1}
\resultdef{secondary.acc.loss.encouraged-tools.wrong-arithmetic.luna}{1.3}
\resultdef{secondary.acc.loss.encouraged-tools.wrong-arithmetic.max}{1.3}
\resultdef{secondary.acc.loss.encouraged-tools.wrong-arithmetic.min}{-0.2}
\resultdef{secondary.acc.loss.encouraged-tools.wrong-arithmetic.qwen}{-0.2}
\resultdef{secondary.acc.loss.encouraged-tools.wrong-arithmetic.sol}{0.9}
\resultdef{secondary.acc.loss.encouraged-tools.wrong-logic.luna}{0.2}
\resultdef{secondary.acc.loss.encouraged-tools.wrong-logic.max}{0.2}
\resultdef{secondary.acc.loss.encouraged-tools.wrong-logic.min}{-0.6}
\resultdef{secondary.acc.loss.encouraged-tools.wrong-logic.qwen}{-0.6}
\resultdef{secondary.acc.loss.encouraged-tools.wrong-logic.sol}{0.0}
\resultdef{secondary.acc.loss.encouraged-tools.wrong-variable.luna}{9.4}
\resultdef{secondary.acc.loss.encouraged-tools.wrong-variable.max}{9.4}
\resultdef{secondary.acc.loss.encouraged-tools.wrong-variable.min}{0.5}
\resultdef{secondary.acc.loss.encouraged-tools.wrong-variable.qwen}{2.4}
\resultdef{secondary.acc.loss.encouraged-tools.wrong-variable.sol}{0.5}
\resultdef{secondary.acc.loss.no-tools.control-flow.luna}{15.9}
\resultdef{secondary.acc.loss.no-tools.control-flow.max}{15.9}
\resultdef{secondary.acc.loss.no-tools.control-flow.min}{0.3}
\resultdef{secondary.acc.loss.no-tools.control-flow.qwen}{4.2}
\resultdef{secondary.acc.loss.no-tools.control-flow.sol}{0.3}
\resultdef{secondary.acc.loss.no-tools.distractors.luna}{2.8}
\resultdef{secondary.acc.loss.no-tools.distractors.max}{2.8}
\resultdef{secondary.acc.loss.no-tools.distractors.min}{-0.1}
\resultdef{secondary.acc.loss.no-tools.distractors.qwen}{0.7}
\resultdef{secondary.acc.loss.no-tools.distractors.sol}{-0.1}
\resultdef{secondary.acc.loss.no-tools.luna}{8.8}
\resultdef{secondary.acc.loss.no-tools.max}{8.8}
\resultdef{secondary.acc.loss.no-tools.maximal-mix.luna}{16.3}
\resultdef{secondary.acc.loss.no-tools.maximal-mix.max}{16.3}
\resultdef{secondary.acc.loss.no-tools.maximal-mix.min}{0.2}
\resultdef{secondary.acc.loss.no-tools.maximal-mix.qwen}{5.3}
\resultdef{secondary.acc.loss.no-tools.maximal-mix.sol}{0.2}
\resultdef{secondary.acc.loss.no-tools.min}{0.2}
\resultdef{secondary.acc.loss.no-tools.off-by-one.luna}{9.3}
\resultdef{secondary.acc.loss.no-tools.off-by-one.max}{9.3}
\resultdef{secondary.acc.loss.no-tools.off-by-one.min}{0.1}
\resultdef{secondary.acc.loss.no-tools.off-by-one.qwen}{1.4}
\resultdef{secondary.acc.loss.no-tools.off-by-one.sol}{0.1}
\resultdef{secondary.acc.loss.no-tools.qwen}{2.3}
\resultdef{secondary.acc.loss.no-tools.rename.luna}{0.0}
\resultdef{secondary.acc.loss.no-tools.rename.max}{0.1}
\resultdef{secondary.acc.loss.no-tools.rename.min}{-0.1}
\resultdef{secondary.acc.loss.no-tools.rename.qwen}{-0.1}
\resultdef{secondary.acc.loss.no-tools.rename.sol}{0.1}
\resultdef{secondary.acc.loss.no-tools.sol}{0.2}
\resultdef{secondary.acc.loss.no-tools.state-and-literals.luna}{8.8}
\resultdef{secondary.acc.loss.no-tools.state-and-literals.max}{8.8}
\resultdef{secondary.acc.loss.no-tools.state-and-literals.min}{0.3}
\resultdef{secondary.acc.loss.no-tools.state-and-literals.qwen}{1.4}
\resultdef{secondary.acc.loss.no-tools.state-and-literals.sol}{0.3}
\resultdef{secondary.acc.loss.no-tools.static-return.luna}{18.3}
\resultdef{secondary.acc.loss.no-tools.static-return.max}{18.3}
\resultdef{secondary.acc.loss.no-tools.static-return.min}{0.1}
\resultdef{secondary.acc.loss.no-tools.static-return.qwen}{5.6}
\resultdef{secondary.acc.loss.no-tools.static-return.sol}{0.1}
\resultdef{secondary.acc.loss.no-tools.wrong-arithmetic.luna}{2.1}
\resultdef{secondary.acc.loss.no-tools.wrong-arithmetic.max}{2.1}
\resultdef{secondary.acc.loss.no-tools.wrong-arithmetic.min}{0.5}
\resultdef{secondary.acc.loss.no-tools.wrong-arithmetic.qwen}{0.8}
\resultdef{secondary.acc.loss.no-tools.wrong-arithmetic.sol}{0.5}
\resultdef{secondary.acc.loss.no-tools.wrong-logic.luna}{2.8}
\resultdef{secondary.acc.loss.no-tools.wrong-logic.max}{2.8}
\resultdef{secondary.acc.loss.no-tools.wrong-logic.min}{0.0}
\resultdef{secondary.acc.loss.no-tools.wrong-logic.qwen}{0.1}
\resultdef{secondary.acc.loss.no-tools.wrong-logic.sol}{0.0}
\resultdef{secondary.acc.loss.no-tools.wrong-variable.luna}{8.6}
\resultdef{secondary.acc.loss.no-tools.wrong-variable.max}{8.6}
\resultdef{secondary.acc.loss.no-tools.wrong-variable.min}{0.2}
\resultdef{secondary.acc.loss.no-tools.wrong-variable.qwen}{2.8}
\resultdef{secondary.acc.loss.no-tools.wrong-variable.sol}{0.2}
\resultdef{secondary.acc.transformed.discretionary-tools.control-flow.luna}{87.5}
\resultdef{secondary.acc.transformed.discretionary-tools.control-flow.max}{99.7}
\resultdef{secondary.acc.transformed.discretionary-tools.control-flow.min}{87.5}
\resultdef{secondary.acc.transformed.discretionary-tools.control-flow.qwen}{97.4}
\resultdef{secondary.acc.transformed.discretionary-tools.control-flow.sol}{99.7}
\resultdef{secondary.acc.transformed.discretionary-tools.distractors.luna}{96.6}
\resultdef{secondary.acc.transformed.discretionary-tools.distractors.max}{99.9}
\resultdef{secondary.acc.transformed.discretionary-tools.distractors.min}{96.6}
\resultdef{secondary.acc.transformed.discretionary-tools.distractors.qwen}{98.8}
\resultdef{secondary.acc.transformed.discretionary-tools.distractors.sol}{99.9}
\resultdef{secondary.acc.transformed.discretionary-tools.luna}{92.0}
\resultdef{secondary.acc.transformed.discretionary-tools.max}{99.6}
\resultdef{secondary.acc.transformed.discretionary-tools.maximal-mix.luna}{86.8}
\resultdef{secondary.acc.transformed.discretionary-tools.maximal-mix.max}{99.6}
\resultdef{secondary.acc.transformed.discretionary-tools.maximal-mix.min}{86.8}
\resultdef{secondary.acc.transformed.discretionary-tools.maximal-mix.qwen}{97.4}
\resultdef{secondary.acc.transformed.discretionary-tools.maximal-mix.sol}{99.6}
\resultdef{secondary.acc.transformed.discretionary-tools.min}{92.0}
\resultdef{secondary.acc.transformed.discretionary-tools.off-by-one.luna}{89.0}
\resultdef{secondary.acc.transformed.discretionary-tools.off-by-one.max}{99.4}
\resultdef{secondary.acc.transformed.discretionary-tools.off-by-one.min}{89.0}
\resultdef{secondary.acc.transformed.discretionary-tools.off-by-one.qwen}{98.3}
\resultdef{secondary.acc.transformed.discretionary-tools.off-by-one.sol}{99.4}
\resultdef{secondary.acc.transformed.discretionary-tools.qwen}{98.0}
\resultdef{secondary.acc.transformed.discretionary-tools.rename.luna}{98.4}
\resultdef{secondary.acc.transformed.discretionary-tools.rename.max}{99.9}
\resultdef{secondary.acc.transformed.discretionary-tools.rename.min}{98.4}
\resultdef{secondary.acc.transformed.discretionary-tools.rename.qwen}{99.3}
\resultdef{secondary.acc.transformed.discretionary-tools.rename.sol}{99.9}
\resultdef{secondary.acc.transformed.discretionary-tools.sol}{99.6}
\resultdef{secondary.acc.transformed.discretionary-tools.state-and-literals.luna}{90.7}
\resultdef{secondary.acc.transformed.discretionary-tools.state-and-literals.max}{99.1}
\resultdef{secondary.acc.transformed.discretionary-tools.state-and-literals.min}{90.7}
\resultdef{secondary.acc.transformed.discretionary-tools.state-and-literals.qwen}{97.5}
\resultdef{secondary.acc.transformed.discretionary-tools.state-and-literals.sol}{99.1}
\resultdef{secondary.acc.transformed.discretionary-tools.static-return.luna}{87.1}
\resultdef{secondary.acc.transformed.discretionary-tools.static-return.max}{99.6}
\resultdef{secondary.acc.transformed.discretionary-tools.static-return.min}{87.1}
\resultdef{secondary.acc.transformed.discretionary-tools.static-return.qwen}{95.7}
\resultdef{secondary.acc.transformed.discretionary-tools.static-return.sol}{99.6}
\resultdef{secondary.acc.transformed.discretionary-tools.subset-max.luna}{98.4}
\resultdef{secondary.acc.transformed.discretionary-tools.subset-max.max}{100.0}
\resultdef{secondary.acc.transformed.discretionary-tools.subset-max.min}{98.4}
\resultdef{secondary.acc.transformed.discretionary-tools.subset-max.qwen}{99.5}
\resultdef{secondary.acc.transformed.discretionary-tools.subset-max.sol}{100.0}
\resultdef{secondary.acc.transformed.discretionary-tools.subset-min.luna}{86.8}
\resultdef{secondary.acc.transformed.discretionary-tools.subset-min.max}{99.1}
\resultdef{secondary.acc.transformed.discretionary-tools.subset-min.min}{86.8}
\resultdef{secondary.acc.transformed.discretionary-tools.subset-min.qwen}{95.7}
\resultdef{secondary.acc.transformed.discretionary-tools.subset-min.sol}{99.1}
\resultdef{secondary.acc.transformed.discretionary-tools.wrong-arithmetic.luna}{98.2}
\resultdef{secondary.acc.transformed.discretionary-tools.wrong-arithmetic.max}{99.5}
\resultdef{secondary.acc.transformed.discretionary-tools.wrong-arithmetic.min}{98.2}
\resultdef{secondary.acc.transformed.discretionary-tools.wrong-arithmetic.qwen}{99.5}
\resultdef{secondary.acc.transformed.discretionary-tools.wrong-arithmetic.sol}{99.3}
\resultdef{secondary.acc.transformed.discretionary-tools.wrong-logic.luna}{97.6}
\resultdef{secondary.acc.transformed.discretionary-tools.wrong-logic.max}{100.0}
\resultdef{secondary.acc.transformed.discretionary-tools.wrong-logic.min}{97.6}
\resultdef{secondary.acc.transformed.discretionary-tools.wrong-logic.qwen}{98.6}
\resultdef{secondary.acc.transformed.discretionary-tools.wrong-logic.sol}{100.0}
\resultdef{secondary.acc.transformed.discretionary-tools.wrong-variable.luna}{90.4}
\resultdef{secondary.acc.transformed.discretionary-tools.wrong-variable.max}{99.8}
\resultdef{secondary.acc.transformed.discretionary-tools.wrong-variable.min}{90.4}
\resultdef{secondary.acc.transformed.discretionary-tools.wrong-variable.qwen}{98.7}
\resultdef{secondary.acc.transformed.discretionary-tools.wrong-variable.sol}{99.8}
\resultdef{secondary.acc.transformed.encouraged-tools.control-flow.luna}{88.2}
\resultdef{secondary.acc.transformed.encouraged-tools.control-flow.max}{99.3}
\resultdef{secondary.acc.transformed.encouraged-tools.control-flow.min}{88.2}
\resultdef{secondary.acc.transformed.encouraged-tools.control-flow.qwen}{98.0}
\resultdef{secondary.acc.transformed.encouraged-tools.control-flow.sol}{99.3}
\resultdef{secondary.acc.transformed.encouraged-tools.distractors.luna}{96.9}
\resultdef{secondary.acc.transformed.encouraged-tools.distractors.max}{99.9}
\resultdef{secondary.acc.transformed.encouraged-tools.distractors.min}{96.9}
\resultdef{secondary.acc.transformed.encouraged-tools.distractors.qwen}{98.8}
\resultdef{secondary.acc.transformed.encouraged-tools.distractors.sol}{99.9}
\resultdef{secondary.acc.transformed.encouraged-tools.luna}{92.7}
\resultdef{secondary.acc.transformed.encouraged-tools.max}{99.6}
\resultdef{secondary.acc.transformed.encouraged-tools.maximal-mix.luna}{87.8}
\resultdef{secondary.acc.transformed.encouraged-tools.maximal-mix.max}{99.3}
\resultdef{secondary.acc.transformed.encouraged-tools.maximal-mix.min}{87.8}
\resultdef{secondary.acc.transformed.encouraged-tools.maximal-mix.qwen}{96.6}
\resultdef{secondary.acc.transformed.encouraged-tools.maximal-mix.sol}{99.3}
\resultdef{secondary.acc.transformed.encouraged-tools.min}{92.7}
\resultdef{secondary.acc.transformed.encouraged-tools.off-by-one.luna}{90.2}
\resultdef{secondary.acc.transformed.encouraged-tools.off-by-one.max}{99.4}
\resultdef{secondary.acc.transformed.encouraged-tools.off-by-one.min}{90.2}
\resultdef{secondary.acc.transformed.encouraged-tools.off-by-one.qwen}{98.2}
\resultdef{secondary.acc.transformed.encouraged-tools.off-by-one.sol}{99.4}
\resultdef{secondary.acc.transformed.encouraged-tools.qwen}{98.3}
\resultdef{secondary.acc.transformed.encouraged-tools.rename.luna}{99.0}
\resultdef{secondary.acc.transformed.encouraged-tools.rename.max}{99.8}
\resultdef{secondary.acc.transformed.encouraged-tools.rename.min}{99.0}
\resultdef{secondary.acc.transformed.encouraged-tools.rename.qwen}{99.4}
\resultdef{secondary.acc.transformed.encouraged-tools.rename.sol}{99.8}
\resultdef{secondary.acc.transformed.encouraged-tools.sol}{99.6}
\resultdef{secondary.acc.transformed.encouraged-tools.state-and-literals.luna}{92.0}
\resultdef{secondary.acc.transformed.encouraged-tools.state-and-literals.max}{99.5}
\resultdef{secondary.acc.transformed.encouraged-tools.state-and-literals.min}{92.0}
\resultdef{secondary.acc.transformed.encouraged-tools.state-and-literals.qwen}{98.5}
\resultdef{secondary.acc.transformed.encouraged-tools.state-and-literals.sol}{99.5}
\resultdef{secondary.acc.transformed.encouraged-tools.static-return.luna}{89.4}
\resultdef{secondary.acc.transformed.encouraged-tools.static-return.max}{99.9}
\resultdef{secondary.acc.transformed.encouraged-tools.static-return.min}{89.4}
\resultdef{secondary.acc.transformed.encouraged-tools.static-return.qwen}{97.5}
\resultdef{secondary.acc.transformed.encouraged-tools.static-return.sol}{99.9}
\resultdef{secondary.acc.transformed.encouraged-tools.subset-max.luna}{99.0}
\resultdef{secondary.acc.transformed.encouraged-tools.subset-max.max}{100.0}
\resultdef{secondary.acc.transformed.encouraged-tools.subset-max.min}{99.0}
\resultdef{secondary.acc.transformed.encouraged-tools.subset-max.qwen}{99.7}
\resultdef{secondary.acc.transformed.encouraged-tools.subset-max.sol}{100.0}
\resultdef{secondary.acc.transformed.encouraged-tools.subset-min.luna}{87.8}
\resultdef{secondary.acc.transformed.encouraged-tools.subset-min.max}{99.1}
\resultdef{secondary.acc.transformed.encouraged-tools.subset-min.min}{87.8}
\resultdef{secondary.acc.transformed.encouraged-tools.subset-min.qwen}{96.6}
\resultdef{secondary.acc.transformed.encouraged-tools.subset-min.sol}{99.1}
\resultdef{secondary.acc.transformed.encouraged-tools.wrong-arithmetic.luna}{98.7}
\resultdef{secondary.acc.transformed.encouraged-tools.wrong-arithmetic.max}{99.7}
\resultdef{secondary.acc.transformed.encouraged-tools.wrong-arithmetic.min}{98.7}
\resultdef{secondary.acc.transformed.encouraged-tools.wrong-arithmetic.qwen}{99.7}
\resultdef{secondary.acc.transformed.encouraged-tools.wrong-arithmetic.sol}{99.1}
\resultdef{secondary.acc.transformed.encouraged-tools.wrong-logic.luna}{98.4}
\resultdef{secondary.acc.transformed.encouraged-tools.wrong-logic.max}{100.0}
\resultdef{secondary.acc.transformed.encouraged-tools.wrong-logic.min}{98.4}
\resultdef{secondary.acc.transformed.encouraged-tools.wrong-logic.qwen}{98.7}
\resultdef{secondary.acc.transformed.encouraged-tools.wrong-logic.sol}{100.0}
\resultdef{secondary.acc.transformed.encouraged-tools.wrong-variable.luna}{89.1}
\resultdef{secondary.acc.transformed.encouraged-tools.wrong-variable.max}{99.5}
\resultdef{secondary.acc.transformed.encouraged-tools.wrong-variable.min}{89.1}
\resultdef{secondary.acc.transformed.encouraged-tools.wrong-variable.qwen}{97.6}
\resultdef{secondary.acc.transformed.encouraged-tools.wrong-variable.sol}{99.5}
\resultdef{secondary.acc.transformed.no-tools.control-flow.luna}{82.3}
\resultdef{secondary.acc.transformed.no-tools.control-flow.max}{99.6}
\resultdef{secondary.acc.transformed.no-tools.control-flow.min}{82.3}
\resultdef{secondary.acc.transformed.no-tools.control-flow.qwen}{94.6}
\resultdef{secondary.acc.transformed.no-tools.control-flow.sol}{99.6}
\resultdef{secondary.acc.transformed.no-tools.distractors.luna}{95.4}
\resultdef{secondary.acc.transformed.no-tools.distractors.max}{100.0}
\resultdef{secondary.acc.transformed.no-tools.distractors.min}{95.4}
\resultdef{secondary.acc.transformed.no-tools.distractors.qwen}{98.1}
\resultdef{secondary.acc.transformed.no-tools.distractors.sol}{100.0}
\resultdef{secondary.acc.transformed.no-tools.luna}{89.4}
\resultdef{secondary.acc.transformed.no-tools.max}{99.8}
\resultdef{secondary.acc.transformed.no-tools.maximal-mix.luna}{81.9}
\resultdef{secondary.acc.transformed.no-tools.maximal-mix.max}{99.8}
\resultdef{secondary.acc.transformed.no-tools.maximal-mix.min}{81.9}
\resultdef{secondary.acc.transformed.no-tools.maximal-mix.qwen}{93.4}
\resultdef{secondary.acc.transformed.no-tools.maximal-mix.sol}{99.8}
\resultdef{secondary.acc.transformed.no-tools.min}{89.4}
\resultdef{secondary.acc.transformed.no-tools.off-by-one.luna}{87.1}
\resultdef{secondary.acc.transformed.no-tools.off-by-one.max}{99.5}
\resultdef{secondary.acc.transformed.no-tools.off-by-one.min}{87.1}
\resultdef{secondary.acc.transformed.no-tools.off-by-one.qwen}{95.8}
\resultdef{secondary.acc.transformed.no-tools.off-by-one.sol}{99.5}
\resultdef{secondary.acc.transformed.no-tools.qwen}{96.4}
\resultdef{secondary.acc.transformed.no-tools.rename.luna}{98.1}
\resultdef{secondary.acc.transformed.no-tools.rename.max}{99.8}
\resultdef{secondary.acc.transformed.no-tools.rename.min}{98.1}
\resultdef{secondary.acc.transformed.no-tools.rename.qwen}{98.8}
\resultdef{secondary.acc.transformed.no-tools.rename.sol}{99.8}
\resultdef{secondary.acc.transformed.no-tools.sol}{99.8}
\resultdef{secondary.acc.transformed.no-tools.state-and-literals.luna}{89.4}
\resultdef{secondary.acc.transformed.no-tools.state-and-literals.max}{99.6}
\resultdef{secondary.acc.transformed.no-tools.state-and-literals.min}{89.4}
\resultdef{secondary.acc.transformed.no-tools.state-and-literals.qwen}{97.4}
\resultdef{secondary.acc.transformed.no-tools.state-and-literals.sol}{99.6}
\resultdef{secondary.acc.transformed.no-tools.static-return.luna}{80.6}
\resultdef{secondary.acc.transformed.no-tools.static-return.max}{99.9}
\resultdef{secondary.acc.transformed.no-tools.static-return.min}{80.6}
\resultdef{secondary.acc.transformed.no-tools.static-return.qwen}{93.7}
\resultdef{secondary.acc.transformed.no-tools.static-return.sol}{99.9}
\resultdef{secondary.acc.transformed.no-tools.subset-max.luna}{98.1}
\resultdef{secondary.acc.transformed.no-tools.subset-max.max}{100.0}
\resultdef{secondary.acc.transformed.no-tools.subset-max.min}{98.1}
\resultdef{secondary.acc.transformed.no-tools.subset-max.qwen}{99.2}
\resultdef{secondary.acc.transformed.no-tools.subset-max.sol}{100.0}
\resultdef{secondary.acc.transformed.no-tools.subset-min.luna}{80.6}
\resultdef{secondary.acc.transformed.no-tools.subset-min.max}{99.5}
\resultdef{secondary.acc.transformed.no-tools.subset-min.min}{80.6}
\resultdef{secondary.acc.transformed.no-tools.subset-min.qwen}{93.4}
\resultdef{secondary.acc.transformed.no-tools.subset-min.sol}{99.5}
\resultdef{secondary.acc.transformed.no-tools.wrong-arithmetic.luna}{97.4}
\resultdef{secondary.acc.transformed.no-tools.wrong-arithmetic.max}{99.5}
\resultdef{secondary.acc.transformed.no-tools.wrong-arithmetic.min}{97.4}
\resultdef{secondary.acc.transformed.no-tools.wrong-arithmetic.qwen}{99.2}
\resultdef{secondary.acc.transformed.no-tools.wrong-arithmetic.sol}{99.5}
\resultdef{secondary.acc.transformed.no-tools.wrong-logic.luna}{94.9}
\resultdef{secondary.acc.transformed.no-tools.wrong-logic.max}{100.0}
\resultdef{secondary.acc.transformed.no-tools.wrong-logic.min}{94.9}
\resultdef{secondary.acc.transformed.no-tools.wrong-logic.qwen}{97.1}
\resultdef{secondary.acc.transformed.no-tools.wrong-logic.sol}{100.0}
\resultdef{secondary.acc.transformed.no-tools.wrong-variable.luna}{89.9}
\resultdef{secondary.acc.transformed.no-tools.wrong-variable.max}{99.8}
\resultdef{secondary.acc.transformed.no-tools.wrong-variable.min}{89.9}
\resultdef{secondary.acc.transformed.no-tools.wrong-variable.qwen}{97.2}
\resultdef{secondary.acc.transformed.no-tools.wrong-variable.sol}{99.8}
\resultdef{secondary.acc.untransformed.discretionary-tools.control-flow.luna}{98.1}
\resultdef{secondary.acc.untransformed.discretionary-tools.control-flow.max}{99.8}
\resultdef{secondary.acc.untransformed.discretionary-tools.control-flow.min}{98.1}
\resultdef{secondary.acc.untransformed.discretionary-tools.control-flow.qwen}{98.9}
\resultdef{secondary.acc.untransformed.discretionary-tools.control-flow.sol}{99.8}
\resultdef{secondary.acc.untransformed.discretionary-tools.distractors.luna}{98.1}
\resultdef{secondary.acc.untransformed.discretionary-tools.distractors.max}{99.8}
\resultdef{secondary.acc.untransformed.discretionary-tools.distractors.min}{98.1}
\resultdef{secondary.acc.untransformed.discretionary-tools.distractors.qwen}{98.9}
\resultdef{secondary.acc.untransformed.discretionary-tools.distractors.sol}{99.8}
\resultdef{secondary.acc.untransformed.discretionary-tools.luna}{98.1}
\resultdef{secondary.acc.untransformed.discretionary-tools.max}{99.8}
\resultdef{secondary.acc.untransformed.discretionary-tools.maximal-mix.luna}{98.1}
\resultdef{secondary.acc.untransformed.discretionary-tools.maximal-mix.max}{99.8}
\resultdef{secondary.acc.untransformed.discretionary-tools.maximal-mix.min}{98.1}
\resultdef{secondary.acc.untransformed.discretionary-tools.maximal-mix.qwen}{98.9}
\resultdef{secondary.acc.untransformed.discretionary-tools.maximal-mix.sol}{99.8}
\resultdef{secondary.acc.untransformed.discretionary-tools.min}{98.1}
\resultdef{secondary.acc.untransformed.discretionary-tools.off-by-one.luna}{96.8}
\resultdef{secondary.acc.untransformed.discretionary-tools.off-by-one.max}{99.2}
\resultdef{secondary.acc.untransformed.discretionary-tools.off-by-one.min}{96.8}
\resultdef{secondary.acc.untransformed.discretionary-tools.off-by-one.qwen}{98.0}
\resultdef{secondary.acc.untransformed.discretionary-tools.off-by-one.sol}{99.2}
\resultdef{secondary.acc.untransformed.discretionary-tools.qwen}{98.9}
\resultdef{secondary.acc.untransformed.discretionary-tools.rename.luna}{98.0}
\resultdef{secondary.acc.untransformed.discretionary-tools.rename.max}{99.7}
\resultdef{secondary.acc.untransformed.discretionary-tools.rename.min}{98.0}
\resultdef{secondary.acc.untransformed.discretionary-tools.rename.qwen}{98.9}
\resultdef{secondary.acc.untransformed.discretionary-tools.rename.sol}{99.7}
\resultdef{secondary.acc.untransformed.discretionary-tools.sol}{99.8}
\resultdef{secondary.acc.untransformed.discretionary-tools.state-and-literals.luna}{98.1}
\resultdef{secondary.acc.untransformed.discretionary-tools.state-and-literals.max}{99.8}
\resultdef{secondary.acc.untransformed.discretionary-tools.state-and-literals.min}{98.1}
\resultdef{secondary.acc.untransformed.discretionary-tools.state-and-literals.qwen}{98.9}
\resultdef{secondary.acc.untransformed.discretionary-tools.state-and-literals.sol}{99.8}
\resultdef{secondary.acc.untransformed.discretionary-tools.static-return.luna}{98.6}
\resultdef{secondary.acc.untransformed.discretionary-tools.static-return.max}{99.6}
\resultdef{secondary.acc.untransformed.discretionary-tools.static-return.min}{98.6}
\resultdef{secondary.acc.untransformed.discretionary-tools.static-return.qwen}{98.9}
\resultdef{secondary.acc.untransformed.discretionary-tools.static-return.sol}{99.6}
\resultdef{secondary.acc.untransformed.discretionary-tools.subset-max.luna}{100.0}
\resultdef{secondary.acc.untransformed.discretionary-tools.subset-max.max}{100.0}
\resultdef{secondary.acc.untransformed.discretionary-tools.subset-max.min}{100.0}
\resultdef{secondary.acc.untransformed.discretionary-tools.subset-max.qwen}{100.0}
\resultdef{secondary.acc.untransformed.discretionary-tools.subset-max.sol}{100.0}
\resultdef{secondary.acc.untransformed.discretionary-tools.subset-min.luna}{96.8}
\resultdef{secondary.acc.untransformed.discretionary-tools.subset-min.max}{99.2}
\resultdef{secondary.acc.untransformed.discretionary-tools.subset-min.min}{96.8}
\resultdef{secondary.acc.untransformed.discretionary-tools.subset-min.qwen}{98.0}
\resultdef{secondary.acc.untransformed.discretionary-tools.subset-min.sol}{99.2}
\resultdef{secondary.acc.untransformed.discretionary-tools.wrong-arithmetic.luna}{100.0}
\resultdef{secondary.acc.untransformed.discretionary-tools.wrong-arithmetic.max}{100.0}
\resultdef{secondary.acc.untransformed.discretionary-tools.wrong-arithmetic.min}{100.0}
\resultdef{secondary.acc.untransformed.discretionary-tools.wrong-arithmetic.qwen}{100.0}
\resultdef{secondary.acc.untransformed.discretionary-tools.wrong-arithmetic.sol}{100.0}
\resultdef{secondary.acc.untransformed.discretionary-tools.wrong-logic.luna}{97.2}
\resultdef{secondary.acc.untransformed.discretionary-tools.wrong-logic.max}{100.0}
\resultdef{secondary.acc.untransformed.discretionary-tools.wrong-logic.min}{97.2}
\resultdef{secondary.acc.untransformed.discretionary-tools.wrong-logic.qwen}{98.6}
\resultdef{secondary.acc.untransformed.discretionary-tools.wrong-logic.sol}{100.0}
\resultdef{secondary.acc.untransformed.discretionary-tools.wrong-variable.luna}{98.1}
\resultdef{secondary.acc.untransformed.discretionary-tools.wrong-variable.max}{100.0}
\resultdef{secondary.acc.untransformed.discretionary-tools.wrong-variable.min}{98.1}
\resultdef{secondary.acc.untransformed.discretionary-tools.wrong-variable.qwen}{99.2}
\resultdef{secondary.acc.untransformed.discretionary-tools.wrong-variable.sol}{100.0}
\resultdef{secondary.acc.untransformed.encouraged-tools.control-flow.luna}{98.9}
\resultdef{secondary.acc.untransformed.encouraged-tools.control-flow.max}{99.8}
\resultdef{secondary.acc.untransformed.encouraged-tools.control-flow.min}{98.9}
\resultdef{secondary.acc.untransformed.encouraged-tools.control-flow.qwen}{99.1}
\resultdef{secondary.acc.untransformed.encouraged-tools.control-flow.sol}{99.8}
\resultdef{secondary.acc.untransformed.encouraged-tools.distractors.luna}{98.9}
\resultdef{secondary.acc.untransformed.encouraged-tools.distractors.max}{99.8}
\resultdef{secondary.acc.untransformed.encouraged-tools.distractors.min}{98.9}
\resultdef{secondary.acc.untransformed.encouraged-tools.distractors.qwen}{99.1}
\resultdef{secondary.acc.untransformed.encouraged-tools.distractors.sol}{99.8}
\resultdef{secondary.acc.untransformed.encouraged-tools.luna}{98.9}
\resultdef{secondary.acc.untransformed.encouraged-tools.max}{99.8}
\resultdef{secondary.acc.untransformed.encouraged-tools.maximal-mix.luna}{98.9}
\resultdef{secondary.acc.untransformed.encouraged-tools.maximal-mix.max}{99.8}
\resultdef{secondary.acc.untransformed.encouraged-tools.maximal-mix.min}{98.9}
\resultdef{secondary.acc.untransformed.encouraged-tools.maximal-mix.qwen}{99.1}
\resultdef{secondary.acc.untransformed.encouraged-tools.maximal-mix.sol}{99.8}
\resultdef{secondary.acc.untransformed.encouraged-tools.min}{98.9}
\resultdef{secondary.acc.untransformed.encouraged-tools.off-by-one.luna}{97.6}
\resultdef{secondary.acc.untransformed.encouraged-tools.off-by-one.max}{99.2}
\resultdef{secondary.acc.untransformed.encouraged-tools.off-by-one.min}{97.6}
\resultdef{secondary.acc.untransformed.encouraged-tools.off-by-one.qwen}{98.8}
\resultdef{secondary.acc.untransformed.encouraged-tools.off-by-one.sol}{99.2}
\resultdef{secondary.acc.untransformed.encouraged-tools.qwen}{99.1}
\resultdef{secondary.acc.untransformed.encouraged-tools.rename.luna}{98.9}
\resultdef{secondary.acc.untransformed.encouraged-tools.rename.max}{99.8}
\resultdef{secondary.acc.untransformed.encouraged-tools.rename.min}{98.9}
\resultdef{secondary.acc.untransformed.encouraged-tools.rename.qwen}{99.1}
\resultdef{secondary.acc.untransformed.encouraged-tools.rename.sol}{99.8}
\resultdef{secondary.acc.untransformed.encouraged-tools.sol}{99.8}
\resultdef{secondary.acc.untransformed.encouraged-tools.state-and-literals.luna}{98.9}
\resultdef{secondary.acc.untransformed.encouraged-tools.state-and-literals.max}{99.8}
\resultdef{secondary.acc.untransformed.encouraged-tools.state-and-literals.min}{98.9}
\resultdef{secondary.acc.untransformed.encouraged-tools.state-and-literals.qwen}{99.1}
\resultdef{secondary.acc.untransformed.encouraged-tools.state-and-literals.sol}{99.8}
\resultdef{secondary.acc.untransformed.encouraged-tools.static-return.luna}{100.0}
\resultdef{secondary.acc.untransformed.encouraged-tools.static-return.max}{100.0}
\resultdef{secondary.acc.untransformed.encouraged-tools.static-return.min}{98.9}
\resultdef{secondary.acc.untransformed.encouraged-tools.static-return.qwen}{98.9}
\resultdef{secondary.acc.untransformed.encouraged-tools.static-return.sol}{100.0}
\resultdef{secondary.acc.untransformed.encouraged-tools.subset-max.luna}{100.0}
\resultdef{secondary.acc.untransformed.encouraged-tools.subset-max.max}{100.0}
\resultdef{secondary.acc.untransformed.encouraged-tools.subset-max.min}{100.0}
\resultdef{secondary.acc.untransformed.encouraged-tools.subset-max.qwen}{100.0}
\resultdef{secondary.acc.untransformed.encouraged-tools.subset-max.sol}{100.0}
\resultdef{secondary.acc.untransformed.encouraged-tools.subset-min.luna}{97.6}
\resultdef{secondary.acc.untransformed.encouraged-tools.subset-min.max}{99.2}
\resultdef{secondary.acc.untransformed.encouraged-tools.subset-min.min}{97.6}
\resultdef{secondary.acc.untransformed.encouraged-tools.subset-min.qwen}{98.1}
\resultdef{secondary.acc.untransformed.encouraged-tools.subset-min.sol}{99.2}
\resultdef{secondary.acc.untransformed.encouraged-tools.wrong-arithmetic.luna}{100.0}
\resultdef{secondary.acc.untransformed.encouraged-tools.wrong-arithmetic.max}{100.0}
\resultdef{secondary.acc.untransformed.encouraged-tools.wrong-arithmetic.min}{99.5}
\resultdef{secondary.acc.untransformed.encouraged-tools.wrong-arithmetic.qwen}{99.5}
\resultdef{secondary.acc.untransformed.encouraged-tools.wrong-arithmetic.sol}{100.0}
\resultdef{secondary.acc.untransformed.encouraged-tools.wrong-logic.luna}{98.6}
\resultdef{secondary.acc.untransformed.encouraged-tools.wrong-logic.max}{100.0}
\resultdef{secondary.acc.untransformed.encouraged-tools.wrong-logic.min}{98.1}
\resultdef{secondary.acc.untransformed.encouraged-tools.wrong-logic.qwen}{98.1}
\resultdef{secondary.acc.untransformed.encouraged-tools.wrong-logic.sol}{100.0}
\resultdef{secondary.acc.untransformed.encouraged-tools.wrong-variable.luna}{98.5}
\resultdef{secondary.acc.untransformed.encouraged-tools.wrong-variable.max}{100.0}
\resultdef{secondary.acc.untransformed.encouraged-tools.wrong-variable.min}{98.5}
\resultdef{secondary.acc.untransformed.encouraged-tools.wrong-variable.qwen}{100.0}
\resultdef{secondary.acc.untransformed.encouraged-tools.wrong-variable.sol}{100.0}
\resultdef{secondary.acc.untransformed.no-tools.control-flow.luna}{98.2}
\resultdef{secondary.acc.untransformed.no-tools.control-flow.max}{99.9}
\resultdef{secondary.acc.untransformed.no-tools.control-flow.min}{98.2}
\resultdef{secondary.acc.untransformed.no-tools.control-flow.qwen}{98.8}
\resultdef{secondary.acc.untransformed.no-tools.control-flow.sol}{99.9}
\resultdef{secondary.acc.untransformed.no-tools.distractors.luna}{98.2}
\resultdef{secondary.acc.untransformed.no-tools.distractors.max}{99.9}
\resultdef{secondary.acc.untransformed.no-tools.distractors.min}{98.2}
\resultdef{secondary.acc.untransformed.no-tools.distractors.qwen}{98.8}
\resultdef{secondary.acc.untransformed.no-tools.distractors.sol}{99.9}
\resultdef{secondary.acc.untransformed.no-tools.luna}{98.2}
\resultdef{secondary.acc.untransformed.no-tools.max}{99.9}
\resultdef{secondary.acc.untransformed.no-tools.maximal-mix.luna}{98.2}
\resultdef{secondary.acc.untransformed.no-tools.maximal-mix.max}{99.9}
\resultdef{secondary.acc.untransformed.no-tools.maximal-mix.min}{98.2}
\resultdef{secondary.acc.untransformed.no-tools.maximal-mix.qwen}{98.8}
\resultdef{secondary.acc.untransformed.no-tools.maximal-mix.sol}{99.9}
\resultdef{secondary.acc.untransformed.no-tools.min}{98.2}
\resultdef{secondary.acc.untransformed.no-tools.off-by-one.luna}{96.4}
\resultdef{secondary.acc.untransformed.no-tools.off-by-one.max}{99.6}
\resultdef{secondary.acc.untransformed.no-tools.off-by-one.min}{96.4}
\resultdef{secondary.acc.untransformed.no-tools.off-by-one.qwen}{97.2}
\resultdef{secondary.acc.untransformed.no-tools.off-by-one.sol}{99.6}
\resultdef{secondary.acc.untransformed.no-tools.qwen}{98.8}
\resultdef{secondary.acc.untransformed.no-tools.rename.luna}{98.1}
\resultdef{secondary.acc.untransformed.no-tools.rename.max}{99.9}
\resultdef{secondary.acc.untransformed.no-tools.rename.min}{98.1}
\resultdef{secondary.acc.untransformed.no-tools.rename.qwen}{98.7}
\resultdef{secondary.acc.untransformed.no-tools.rename.sol}{99.9}
\resultdef{secondary.acc.untransformed.no-tools.sol}{99.9}
\resultdef{secondary.acc.untransformed.no-tools.state-and-literals.luna}{98.2}
\resultdef{secondary.acc.untransformed.no-tools.state-and-literals.max}{99.9}
\resultdef{secondary.acc.untransformed.no-tools.state-and-literals.min}{98.2}
\resultdef{secondary.acc.untransformed.no-tools.state-and-literals.qwen}{98.8}
\resultdef{secondary.acc.untransformed.no-tools.state-and-literals.sol}{99.9}
\resultdef{secondary.acc.untransformed.no-tools.static-return.luna}{98.9}
\resultdef{secondary.acc.untransformed.no-tools.static-return.max}{100.0}
\resultdef{secondary.acc.untransformed.no-tools.static-return.min}{98.9}
\resultdef{secondary.acc.untransformed.no-tools.static-return.qwen}{99.3}
\resultdef{secondary.acc.untransformed.no-tools.static-return.sol}{100.0}
\resultdef{secondary.acc.untransformed.no-tools.subset-max.luna}{99.5}
\resultdef{secondary.acc.untransformed.no-tools.subset-max.max}{100.0}
\resultdef{secondary.acc.untransformed.no-tools.subset-max.min}{99.5}
\resultdef{secondary.acc.untransformed.no-tools.subset-max.qwen}{100.0}
\resultdef{secondary.acc.untransformed.no-tools.subset-max.sol}{100.0}
\resultdef{secondary.acc.untransformed.no-tools.subset-min.luna}{96.4}
\resultdef{secondary.acc.untransformed.no-tools.subset-min.max}{99.6}
\resultdef{secondary.acc.untransformed.no-tools.subset-min.min}{96.4}
\resultdef{secondary.acc.untransformed.no-tools.subset-min.qwen}{97.2}
\resultdef{secondary.acc.untransformed.no-tools.subset-min.sol}{99.6}
\resultdef{secondary.acc.untransformed.no-tools.wrong-arithmetic.luna}{99.5}
\resultdef{secondary.acc.untransformed.no-tools.wrong-arithmetic.max}{100.0}
\resultdef{secondary.acc.untransformed.no-tools.wrong-arithmetic.min}{99.5}
\resultdef{secondary.acc.untransformed.no-tools.wrong-arithmetic.qwen}{100.0}
\resultdef{secondary.acc.untransformed.no-tools.wrong-arithmetic.sol}{100.0}
\resultdef{secondary.acc.untransformed.no-tools.wrong-logic.luna}{97.7}
\resultdef{secondary.acc.untransformed.no-tools.wrong-logic.max}{100.0}
\resultdef{secondary.acc.untransformed.no-tools.wrong-logic.min}{97.2}
\resultdef{secondary.acc.untransformed.no-tools.wrong-logic.qwen}{97.2}
\resultdef{secondary.acc.untransformed.no-tools.wrong-logic.sol}{100.0}
\resultdef{secondary.acc.untransformed.no-tools.wrong-variable.luna}{98.5}
\resultdef{secondary.acc.untransformed.no-tools.wrong-variable.max}{100.0}
\resultdef{secondary.acc.untransformed.no-tools.wrong-variable.min}{98.5}
\resultdef{secondary.acc.untransformed.no-tools.wrong-variable.qwen}{100.0}
\resultdef{secondary.acc.untransformed.no-tools.wrong-variable.sol}{100.0}
\resultdef{secondary.diff.ci.luna}{[+1.6, +3.6]}
\resultdef{secondary.diff.ci.qwen}{[+0.6, +2.3]}
\resultdef{secondary.diff.ci.sol}{[$-$0.3, +0.2]}
\resultdef{secondary.diff.control-flow.ci.luna}{[+2.7, +7.6]}
\resultdef{secondary.diff.control-flow.ci.qwen}{[+1.3, +4.9]}
\resultdef{secondary.diff.control-flow.ci.sol}{[$-$0.6, +0.2]}
\resultdef{secondary.diff.control-flow.luna}{+5.2}
\resultdef{secondary.diff.control-flow.max}{+5.2}
\resultdef{secondary.diff.control-flow.min}{$-$0.2}
\resultdef{secondary.diff.control-flow.qwen}{+3.1}
\resultdef{secondary.diff.control-flow.sol}{$-$0.2}
\resultdef{secondary.diff.distractors.ci.luna}{[$-$0.3, +1.8]}
\resultdef{secondary.diff.distractors.ci.qwen}{[$-$0.6, +1.3]}
\resultdef{secondary.diff.distractors.ci.sol}{[$-$0.2, +0.2]}
\resultdef{secondary.diff.distractors.luna}{+0.7}
\resultdef{secondary.diff.distractors.max}{+0.7}
\resultdef{secondary.diff.distractors.min}{+0.0}
\resultdef{secondary.diff.distractors.qwen}{+0.3}
\resultdef{secondary.diff.distractors.sol}{+0.0}
\resultdef{secondary.diff.luna}{+2.6}
\resultdef{secondary.diff.max}{+2.6}
\resultdef{secondary.diff.maximal-mix.ci.luna}{[+2.9, +7.4]}
\resultdef{secondary.diff.maximal-mix.ci.qwen}{[+1.4, +4.4]}
\resultdef{secondary.diff.maximal-mix.ci.sol}{[$-$0.9, +0.2]}
\resultdef{secondary.diff.maximal-mix.luna}{+5.2}
\resultdef{secondary.diff.maximal-mix.max}{+5.2}
\resultdef{secondary.diff.maximal-mix.min}{$-$0.3}
\resultdef{secondary.diff.maximal-mix.qwen}{+2.9}
\resultdef{secondary.diff.maximal-mix.sol}{$-$0.3}
\resultdef{secondary.diff.min}{$-$0.1}
\resultdef{secondary.diff.off-by-one.ci.luna}{[$-$0.3, +4.2]}
\resultdef{secondary.diff.off-by-one.ci.qwen}{[$-$1.2, +2.7]}
\resultdef{secondary.diff.off-by-one.ci.sol}{[$-$0.3, +1.2]}
\resultdef{secondary.diff.off-by-one.luna}{+1.9}
\resultdef{secondary.diff.off-by-one.max}{+1.9}
\resultdef{secondary.diff.off-by-one.min}{+0.2}
\resultdef{secondary.diff.off-by-one.qwen}{+0.8}
\resultdef{secondary.diff.off-by-one.sol}{+0.2}
\resultdef{secondary.diff.qwen}{+1.5}
\resultdef{secondary.diff.rename-distractors.subset-max.luna}{+0.7}
\resultdef{secondary.diff.rename-distractors.subset-max.max}{+0.7}
\resultdef{secondary.diff.rename-distractors.subset-max.min}{+0.1}
\resultdef{secondary.diff.rename-distractors.subset-max.qwen}{+0.3}
\resultdef{secondary.diff.rename-distractors.subset-max.sol}{+0.1}
\resultdef{secondary.diff.rename-distractors.subset-min.luna}{+0.1}
\resultdef{secondary.diff.rename-distractors.subset-min.max}{+0.3}
\resultdef{secondary.diff.rename-distractors.subset-min.min}{+0.0}
\resultdef{secondary.diff.rename-distractors.subset-min.qwen}{+0.3}
\resultdef{secondary.diff.rename-distractors.subset-min.sol}{+0.0}
\resultdef{secondary.diff.rename.ci.luna}{[$-$0.7, +0.9]}
\resultdef{secondary.diff.rename.ci.qwen}{[$-$0.6, +1.1]}
\resultdef{secondary.diff.rename.ci.sol}{[+0.0, +0.3]}
\resultdef{secondary.diff.rename.luna}{+0.1}
\resultdef{secondary.diff.rename.max}{+0.3}
\resultdef{secondary.diff.rename.min}{+0.1}
\resultdef{secondary.diff.rename.qwen}{+0.3}
\resultdef{secondary.diff.rename.sol}{+0.1}
\resultdef{secondary.diff.sol}{$-$0.1}
\resultdef{secondary.diff.state-and-literals.ci.luna}{[+0.2, +3.5]}
\resultdef{secondary.diff.state-and-literals.ci.qwen}{[$-$0.3, +2.0]}
\resultdef{secondary.diff.state-and-literals.ci.sol}{[$-$0.5, +0.5]}
\resultdef{secondary.diff.state-and-literals.luna}{+1.8}
\resultdef{secondary.diff.state-and-literals.max}{+1.8}
\resultdef{secondary.diff.state-and-literals.min}{+0.0}
\resultdef{secondary.diff.state-and-literals.qwen}{+0.8}
\resultdef{secondary.diff.state-and-literals.sol}{+0.0}
\resultdef{secondary.diff.static-return.ci.luna}{[+5.1, +10.3]}
\resultdef{secondary.diff.static-return.ci.qwen}{[+2.2, +6.1]}
\resultdef{secondary.diff.static-return.ci.sol}{[$-$0.3, +0.4]}
\resultdef{secondary.diff.static-return.luna}{+7.7}
\resultdef{secondary.diff.static-return.max}{+7.7}
\resultdef{secondary.diff.static-return.min}{+0.0}
\resultdef{secondary.diff.static-return.qwen}{+4.1}
\resultdef{secondary.diff.static-return.sol}{+0.0}
\resultdef{secondary.diff.subset-max.luna}{+7.7}
\resultdef{secondary.diff.subset-max.max}{+7.7}
\resultdef{secondary.diff.subset-max.min}{+0.2}
\resultdef{secondary.diff.subset-max.qwen}{+4.1}
\resultdef{secondary.diff.subset-max.sol}{+0.2}
\resultdef{secondary.diff.subset-min.luna}{$-$0.8}
\resultdef{secondary.diff.subset-min.max}{+0.3}
\resultdef{secondary.diff.subset-min.min}{$-$0.8}
\resultdef{secondary.diff.subset-min.qwen}{+0.3}
\resultdef{secondary.diff.subset-min.sol}{$-$0.4}
\resultdef{secondary.diff.wrong-arithmetic.ci.luna}{[$-$0.3, +2.1]}
\resultdef{secondary.diff.wrong-arithmetic.ci.qwen}{[+0.0, +2.3]}
\resultdef{secondary.diff.wrong-arithmetic.ci.sol}{[$-$1.1, +0.0]}
\resultdef{secondary.diff.wrong-arithmetic.luna}{+0.9}
\resultdef{secondary.diff.wrong-arithmetic.max}{+1.0}
\resultdef{secondary.diff.wrong-arithmetic.min}{$-$0.4}
\resultdef{secondary.diff.wrong-arithmetic.qwen}{+1.0}
\resultdef{secondary.diff.wrong-arithmetic.sol}{$-$0.4}
\resultdef{secondary.diff.wrong-logic.ci.luna}{[+0.7, +4.5]}
\resultdef{secondary.diff.wrong-logic.ci.qwen}{[$-$2.3, +3.2]}
\resultdef{secondary.diff.wrong-logic.ci.sol}{[+0.0, +0.0]}
\resultdef{secondary.diff.wrong-logic.luna}{+2.6}
\resultdef{secondary.diff.wrong-logic.max}{+2.6}
\resultdef{secondary.diff.wrong-logic.min}{+0.0}
\resultdef{secondary.diff.wrong-logic.qwen}{+0.7}
\resultdef{secondary.diff.wrong-logic.sol}{+0.0}
\resultdef{secondary.diff.wrong-variable.ci.luna}{[$-$2.5, +0.9]}
\resultdef{secondary.diff.wrong-variable.ci.qwen}{[$-$0.8, +1.6]}
\resultdef{secondary.diff.wrong-variable.ci.sol}{[$-$1.2, +0.2]}
\resultdef{secondary.diff.wrong-variable.luna}{$-$0.8}
\resultdef{secondary.diff.wrong-variable.max}{+0.5}
\resultdef{secondary.diff.wrong-variable.min}{$-$0.8}
\resultdef{secondary.diff.wrong-variable.qwen}{+0.5}
\resultdef{secondary.diff.wrong-variable.sol}{$-$0.3}
\resultdef{secondary.disc-minus-enc.transformed.ci.luna}{[$-$1.4, $-$0.1]}
\resultdef{secondary.disc-minus-enc.transformed.ci.qwen}{[$-$0.6, +0.2]}
\resultdef{secondary.disc-minus-enc.transformed.ci.sol}{[$-$0.2, +0.3]}
\resultdef{secondary.disc-minus-enc.transformed.luna}{$-$0.7}
\resultdef{secondary.disc-minus-enc.transformed.max}{+0.0}
\resultdef{secondary.disc-minus-enc.transformed.min}{$-$0.7}
\resultdef{secondary.disc-minus-enc.transformed.qwen}{$-$0.2}
\resultdef{secondary.disc-minus-enc.transformed.sol}{+0.0}
\resultdef{secondary.disc-minus-enc.untransformed.ci.luna}{[$-$1.6, $-$0.2]}
\resultdef{secondary.disc-minus-enc.untransformed.ci.qwen}{[$-$0.7, +0.3]}
\resultdef{secondary.disc-minus-enc.untransformed.ci.sol}{[$-$0.2, +0.0]}
\resultdef{secondary.disc-minus-enc.untransformed.luna}{$-$0.8}
\resultdef{secondary.disc-minus-enc.untransformed.max}{$-$0.1}
\resultdef{secondary.disc-minus-enc.untransformed.min}{$-$0.8}
\resultdef{secondary.disc-minus-enc.untransformed.qwen}{$-$0.2}
\resultdef{secondary.disc-minus-enc.untransformed.sol}{$-$0.1}
\resultdef{secondary.disc-minus-no.transformed.ci.luna}{[+1.7, +3.4]}
\resultdef{secondary.disc-minus-no.transformed.ci.qwen}{[+1.1, +2.2]}
\resultdef{secondary.disc-minus-no.transformed.ci.sol}{[$-$0.3, +0.0]}
\resultdef{secondary.disc-minus-no.transformed.luna}{+2.6}
\resultdef{secondary.disc-minus-no.transformed.max}{+2.6}
\resultdef{secondary.disc-minus-no.transformed.min}{$-$0.1}
\resultdef{secondary.disc-minus-no.transformed.qwen}{+1.6}
\resultdef{secondary.disc-minus-no.transformed.sol}{$-$0.1}
\resultdef{secondary.disc-minus-no.untransformed.ci.luna}{[$-$1.0, +0.7]}
\resultdef{secondary.disc-minus-no.untransformed.ci.qwen}{[$-$0.3, +0.7]}
\resultdef{secondary.disc-minus-no.untransformed.ci.sol}{[$-$0.4, +0.0]}
\resultdef{secondary.disc-minus-no.untransformed.luna}{$-$0.1}
\resultdef{secondary.disc-minus-no.untransformed.max}{+0.2}
\resultdef{secondary.disc-minus-no.untransformed.min}{$-$0.2}
\resultdef{secondary.disc-minus-no.untransformed.qwen}{+0.2}
\resultdef{secondary.disc-minus-no.untransformed.sol}{$-$0.2}
\resultdef{secondary.effect.transformed.ci.luna}{[+2.5, +4.1]}
\resultdef{secondary.effect.transformed.ci.qwen}{[+1.3, +2.4]}
\resultdef{secondary.effect.transformed.ci.sol}{[$-$0.4, +0.0]}
\resultdef{secondary.effect.transformed.control-flow.ci.luna}{[+3.5, +8.3]}
\resultdef{secondary.effect.transformed.control-flow.ci.qwen}{[+1.9, +5.1]}
\resultdef{secondary.effect.transformed.control-flow.ci.sol}{[$-$0.7, +0.0]}
\resultdef{secondary.effect.transformed.control-flow.luna}{+5.9}
\resultdef{secondary.effect.transformed.control-flow.max}{+5.9}
\resultdef{secondary.effect.transformed.control-flow.min}{$-$0.2}
\resultdef{secondary.effect.transformed.control-flow.qwen}{+3.4}
\resultdef{secondary.effect.transformed.control-flow.sol}{$-$0.2}
\resultdef{secondary.effect.transformed.distractors.ci.luna}{[+0.5, +2.5]}
\resultdef{secondary.effect.transformed.distractors.ci.qwen}{[$-$0.1, +1.6]}
\resultdef{secondary.effect.transformed.distractors.ci.sol}{[$-$0.3, +0.0]}
\resultdef{secondary.effect.transformed.distractors.luna}{+1.5}
\resultdef{secondary.effect.transformed.distractors.max}{+1.5}
\resultdef{secondary.effect.transformed.distractors.min}{$-$0.1}
\resultdef{secondary.effect.transformed.distractors.qwen}{+0.7}
\resultdef{secondary.effect.transformed.distractors.sol}{$-$0.1}
\resultdef{secondary.effect.transformed.luna}{+3.4}
\resultdef{secondary.effect.transformed.max}{+3.4}
\resultdef{secondary.effect.transformed.maximal-mix.ci.luna}{[+3.6, +8.0]}
\resultdef{secondary.effect.transformed.maximal-mix.ci.qwen}{[+1.9, +4.7]}
\resultdef{secondary.effect.transformed.maximal-mix.ci.sol}{[$-$0.9, +0.0]}
\resultdef{secondary.effect.transformed.maximal-mix.luna}{+5.9}
\resultdef{secondary.effect.transformed.maximal-mix.max}{+5.9}
\resultdef{secondary.effect.transformed.maximal-mix.min}{$-$0.4}
\resultdef{secondary.effect.transformed.maximal-mix.qwen}{+3.2}
\resultdef{secondary.effect.transformed.maximal-mix.sol}{$-$0.4}
\resultdef{secondary.effect.transformed.min}{$-$0.2}
\resultdef{secondary.effect.transformed.off-by-one.ci.luna}{[+1.1, +5.1]}
\resultdef{secondary.effect.transformed.off-by-one.ci.qwen}{[+1.2, +3.8]}
\resultdef{secondary.effect.transformed.off-by-one.ci.sol}{[$-$0.4, +0.0]}
\resultdef{secondary.effect.transformed.off-by-one.luna}{+3.1}
\resultdef{secondary.effect.transformed.off-by-one.max}{+3.1}
\resultdef{secondary.effect.transformed.off-by-one.min}{$-$0.2}
\resultdef{secondary.effect.transformed.off-by-one.qwen}{+2.4}
\resultdef{secondary.effect.transformed.off-by-one.sol}{$-$0.2}
\resultdef{secondary.effect.transformed.qwen}{+1.8}
\resultdef{secondary.effect.transformed.rename.ci.luna}{[+0.2, +1.6]}
\resultdef{secondary.effect.transformed.rename.ci.qwen}{[+0.0, +1.3]}
\resultdef{secondary.effect.transformed.rename.ci.sol}{[+0.0, +0.0]}
\resultdef{secondary.effect.transformed.rename.luna}{+0.9}
\resultdef{secondary.effect.transformed.rename.max}{+0.9}
\resultdef{secondary.effect.transformed.rename.min}{+0.0}
\resultdef{secondary.effect.transformed.rename.qwen}{+0.6}
\resultdef{secondary.effect.transformed.rename.sol}{+0.0}
\resultdef{secondary.effect.transformed.sol}{$-$0.2}
\resultdef{secondary.effect.transformed.state-and-literals.ci.luna}{[+0.9, +4.2]}
\resultdef{secondary.effect.transformed.state-and-literals.ci.qwen}{[+0.2, +2.1]}
\resultdef{secondary.effect.transformed.state-and-literals.ci.sol}{[$-$0.7, +0.4]}
\resultdef{secondary.effect.transformed.state-and-literals.luna}{+2.5}
\resultdef{secondary.effect.transformed.state-and-literals.max}{+2.5}
\resultdef{secondary.effect.transformed.state-and-literals.min}{$-$0.1}
\resultdef{secondary.effect.transformed.state-and-literals.qwen}{+1.1}
\resultdef{secondary.effect.transformed.state-and-literals.sol}{$-$0.1}
\resultdef{secondary.effect.transformed.static-return.ci.luna}{[+6.3, +11.3]}
\resultdef{secondary.effect.transformed.static-return.ci.qwen}{[+2.4, +5.3]}
\resultdef{secondary.effect.transformed.static-return.ci.sol}{[$-$0.3, +0.4]}
\resultdef{secondary.effect.transformed.static-return.luna}{+8.8}
\resultdef{secondary.effect.transformed.static-return.max}{+8.8}
\resultdef{secondary.effect.transformed.static-return.min}{+0.0}
\resultdef{secondary.effect.transformed.static-return.qwen}{+3.8}
\resultdef{secondary.effect.transformed.static-return.sol}{+0.0}
\resultdef{secondary.effect.transformed.subset-max.luna}{+8.8}
\resultdef{secondary.effect.transformed.subset-max.max}{+8.8}
\resultdef{secondary.effect.transformed.subset-max.min}{+0.0}
\resultdef{secondary.effect.transformed.subset-max.qwen}{+3.8}
\resultdef{secondary.effect.transformed.subset-max.sol}{+0.0}
\resultdef{secondary.effect.transformed.subset-min.luna}{$-$0.8}
\resultdef{secondary.effect.transformed.subset-min.max}{+0.5}
\resultdef{secondary.effect.transformed.subset-min.min}{$-$0.8}
\resultdef{secondary.effect.transformed.subset-min.qwen}{+0.5}
\resultdef{secondary.effect.transformed.subset-min.sol}{$-$0.4}
\resultdef{secondary.effect.transformed.subsets-ci-positive.luna}{9}
\resultdef{secondary.effect.transformed.subsets-ci-positive.max}{9}
\resultdef{secondary.effect.transformed.subsets-ci-positive.min}{0}
\resultdef{secondary.effect.transformed.subsets-ci-positive.qwen}{6}
\resultdef{secondary.effect.transformed.subsets-ci-positive.sol}{0}
\resultdef{secondary.effect.transformed.subsets-positive.luna}{9}
\resultdef{secondary.effect.transformed.subsets-positive.max}{10}
\resultdef{secondary.effect.transformed.subsets-positive.min}{0}
\resultdef{secondary.effect.transformed.subsets-positive.qwen}{10}
\resultdef{secondary.effect.transformed.subsets-positive.sol}{0}
\resultdef{secondary.effect.transformed.wrong-arithmetic.ci.luna}{[+0.2, +2.6]}
\resultdef{secondary.effect.transformed.wrong-arithmetic.ci.qwen}{[$-$0.1, +1.1]}
\resultdef{secondary.effect.transformed.wrong-arithmetic.ci.sol}{[$-$1.1, +0.0]}
\resultdef{secondary.effect.transformed.wrong-arithmetic.luna}{+1.4}
\resultdef{secondary.effect.transformed.wrong-arithmetic.max}{+1.4}
\resultdef{secondary.effect.transformed.wrong-arithmetic.min}{$-$0.4}
\resultdef{secondary.effect.transformed.wrong-arithmetic.qwen}{+0.5}
\resultdef{secondary.effect.transformed.wrong-arithmetic.sol}{$-$0.4}
\resultdef{secondary.effect.transformed.wrong-logic.ci.luna}{[+2.2, +5.1]}
\resultdef{secondary.effect.transformed.wrong-logic.ci.qwen}{[+0.5, +2.8]}
\resultdef{secondary.effect.transformed.wrong-logic.ci.sol}{[+0.0, +0.0]}
\resultdef{secondary.effect.transformed.wrong-logic.luna}{+3.5}
\resultdef{secondary.effect.transformed.wrong-logic.max}{+3.5}
\resultdef{secondary.effect.transformed.wrong-logic.min}{+0.0}
\resultdef{secondary.effect.transformed.wrong-logic.qwen}{+1.6}
\resultdef{secondary.effect.transformed.wrong-logic.sol}{+0.0}
\resultdef{secondary.effect.transformed.wrong-variable.ci.luna}{[$-$2.1, +0.5]}
\resultdef{secondary.effect.transformed.wrong-variable.ci.qwen}{[$-$0.8, +1.6]}
\resultdef{secondary.effect.transformed.wrong-variable.ci.sol}{[$-$1.2, +0.2]}
\resultdef{secondary.effect.transformed.wrong-variable.luna}{$-$0.8}
\resultdef{secondary.effect.transformed.wrong-variable.max}{+0.5}
\resultdef{secondary.effect.transformed.wrong-variable.min}{$-$0.8}
\resultdef{secondary.effect.transformed.wrong-variable.qwen}{+0.5}
\resultdef{secondary.effect.transformed.wrong-variable.sol}{$-$0.3}
\resultdef{secondary.effect.untransformed.ci.luna}{[+0.2, +1.4]}
\resultdef{secondary.effect.untransformed.ci.qwen}{[$-$0.3, +1.1]}
\resultdef{secondary.effect.untransformed.ci.sol}{[$-$0.3, +0.0]}
\resultdef{secondary.effect.untransformed.control-flow.ci.luna}{[+0.2, +1.3]}
\resultdef{secondary.effect.untransformed.control-flow.ci.qwen}{[$-$0.3, +1.0]}
\resultdef{secondary.effect.untransformed.control-flow.ci.sol}{[$-$0.3, +0.0]}
\resultdef{secondary.effect.untransformed.control-flow.luna}{+0.7}
\resultdef{secondary.effect.untransformed.control-flow.max}{+0.7}
\resultdef{secondary.effect.untransformed.control-flow.min}{$-$0.1}
\resultdef{secondary.effect.untransformed.control-flow.qwen}{+0.3}
\resultdef{secondary.effect.untransformed.control-flow.sol}{$-$0.1}
\resultdef{secondary.effect.untransformed.distractors.ci.luna}{[+0.2, +1.3]}
\resultdef{secondary.effect.untransformed.distractors.ci.qwen}{[$-$0.3, +1.0]}
\resultdef{secondary.effect.untransformed.distractors.ci.sol}{[$-$0.3, +0.0]}
\resultdef{secondary.effect.untransformed.distractors.luna}{+0.7}
\resultdef{secondary.effect.untransformed.distractors.max}{+0.7}
\resultdef{secondary.effect.untransformed.distractors.min}{$-$0.1}
\resultdef{secondary.effect.untransformed.distractors.qwen}{+0.3}
\resultdef{secondary.effect.untransformed.distractors.sol}{$-$0.1}
\resultdef{secondary.effect.untransformed.luna}{+0.7}
\resultdef{secondary.effect.untransformed.max}{+0.7}
\resultdef{secondary.effect.untransformed.maximal-mix.ci.luna}{[+0.2, +1.3]}
\resultdef{secondary.effect.untransformed.maximal-mix.ci.qwen}{[$-$0.3, +1.1]}
\resultdef{secondary.effect.untransformed.maximal-mix.ci.sol}{[$-$0.3, +0.0]}
\resultdef{secondary.effect.untransformed.maximal-mix.luna}{+0.7}
\resultdef{secondary.effect.untransformed.maximal-mix.max}{+0.7}
\resultdef{secondary.effect.untransformed.maximal-mix.min}{$-$0.1}
\resultdef{secondary.effect.untransformed.maximal-mix.qwen}{+0.3}
\resultdef{secondary.effect.untransformed.maximal-mix.sol}{$-$0.1}
\resultdef{secondary.effect.untransformed.min}{$-$0.1}
\resultdef{secondary.effect.untransformed.off-by-one.ci.luna}{[+0.0, +2.7]}
\resultdef{secondary.effect.untransformed.off-by-one.ci.qwen}{[+0.0, +3.7]}
\resultdef{secondary.effect.untransformed.off-by-one.ci.sol}{[$-$1.3, +0.0]}
\resultdef{secondary.effect.untransformed.off-by-one.luna}{+1.2}
\resultdef{secondary.effect.untransformed.off-by-one.max}{+1.6}
\resultdef{secondary.effect.untransformed.off-by-one.min}{$-$0.4}
\resultdef{secondary.effect.untransformed.off-by-one.qwen}{+1.6}
\resultdef{secondary.effect.untransformed.off-by-one.sol}{$-$0.4}
\resultdef{secondary.effect.untransformed.qwen}{+0.3}
\resultdef{secondary.effect.untransformed.rename.ci.luna}{[+0.2, +1.4]}
\resultdef{secondary.effect.untransformed.rename.ci.qwen}{[$-$0.3, +1.1]}
\resultdef{secondary.effect.untransformed.rename.ci.sol}{[$-$0.3, +0.0]}
\resultdef{secondary.effect.untransformed.rename.luna}{+0.8}
\resultdef{secondary.effect.untransformed.rename.max}{+0.8}
\resultdef{secondary.effect.untransformed.rename.min}{$-$0.1}
\resultdef{secondary.effect.untransformed.rename.qwen}{+0.3}
\resultdef{secondary.effect.untransformed.rename.sol}{$-$0.1}
\resultdef{secondary.effect.untransformed.sol}{$-$0.1}
\resultdef{secondary.effect.untransformed.state-and-literals.ci.luna}{[+0.2, +1.3]}
\resultdef{secondary.effect.untransformed.state-and-literals.ci.qwen}{[$-$0.3, +1.0]}
\resultdef{secondary.effect.untransformed.state-and-literals.ci.sol}{[$-$0.3, +0.0]}
\resultdef{secondary.effect.untransformed.state-and-literals.luna}{+0.7}
\resultdef{secondary.effect.untransformed.state-and-literals.max}{+0.7}
\resultdef{secondary.effect.untransformed.state-and-literals.min}{$-$0.1}
\resultdef{secondary.effect.untransformed.state-and-literals.qwen}{+0.3}
\resultdef{secondary.effect.untransformed.state-and-literals.sol}{$-$0.1}
\resultdef{secondary.effect.untransformed.static-return.ci.luna}{[+0.0, +2.3]}
\resultdef{secondary.effect.untransformed.static-return.ci.qwen}{[$-$1.5, +0.7]}
\resultdef{secondary.effect.untransformed.static-return.ci.sol}{[+0.0, +0.0]}
\resultdef{secondary.effect.untransformed.static-return.luna}{+1.1}
\resultdef{secondary.effect.untransformed.static-return.max}{+1.1}
\resultdef{secondary.effect.untransformed.static-return.min}{$-$0.4}
\resultdef{secondary.effect.untransformed.static-return.qwen}{$-$0.4}
\resultdef{secondary.effect.untransformed.static-return.sol}{+0.0}
\resultdef{secondary.effect.untransformed.subset-max.luna}{+1.2}
\resultdef{secondary.effect.untransformed.subset-max.max}{+1.6}
\resultdef{secondary.effect.untransformed.subset-max.min}{+0.0}
\resultdef{secondary.effect.untransformed.subset-max.qwen}{+1.6}
\resultdef{secondary.effect.untransformed.subset-max.sol}{+0.0}
\resultdef{secondary.effect.untransformed.subset-min.luna}{+0.0}
\resultdef{secondary.effect.untransformed.subset-min.max}{+0.0}
\resultdef{secondary.effect.untransformed.subset-min.min}{$-$0.5}
\resultdef{secondary.effect.untransformed.subset-min.qwen}{$-$0.5}
\resultdef{secondary.effect.untransformed.subset-min.sol}{$-$0.4}
\resultdef{secondary.effect.untransformed.wrong-arithmetic.ci.luna}{[+0.0, +1.5]}
\resultdef{secondary.effect.untransformed.wrong-arithmetic.ci.qwen}{[$-$1.6, +0.0]}
\resultdef{secondary.effect.untransformed.wrong-arithmetic.ci.sol}{[+0.0, +0.0]}
\resultdef{secondary.effect.untransformed.wrong-arithmetic.luna}{+0.5}
\resultdef{secondary.effect.untransformed.wrong-arithmetic.max}{+0.5}
\resultdef{secondary.effect.untransformed.wrong-arithmetic.min}{$-$0.5}
\resultdef{secondary.effect.untransformed.wrong-arithmetic.qwen}{$-$0.5}
\resultdef{secondary.effect.untransformed.wrong-arithmetic.sol}{+0.0}
\resultdef{secondary.effect.untransformed.wrong-logic.ci.luna}{[$-$0.5, +2.9]}
\resultdef{secondary.effect.untransformed.wrong-logic.ci.qwen}{[$-$1.4, +3.8]}
\resultdef{secondary.effect.untransformed.wrong-logic.ci.sol}{[+0.0, +0.0]}
\resultdef{secondary.effect.untransformed.wrong-logic.luna}{+0.9}
\resultdef{secondary.effect.untransformed.wrong-logic.max}{+0.9}
\resultdef{secondary.effect.untransformed.wrong-logic.min}{+0.0}
\resultdef{secondary.effect.untransformed.wrong-logic.qwen}{+0.9}
\resultdef{secondary.effect.untransformed.wrong-logic.sol}{+0.0}
\resultdef{secondary.effect.untransformed.wrong-variable.ci.luna}{[$-$1.1, +1.1]}
\resultdef{secondary.effect.untransformed.wrong-variable.ci.qwen}{[+0.0, +0.0]}
\resultdef{secondary.effect.untransformed.wrong-variable.ci.sol}{[+0.0, +0.0]}
\resultdef{secondary.effect.untransformed.wrong-variable.luna}{+0.0}
\resultdef{secondary.effect.untransformed.wrong-variable.max}{+0.0}
\resultdef{secondary.effect.untransformed.wrong-variable.min}{+0.0}
\resultdef{secondary.effect.untransformed.wrong-variable.qwen}{+0.0}
\resultdef{secondary.effect.untransformed.wrong-variable.sol}{+0.0}
\resultdef{secondary.enc-minus-disc.transformed.luna}{+0.7}
\resultdef{secondary.enc-minus-disc.transformed.max}{+0.7}
\resultdef{secondary.enc-minus-disc.transformed.min}{+0.0}
\resultdef{secondary.enc-minus-disc.transformed.qwen}{+0.2}
\resultdef{secondary.enc-minus-disc.transformed.sol}{+0.0}
\resultdef{secondary.enc-minus-disc.untransformed.luna}{+0.8}
\resultdef{secondary.enc-minus-disc.untransformed.max}{+0.8}
\resultdef{secondary.enc-minus-disc.untransformed.min}{+0.1}
\resultdef{secondary.enc-minus-disc.untransformed.qwen}{+0.2}
\resultdef{secondary.enc-minus-disc.untransformed.sol}{+0.1}
\resultdef{secondary.error-reduction.transformed.control-flow.luna}{33}
\resultdef{secondary.error-reduction.transformed.control-flow.max}{64}
\resultdef{secondary.error-reduction.transformed.control-flow.min}{-60}
\resultdef{secondary.error-reduction.transformed.control-flow.qwen}{64}
\resultdef{secondary.error-reduction.transformed.control-flow.sol}{-60}
\resultdef{secondary.error-reduction.transformed.distractors.luna}{32}
\resultdef{secondary.error-reduction.transformed.distractors.max}{35}
\resultdef{secondary.error-reduction.transformed.distractors.min}{32}
\resultdef{secondary.error-reduction.transformed.distractors.qwen}{35}
\resultdef{secondary.error-reduction.transformed.luna}{32}
\resultdef{secondary.error-reduction.transformed.max}{51}
\resultdef{secondary.error-reduction.transformed.maximal-mix.luna}{33}
\resultdef{secondary.error-reduction.transformed.maximal-mix.max}{49}
\resultdef{secondary.error-reduction.transformed.maximal-mix.min}{-167}
\resultdef{secondary.error-reduction.transformed.maximal-mix.qwen}{49}
\resultdef{secondary.error-reduction.transformed.maximal-mix.sol}{-167}
\resultdef{secondary.error-reduction.transformed.min}{-67}
\resultdef{secondary.error-reduction.transformed.off-by-one.luna}{24}
\resultdef{secondary.error-reduction.transformed.off-by-one.max}{58}
\resultdef{secondary.error-reduction.transformed.off-by-one.min}{-33}
\resultdef{secondary.error-reduction.transformed.off-by-one.qwen}{58}
\resultdef{secondary.error-reduction.transformed.off-by-one.sol}{-33}
\resultdef{secondary.error-reduction.transformed.qwen}{51}
\resultdef{secondary.error-reduction.transformed.rename.luna}{45}
\resultdef{secondary.error-reduction.transformed.rename.max}{50}
\resultdef{secondary.error-reduction.transformed.rename.min}{0}
\resultdef{secondary.error-reduction.transformed.rename.qwen}{50}
\resultdef{secondary.error-reduction.transformed.rename.sol}{0}
\resultdef{secondary.error-reduction.transformed.sol}{-67}
\resultdef{secondary.error-reduction.transformed.state-and-literals.luna}{24}
\resultdef{secondary.error-reduction.transformed.state-and-literals.max}{44}
\resultdef{secondary.error-reduction.transformed.state-and-literals.min}{-20}
\resultdef{secondary.error-reduction.transformed.state-and-literals.qwen}{44}
\resultdef{secondary.error-reduction.transformed.state-and-literals.sol}{-20}
\resultdef{secondary.error-reduction.transformed.static-return.luna}{46}
\resultdef{secondary.error-reduction.transformed.static-return.max}{60}
\resultdef{secondary.error-reduction.transformed.static-return.min}{0}
\resultdef{secondary.error-reduction.transformed.static-return.qwen}{60}
\resultdef{secondary.error-reduction.transformed.static-return.sol}{0}
\resultdef{secondary.error-reduction.transformed.wrong-arithmetic.luna}{52}
\resultdef{secondary.error-reduction.transformed.wrong-arithmetic.max}{63}
\resultdef{secondary.error-reduction.transformed.wrong-arithmetic.min}{-80}
\resultdef{secondary.error-reduction.transformed.wrong-arithmetic.qwen}{63}
\resultdef{secondary.error-reduction.transformed.wrong-arithmetic.sol}{-80}
\resultdef{secondary.error-reduction.transformed.wrong-logic.luna}{69}
\resultdef{secondary.error-reduction.transformed.wrong-logic.max}{69}
\resultdef{secondary.error-reduction.transformed.wrong-logic.min}{55}
\resultdef{secondary.error-reduction.transformed.wrong-logic.qwen}{55}
\resultdef{secondary.error-reduction.transformed.wrong-variable.luna}{-8}
\resultdef{secondary.error-reduction.transformed.wrong-variable.max}{16}
\resultdef{secondary.error-reduction.transformed.wrong-variable.min}{-200}
\resultdef{secondary.error-reduction.transformed.wrong-variable.qwen}{16}
\resultdef{secondary.error-reduction.transformed.wrong-variable.sol}{-200}
\resultdef{secondary.error-reduction.untransformed.control-flow.luna}{41}
\resultdef{secondary.error-reduction.untransformed.control-flow.max}{41}
\resultdef{secondary.error-reduction.untransformed.control-flow.min}{-100}
\resultdef{secondary.error-reduction.untransformed.control-flow.qwen}{27}
\resultdef{secondary.error-reduction.untransformed.control-flow.sol}{-100}
\resultdef{secondary.error-reduction.untransformed.distractors.luna}{41}
\resultdef{secondary.error-reduction.untransformed.distractors.max}{41}
\resultdef{secondary.error-reduction.untransformed.distractors.min}{-100}
\resultdef{secondary.error-reduction.untransformed.distractors.qwen}{27}
\resultdef{secondary.error-reduction.untransformed.distractors.sol}{-100}
\resultdef{secondary.error-reduction.untransformed.luna}{41}
\resultdef{secondary.error-reduction.untransformed.max}{41}
\resultdef{secondary.error-reduction.untransformed.maximal-mix.luna}{41}
\resultdef{secondary.error-reduction.untransformed.maximal-mix.max}{41}
\resultdef{secondary.error-reduction.untransformed.maximal-mix.min}{-100}
\resultdef{secondary.error-reduction.untransformed.maximal-mix.qwen}{27}
\resultdef{secondary.error-reduction.untransformed.maximal-mix.sol}{-100}
\resultdef{secondary.error-reduction.untransformed.min}{-100}
\resultdef{secondary.error-reduction.untransformed.off-by-one.luna}{33}
\resultdef{secondary.error-reduction.untransformed.off-by-one.max}{57}
\resultdef{secondary.error-reduction.untransformed.off-by-one.min}{-100}
\resultdef{secondary.error-reduction.untransformed.off-by-one.qwen}{57}
\resultdef{secondary.error-reduction.untransformed.off-by-one.sol}{-100}
\resultdef{secondary.error-reduction.untransformed.qwen}{27}
\resultdef{secondary.error-reduction.untransformed.rename.luna}{41}
\resultdef{secondary.error-reduction.untransformed.rename.max}{41}
\resultdef{secondary.error-reduction.untransformed.rename.min}{-100}
\resultdef{secondary.error-reduction.untransformed.rename.qwen}{27}
\resultdef{secondary.error-reduction.untransformed.rename.sol}{-100}
\resultdef{secondary.error-reduction.untransformed.sol}{-100}
\resultdef{secondary.error-reduction.untransformed.state-and-literals.luna}{41}
\resultdef{secondary.error-reduction.untransformed.state-and-literals.max}{41}
\resultdef{secondary.error-reduction.untransformed.state-and-literals.min}{-100}
\resultdef{secondary.error-reduction.untransformed.state-and-literals.qwen}{27}
\resultdef{secondary.error-reduction.untransformed.state-and-literals.sol}{-100}
\resultdef{secondary.error-reduction.untransformed.static-return.luna}{100}
\resultdef{secondary.error-reduction.untransformed.static-return.max}{100}
\resultdef{secondary.error-reduction.untransformed.static-return.min}{-50}
\resultdef{secondary.error-reduction.untransformed.static-return.qwen}{-50}
\resultdef{secondary.error-reduction.untransformed.wrong-arithmetic.luna}{100}
\resultdef{secondary.error-reduction.untransformed.wrong-logic.luna}{40}
\resultdef{secondary.error-reduction.untransformed.wrong-logic.max}{40}
\resultdef{secondary.error-reduction.untransformed.wrong-logic.min}{33}
\resultdef{secondary.error-reduction.untransformed.wrong-logic.qwen}{33}
\resultdef{secondary.error-reduction.untransformed.wrong-variable.luna}{0}
\resultdef{secondary.extra-correct.minutes.luna}{19.6}
\resultdef{secondary.extra-correct.minutes.max}{260.0}
\resultdef{secondary.extra-correct.minutes.min}{19.6}
\resultdef{secondary.extra-correct.minutes.qwen}{260.0}
\resultdef{secondary.extra-correct.usd.luna}{0.04}
\resultdef{secondary.extra-correct.usd.max}{1.01}
\resultdef{secondary.extra-correct.usd.min}{0.04}
\resultdef{secondary.extra-correct.usd.qwen}{1.01}
\resultdef{secondary.slope.debugger.transformed.ci.qwen}{[+1.5, +10.1]}
\resultdef{secondary.slope.debugger.transformed.ci.sol}{[+0.0, +0.8]}
\resultdef{secondary.slope.debugger.transformed.luna}{+52.3}
\resultdef{secondary.slope.debugger.transformed.max}{+52.3}
\resultdef{secondary.slope.debugger.transformed.min}{+0.3}
\resultdef{secondary.slope.debugger.transformed.placebo-miss.luna}{17.9}
\resultdef{secondary.slope.debugger.transformed.placebo-miss.max}{17.9}
\resultdef{secondary.slope.debugger.transformed.placebo-miss.min}{0.0}
\resultdef{secondary.slope.debugger.transformed.placebo-miss.qwen}{5.8}
\resultdef{secondary.slope.debugger.transformed.placebo-miss.sol}{0.0}
\resultdef{secondary.slope.debugger.transformed.qwen}{+5.7}
\resultdef{secondary.slope.debugger.transformed.sol}{+0.3}
\resultdef{secondary.slope.debugger.untransformed.ci.qwen}{[$-$2.1, +0.7]}
\resultdef{secondary.slope.debugger.untransformed.placebo-miss.qwen}{0.3}
\resultdef{secondary.slope.debugger.untransformed.qwen}{$-$0.7}
\resultdef{secondary.slope.probe.transformed.ci.luna}{[+6.9, +17.4]}
\resultdef{secondary.slope.probe.transformed.ci.qwen}{[+3.5, +8.0]}
\resultdef{secondary.slope.probe.transformed.ci.sol}{[+0.0, +0.2]}
\resultdef{secondary.slope.probe.transformed.luna}{+12.1}
\resultdef{secondary.slope.probe.transformed.max}{+12.1}
\resultdef{secondary.slope.probe.transformed.min}{+0.1}
\resultdef{secondary.slope.probe.transformed.placebo-miss.luna}{6.1}
\resultdef{secondary.slope.probe.transformed.placebo-miss.max}{6.1}
\resultdef{secondary.slope.probe.transformed.placebo-miss.min}{2.0}
\resultdef{secondary.slope.probe.transformed.placebo-miss.qwen}{3.1}
\resultdef{secondary.slope.probe.transformed.placebo-miss.sol}{2.0}
\resultdef{secondary.slope.probe.transformed.qwen}{+5.6}
\resultdef{secondary.slope.probe.transformed.sol}{+0.1}
\resultdef{secondary.slope.probe.untransformed.ci.qwen}{[$-$1.0, +5.5]}
\resultdef{secondary.slope.probe.untransformed.luna}{+3.2}
\resultdef{secondary.slope.probe.untransformed.max}{+3.2}
\resultdef{secondary.slope.probe.untransformed.min}{$-$1.1}
\resultdef{secondary.slope.probe.untransformed.placebo-miss.luna}{14.9}
\resultdef{secondary.slope.probe.untransformed.placebo-miss.max}{71.2}
\resultdef{secondary.slope.probe.untransformed.placebo-miss.min}{3.2}
\resultdef{secondary.slope.probe.untransformed.placebo-miss.qwen}{3.2}
\resultdef{secondary.slope.probe.untransformed.placebo-miss.sol}{71.2}
\resultdef{secondary.slope.probe.untransformed.qwen}{+1.9}
\resultdef{secondary.slope.probe.untransformed.sol}{$-$1.1}
\resultdef{tool.any-debugger.b.luna}{0.0}
\resultdef{tool.any-debugger.b.max}{0.9}
\resultdef{tool.any-debugger.b.min}{0.0}
\resultdef{tool.any-debugger.b.qwen}{0.9}
\resultdef{tool.any-debugger.b.sol}{0.0}
\resultdef{tool.any-debugger.c.luna}{100.0}
\resultdef{tool.any-debugger.c.max}{100.0}
\resultdef{tool.any-debugger.c.min}{20.2}
\resultdef{tool.any-debugger.c.qwen}{20.2}
\resultdef{tool.any-debugger.c.sol}{100.0}
\resultdef{tool.any-debugger.e.luna}{0.3}
\resultdef{tool.any-debugger.e.max}{6.2}
\resultdef{tool.any-debugger.e.min}{0.2}
\resultdef{tool.any-debugger.e.qwen}{6.2}
\resultdef{tool.any-debugger.e.sol}{0.2}
\resultdef{tool.any-debugger.f.luna}{100.0}
\resultdef{tool.any-debugger.f.max}{100.0}
\resultdef{tool.any-debugger.f.min}{41.7}
\resultdef{tool.any-debugger.f.qwen}{41.7}
\resultdef{tool.any-debugger.f.sol}{100.0}
\resultdef{tool.breakpoints.b.qwen}{31.5}
\resultdef{tool.breakpoints.c.luna}{14.6}
\resultdef{tool.breakpoints.c.max}{22.6}
\resultdef{tool.breakpoints.c.min}{14.6}
\resultdef{tool.breakpoints.c.qwen}{22.6}
\resultdef{tool.breakpoints.c.sol}{20.1}
\resultdef{tool.breakpoints.e.luna}{18.4}
\resultdef{tool.breakpoints.e.max}{26.1}
\resultdef{tool.breakpoints.e.min}{18.4}
\resultdef{tool.breakpoints.e.qwen}{26.1}
\resultdef{tool.breakpoints.e.sol}{23.1}
\resultdef{tool.breakpoints.f.luna}{17.5}
\resultdef{tool.breakpoints.f.max}{26.1}
\resultdef{tool.breakpoints.f.min}{17.5}
\resultdef{tool.breakpoints.f.qwen}{26.1}
\resultdef{tool.breakpoints.f.sol}{21.3}
\resultdef{tool.debugger-actions.b.luna}{0.0}
\resultdef{tool.debugger-actions.b.max}{0.1}
\resultdef{tool.debugger-actions.b.min}{0.0}
\resultdef{tool.debugger-actions.b.qwen}{0.1}
\resultdef{tool.debugger-actions.b.sol}{0.0}
\resultdef{tool.debugger-actions.c.luna}{11.6}
\resultdef{tool.debugger-actions.c.max}{11.6}
\resultdef{tool.debugger-actions.c.min}{1.5}
\resultdef{tool.debugger-actions.c.qwen}{1.5}
\resultdef{tool.debugger-actions.c.sol}{8.0}
\resultdef{tool.debugger-actions.e.luna}{0.0}
\resultdef{tool.debugger-actions.e.max}{0.4}
\resultdef{tool.debugger-actions.e.min}{0.0}
\resultdef{tool.debugger-actions.e.qwen}{0.4}
\resultdef{tool.debugger-actions.e.sol}{0.0}
\resultdef{tool.debugger-actions.f.luna}{12.5}
\resultdef{tool.debugger-actions.f.max}{12.5}
\resultdef{tool.debugger-actions.f.min}{3.1}
\resultdef{tool.debugger-actions.f.qwen}{3.1}
\resultdef{tool.debugger-actions.f.sol}{8.4}
\resultdef{tool.inspection.b.qwen}{8.2}
\resultdef{tool.inspection.c.luna}{18.8}
\resultdef{tool.inspection.c.max}{19.7}
\resultdef{tool.inspection.c.min}{13.4}
\resultdef{tool.inspection.c.qwen}{13.4}
\resultdef{tool.inspection.c.sol}{19.7}
\resultdef{tool.inspection.e.luna}{13.2}
\resultdef{tool.inspection.e.max}{17.9}
\resultdef{tool.inspection.e.min}{11.6}
\resultdef{tool.inspection.e.qwen}{11.6}
\resultdef{tool.inspection.e.sol}{17.9}
\resultdef{tool.inspection.f.luna}{16.0}
\resultdef{tool.inspection.f.max}{20.2}
\resultdef{tool.inspection.f.min}{12.4}
\resultdef{tool.inspection.f.qwen}{12.4}
\resultdef{tool.inspection.f.sol}{20.2}
\resultdef{tool.no-tool.b.luna}{91.2}
\resultdef{tool.no-tool.b.max}{91.2}
\resultdef{tool.no-tool.b.min}{0.0}
\resultdef{tool.no-tool.b.qwen}{0.0}
\resultdef{tool.no-tool.b.sol}{13.3}
\resultdef{tool.no-tool.c.luna}{0.0}
\resultdef{tool.no-tool.c.max}{0.1}
\resultdef{tool.no-tool.c.min}{0.0}
\resultdef{tool.no-tool.c.qwen}{0.1}
\resultdef{tool.no-tool.c.sol}{0.0}
\resultdef{tool.no-tool.e.luna}{87.9}
\resultdef{tool.no-tool.e.max}{87.9}
\resultdef{tool.no-tool.e.min}{0.2}
\resultdef{tool.no-tool.e.qwen}{0.2}
\resultdef{tool.no-tool.e.sol}{1.4}
\resultdef{tool.no-tool.f.luna}{0.0}
\resultdef{tool.no-tool.f.max}{0.2}
\resultdef{tool.no-tool.f.min}{0.0}
\resultdef{tool.no-tool.f.qwen}{0.2}
\resultdef{tool.no-tool.f.sol}{0.0}
\resultdef{tool.probe-only.b.luna}{8.8}
\resultdef{tool.probe-only.b.max}{99.1}
\resultdef{tool.probe-only.b.min}{8.8}
\resultdef{tool.probe-only.b.qwen}{99.1}
\resultdef{tool.probe-only.b.sol}{86.7}
\resultdef{tool.probe-only.c.luna}{0.0}
\resultdef{tool.probe-only.c.max}{79.8}
\resultdef{tool.probe-only.c.min}{0.0}
\resultdef{tool.probe-only.c.qwen}{79.8}
\resultdef{tool.probe-only.c.sol}{0.0}
\resultdef{tool.probe-only.e.luna}{11.8}
\resultdef{tool.probe-only.e.max}{98.4}
\resultdef{tool.probe-only.e.min}{11.8}
\resultdef{tool.probe-only.e.qwen}{93.6}
\resultdef{tool.probe-only.e.sol}{98.4}
\resultdef{tool.probe-only.f.luna}{0.0}
\resultdef{tool.probe-only.f.max}{58.1}
\resultdef{tool.probe-only.f.min}{0.0}
\resultdef{tool.probe-only.f.qwen}{58.1}
\resultdef{tool.probe-only.f.sol}{0.0}
\resultdef{tool.probe-runs.b.luna}{0.1}
\resultdef{tool.probe-runs.b.max}{1.8}
\resultdef{tool.probe-runs.b.min}{0.1}
\resultdef{tool.probe-runs.b.qwen}{1.8}
\resultdef{tool.probe-runs.b.sol}{0.9}
\resultdef{tool.probe-runs.c.luna}{1.1}
\resultdef{tool.probe-runs.c.max}{2.3}
\resultdef{tool.probe-runs.c.min}{1.1}
\resultdef{tool.probe-runs.c.qwen}{2.3}
\resultdef{tool.probe-runs.c.sol}{1.1}
\resultdef{tool.probe-runs.e.luna}{0.2}
\resultdef{tool.probe-runs.e.max}{2.2}
\resultdef{tool.probe-runs.e.min}{0.2}
\resultdef{tool.probe-runs.e.qwen}{2.2}
\resultdef{tool.probe-runs.e.sol}{1.0}
\resultdef{tool.probe-runs.f.luna}{1.2}
\resultdef{tool.probe-runs.f.max}{2.7}
\resultdef{tool.probe-runs.f.min}{1.1}
\resultdef{tool.probe-runs.f.qwen}{2.7}
\resultdef{tool.probe-runs.f.sol}{1.1}
\resultdef{tool.stepping.b.qwen}{17.8}
\resultdef{tool.stepping.c.luna}{45.8}
\resultdef{tool.stepping.c.max}{45.8}
\resultdef{tool.stepping.c.min}{27.3}
\resultdef{tool.stepping.c.qwen}{27.3}
\resultdef{tool.stepping.c.sol}{45.5}
\resultdef{tool.stepping.e.luna}{27.6}
\resultdef{tool.stepping.e.max}{27.6}
\resultdef{tool.stepping.e.min}{19.2}
\resultdef{tool.stepping.e.qwen}{19.2}
\resultdef{tool.stepping.e.sol}{23.1}
\resultdef{tool.stepping.f.luna}{41.7}
\resultdef{tool.stepping.f.max}{44.0}
\resultdef{tool.stepping.f.min}{22.9}
\resultdef{tool.stepping.f.qwen}{22.9}
\resultdef{tool.stepping.f.sol}{44.0}

%% file: figures/study-overview.tex
\definecolor{ovP}{RGB}{213,230,246}
\definecolor{ovT}{RGB}{251,225,198}
\definecolor{ovcheck}{RGB}{38,135,70}
\newcommand{\ovcard}[5]{%
  \filldraw[fill=white, draw=black!50, line width=0.4pt, rounded corners=0.8pt] (#1,#2) rectangle (#1+#3,#2-#4);
  \fill[#5, rounded corners=0.6pt] (#1+0.25,#2-0.25) rectangle (#1+#3-0.25,#2-4.4);
  \draw[black!45, line width=0.3pt] (#1,#2-4.4) -- (#1+#3,#2-4.4);}
\newcommand{\ovstack}[5]{\ovcard{#1+4}{#2+4}{#3}{#4}{#5}\ovcard{#1+2}{#2+2}{#3}{#4}{#5}\ovcard{#1}{#2}{#3}{#4}{#5}}
\newcommand{\ovcode}[5]{%
  \foreach \ind/\len [count=\n] in {#5} {%
    \ifnum\len>0
      \draw[black!22, line width={#4*0.3pt}, line cap=round] (#1+2.6+\ind*#3, #2-7.4-\n*#4+#4) -- ++(\len*#3, 0);
    \fi}}
\newcommand{\ovhl}[5]{\fill[faulthl] (#1+0.5, #2-7.4-#5*#4+#4+#4*0.45) rectangle (#1+#3-0.5, #2-7.4-#5*#4+#4-#4*0.45);}
\newcommand{\ovtext}[8]{%
  \draw[black!22, line width={#4*0.3pt}, line cap=round] (#1+2.6+#6*#3, #2-7.4-#5*#4+#4) -- ++(#7*#3, 0);
  \node[anchor=west, inner sep=0pt, font=\fontfamily{lmtt}\fontsize{4.4}{4.4}\selectfont]
    at (#1+2.6+#6*#3+#7*#3+0.9, #2-7.4-#5*#4+#4) {#8};}
\begin{tikzpicture}[x=1pt, y=1pt, font=\sffamily,
  panel/.style={draw=black!30, line width=0.5pt, rounded corners=3pt, fill=white},
  num/.style={circle, fill=black, text=white, inner sep=0pt, minimum size=8.4pt,
    font=\sffamily\bfseries\fontsize{5.8}{6}\selectfont},
  ttl/.style={anchor=base west, yshift=-2.65pt, inner sep=0pt, font=\sffamily\bfseries\boldmath\fontsize{7.4}{8.5}\selectfont},
  txt/.style={inner sep=0pt, font=\sffamily\fontsize{6}{7.3}\selectfont},
  lbl/.style={inner sep=0pt, font=\sffamily\bfseries\boldmath\fontsize{6.2}{7}\selectfont},
  sub/.style={inner sep=0pt, font=\sffamily\fontsize{5.4}{6}\selectfont},
  flow/.style={-{Triangle[length=3.4pt, width=5.2pt]}, line width=1.4pt, black!55},
]
\draw[rounded corners=4pt, fill=black!3, draw=black!20, line width=0.4pt] (0,0) rectangle (516,-186);
\draw[panel] (5,-5) rectangle (103,-98);
\draw[panel] (112,-5) rectangle (239,-98);
\draw[panel] (248,-5) rectangle (403,-98);
\draw[panel] (412,-5) rectangle (511,-98);
\draw[panel] (5,-107) rectangle (151,-181);
\draw[panel] (160,-107) rectangle (511,-181);
\draw[flow] (103.8,-51.5) -- (111.4,-51.5);
\draw[flow] (239.8,-51.5) -- (247.4,-51.5);
\draw[flow] (403.8,-51.5) -- (411.4,-51.5);
\draw[flow] (461.5,-98.8) -- (461.5,-106.4);
\draw[flow] (159.2,-144) -- (151.6,-144);
\node[num] at (14,-14) {1};   \node[ttl] at (21,-14) {Reference programs};
\node[num] at (121,-14) {2};  \node[ttl] at (128,-14) {Fault seeding};
\node[num] at (257,-14) {3};  \node[ttl] at (264,-14) {Transformation of $P$ and $M$};
\node[num] at (421,-14) {4};  \node[ttl] at (428,-14) {Admission};
\node[num] at (169,-116) {5}; \node[ttl] at (176,-116) {Evaluation};
\node[num] at (14,-116) {6};  \node[ttl] at (21,-116) {Outcomes};
\ovcard{10.5}{-28.5}{23}{33}{black!8}
\draw[black!45, line width=1.08pt, line cap=round] (13.1,-35.9) -- ++(9.6,0);
\ovcode{10.5}{-28.5}{0.8}{3.6}{0/0, 0/0, 0/21, 0/21, 0/20, 0/21, 0/13}
\ovcard{40.5}{-28.5}{27}{33}{ovP}
\ovcode{40.5}{-28.5}{0.8}{3.6}{0/13, 2/21, 4/12, 4/12, 5/0, 5/11, 4/8}
\ovtext{40.5}{-28.5}{0.8}{3.6}{5}{5}{5}{\textcolor{black!70}{v - mi}}
\ovcard{74.5}{-28.5}{23}{33}{black!8}
\foreach \len [count=\n] in {10,13,9,12,11,10} {%
  \draw[black!22, line width=1.08pt, line cap=round] (77.1,-35.9-\n*3.6+3.6) -- ++(\len*0.8,0);
  \node[inner sep=0pt, text=ovcheck] at (93.9,-35.9-\n*3.6+3.6) {\fontsize{3.4}{3.4}\selectfont\faCheck};}
\node[sub, anchor=north] at (22,-64.5) {Description};
\node[sub, anchor=north] at (54,-64.5) {Program $P$};
\node[sub, anchor=north] at (86,-64.5) {Tests};
\node[txt, align=center, anchor=north] at (54,-75) {150 from LeetCodeDataset\\50 Easy \textperiodcentered{} 50 Medium \textperiodcentered{} 50 Hard};
\node[inner sep=0pt] at (120.5,-44.7) {\fontsize{11}{11}\selectfont\faBug};
\node[txt, align=left, anchor=north west] at (129,-27.5) {\textbullet\ Off-by-one\\\textbullet\ Wrong arithmetic\\\textbullet\ Wrong logic\\\textbullet\ Wrong variable\\\textbullet\ Static return};
\node[draw=black!25, fill=black!4, rounded corners=1pt, inner sep=1.6pt, font=\fontfamily{lmtt}\fontsize{5.6}{6}\selectfont]
  at (143,-80.2) {\textcolor{black!55}{v - mi}\ $\rightarrow$\ \textcolor{red!60!black}{v + mi}};
\ovstack{184}{-28.5}{40}{36}{ovP}
\ovhl{184}{-28.5}{40}{4.2}{5}
\ovcode{184}{-28.5}{1.0}{4.2}{0/15, 2/28, 4/16, 4/16, 6/0, 6/15, 4/10}
\ovtext{184}{-28.5}{1.0}{4.2}{5}{6}{15}{\textcolor{red!60!black}{v + mi}}
\node[txt, align=center, anchor=north] at (205,-70.5) {\textbf{Faulty programs \boldmath$M$}\\one fault per type\\each fails a test};
\node[txt, align=left, anchor=north west] at (254,-35.1) {\textbullet\ Rename\\\textbullet\ State \& literals\\\textbullet\ Control flow\\\textbullet\ Distractors\\\textbullet\ Maximal mix};
\ovstack{312}{-28.5}{35}{51}{ovT}
\ovcode{312}{-28.5}{0.8}{4.0}{0/15, 2/28, 4/22, 4/17, 4/16, 6/0, 6/27, 6/13, 6/20, 6/12, 4/10}
\ovtext{312}{-28.5}{0.8}{4.0}{6}{6}{11}{\textcolor{black!70}{v - mi}}
\ovstack{356}{-28.5}{35}{51}{ovT}
\ovhl{356}{-28.5}{35}{4.0}{6}
\ovcode{356}{-28.5}{0.8}{4.0}{0/15, 2/28, 4/22, 4/17, 4/16, 6/0, 6/27, 6/13, 6/20, 6/12, 4/10}
\ovtext{356}{-28.5}{0.8}{4.0}{6}{6}{11}{\textcolor{red!60!black}{v + mi}}
\node[lbl] at (331,-86.5) {$P'$};
\node[lbl] at (375,-86.5) {$M'$};
\node[txt, align=left, anchor=north west] at (417,-23) {\textcolor{ovcheck}{\faCheck}\ $P'$ passes all tests\\\textcolor{ovcheck}{\faCheck}\ $M'$ fails like $M$\\\textcolor{ovcheck}{\faCheck}\ restoring $L'$ makes $M'$ pass};
\ovcard{431.5}{-47}{26}{29}{ovP}
\ovhl{431.5}{-47}{26}{3.0}{5}
\ovcode{431.5}{-47}{0.66}{3.0}{0/15, 2/28, 4/16, 4/16, 6/22, 6/15, 4/10}
\ovcard{465.5}{-47}{26}{41}{ovT}
\ovhl{465.5}{-47}{26}{3.0}{6}
\ovcode{465.5}{-47}{0.66}{3.0}{0/15, 2/28, 4/22, 4/17, 4/16, 6/18, 6/27, 6/13, 6/20, 6/12, 4/10}
\draw[red!60!black, densely dashed, line width=0.5pt] (457.5,-66.4) -- (465.5,-69.4);
\node[txt, anchor=east] at (430,-66.4) {$L$};
\node[txt, anchor=west] at (493,-69.4) {$L'$};
\node[lbl] at (461.5,-92.5) {Matched pairs $(M, M')$};
\fill[black!5] (228,-123) rectangle (444,-141);
\fill[ovP] (168,-141) rectangle (228,-157);
\fill[ovT] (168,-157) rectangle (228,-173);
\draw[black!35, line width=0.4pt] (228,-123) rectangle (444,-173) (168,-141) rectangle (228,-173)
  (168,-157) -- (444,-157) (228,-141) -- (444,-141) (300,-123) -- (300,-173) (372,-123) -- (372,-173);
\node[lbl] at (264,-128.5) {No-tools};
\node[lbl] at (336,-128.5) {Discretionary-tools};
\node[lbl] at (408,-128.5) {Encouraged-tools};
\node[sub] at (264,-136) {{\fontfamily{lmtt}\selectfont report\_bug} only};
\node[sub] at (336,-136) {+ {\fontfamily{lmtt}\selectfont run\_probe} and debugger};
\node[sub] at (408,-136) {+ instruction to use them};
\node[lbl] at (198,-149) {Untransformed $M$};
\node[lbl] at (198,-165) {Transformed $M'$};
\foreach \x/\y/\c in {264/-149/A, 336/-149/B, 408/-149/C, 264/-165/D, 336/-165/E, 408/-165/F}
  \node[font=\sffamily\bfseries\fontsize{8.5}{9}\selectfont] at (\x,\y) {\c};
\node[inner sep=0pt] at (477,-140) {\fontsize{13}{13}\selectfont\faRobot};
\node[txt, align=center, anchor=north] at (477,-150) {Sol 6.1 \textperiodcentered{} Luna 6\\Qwen 3.8};
\node[inner sep=0pt] at (33.6,-151.5) {\fontsize{15}{15}\selectfont\faChartBar};
\node[txt, align=left, anchor=north west] at (45.6,-130.3) {\textbullet\ Top-1 accuracy,\ \ $G = D - A$\\\textbullet\ $\Delta = C - A$,\ \ $\Delta' = F - D$\\\textbullet\ $I = \Delta' - \Delta$\\\textbullet\ tokens, time, and cost\\\textbullet\ B and E vs.\ other settings\\\textbullet\ tool use, slope $\beta_1$};
\end{tikzpicture}

%% file: sections/01-introduction.tex
\section{Introduction}
\label{sec:introduction}

Fault localization is central to debugging, requiring a developer or coding agent to identify where a program deviates from its intended behavior~\cite{parnin_2011_debugging,kochhar_2016_expectations,pearson_2017_localization}. Rather than reasoning about execution from source code alone, a coding agent can use \emph{debugger assistance} by running the program and stepping through it in an interactive debugger. Earlier large language models (LLMs) struggled to trace execution statically, predicting 46 to 81\% of outputs on 3-to-13-line Python functions in CRUXEval~\cite{gu_2024_cruxeval,chen_2025_runtime_reasoning}. Debugging systems therefore supplied runtime evidence through recorded traces~\cite{chen_2024_self_debug,zhong_2024_ldb} or interactive debuggers~\cite{levin_2025_chatdbg,yuan_2025_debug_gym,wang_2026_inspectcoder,sun_2026_debugharness}. Yet operating a stateful debugger proved difficult for earlier agents, and direct \texttt{pdb} access often lowered their repair rates~\cite{yuan_2025_debug_gym,wang_2026_inspectcoder}. Recent reasoning models have changed both sides of this trade-off: \texttt{pdb} improved their repair success in debug-gym~\cite{yuan_2025_debug_gym}, while on CRUXEval they predict 93 to 98\% of outputs from source code alone~\cite{lam_2025_codecrash}. On obfuscated LeetCode bugs, optional probes and debugging raised Qwen~3.6's repair accuracy from 78.0\% to 94.8\%, whereas GPT-5.4 moved only from 91.3\% to 91.7\%~\cite{daxerer_2026_thesis}.

This rapid progress in model capabilities raises three questions: how accurately the latest generation of frontier models localize faults from source code alone, how sensitive that accuracy is to behavior-preserving transformations, and how much debugger assistance still improves outcomes on top of static reasoning. However, three gaps in prior evaluations leave these questions open. First, most agentic debugging studies measure end-to-end program repair~\cite{zhong_2024_ldb,yuan_2025_debug_gym,wang_2026_inspectcoder,sun_2026_debugharness}, which confounds fault localization with patch generation because correct patches can arise from imprecise locations~\cite{hossain_2024_localization_repair} and plausible patches can overfit to tests~\cite{smith_2015_overfitting,qi_2015_plausibility,wang_2026_swebench_solved}. Second, high static accuracy on public benchmarks can reflect training-data exposure~\cite{riddell_2024_contamination,ramos_2025_memorization}. Behavior-preserving source transformations reduce verbatim overlap with pretraining corpora~\cite{cao_2025_codecleaner} and caused earlier static LLM localizers to miss 78\% of the faults they had localized in untransformed programs~\cite{haroon_2026_localizability}, but prior work has not measured how much such transformations still affect frontier models or whether debugger assistance recovers any lost accuracy. Third, prior studies either enforce fixed runtime instrumentation~\cite{zhong_2024_ldb,yang_2026_semloc} or evaluate a single combined tool loop~\cite{yuan_2025_debug_gym,wang_2026_inspectcoder,daxerer_2026_thesis}, without separating encouraged from optional tool use, comparing the costs of both~\cite{kapoor_2025_agents,qian_2025_smart}, or tracking whether agents given a choice step through the program in the debugger or simply run it.

We address all three gaps in a controlled experiment on self-contained algorithmic programs, the regime that originally motivated runtime feedback~\cite{gu_2024_cruxeval,chen_2024_self_debug,zhong_2024_ldb}. From 150 LeetCodeDataset problems~\cite{xia_2025_leetcodedataset}, each with a problem description, a verified Python \emph{reference program}, and an executable test suite, we seed \result{programs.faulty.admitted}{610} single-line faults across five fault types (Fig.~\ref{fig:study-overview}). We rewrite each reference and faulty program with five fixed recipes of behavior-preserving transformations, ranging from variable renaming and dead-code distractors to restructuring loops and conditions (\result{programs.transformed.admitted}{3{,}022} transformed programs, Fig.~\ref{fig:example}). We admit only matched pairs where the transformation preserves the reference program's passing behavior, reproduces the faulty program's exact test failure, and keeps the fault restorable at the tracked line. Agents driven by Sol~6.1, Luna~6, and Qwen~3.8 then localize each fault across \result{facts.episodes.kept}{65{,}376} agent runs (\emph{episodes}) in three tool-use settings: \emph{no-tools} (problem description and source code alone), \emph{discretionary-tools} (adding \texttt{run\_probe} alongside a pdb-style debugger for optional use), and \emph{encouraged-tools} (providing the same tools and instructing the agent to gather runtime evidence before deciding).

\begin{figure}[t]
\centering
\figcode
\renewcommand{\arraystretch}{0.92}\setlength{\tabcolsep}{1pt}
\begin{minipage}[t]{0.455\columnwidth}
{\normalfont\footnotesize Untransformed}\\[2pt]
\begin{tabular}{@{}rcl@{}}
\cln{1}{}{class Solution:}
\cln{2}{}{\ \ def maxProfit(self, prices):}
\cln{3}{}{\ \ \ \ ans, mi = 0, inf}
\cln{4}{}{\ \ \ \ for v in prices:}
\cln{5}{\color{gray}-}{\color{gray}\ \ \ \ \ \ ans = max(ans, v - mi)}
\rowcolor{faulthl}\cln{5}{+}{\ \ \ \ \ \ ans = max(ans, v + mi)}
\cln{6}{}{\ \ \ \ \ \ mi = min(mi, v)}
\cln{7}{}{\ \ \ \ return ans}
\end{tabular}
\end{minipage}\hfill
\begin{minipage}[t]{0.525\columnwidth}
{\normalfont\footnotesize Transformed (State and literals)}\\[2pt]
\begin{tabular}{@{}rcl@{}}
\cln{1}{}{class Solution:}
\cln{2}{}{\ \ def maxProfit(self, prices):}
\cln{3}{}{\ \ \ \ \_frag\_0 = (7 - 7), inf}
\cln{4}{}{\ \ \ \ ans, mi = \_frag\_0}
\cln{5}{}{\ \ \ \ for v in prices:}
\cln{6}{\color{gray}-}{\color{gray}\ \ \ \ \ \ \_frag\_1 = (v - mi)}
\rowcolor{faulthl}\cln{6}{+}{\ \ \ \ \ \ \_frag\_1 = (v + mi)}
\cln{7}{}{\ \ \ \ \ \ \_frag\_2 = max(ans, \_frag\_1)}
\cln{8}{}{\ \ \ \ \ \ ans = \_frag\_2}
\cln{9}{}{\ \ \ \ \ \ \_frag\_3 = min(mi, v)}
\cln{10}{}{\ \ \ \ \ \ mi = \_frag\_3}
\cln{11}{}{\ \ \ \ return ans}
\end{tabular}
\end{minipage}
\caption{A seeded fault in an untransformed faulty program (left) and in its transformed counterpart (right). Lines marked $-$ belong to the reference programs, and lines marked $+$ to the faulty programs. The \emph{State and literals} recipe rewrites literals and splits the faulty statement across lines 6 to~8, moving the mutated expression from line~5 to line~6.}
\label{fig:example}
\end{figure}

From source code alone, all three agents approached the accuracy ceiling, localizing \result{primary.acc.untransformed.no-tools.min}{98.2} to \result{primary.acc.untransformed.no-tools.max}{99.9}\% of faults in untransformed programs and \result{primary.acc.transformed.no-tools.luna}{96.7}\% (Luna~6), \result{primary.acc.transformed.no-tools.qwen}{97.9}\% (Qwen~3.8), and \result{primary.acc.transformed.no-tools.sol}{99.9}\% (Sol~6.1) in transformed programs. Variable renaming and distractor insertion barely changed static accuracy, whereas restructuring loops and wrapping conditions in double negations lowered it by up to \result{primary.acc.loss.no-tools.maximal-mix.luna}{3.8} percentage points in a single recipe, where most new errors blamed doubly negated conditions. Encouraging debugger assistance raised transformed accuracy to \result{primary.acc.transformed.encouraged-tools.luna}{98.8}\% for Luna~6 (\result{primary.effect.transformed.luna}{+2.1} points, removing \result{primary.error-reduction.transformed.luna}{63}\% of its remaining errors) and \result{primary.acc.transformed.encouraged-tools.qwen}{99.2}\% for Qwen~3.8 (\result{primary.effect.transformed.qwen}{+1.3} points, removing \result{primary.error-reduction.transformed.qwen}{61}\%), with positive effects across every recipe and fault type, while Sol~6.1 stayed at \result{primary.acc.transformed.encouraged-tools.sol}{99.9}\%. These gains were \result{primary.diff.luna}{+1.4} and \result{primary.diff.qwen}{+1.0} percentage points larger on transformed programs than on their untransformed counterparts, closing the transformation gap to within \result{primary.acc.gap.encouraged-tools.min}{$-$0.1} to \result{primary.acc.gap.encouraged-tools.max}{+0.1} points. However, encouraged assistance multiplied token use by \result{cost.ratio.tokens.transformed.min}{14.7} to \result{cost.ratio.tokens.transformed.max}{27.0}, wall time by \result{cost.ratio.wall.transformed.min}{5.3} to \result{cost.ratio.wall.transformed.max}{7.7}, and monetary cost by \result{cost.ratio.usd.transformed.min}{3.4} to \result{cost.ratio.usd.transformed.max}{5.0}.

When debugger assistance was optional, the agents rarely stepped through code in the interactive debugger. Under discretionary-tools, they took debugger actions in only \result{tool.any-debugger.e.min}{0.2} to \result{tool.any-debugger.e.max}{6.2}\% of episodes on transformed programs (against \result{tool.any-debugger.f.min}{41.7} to \result{tool.any-debugger.f.max}{100.0}\% when encouraged) and instead ran execution probes or answered from source code alone. Even so, Luna~6 and Qwen~3.8 gained \result{primary.disc-minus-no.transformed.luna}{+1.5} and \result{primary.disc-minus-no.transformed.qwen}{+1.3} points over no-tools, and Qwen~3.8 ran more probes on the faults where debugger assistance helped most. Encouraging debugger use added only \result{primary.enc-minus-disc.transformed.min}{+0.0} to \result{primary.enc-minus-disc.transformed.max}{+0.6} points over optional access. In this evaluation, coding agents rarely needed debugger assistance, and when given the choice, they obtained most of its benefit by running the program rather than stepping through it.

This paper makes the following contributions:
\begin{itemize}
  \item \textbf{Static localization across transformations:} We evaluate fault localization accuracy across \result{programs.faulty.admitted}{610} untransformed and \result{programs.transformed.admitted}{3{,}022} transformed programs spanning five fault types and five transformation recipes (Section~\ref{sec:results-rq1}).
  \item \textbf{Effect and cost of debugger assistance:} We measure the accuracy gain, error reduction, and resource cost of encouraged debugger assistance across original and transformed programs and three models (Section~\ref{sec:results-rq2}).
  \item \textbf{Discretionary versus encouraged use:} We quantify how often agents choose execution probes over the interactive debugger and how much of the accuracy benefit optional tool use retains at lower cost (Section~\ref{sec:results-rq3}).
  \item \textbf{Replication package:} We release our pipeline, programs, \result{facts.episodes.kept}{65{,}376} agent traces, and analysis scripts~\cite{anonymous_2026_replication}.
\end{itemize}

%% file: sections/02-related-work.tex
\section{Related Work}
\label{sec:related}

\paragraph{LLM-based fault localization}
Most LLM-based fault localization pipelines use fixed execution rules to filter candidates before the model reads the code. AutoFL lets an LLM retrieve the classes and methods covered by failing tests on Defects4J and BugsInPy~\cite{kang_2024_autofl}, FlexFL narrows candidate methods with traditional spectrum-based and information-retrieval rankings before an LLM inspects their source~\cite{xu_2025_flexfl}, and SieveFL combines failing-test coverage with semantic similarity to prune 79\% of candidate methods and 49\% of input tokens at unchanged Top-1 accuracy~\cite{farzandway_2026_sievefl}. SemLoc localizes faulty lines by running instrumented code as a fixed pipeline step to check model-inferred semantic constraints~\cite{yang_2026_semloc}. In PingFL, the agent inserts print statements into suspicious methods and reruns failing tests~\cite{qin_2025_pingfl}. Apart from SemLoc, these approaches rank methods rather than lines or statements, and none compares lightweight script execution against interactive breakpoint stepping.

\paragraph{Source transformations and benchmark familiarity}
Benchmark designers counter training-data exposure by selecting problems released after model training cutoffs, as in DebugBench~\cite{tian_2024_debugbench} and LiveCodeBench~\cite{jain_2025_livecodebench}, or by rewriting existing benchmarks. Rewrites either apply behavior-preserving code transformations, as in CodeCleaner~\cite{cao_2025_codecleaner}, or wrap the same algorithms in new problem contexts, as in DyCodeEval~\cite{chen_2025_dycodeeval}. Even after surface rewrites, semantic overlap with pretraining data can remain and raise benchmark scores~\cite{spiesberger_2026_soft_contamination}. Prior studies also use behavior-preserving transformations to test the robustness of static fault localization, program repair, and code reasoning. Haroon et al.\ evaluate line-level localization by ten models given only the source code and specification across 750,013 fault-localization tasks~\cite{haroon_2026_localizability}. Li et al.\ transform 438 single-function Defects4J bugs and observe 49\% fewer correct patches, with many edits targeting inserted dead code or rewritten predicates~\cite{li_2025_generalizability}, and HEJ-Robust observes its largest Java repair drops after local variable renaming~\cite{rabbi_2026_hej_robust}. Conversely, variable renaming can raise output-prediction accuracy on LiveCodeBench~\cite{orvalho_2025_robustness} and refactoring can improve code completion~\cite{cao_2024_countermeasures}. On 16 bootstrap-filtered LeetCode programs, Daxerer finds that combined structural-lexical obfuscation lowers judge-approved bug fixing much more for Qwen~3.6 than for GPT-5.4 and that optional runtime tools recover Qwen~3.6's performance~\cite{daxerer_2026_thesis}. Whereas prior work evaluates static predictions or repair proposals under a single setup, we isolate fault localization on matched pairs across five transformation recipes and three tool-use settings.

\paragraph{Runtime evidence in LLM-based debugging}
Debugging pipelines supply LLMs with runtime information at different levels of agent control. Fixed pipelines execute the program outside the model's control and append the resulting outputs or traces to the prompt: Self-Debugging feeds back unit-test results alongside model-generated code explanations~\cite{chen_2024_self_debug}, and LDB instruments basic-block boundaries to record intermediate variable values before the model refines a generated program~\cite{zhong_2024_ldb}. Agentic systems let the model issue debugger commands directly, but must guard against invalid commands and verbose outputs. ChatDBG lets GPT-4 query \texttt{pdb}, \texttt{lldb}, or \texttt{gdb} to explain crashes and propose fixes~\cite{levin_2025_chatdbg}. In debug-gym, exposing \texttt{pdb} on SWE-bench-Lite lowered repair rates for GPT-4o and GPT-4o-mini but improved them for reasoning models such as o1-preview, o3-mini, and Claude~3.7 Sonnet~\cite{yuan_2025_debug_gym}. InspectCoder interposes a state-tracking middleware between the agent and \texttt{pdb} on BigCodeBench and LiveCodeBench, where direct \texttt{pdb} interaction without the middleware lowered repair success from 67.9\% to 25.7\%~\cite{wang_2026_inspectcoder}. DebugHarness combines \texttt{gdb}, deterministic replay, and script-based output distillation to patch C and C++ vulnerabilities~\cite{sun_2026_debugharness}. These studies evaluate end-to-end repair or explanation quality under a single tool configuration, whereas we isolate fault localization and compare no, optional, and encouraged debugger assistance across original and transformed programs.

%% file: sections/03-study-design.tex
\section{Study Design}
\label{sec:design}

Fig.~\ref{fig:study-overview} summarizes our study design. We seed single-line faults into reference programs from a public dataset and rewrite each resulting faulty program with behavior-preserving transformations. Agents localize the faults in both representations (untransformed and transformed) under three tool-use settings: no-tools, discretionary-tools, and encouraged-tools. Crossing the two representations with the three settings yields Conditions A to C on untransformed programs and Conditions D to F on transformed programs. A \emph{localization case} pairs a faulty program with the natural-language \emph{problem description} of the underlying LeetCode task, and an \emph{episode} is a single agent run on a case under a given tool-use setting.

Both discretionary-tools and encouraged-tools provide debugger assistance. We compare Top-1 accuracy, episode cost, and tool use across settings. Defining the \emph{no-tools change} ($G$) as transformed minus untransformed static accuracy and the \emph{assistance effect} ($\Delta, \Delta'$) as encouraged-tools minus no-tools accuracy, we address three research questions:

\begin{keeptogether}
\begin{itemize}
  \item \textbf{RQ1 (Static localization and transformation sensitivity):} How accurately do coding agents localize faults from source code alone, and how much do behavior-preserving source transformations change this accuracy?
  \item \textbf{RQ2 (Debugger effect):} How much does encouraged debugger assistance improve localization accuracy on un-/transformed programs, and at what cost?
  \item \textbf{RQ3 (Discretionary use):} How often do agents use debugger assistance when it is optional, and how does this affect accuracy relative to no and encouraged assistance?
\end{itemize}
\end{keeptogether}

\subsection{Localization Cases}
\label{sec:design-tasks}

\paragraph{Source programs}
We draw reference programs from LeetCodeDataset~\cite{xia_2025_leetcodedataset}, which pairs LeetCode problems with natural-language descriptions, verified Python reference solutions, and executable tests. Self-contained algorithmic programs, which prior runtime-debugging and transformation studies also evaluate~\cite{zhong_2024_ldb,tian_2024_debugbench,haroon_2026_localizability}, provide the dense test suites needed to verify behavior preservation and identical test failures across transformations. We sample from the dataset's training split, which contains problems released before July~2024 according to the authors' estimates and is intended for training code models~\cite{xia_2025_leetcodedataset}, drawing 50 reference programs at random from each of the Easy, Medium, and Hard difficulty groups (150 in total, Table~\ref{tab:programs}).
\begin{table}[t]
\centering
\caption{Constructed programs. Programs gives admitted out of attempted. LOC and CC give medians with interquartile ranges, and for transformed programs the median change from the untransformed faulty program, in percent for LOC. Kept is the median share of preserved four-token sequences.}
\label{tab:programs}
\footnotesize
\setlength{\tabcolsep}{3pt}
\newcommand{\iqr}[1]{{\scriptsize[#1]}}
\begin{tabular}{@{}l r@{\,/\,}l r@{\hspace{3pt}}l r@{\hspace{3pt}}l r@{}}
\toprule
 & \multicolumn{2}{c}{Programs} & \multicolumn{2}{c}{LOC} & \multicolumn{2}{c}{CC} & Kept \\
\midrule
\csname @@input\endcsname generated/table-programs-rows.tex
\bottomrule
\end{tabular}
\end{table}

\paragraph{Fault seeding}
From each reference program $P$, we derive up to five faulty programs $M$, one per fault type. Each fault is a single edit to one expression on one line. \emph{Off-by-one} changes a numeric literal by one, and \emph{wrong arithmetic} replaces an arithmetic operator. \emph{Wrong logic} swaps \texttt{and} and \texttt{or}, changes a comparison operator, flips a Boolean literal, or adds or removes a negation. \emph{Wrong variable} replaces a variable with another variable in scope, and \emph{static return} replaces a returned expression with a constant such as \texttt{0} or \texttt{None}. Except for wrong variable, these fault types adapt mutation operators from PIT~\cite{coles_2016_pit,pit_mutators}. Wrong variable corresponds to the changed-identifier pattern that Kamienski et al.\ report among single-statement bugs in Python projects~\cite{kamienski_2021_pysstubs}. For each reference program and fault type, we apply a randomly chosen edit of that type that fails at least one test and record the edited line~$L$, or skip the fault type if no candidate edit qualifies.

\paragraph{Transformations}
We rewrite every faulty program with five fixed recipes, each consisting of an ordered sequence of behavior-preserving operations. \emph{Rename} replaces variable names with single letters. \emph{State and literals} routes attribute access through \texttt{getattr} calls, rewrites numeric and Boolean literals as equivalent expressions, and splits expressions into intermediate assignments. \emph{Control flow} converts \texttt{for} loops into explicit iterator loops, expands conditions with double negations, inserts dead code, and splits expressions. \emph{Distractors} wraps statements in opaque predicates that always evaluate to true, inserts dead code and generic comments such as \texttt{\# WARNING: do not modify}, and splits expressions. \emph{Maximal mix} chains 14 steps of these operations and additionally extracts statement blocks into helper functions and constructs strings from character codes. Fig.~\ref{fig:example} illustrates the State and literals recipe applied to a faulty program. Prior studies use similar operations to evaluate fault localization~\cite{haroon_2026_localizability} and program repair~\cite{li_2025_generalizability} on rewritten code, to test whether LLMs reason about program semantics~\cite{orvalho_2025_robustness,maveli_2025_equivalence}, and to reduce overlap with training corpora~\cite{cao_2025_codecleaner,zhang_2025_unseen_horizons}.

\paragraph{Admission}
We apply each recipe to $P$ and $M$ alike to obtain $P'$ and $M'$, track the mutated expression to identify its new line~$L'$ in~$M'$, and admit $(M, M')$ as a \emph{matched pair} only if $P'$ passes all tests, $M'$ fails on the exact same test assertion as~$M$, and restoring the original expression at~$L'$ causes $M'$ to pass all tests. On the dataset's test suites, which have a median of \result{programs.sampled.test-assertions.median}{99} assertions per problem, these checks verify that the recipe preserves tested behavior, that the fault causes the same failure, and that line~$L'$ holds the fault.

Table~\ref{tab:programs} characterizes the constructed programs using non-empty lines of code including comments (LOC), McCabe's cyclomatic complexity (CC)~\cite{mccabe_1976_complexity}, and the share of untransformed four-token sequences preserved after transformation (Kept), adapting CodeCleaner's n-gram overlap ratio~\cite{cao_2025_codecleaner}. In total, admission retained \result{programs.reference.admitted}{144} of \result{programs.reference.attempted}{150} reference programs, \result{programs.faulty.admitted}{610} of \result{programs.faulty.attempted}{750} seeded faults (\result{programs.faulty.admitted.wrong-arithmetic}{104} to \result{programs.faulty.admitted.static-return}{141} per fault type), and \result{programs.transformed.admitted}{3{,}022} of \result{programs.transformed.attempted}{3{,}050} transformed programs. Admitted reference programs have median LOC of \result{programs.reference.loc.median.easy}{8}, \result{programs.reference.loc.median.medium}{11}, and \result{programs.reference.loc.median.hard}{19} and median CC of \result{programs.reference.cc.median.easy}{3}, \result{programs.reference.cc.median.medium}{4}, and \result{programs.reference.cc.median.hard}{8} across Easy, Medium, and Hard. Across recipes, median changes ranged from +\result{programs.transformed.loc-change-percent.median.rename}{0}\% (Rename) to +\result{programs.transformed.loc-change-percent.median.maximal-mix}{450}\% (Maximal mix) in LOC and +\result{programs.transformed.cc-change.median.rename}{0} to +\result{programs.transformed.cc-change.median.distractors}{11} (Distractors) in CC, while median Kept ranged from \result{programs.transformed.kept.median.maximal-mix}{37}\% (Maximal mix) to \result{programs.transformed.kept.median.distractors}{71}\% (Distractors).

\subsection{Agent Interaction}
\label{sec:design-workflows}

\paragraph{Task}
Each episode begins with a system prompt and a task message (Listing~\ref{lst:prompts}). The task message provides the problem description, the faulty program with 1-based line numbers, and shared supporting code from the dataset, such as standard imports and data-structure definitions. It concludes by instructing the agent to report the single most likely affected line. The agent submits its answer by calling \texttt{report\_bug} with the line number and an explanation. Evaluation tests are withheld from the agent in all settings.

\paragraph{Tool-use settings}
Each setting adds one element to the previous one. In no-tools, \texttt{report\_bug} is the only available tool, so the agent must localize the fault from the problem description and source code alone. Discretionary-tools adds tools to execute and debug the program along with their documentation. Using \texttt{run\_probe}, the agent can execute a custom Python script, for example one that invokes the target method on chosen inputs, and observe its output and any raised exception. The agent can also launch such a script in an interactive debugger, set breakpoints on program lines, and step into, over, or out of function calls. Whenever execution pauses, the agent can inspect local variables, the call stack, program output, and the values of arbitrary Python expressions. Encouraged-tools provides the same toolset and appends a two-sentence instruction directing the agent to gather runtime evidence before deciding (Listing~\ref{lst:prompts}).

\begin{lstlisting}[float=t, caption={Prompt text shared by all tool-use settings, with the instruction that encouraged-tools appends to the system prompt.}, label={lst:prompts}]
(*\textrm{\textit{System prompt:}}*)
You are a Python debugging agent. Find the injected defect in the supplied implementation. You do not have access to hidden evaluation tests.
(*\textrm{\textit{Appended in encouraged-tools:}}*)
Actively use the probe and debugger tools to gather runtime evidence before deciding. Inspect relevant state and execution flow.
(*\textrm{\textit{End of task message:}}*)
The implementation contains one injected bug. Locate it and call `report_bug` exactly once. Provide the single most likely affected line's 1-based line number.
\end{lstlisting}

\paragraph{Prompt design}
Aside from the appended instruction in encouraged-tools, the prompt text is identical across all settings, avoiding prompt-wording variations that can alter LLM performance~\cite{mizrahi_2024_multi_prompt,sclar_2024_format_sensitivity}. The prompts specify the task and required output format without prescribing a debugging strategy, leaving tool selection to the model. A 100-episode pilot run verified that the agents invoked the tools and called \texttt{report\_bug} as specified. Following reporting guidelines for LLM studies~\cite{korn_2026_prompt_reporting,baltes_2025_llm_guidelines}, our replication package~\cite{anonymous_2026_replication} provides the tool descriptions and full prompts.

\paragraph{Models and limits}
We evaluate Sol~6.1, Luna~6, and Qwen~3.8, running two independent episodes per model, case, and setting.
Sol~6.1 and Luna~6 are accessed through OpenAI's Responses API under the identifiers \texttt{gpt-6.1-sol} and \texttt{gpt-6-luna} at the provider's default reasoning effort (medium). Qwen~3.8 runs from its 27-billion-parameter FP8 checkpoint (\texttt{Qwen/Qwen3.8-27B-FP8}) on our own vLLM server with thinking enabled at the model's default reasoning effort (xhigh). All models use default sampling parameters. We collected all episodes between October~6 and~14, 2026.
Each untransformed case is evaluated once and shared across its matched pairs (\result{facts.episodes.planned}{65{,}376} episodes in total). An episode terminates when the agent calls \texttt{report\_bug}. If an agent responds without calling any tool or reaches 50 tool rounds, where a round is one model response and its requested tool calls, it receives one final prompt requiring a \texttt{report\_bug} call. Across conditions, at most \result{cost.forced-answer.f.max}{0.3}\% of episodes needed this fallback, and at most \result{cost.no-answer.f.max}{0.1}\% ended without an answer.

\subsection{Measurements}
\label{sec:design-measurements}

\paragraph{Localization}
We grade each reported line under two criteria. Statement-level grading accepts any line belonging to the faulty statement, which a transformation recipe can split across multiple lines, as in lines 6 to~8 of Fig.~\ref{fig:example}. Recipes split the faulty statement in \result{programs.transformed.spread-share}{47}\% of transformed cases. Line-level grading accepts only the exact line holding the mutated expression ($L$ or~$L'$). We measure Top-1 accuracy as the share of episodes that successfully localize the fault, counting episodes without an answer as unsuccessful.

\paragraph{Tool use and cost}
Beyond \texttt{report\_bug}, agents can invoke two categories of runtime tools. Probe runs execute agent-written scripts through \texttt{run\_probe} without the debugger. Debugger actions encompass all calls that start or stop a debug session, manage breakpoints, step through execution, or inspect the paused program state. For every episode, we record total tokens, wall-clock time, and monetary cost at the providers' list prices as of October~7, 2026, accounting for cached-input rates. For self-hosted Qwen~3.8, we estimate monetary cost from DeepInfra's list prices~\cite{deepinfra_qwen_pricing}.

\subsection{Analysis}
\label{sec:design-analysis}

We analyze each model separately. For a case~$X$, let $a_{\mathrm{no}}(X)$, $a_{\mathrm{disc}}(X)$, and $a_{\mathrm{enc}}(X)$ denote Top-1 accuracy under no-tools, discretionary-tools, and encouraged-tools.

\paragraph{RQ1 and RQ2}
To measure how transformations affect static accuracy, the no-tools change ($G$) averages the per-pair difference over all matched pairs~$\mathcal{E}$:
\[
G = \frac{1}{|\mathcal{E}|}\sum_{(M,M')\in\mathcal{E}} \bigl(a_{\mathrm{no}}(M') - a_{\mathrm{no}}(M)\bigr),
\]
while the assistance effects $\Delta$ and $\Delta'$ average $a_{\mathrm{enc}}(M) - a_{\mathrm{no}}(M)$ and $a_{\mathrm{enc}}(M') - a_{\mathrm{no}}(M')$ over the same pairs~$\mathcal{E}$.
RQ1 examines $a_{\mathrm{no}}(M)$, $a_{\mathrm{no}}(M')$, and $G$ across all programs and within each transformation recipe and fault type, along with the error patterns on transformed programs. Because an agent can improve by at most its no-tools error rate, RQ2 examines $\Delta$ and $\Delta'$ alongside the share of no-tools errors they remove, $\Delta / (1 - \bar{a}_{\mathrm{no}})$ and $\Delta' / (1 - \bar{a}'_{\mathrm{no}})$. It also evaluates the interaction difference $I = \Delta' - \Delta$, which is positive when debugger assistance yields a larger gain on transformed than on untransformed programs, and the ratios of mean tokens, wall time, and cost between encouraged-tools and no-tools.

\paragraph{RQ3}
RQ3 investigates how often agents take debugger actions and run probes under discretionary-tools and compares their debugger use with encouraged-tools, which checks that the instruction changes agent behavior. It then compares discretionary-tools against no-tools and encouraged-tools through episode cost and the mean per-case accuracy differences $a_{\mathrm{disc}}(X) - a_{\mathrm{no}}(X)$ and $a_{\mathrm{disc}}(X) - a_{\mathrm{enc}}(X)$ across the cases of each representation. To test whether agents invoke tools more often on cases where debugger assistance raises accuracy more, let $u(X)$ denote the mean number of probe runs, or of debugger actions, across the discretionary-tools episodes of case~$X$. Let $r(X)$ be the fractional rank of $u(X)$ across all cases for a model and representation, scaled from 0 to 1 with average ranks for ties. For each model, representation, and tool category, we fit
\[
a_{\mathrm{enc}}(X) - a_{\mathrm{no}}(X) = \beta_0 + \beta_1\, r(X) + \varepsilon(X)
\]
by ordinary least squares. The slope $\beta_1$ estimates the fitted difference in the assistance effect between the bottom and top of the tool-use ranking. Using ranks prevents a few cases with many debugger actions from dominating the fit.

\paragraph{Uncertainty}
We report 95\% percentile confidence intervals from a nonparametric cluster bootstrap with 2{,}000 resamples~\cite{davison_1997_bootstrap}. Because cases derived from the same reference program share its problem specification and underlying algorithm, their outcomes are likely correlated, and treating them as independent observations can understate uncertainty~\cite{miller_2024_errorbars}. Each bootstrap resample therefore draws whole reference programs with replacement, stratified by difficulty group, and retains all associated cases and episodes, capturing both program-level sampling variation and episode-to-episode variance within each case~\cite{davison_1997_bootstrap}. Every resample recomputes all summary statistics, tool-use ranks, and RQ3 regressions.

Under discretionary-tools, tool invocations can be concentrated in a small number of reference programs. A bootstrap resample that draws none of those programs leaves $\beta_1$ undefined, and computing an interval from the remaining resamples is invalid~\cite{cameron_2015_guide}. Because the number of clusters with tool use required for valid inference depends on the empirical distribution~\cite{mackinnon_2017_sizes,mackinnon_2023_guide}, we validate each $\beta_1$ interval using 1{,}000 cluster-level placebo datasets~\cite{bertrand_2004_trust}. Each placebo dataset keeps localization outcomes fixed and reassigns observed tool use within each difficulty group to randomly selected reference programs from the same group in proportion to their case counts, making tool use independent of the assistance effect within each group. We report the confidence interval for $\beta_1$ only if $\beta_1$ is defined in all 2{,}000 bootstrap resamples and the share of the 1{,}000 placebo intervals that miss the mean placebo slope is not significantly above 5\% under a one-sided exact binomial test at the 5\% level. Otherwise, Table~\ref{tab:discretionary} marks $\beta_1$ with a dagger ($^\dagger$) or shows a dash when fewer than two reference programs have such tool use.

%% file: sections/04-results.tex
\section{Results}
\label{sec:results}

\begin{table*}[t]
\centering
\caption{Top-1 localization accuracy in percent under statement-level grading. No-tools change is transformed minus untransformed static accuracy ($G$), Effect is encouraged-tools minus no-tools ($\Delta$ and $\Delta'$ on un-/transformed programs), and Difference is $I = \Delta' - \Delta$, all in percentage points. Untransformed accuracies weight each program by its number of matched transformed versions, as $G$ and $\Delta$ do. Brackets give 95\% intervals. Fig.~\ref{fig:effects} breaks $G$ and $\Delta'$ down by recipe and fault type.}
\label{tab:accuracy}
\footnotesize
\begin{tabular}{@{}llcccccc@{}}
\toprule
Model & Programs & No-tools & Discretionary & Encouraged & No-tools change ($G$) & Effect ($\Delta, \Delta'$) & Difference ($I = \Delta' - \Delta$) \\
\midrule
\csname @@input\endcsname generated/table-accuracy-rows.tex
\bottomrule
\end{tabular}
\end{table*}

\begin{figure*}[t]
\centering
\includegraphics[width=\textwidth]{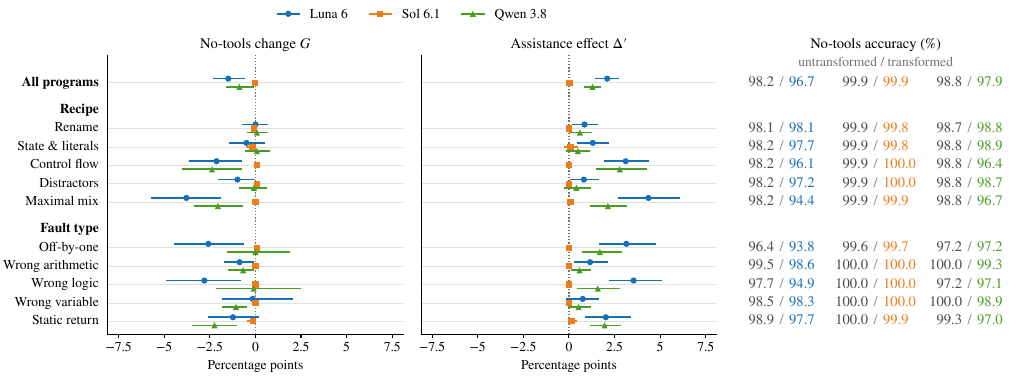}
\caption{No-tools change $G$ (left, transformed minus untransformed accuracy under no-tools) and assistance effect $\Delta'$ on transformed programs (right, encouraged-tools minus no-tools) by recipe and fault type, per model, in percentage points, with 95\% intervals. The rightmost column gives no-tools accuracy in percent on the matched untransformed (in gray) and transformed programs.}
\label{fig:effects}
\end{figure*}

\begin{table*}[t]
\centering
\caption{Episode cost. Tokens and wall time are means per episode, and monetary cost is the mean cost of 1,000 episodes, estimated for Qwen~3.8 from a provider's list prices. The last two columns divide encouraged-tools by no-tools.}
\label{tab:cost}
\footnotesize
\begin{tabular}{@{}llcccccccc@{}}
\toprule
 & & \multicolumn{3}{c}{Untransformed} & \multicolumn{3}{c}{Transformed} & \multicolumn{2}{c}{Encouraged / no-tools} \\
\cmidrule(lr){3-5}\cmidrule(lr){6-8}\cmidrule(l){9-10}
Model & Measure & No-tools & Discr. & Encour. & No-tools & Discr. & Encour. & Untr. & Transf. \\
\midrule
\csname @@input\endcsname generated/table-cost-rows.tex
\bottomrule
\end{tabular}
\end{table*}

\subsection{RQ1: Static Localization and Transformation Sensitivity}
\label{sec:results-rq1}

Without tools, all three agents operated near the accuracy ceiling, localizing \result{primary.acc.untransformed.no-tools.luna}{98.2}\% (Luna~6), \result{primary.acc.untransformed.no-tools.qwen}{98.8}\% (Qwen~3.8), and \result{primary.acc.untransformed.no-tools.sol}{99.9}\% (Sol~6.1) of faults in untransformed programs, and \result{primary.acc.transformed.no-tools.luna}{96.7}\%, \result{primary.acc.transformed.no-tools.qwen}{97.9}\%, and \result{primary.acc.transformed.no-tools.sol}{99.9}\% in transformed programs (Table~\ref{tab:accuracy}). Behavior-preserving transformations changed overall static accuracy by $G =$~\result{primary.acc.gap.no-tools.luna}{$-$1.5}~\result{primary.acc.gap.no-tools.ci.luna}{[$-$2.3, $-$0.6]} percentage points for Luna~6 and \result{primary.acc.gap.no-tools.qwen}{$-$0.9}~\result{primary.acc.gap.no-tools.ci.qwen}{[$-$1.6, $-$0.1]} points for Qwen~3.8, while Sol~6.1 remained unchanged ($G =$~\result{primary.acc.gap.no-tools.sol}{+0.0}~\result{primary.acc.gap.no-tools.ci.sol}{[$-$0.1, +0.0]} points). Across the five fault types, untransformed no-tools accuracy was lowest on off-by-one (\result{primary.acc.untransformed.no-tools.off-by-one.luna}{96.4}\% for Luna~6 and \result{primary.acc.untransformed.no-tools.off-by-one.qwen}{97.2}\% for Qwen~3.8) and wrong-logic faults (\result{primary.acc.untransformed.no-tools.wrong-logic.luna}{97.7}\% and \result{primary.acc.untransformed.no-tools.wrong-logic.qwen}{97.2}\%, Fig.~\ref{fig:effects}). Transformations lowered Luna~6's static accuracy most on those two fault types (to \result{primary.acc.transformed.no-tools.off-by-one.luna}{93.8}\% and \result{primary.acc.transformed.no-tools.wrong-logic.luna}{94.9}\%, $G =$~\result{primary.acc.gap.no-tools.off-by-one.luna}{$-$2.6} and \result{primary.acc.gap.no-tools.wrong-logic.luna}{$-$2.8} points) and Qwen~3.8's most on static-return faults (from \result{primary.acc.untransformed.no-tools.static-return.qwen}{99.3}\% to \result{primary.acc.transformed.no-tools.static-return.qwen}{97.0}\%, $G =$~\result{primary.acc.gap.no-tools.static-return.qwen}{$-$2.2} points).

Across the five transformation recipes, the static accuracy loss concentrated in Control flow and Maximal mix, the two recipes that rewrite \texttt{for} loops into explicit iterator loops and wrap conditions in double negations (Fig.~\ref{fig:effects}). Without tools, these two recipes lowered Luna~6's accuracy from \result{primary.acc.untransformed.no-tools.maximal-mix.luna}{98.2}\% on the untransformed counterparts to \result{primary.acc.transformed.no-tools.control-flow.luna}{96.1}\% and \result{primary.acc.transformed.no-tools.maximal-mix.luna}{94.4}\% ($G =$~\result{primary.acc.gap.no-tools.control-flow.luna}{$-$2.1}~\result{primary.acc.gap.no-tools.control-flow.ci.luna}{[$-$3.7, $-$0.7]} and \result{primary.acc.gap.no-tools.maximal-mix.luna}{$-$3.8}~\result{primary.acc.gap.no-tools.maximal-mix.ci.luna}{[$-$5.7, $-$1.9]} points) and Qwen~3.8's accuracy from \result{primary.acc.untransformed.no-tools.control-flow.qwen}{98.8}\% to \result{primary.acc.transformed.no-tools.control-flow.qwen}{96.4}\% and \result{primary.acc.transformed.no-tools.maximal-mix.qwen}{96.7}\% (\result{primary.acc.gap.no-tools.control-flow.qwen}{$-$2.4}~\result{primary.acc.gap.no-tools.control-flow.ci.qwen}{[$-$4.0, $-$0.8]} and \result{primary.acc.gap.no-tools.maximal-mix.qwen}{$-$2.1}~\result{primary.acc.gap.no-tools.maximal-mix.ci.qwen}{[$-$3.3, $-$0.7]} points). By contrast, Rename (preserving \result{programs.transformed.kept.median.rename}{40}\% of four-token sequences, Table~\ref{tab:programs}), State and literals, and Distractors (increasing line count by \result{programs.transformed.loc-change-percent.median.distractors}{400}\% and cyclomatic complexity by \result{programs.transformed.cc-change.median.distractors}{11}) changed no-tools accuracy by only \result{primary.acc.gap.no-tools.distractors.luna}{$-$1.0} to \result{primary.acc.gap.no-tools.rename.qwen}{+0.1} points across all three models, with every 95\% interval including zero.

Inspecting the failed no-tools episodes on Control flow and Maximal mix shows that \result{facts.primary.wrong.counterpart-solved.combined.luna}{91} of Luna~6's \result{facts.primary.wrong.wrong-answers.combined.luna}{116} wrong answers and \result{facts.primary.wrong.counterpart-solved.combined.qwen}{66} of Qwen~3.8's \result{facts.primary.wrong.wrong-answers.combined.qwen}{79} wrong answers occurred on faults that the same agent localized in both no-tools episodes on the untransformed program. Among these \emph{new errors}, \result{facts.primary.wrong.double-negation-share.combined.luna}{71.4}\% for Luna~6 and \result{facts.primary.wrong.double-negation-share.combined.qwen}{68.2}\% for Qwen~3.8 named a condition that the recipe had wrapped in a double negation, even though doubly negated conditions make up only \result{programs.transformed.double-negation-share.control-flow}{7.8}\% of the lines in Control flow programs and \result{programs.transformed.double-negation-share.maximal-mix}{4.4}\% in Maximal mix programs.

\rqanswer{1}{Without tools, the agents localized \result{primary.acc.untransformed.no-tools.min}{98.2} to \result{primary.acc.untransformed.no-tools.max}{99.9}\% of faults in untransformed and \result{primary.acc.transformed.no-tools.min}{96.7} to \result{primary.acc.transformed.no-tools.max}{99.9}\% in transformed programs ($G =$~\result{primary.acc.gap.no-tools.luna}{$-$1.5} percentage points for Luna~6, \result{primary.acc.gap.no-tools.qwen}{$-$0.9} for Qwen~3.8, and \result{primary.acc.gap.no-tools.sol}{+0.0} for Sol~6.1). Variable renaming and dead-code distractors changed static accuracy by at most \result{primary.acc.gap.no-tools.distractors.luna}{$-$1.0} to \result{primary.acc.gap.no-tools.rename.qwen}{+0.1} points, whereas restructuring loops and doubly negating conditions lowered it by \result{primary.acc.loss.no-tools.control-flow.luna}{2.1} to \result{primary.acc.loss.no-tools.maximal-mix.luna}{3.8} points for Luna~6 and Qwen~3.8, with \result{facts.primary.wrong.double-negation-share.combined.qwen}{68.2} to \result{facts.primary.wrong.double-negation-share.combined.luna}{71.4}\% of new errors blaming doubly negated conditions.}

\subsection{RQ2: Debugger Effect}
\label{sec:results-rq2}

Encouraging debugger assistance improved accuracy for the two models not already at 99.9\% and closed almost the entire static transformation gap (Table~\ref{tab:accuracy}). On untransformed programs, Luna~6 gained $\Delta =$~\result{primary.effect.untransformed.luna}{+0.7}~\result{primary.effect.untransformed.ci.luna}{[+0.2, +1.4]} percentage points (\result{primary.error-reduction.untransformed.luna}{41}\% of its no-tools errors) and Qwen~3.8 gained \result{primary.effect.untransformed.qwen}{+0.3}~\result{primary.effect.untransformed.ci.qwen}{[$-$0.3, +1.1]} points (\result{primary.error-reduction.untransformed.qwen}{27}\%). On transformed programs, where no-tools error rates were higher, their assistance effects rose to $\Delta' =$~\result{primary.effect.transformed.luna}{+2.1}~\result{primary.effect.transformed.ci.luna}{[+1.4, +2.8]} and \result{primary.effect.transformed.qwen}{+1.3}~\result{primary.effect.transformed.ci.qwen}{[+0.8, +1.8]} points, removing \result{primary.error-reduction.transformed.luna}{63}\% and \result{primary.error-reduction.transformed.qwen}{61}\% of their no-tools errors. These larger gains ($I = \Delta' - \Delta$ of \result{primary.diff.luna}{+1.4}~\result{primary.diff.ci.luna}{[+0.6, +2.1]} and \result{primary.diff.qwen}{+1.0}~\result{primary.diff.ci.qwen}{[+0.1, +1.7]} points) left encouraged-tools accuracy within \result{primary.acc.gap.encouraged-tools.min}{$-$0.1} to \result{primary.acc.gap.encouraged-tools.max}{+0.1} points between representations. Sol~6.1 did not improve in either representation ($\Delta =$~\result{primary.effect.untransformed.sol}{$-$0.1}~\result{primary.effect.untransformed.ci.sol}{[$-$0.3, +0.0]}, $\Delta' =$~\result{primary.effect.transformed.sol}{+0.0}~\result{primary.effect.transformed.ci.sol}{[+0.0, +0.1]}, and $I =$~\result{primary.diff.sol}{+0.1}~\result{primary.diff.ci.sol}{[+0.0, +0.3]} points).

For both Luna~6 and Qwen~3.8, $\Delta'$ was positive in all five transformation recipes and all five fault types, with 95\% intervals strictly above zero in \result{primary.effect.transformed.subsets-ci-positive.luna}{9} of the ten subsets for Luna~6 and \result{primary.effect.transformed.subsets-ci-positive.qwen}{6} of ten for Qwen~3.8 (Fig.~\ref{fig:effects}). Across recipes, the largest gains occurred on Control flow and Maximal mix, where encouraged assistance raised accuracy by \result{primary.effect.transformed.control-flow.luna}{+3.1} and \result{primary.effect.transformed.maximal-mix.luna}{+4.4} points for Luna~6 (removing \result{primary.error-reduction.transformed.control-flow.luna}{79}\% and \result{primary.error-reduction.transformed.maximal-mix.luna}{78}\% of the faults missed without tools, $I =$~\result{primary.diff.control-flow.luna}{+2.4} and \result{primary.diff.maximal-mix.luna}{+3.6}) and by \result{primary.effect.transformed.control-flow.qwen}{+2.8} and \result{primary.effect.transformed.maximal-mix.qwen}{+2.1} points for Qwen~3.8 (removing \result{primary.error-reduction.transformed.control-flow.qwen}{77}\% and \result{primary.error-reduction.transformed.maximal-mix.qwen}{65}\%, $I =$~\result{primary.diff.control-flow.qwen}{+2.5} and \result{primary.diff.maximal-mix.qwen}{+1.8}). Rename, State and literals, and Distractors still gained \result{primary.effect.transformed.subset-min.luna}{+0.8} to \result{primary.effect.transformed.state-and-literals.luna}{+1.3} points for Luna~6 and \result{primary.effect.transformed.subset-min.qwen}{+0.4} to \result{primary.effect.transformed.rename.qwen}{+0.6} for Qwen~3.8 ($I =$~\result{primary.diff.distractors.luna}{+0.1} to \result{primary.diff.state-and-literals.luna}{+0.6}). Across fault types, $\Delta'$ was largest on off-by-one (\result{primary.effect.transformed.off-by-one.luna}{+3.1} and \result{primary.effect.transformed.off-by-one.qwen}{+1.7} points), wrong-logic (\result{primary.effect.transformed.wrong-logic.luna}{+3.5} and \result{primary.effect.transformed.wrong-logic.qwen}{+1.6}), and static-return faults (\result{primary.effect.transformed.static-return.luna}{+2.0} for both models). Across all five fault types, encouraged assistance removed \result{primary.error-reduction.transformed.off-by-one.luna}{51} to \result{primary.error-reduction.transformed.static-return.luna}{87}\% of Luna~6's and \result{primary.error-reduction.transformed.wrong-variable.qwen}{50} to \result{primary.error-reduction.transformed.wrong-arithmetic.qwen}{86}\% of Qwen~3.8's no-tools errors.

Encouraging debugger assistance multiplied the resource cost of an episode in both representations (Table~\ref{tab:cost}). Without tools, episodes averaged \result{cost.tokens-thousands.a.min}{1.3} to \result{cost.tokens-thousands.a.max}{3.3} thousand tokens and \result{cost.wall-seconds.a.min}{4.9} to \result{cost.wall-seconds.a.max}{37.9}\,s on untransformed programs, and \result{cost.tokens-thousands.d.min}{1.7} to \result{cost.tokens-thousands.d.max}{5.0} thousand tokens and \result{cost.wall-seconds.d.min}{5.2} to \result{cost.wall-seconds.d.max}{66.4}\,s on transformed programs, costing \result{cost.usd-per-1000.a.min}{0.22} to \result{cost.usd-per-1000.d.max}{7.65} US dollars per 1,000 episodes. Under encouraged-tools, token usage rose to \result{cost.tokens-thousands.c.min}{25.0} to \result{cost.tokens-thousands.c.max}{41.8} thousand on untransformed and \result{cost.tokens-thousands.f.min}{32.2} to \result{cost.tokens-thousands.f.max}{73.5} thousand on transformed programs (\result{cost.ratio.tokens.untransformed.min}{11.2} to \result{cost.ratio.tokens.untransformed.max}{29.7} and \result{cost.ratio.tokens.transformed.min}{14.7} to \result{cost.ratio.tokens.transformed.max}{27.0} times no-tools), and wall time rose to \result{cost.wall-seconds.c.min}{37.8} to \result{cost.wall-seconds.c.max}{213.6}\,s and \result{cost.wall-seconds.f.min}{39.9} to \result{cost.wall-seconds.f.max}{350.4}\,s (\result{cost.ratio.wall.untransformed.min}{5.6} to \result{cost.ratio.wall.untransformed.max}{7.7} and \result{cost.ratio.wall.transformed.min}{5.3} to \result{cost.ratio.wall.transformed.max}{7.7} times). Monetary cost rose to \result{cost.usd-per-1000.c.min}{1.26} to \result{cost.usd-per-1000.c.max}{15.68} and \result{cost.usd-per-1000.f.min}{1.60} to \result{cost.usd-per-1000.f.max}{26.06} US dollars per 1,000 episodes (\result{cost.ratio.usd.untransformed.min}{3.3} to \result{cost.ratio.usd.untransformed.max}{5.7} and \result{cost.ratio.usd.transformed.min}{3.4} to \result{cost.ratio.usd.transformed.max}{5.0} times, smaller than the token ratios because multi-turn tool calls reuse cached prompt prefixes). Sol~6.1 used \result{cost.ratio.tokens.transformed.sol}{18.4} to \result{cost.ratio.tokens.untransformed.sol}{19.8} times as many tokens and \result{cost.ratio.wall.transformed.sol}{7.7} times the wall time for no accuracy gain.

\rqanswer{2}{Encouraged debugger assistance raised accuracy for Luna~6 and Qwen~3.8 by $\Delta =$~\result{primary.effect.untransformed.luna}{+0.7} and \result{primary.effect.untransformed.qwen}{+0.3} percentage points on untransformed programs (removing \result{primary.error-reduction.untransformed.luna}{41}\% and \result{primary.error-reduction.untransformed.qwen}{27}\% of no-tools errors) and by $\Delta' =$~\result{primary.effect.transformed.luna}{+2.1} and \result{primary.effect.transformed.qwen}{+1.3} points on transformed programs (\result{primary.error-reduction.transformed.luna}{63}\% and \result{primary.error-reduction.transformed.qwen}{61}\%), while Sol~6.1 stayed at \result{primary.acc.transformed.no-tools.sol}{99.9}\%. Encouraged tools closed the static transformation gap to \result{primary.acc.gap.encouraged-tools.min}{$-$0.1} to \result{primary.acc.gap.encouraged-tools.max}{+0.1} points ($I =$~\result{primary.diff.luna}{+1.4} and \result{primary.diff.qwen}{+1.0} points, up to \result{primary.diff.subset-max.max}{+3.6} on individual recipes), with the largest gains where control-flow rewrites had lowered static accuracy. On transformed programs, it multiplied tokens by \result{cost.ratio.tokens.transformed.min}{14.7} to \result{cost.ratio.tokens.transformed.max}{27.0}, wall time by \result{cost.ratio.wall.transformed.min}{5.3} to \result{cost.ratio.wall.transformed.max}{7.7}, and cost by \result{cost.ratio.usd.transformed.min}{3.4} to \result{cost.ratio.usd.transformed.max}{5.0}.}

\subsection{RQ3: Discretionary Use}
\label{sec:results-rq3}

When debugger assistance was optional, the agents rarely used the interactive debugger, taking debugger actions in \result{tool.any-debugger.b.min}{0.0} to \result{tool.any-debugger.b.max}{0.9}\% of episodes on untransformed programs and \result{tool.any-debugger.e.min}{0.2} to \result{tool.any-debugger.e.max}{6.2}\% on transformed programs (Table~\ref{tab:tool-use}). Instead, Sol~6.1 and Qwen~3.8 ran execution probes without debugger actions in \result{tool.probe-only.b.sol}{86.7} and \result{tool.probe-only.b.qwen}{99.1}\% of untransformed episodes and \result{tool.probe-only.e.sol}{98.4} and \result{tool.probe-only.e.qwen}{93.6}\% of transformed episodes. Luna~6 called no tool in \result{tool.no-tool.b.luna}{91.2}\% of untransformed and \result{tool.no-tool.e.luna}{87.9}\% of transformed discretionary episodes and ran only probes in \result{tool.probe-only.b.luna}{8.8} and \result{tool.probe-only.e.luna}{11.8}\%. Across the executed probe scripts on transformed programs, \result{facts.probe.calls-program-share.transformed.discretionary-tools.qwen}{57.9}\% (Qwen~3.8) to \result{facts.probe.calls-program-share.transformed.discretionary-tools.sol}{99.8}\% (Sol~6.1) called the program under test, and \result{facts.probe.checks-expected-share.transformed.discretionary-tools.sol}{1.2}\% (Sol~6.1) to \result{facts.probe.checks-expected-share.transformed.discretionary-tools.qwen}{25.6}\% (Qwen~3.8) checked the return value via an in-script assertion or equality test instead of only printing outputs to inspect. Encouraging tool use raised the share of episodes with a debugger action to \result{tool.any-debugger.c.min}{20.2} to \result{tool.any-debugger.c.max}{100.0}\% on untransformed programs and \result{tool.any-debugger.f.min}{41.7} to \result{tool.any-debugger.f.max}{100.0}\% on transformed programs, with Luna~6 and Sol~6.1 averaging \result{tool.debugger-actions.f.luna}{12.5} and \result{tool.debugger-actions.f.sol}{8.4} debugger actions per transformed episode while Qwen~3.8 still ran only probes in \result{tool.probe-only.f.qwen}{58.1}\%. Among encouraged debugger actions on transformed programs, stepping accounted for \result{tool.stepping.f.luna}{41.7}\% (Luna~6) and \result{tool.stepping.f.sol}{44.0}\% (Sol~6.1) versus \result{tool.stepping.f.qwen}{22.9}\% (Qwen~3.8), breakpoints for \result{tool.breakpoints.f.luna}{17.5}\% to \result{tool.breakpoints.f.qwen}{26.1}\%, and state inspection for \result{tool.inspection.f.qwen}{12.4}\% to \result{tool.inspection.f.sol}{20.2}\% of debugger actions.

\begin{table}[t]
\centering
\caption{Tool use in the tool-enabled settings on untransformed and transformed programs. The first two rows per model give the shares of episodes with at least one debugger action and of episodes with probe runs but no debugger action. The next two give the mean numbers of debugger actions and probe runs per episode. The last three give the shares of debugger actions that manage breakpoints, step, and inspect state.}
\label{tab:tool-use}
\footnotesize
\setlength{\tabcolsep}{4pt}
\begin{tabular}{@{}llcccc@{}}
\toprule
 & & \multicolumn{2}{c}{Discretionary} & \multicolumn{2}{c}{Encouraged} \\
\cmidrule(lr){3-4}\cmidrule(l){5-6}
Model & Measure & Untr. & Transf. & Untr. & Transf. \\
\midrule
\csname @@input\endcsname generated/table-tool-use-rows.tex
\bottomrule
\end{tabular}
\end{table}

Although the agents rarely invoked the debugger under discretionary-tools, Luna~6 and Qwen~3.8 obtained most of their encouraged-tools accuracy gains at a fraction of the cost. On transformed programs, discretionary-tools reached \result{primary.acc.transformed.discretionary-tools.luna}{98.2}\% accuracy for Luna~6, \result{primary.acc.transformed.discretionary-tools.qwen}{99.1}\% for Qwen~3.8, and \result{primary.acc.transformed.discretionary-tools.sol}{99.9}\% for Sol~6.1 (Table~\ref{tab:accuracy}). For Luna~6 and Qwen~3.8, this raised accuracy over no-tools by \result{primary.disc-minus-no.transformed.luna}{+1.5}~\result{primary.disc-minus-no.transformed.ci.luna}{[+0.9, +2.2]} and \result{primary.disc-minus-no.transformed.qwen}{+1.3}~\result{primary.disc-minus-no.transformed.ci.qwen}{[+0.8, +1.8]} points, within \result{primary.disc-minus-enc.transformed.luna}{$-$0.6}~\result{primary.disc-minus-enc.transformed.ci.luna}{[$-$1.0, $-$0.2]} and \result{primary.disc-minus-enc.transformed.qwen}{+0.0}~\result{primary.disc-minus-enc.transformed.ci.qwen}{[$-$0.3, +0.2]} points of encouraged-tools (Table~\ref{tab:discretionary}). On untransformed programs, where baseline accuracy was already \result{primary.acc.untransformed.no-tools.min}{98.2} to \result{primary.acc.untransformed.no-tools.max}{99.9}\%, discretionary-tools differed from no-tools by \result{primary.disc-minus-no.untransformed.min}{$-$0.2} to \result{primary.disc-minus-no.untransformed.max}{+0.2} points and from encouraged-tools by \result{primary.disc-minus-enc.untransformed.min}{$-$0.8} to \result{primary.disc-minus-enc.untransformed.max}{$-$0.1} points. Like encouraged-tools, discretionary-tools accuracy on transformed programs differed from the untransformed counterparts by only \result{primary.acc.gap.discretionary-tools.min}{+0.1} to \result{primary.acc.gap.discretionary-tools.max}{+0.2} points. Meanwhile, discretionary-tools episodes on transformed programs used \result{cost.tokens-thousands.e.luna}{3.3} thousand tokens (\result{cost.wall-seconds.e.luna}{6.4}\,s, \result{cost.usd-per-1000.e.luna}{0.38} US dollars per 1,000 episodes) for Luna~6, \result{cost.tokens-thousands.e.sol}{4.9} thousand tokens for Sol~6.1, and \result{cost.tokens-thousands.e.qwen}{38.3} thousand tokens (\result{cost.wall-seconds.e.qwen}{245.0}\,s, \result{cost.usd-per-1000.e.qwen}{18.92} US dollars) for Qwen~3.8, compared with \result{cost.tokens-thousands.f.luna}{55.0}, \result{cost.tokens-thousands.f.sol}{32.2}, and \result{cost.tokens-thousands.f.qwen}{73.5} thousand under encouraged-tools (Table~\ref{tab:cost}).

Qwen~3.8 also ran more probes on the cases where debugger assistance raised its accuracy more. Regressing the per-case assistance effect on the discretionary probe-run rank gives a slope $\beta_1$ of \result{primary.slope.probe.transformed.qwen}{+5.3}~\result{primary.slope.probe.transformed.ci.qwen}{[+3.2, +7.5]} percentage points from the bottom to the top of the probe-use ranking on transformed programs (placebo miss rate \result{primary.slope.probe.transformed.placebo-miss.qwen}{4.2}\%) and \result{primary.slope.probe.untransformed.qwen}{+1.9}~\result{primary.slope.probe.untransformed.ci.qwen}{[$-$1.0, +5.5]} points on untransformed programs (Table~\ref{tab:discretionary}). Luna~6 made probe runs on \result{count.probe-cases.transformed.luna}{596} of the \result{count.cases.transformed.luna}{3{,}022} transformed cases, with a slope estimate of \result{primary.slope.probe.transformed.luna}{+15.0} points that is reported without an interval because \result{primary.slope.probe.transformed.placebo-miss.luna}{6.6}\% of its placebo intervals missed the target. By debugger-action rank on transformed programs, Luna~6 and Qwen~3.8 had descriptive slopes of \result{primary.slope.debugger.transformed.luna}{+66.7} points (across \result{count.debugger-cases.transformed.luna}{17} cases, placebo miss rate \result{primary.slope.debugger.transformed.placebo-miss.luna}{36.2}\%) and \result{primary.slope.debugger.transformed.qwen}{+2.7} points (miss rate \result{primary.slope.debugger.transformed.placebo-miss.qwen}{6.7}\%). Because encouraged-tools raised Sol~6.1's accuracy over no-tools on \result{facts.primary.reversal.better.transformed.sol}{2} transformed cases and lowered it on \result{facts.primary.reversal.worse.transformed.sol}{1}, its probe and debugger slopes stayed near zero.

\rqanswer{3}{When assistance was optional, the agents took debugger actions in \result{tool.any-debugger.b.min}{0.0} to \result{tool.any-debugger.b.max}{0.9}\% of untransformed and \result{tool.any-debugger.e.min}{0.2} to \result{tool.any-debugger.e.max}{6.2}\% of transformed episodes. Sol~6.1 and Qwen~3.8 mostly ran probes without the debugger, and Luna~6 called no tool in \result{tool.no-tool.e.luna}{87.9}\% of transformed episodes. Even so, optional assistance raised transformed accuracy over no-tools by \result{primary.disc-minus-no.transformed.luna}{+1.5} points for Luna~6 and \result{primary.disc-minus-no.transformed.qwen}{+1.3} for Qwen~3.8 (versus \result{primary.effect.transformed.luna}{+2.1} and \result{primary.effect.transformed.qwen}{+1.3} when encouraged) using \result{cost.tokens-thousands.e.luna}{3.3} and \result{cost.tokens-thousands.e.qwen}{38.3} thousand tokens instead of \result{cost.tokens-thousands.f.luna}{55.0} and \result{cost.tokens-thousands.f.qwen}{73.5} thousand, and Qwen~3.8 ran more probes where assistance helped more.}
\begin{table}[t]
\centering
\caption{Discretionary use. Mean per-case accuracy differences between discretionary-tools and the other two settings, and the slopes $\beta_1$ of the per-case assistance effect on the rank of probe runs and of debugger actions under discretionary-tools (Section~\ref{sec:design-analysis}), all in percentage points. Brackets give 95\% intervals. A dagger marks a slope without an interval and a dash a slope that is not reported.}
\label{tab:discretionary}
\scriptsize
\setlength{\tabcolsep}{1.3pt}
\begin{tabular}{@{}lcccc@{}}
\toprule
 & \multicolumn{2}{c}{Discretionary minus} & \multicolumn{2}{c}{Slope $\beta_1$ by} \\
\cmidrule(lr){2-3}\cmidrule(lr){4-5}
 & No-tools & Encouraged & Probe runs & Debugger \\
\midrule
\csname @@input\endcsname generated/table-discretionary-rows.tex
\bottomrule
\end{tabular}
\end{table}

\subsection{Robustness}
\label{sec:results-robustness}

To test sensitivity to our grading criterion, we re-evaluated all episodes under line-level grading, which accepts only the tracked line~$L'$ holding the edited expression rather than any line of a split faulty statement, such as lines 7 and~8 in Fig.~\ref{fig:example} (Section~\ref{sec:design-measurements}). Naming another line of a split faulty statement accounted for \result{facts.statement-only.share.luna}{7.3}\% of Luna~6's no-tools answers on transformed programs and \result{facts.statement-only.share.sol}{0.1}\% of Sol~6.1's. Line-level grading lowered no-tools accuracy on transformed programs to \result{secondary.acc.transformed.no-tools.luna}{89.4}\% for Luna~6 and \result{secondary.acc.transformed.no-tools.qwen}{96.4}\% for Qwen~3.8 ($G =$~\result{secondary.acc.gap.no-tools.luna}{$-$8.8} and \result{secondary.acc.gap.no-tools.qwen}{$-$2.3} points), raising their $\Delta'$ estimates to \result{secondary.effect.transformed.luna}{+3.4} and \result{secondary.effect.transformed.qwen}{+1.8} points and $I$ to \result{secondary.diff.luna}{+2.6}~\result{secondary.diff.ci.luna}{[+1.6, +3.6]} and \result{secondary.diff.qwen}{+1.5}~\result{secondary.diff.ci.qwen}{[+0.6, +2.3]} points.

To check whether answers accepted only by statement-level grading diagnosed the seeded fault, we conducted a blinded, LLM-assisted audit (verified by an author) of 45 such answers per model across the three settings (which for Sol~6.1 were all of them) alongside 20 line-level successes per model. The agent's explanation identified the seeded fault in \result{audit.statement-only.diagnosed}{133} of the \result{audit.statement-only.sampled}{135} statement-only answers and in \result{audit.exact-line.diagnosed}{59} of the \result{audit.exact-line.sampled}{60} line-level successes, showing that statement-only answers diagnosed the fault as often as line-level successes did.

%% file: sections/05-discussion.tex
\section{Discussion}
\label{sec:discussion}

\paragraph{Fault localization at the accuracy ceiling}
Without tools, the three agents localized \result{primary.acc.untransformed.no-tools.min}{98.2} to \result{primary.acc.untransformed.no-tools.max}{99.9}\% of faults in untransformed and \result{primary.acc.transformed.no-tools.min}{96.7} to \result{primary.acc.transformed.no-tools.max}{99.9}\% in transformed programs, and neither replacing most four-token sequences in Rename nor quadrupling line count in Distractors lowered static accuracy by more than \result{primary.acc.loss.no-tools.distractors.luna}{1.0} percentage point (Section~\ref{sec:results-rq1}). This contrasts with earlier evaluations in which models without runtime feedback left substantial shares of faulty programs unrepaired~\cite{zhong_2024_ldb,tian_2024_debugbench} and static localizers lost 78\% of previously localized faults after behavior-preserving transformations~\cite{haroon_2026_localizability}. Three differences in Haroon et al.'s setup may account for their larger losses: their metric conditions on solving the original program, their programs average 250 lines, and most of their models date from March~2025. Meanwhile, unlike earlier agents whose repair rates dropped under direct \texttt{pdb} access or unfiltered output~\cite{yuan_2025_debug_gym,wang_2026_inspectcoder}, encouraged debugger use never caused an accuracy collapse. Even when Luna~6 and Sol~6.1 used the interactive debugger in \result{tool.any-debugger.f.luna}{100.0}\% of transformed episodes, forced-answer and no-answer rates stayed at or below \result{cost.forced-answer.f.max}{0.3}\% and \result{cost.no-answer.f.max}{0.1}\%, Sol~6.1 remained at \result{primary.acc.transformed.encouraged-tools.sol}{99.9}\%, and Luna~6 and Qwen~3.8 removed \result{primary.error-reduction.transformed.luna}{63}\% and \result{primary.error-reduction.transformed.qwen}{61}\% of their remaining errors. With structured tool interfaces and current reasoning models, the main limitation of debugger assistance on short, well-specified programs is no longer tool reliability, but that static reasoning already solves nearly every case.

\paragraph{Syntactic rewrites versus familiarity}
The static accuracy loss from transformation did not follow how much surface code a recipe changed. Without tools, Rename did not lower accuracy despite preserving only \result{programs.transformed.kept.median.rename}{40}\% of four-token sequences, whereas Control flow and Maximal mix lowered accuracy by \result{primary.acc.loss.no-tools.control-flow.luna}{2.1} to \result{primary.acc.loss.no-tools.maximal-mix.luna}{3.8} points, with \result{facts.primary.wrong.double-negation-share.combined.luna}{71.4}\% (Luna~6) and \result{facts.primary.wrong.double-negation-share.combined.qwen}{68.2}\% (Qwen~3.8) of their new errors naming a doubly negated condition (Section~\ref{sec:results-rq1}). Because our wrong-logic faults add or remove a negation and the prompt states that the program contains one injected bug, a static reader can mistake a doubly negated condition for the seeded fault, much as LLMs fail code-equivalence checks under simple syntactic rewrites~\cite{wei_2025_equibench,oza_2026_cetbench} and follow misleading comments~\cite{lam_2025_codecrash,wu_2026_alibi}. Running the program shows that the condition evaluates as intended, which explains why both tool-enabled settings closed almost the entire accuracy gap between representations. Before attributing an accuracy drop on transformed code to reduced training-data familiarity, benchmark designers should therefore check whether new errors cluster on specific syntactic constructs.

\paragraph{Program execution versus interactive stepping}
When debugger assistance was optional, the agents obtained most of its accuracy benefit while rarely invoking the interactive debugger. Instructing agents to use the tools raised the share of transformed episodes with a debugger action from \result{tool.any-debugger.e.min}{0.2} to \result{tool.any-debugger.e.max}{6.2}\% up to \result{tool.any-debugger.f.min}{41.7} to \result{tool.any-debugger.f.max}{100.0}\%, but raised accuracy by only \result{primary.enc-minus-disc.transformed.luna}{+0.6} points for Luna~6 and \result{primary.enc-minus-disc.transformed.qwen}{+0.0} points for Qwen~3.8 (Section~\ref{sec:results-rq3}). On short programs, running \texttt{run\_probe} on chosen inputs appears to supply most of the runtime evidence, paralleling the gains from print debugging in PingFL~\cite{qin_2025_pingfl}. In an exploratory post-hoc split of transformed discretionary-tools episodes, probe-only runs accounted for \result{facts.primary.gain-by-use.discretionary-tools.probes-only.transformed.qwen}{+1.1} of Qwen~3.8's \result{primary.disc-minus-no.transformed.qwen}{+1.3}-point gain and \result{facts.primary.gain-by-use.discretionary-tools.probes-only.transformed.luna}{+0.7} of Luna~6's \result{primary.disc-minus-no.transformed.luna}{+1.5} points, with no-tool runs contributing another \result{facts.primary.gain-by-use.discretionary-tools.no-tool.transformed.luna}{+0.7} points for Luna~6. Because agents choose tools in response to the fault and earlier outputs, conditioning on realized tool use mixes tool effects with selection bias, and extra model turns can also aid accuracy~\cite{kapoor_2025_agents}. Isolating program execution from interactive debugging therefore requires a dedicated execution-only setting.

\paragraph{Cost-effectiveness and model selection}
Across models and settings, how often an agent invoked tools did not match how much accuracy it gained from them (Tables~\ref{tab:accuracy} to~\ref{tab:discretionary}). Sol~6.1 ran probes in \result{tool.probe-only.e.sol}{98.4}\% of its discretionary episodes on transformed programs even though it already localized \result{primary.acc.transformed.no-tools.sol}{99.9}\% of the faults without tools, consistent with reports that LLMs call tools when internal knowledge suffices~\cite{qian_2025_smart}. Conversely, Luna~6 called no tool in \result{tool.no-tool.e.luna}{87.9}\% of its discretionary episodes yet retained \result{primary.disc-minus-no.transformed.luna}{+1.5} of its \result{primary.effect.transformed.luna}{+2.1}-point gain at \result{cost.tokens-thousands.e.luna}{3.3} thousand tokens per episode versus \result{cost.tokens-thousands.f.luna}{55.0} thousand under encouraged-tools. Qwen~3.8 matched its encouraged accuracy (\result{primary.disc-minus-no.transformed.qwen}{+1.3} points) at roughly half the tokens (\result{cost.tokens-thousands.e.qwen}{38.3} versus \result{cost.tokens-thousands.f.qwen}{73.5} thousand) while running more probes on the faults where assistance helped more. Moreover, Sol~6.1 without tools (\result{primary.acc.transformed.no-tools.sol}{99.9}\% accuracy, \result{cost.wall-seconds.d.sol}{5.2}\,s, \result{cost.usd-per-1000.d.sol}{3.41} US dollars per 1,000 transformed episodes) was both more accurate and faster than Luna~6 or Qwen~3.8 with encouraged debugging (\result{primary.acc.transformed.encouraged-tools.luna}{98.8}\% in \result{cost.wall-seconds.f.luna}{47.0}\,s and \result{primary.acc.transformed.encouraged-tools.qwen}{99.2}\% in \result{cost.wall-seconds.f.qwen}{350.4}\,s). For agent builders, multi-turn debugging scaffolding on self-contained code cannot substitute for a stronger base model and is most cost-effective when left optional rather than forced by prompt instructions. Because most executed probe scripts did not check expected outputs even though every problem description includes example input-output pairs (Section~\ref{sec:results-rq3}), dedicated tools that run specification examples automatically could further reduce probing overhead.
\paragraph{Grading after transformations}
Statement-level and line-level grading differ only where a transformation recipe splits the faulty statement across lines (\result{programs.transformed.spread-share}{47}\% of transformed cases). In our audit, answers accepted only by statement-level grading diagnosed the seeded fault as often as exact-line successes, and the share of such answers varied from \result{facts.statement-only.share.sol}{0.1}\% for Sol~6.1 to \result{facts.statement-only.share.luna}{7.3}\% for Luna~6 (Section~\ref{sec:results-robustness}). Strict line-level grading would count these valid diagnoses as failures, inflate $|G|$ and $I$, and make model comparisons depend on which fragment of a split statement each model reports. Evaluations on transformed code should therefore track the full span of the faulty statement across rewrites and report gains relative to the no-tools error rate alongside percentage-point effects when baseline accuracy is near the ceiling.

%% file: sections/06-threats.tex
\section{Threats to Validity}
\label{sec:threats}

\paragraph{Comparing representations}
Like prior studies~\cite{cao_2025_codecleaner,li_2025_generalizability,zhang_2025_unseen_horizons}, we use behavior-preserving transformations to reduce surface overlap with pretraining corpora. However, Kept measures four-token overlap between program versions rather than against a model's training corpus, transformed programs remain semantic duplicates of public solutions~\cite{spiesberger_2026_soft_contamination}, and the problem description stays unchanged because fault localization requires a specification of intended behavior~\cite{haroon_2026_localizability}. An agent can therefore still recognize an underlying LeetCode problem seen during training~\cite{jain_2025_livecodebench}. Residual familiarity makes the two representations more alike and attenuates any part of $G$ or $I$ driven by unfamiliarity, so we interpret $G$ and $I$ as sensitivity to the transformation recipes rather than to unfamiliarity alone.

\paragraph{Measuring the assistance effect}
Compared with no-tools, encouraged-tools adds program execution, the interactive debugger, an explicit tool-use instruction, and additional model calls, so $\Delta$ and $\Delta'$ measure their combined effect. Top-1 accuracy grades the reported line number rather than the explanation, and LLMs can reach correct answers from flawed reasoning~\cite{orvalho_2025_robustness}. Coincidental hits would bias $\Delta$ or $\Delta'$ only if they occurred at different rates across settings, and our blinded audit found valid fault diagnoses in \result{audit.exact-line.diagnosed}{59} of \result{audit.exact-line.sampled}{60} exact-line and \result{audit.statement-only.diagnosed}{133} of \result{audit.statement-only.sampled}{135} statement-only successes (Section~\ref{sec:results-robustness}). The prompt asks for the ``most likely affected line'' (Listing~\ref{lst:prompts}), which an agent could interpret as a downstream line where the fault manifests, though all settings share this prompt. Finally, monetary costs depend on provider pricing and cached-input discounts, and our DeepInfra-based estimate for self-hosted Qwen~3.8 is an external proxy, so we report token counts and wall time alongside dollar costs for every condition.

\paragraph{Discretionary use}
An agent may invoke a tool without expecting an accuracy gain~\cite{qian_2025_smart}, for example because a program looks complex. The RQ3 slope $\beta_1$ therefore measures whether discretionary tool use is associated with larger per-case assistance effects, not why the agent used a tool. With two episodes per case and setting, per-case noise can attenuate $\beta_1$ toward zero, so we draw no case-level conclusions.

\paragraph{Generalization}
We evaluate short Python programs for precisely specified algorithmic problems because constructing verified matched pairs requires a problem description, a reference solution, and an executable test suite. Our findings do not generalize directly to multi-file repositories, incomplete specifications, or multi-statement defects, where the relevant state may not fit in the model's context and interactive debugging has shown larger gains~\cite{yuan_2025_debug_gym,sun_2026_debugharness}. Four of our five fault types instantiate single-statement bug patterns that account for a quarter of the 73,013 Python bugs mined by Kamienski et al.~\cite{kamienski_2021_pysstubs}, and 73\% of 357 real Java faults in Just et al.\ were coupled to mutants of common operators~\cite{just_2014_mutants}.

\paragraph{Reproducibility}
Commercial API updates can alter model behavior over time~\cite{baltes_2025_llm_guidelines}. We record model identifiers, reasoning settings, and collection dates (Section~\ref{sec:design-workflows}), release all prompts, traces, and analysis scripts~\cite{anonymous_2026_replication}, and serve Qwen~3.8 ourselves from fixed open-weights checkpoints.

%% file: sections/07-conclusion.tex
\section{Conclusion}
\label{sec:conclusion}

Across \result{programs.faulty.admitted}{610} untransformed and \result{programs.transformed.admitted}{3{,}022} transformed Python programs (\result{facts.episodes.kept}{65{,}376} episodes), we evaluated how accurately coding agents localize single-line faults from source code alone and how much debugger assistance improves outcomes. Without tools, agents based on Sol~6.1, Luna~6, and Qwen~3.8 already approached the accuracy ceiling, localizing \result{primary.acc.untransformed.no-tools.min}{98.2} to \result{primary.acc.untransformed.no-tools.max}{99.9}\% of faults in untransformed and \result{primary.acc.transformed.no-tools.min}{96.7} to \result{primary.acc.transformed.no-tools.max}{99.9}\% in transformed programs. Variable renaming and distractors barely changed static accuracy, whereas restructured loops and doubly negated conditions lowered it for Luna~6 and Qwen~3.8. Encouraged debugger assistance removed \result{primary.error-reduction.transformed.luna}{63}\% and \result{primary.error-reduction.transformed.qwen}{61}\% of their remaining transformed errors (\result{primary.effect.transformed.luna}{+2.1} and \result{primary.effect.transformed.qwen}{+1.3} percentage points) and closed the transformation gap, while multiplying token use by \result{cost.ratio.tokens.transformed.min}{14.7} to \result{cost.ratio.tokens.transformed.max}{27.0} and wall time by \result{cost.ratio.wall.transformed.min}{5.3} to \result{cost.ratio.wall.transformed.max}{7.7}. When debugger assistance was optional, the agents used the interactive debugger in only \result{tool.any-debugger.e.min}{0.2} to \result{tool.any-debugger.e.max}{6.2}\% of transformed episodes and instead ran execution probes or answered from source code alone, yet kept most of the accuracy benefit at a fraction of the token cost. Future work should isolate program execution from interactive debugging with an execution-only setting and evaluate larger, less precisely specified repository bugs where static reasoning leaves more room for state inspection.

\section*{Data Availability}
Our replication package~\cite{anonymous_2026_replication} provides our pipeline, programs, prompts, episode traces, and analysis scripts.

%% file: sections/08-acknowledgment.tex
\newcommand{\sccfundingfull}{The research reported in this paper has been partly funded by the Federal Ministry for Innovation, Mobility and Infrastructure (BMIMI), the Federal Ministry for Economy, Energy and Tourism (BMWET), and the State of Upper Austria in the frame of the SCCH competence center INTEGRATE (FFG grant no.~892418) in the COMET - Competence Centers for Excellent Technologies Programme managed by Austrian Research Promotion Agency FFG.}
\newcommand{\sccfundingshort}{The research reported in this paper has been partly funded by BMIMI, BMWET, and the State of Upper Austria in the frame of the SCCH competence center INTEGRATE (FFG grant no.~892418) part of the FFG COMET Competence Centers for Excellent Technologies Programme.}
\newcommand{\sccfundingveryshort}{The research was supported by the Austrian ministries BMIMI, BMWET and the State of Upper Austria in the frame of the SCCH COMET competence center INTEGRATE (FFG 892418).}
\newcommand{\aidisclosure}{We used OpenAI GPT, Anthropic Claude, and Google Gemini models to assist with writing, coding, and literature search. The authors reviewed and refined the generated material and take full responsibility for the final manuscript.}

\ifcameraready
\section*{Acknowledgment}
\sccfundingfull{} \aidisclosure
\else
\section*{Use of Generative AI}
\aidisclosure
\fi